%% file: maincurrent.tex
\documentclass[a4paper, 12pt]{article} 

\usepackage{dcolumn}
\usepackage[margin=1in]{geometry}
\usepackage[utf8]{inputenc}
\usepackage{graphicx}
\usepackage{subfigure}
\usepackage{authblk}

\usepackage[style=nature,backend=biber,sortcites=true]{biblatex}
\AtEveryBibitem{%
 \clearfield{url}%
 \clearfield{month}%
 \clearfield{issn}%
 \clearfield{doi}
 \clearfield{address}
}

\newcommand{\beginsupplement}{%
        \setcounter{table}{0}
        \renewcommand{\thetable}{S\arabic{table}}%
        \setcounter{figure}{0}
        \renewcommand{\thefigure}{S\arabic{figure}}%
     }

\usepackage{xcite}

\usepackage{longtable} 
\usepackage{amsmath}
\usepackage{amssymb}
\usepackage{newtxmath}
\DeclareMathAlphabet{\mathpzc}{T1}{pzc}{m}{it}

\usepackage{bm}

\usepackage[labelfont=bf]{caption}
\usepackage{floatrow}
\usepackage[dvipsnames]{xcolor}
\definecolor{dkgreen}{rgb}{0,0.6,0}
\definecolor{gray}{rgb}{0.5,0.5,0.5}

\usepackage{xurl}
\usepackage[utf8]{inputenc}
\usepackage{changepage}

\title{
   Nested Skills in Labor Ecosystems: A Hidden Dimension of Human Capital
}

\author[1,2,3]{Moh Hosseinioun}
\author[4]{Frank Neffke}
\author[5]{Letian (LT) Zhang}
\author[1,2,6]{Hyejin Youn \thanks{Correspondence can be sent to hyejin.youn@kellogg.northwestern.edu.}}

\affil[1]{Kellogg School of Management, Northwestern University, Evanston, IL, USA}
\affil[2]{Northwestern Institute on Complex Systems, Evanston, IL, USA}
\affil[3]{Department of Information and Decision Sciences, University of Illinois, Chicago, IL, USA}
\affil[4]{Complexity Science Hub Vienna, Vienna, Austria}
\affil[5]{Harvard Business School, Harvard University, Cambridge, MA, USA}
\affil[6]{Santa Fe Institute, Santa Fe, NM, USA}

\date{\today}

\begin{document}

\maketitle
\thispagestyle{empty}

\vspace{-0.5cm}
\begin{abstract} \label{sec: abstract}
Modern economies, characterized by their vast output of goods and services, operate through globally interconnected networks. 
As economies become more complex, so do these networks, coordinating increasingly diverse portfolios of specialized efforts and knowledge. 
In this study, we analyze U.S. survey data (2005--2019) to infer an underlying interdependency tree within the fabric of skill portfolios. Hierarchically constructed, this skill tree starts from widely needed, foundational abilities, constituting the root, and extends to highly specialized, niche skills required by select jobs at the extremities. 
The directionality is defined by the asymmetrical conditional probabilities of the presence of one skill given the existence of another. 
Examining 70 million job transitions in resumes and national surveys, we observe that individuals tend to delve deeper into these nested specialization paths as they ascend the career ladder to enjoy higher wage premiums.  
Nevertheless, we find the role of foundational skills for such ascent remains pivotal; without reinforcing them, the anticipated wage premiums may vanish. 
Hence, we further differentiate \textit{nested} skills from others, with the former building on common prerequisites while the latter does not, 
and analyze disparities in these skill gaps across different genders and racial/ethnic groups. Our analysis reveals a growing and concerning fragmentation in the divide between these two skill groups over the past two decades, suggesting further polarization within the job landscape \cite{Autor2013}.
Our findings highlight the critical role of robust foundational skills as a stepping stone to specialization and the economic advantages it can confer, reinforcing the need for balanced skill development strategies in complex economies \cite{Althobaiti2022}.
\end{abstract}
\newpage
\pagenumbering{arabic}

\section*{Introduction}

Complexity and specialization are foundational to the narrative of economic growth and innovation \cite{Carneiro1986, henrich2015secret, richerson1999complex, MitchellMelanie2009}. As society advances, creating and maintaining sophisticated goods, services, and infrastructure, these socio-economic complexities have surpassed what individuals can embody and manage on their own \cite{BenJ2009, Gamble2002}. It is no longer feasible for individuals to master universal expertise across all areas. For economies, this means developing deep divisions of labor and knowledge that first distribute knowledge across people and then coordinate this distributed knowledge in teams, firms, and value chains \cite{BeckerG.S1992TDoL, Hidalgo2015, Azoulay2020, Pichler2023}.  For individuals, this means specializing, and deciding which skills to acquire over long educational and work trajectories has become increasingly important \cite{Acemoglu2020}. As such, human capital is far from an isolated entity but an interdependent ecosystem of skills and knowledge in economies.

This leads to research questions: What does the structure of these interdependencies look like? And, more importantly, what implications does this nested structure carry? Division of labor, division of knowledge, and the existence of such an interdependency web are not in doubt as they manifest in education and career paths in a way we experience every day, shaping social and economic systems \cite{Autor2013, SchwabeHenrik2020Awsa}. However, though the framework may seem intuitive, it is essential to note that the hierarchical layout of skills reflected in job roles has often been assumed rather than empirically evidenced.

Emerging research aimed at understanding the network architecture of human capital has yielded insights into the detailed tasks that individuals perform at work and the skills they require to do so \cite{AndersonKatharineA2017Snam, Borner2018, Alabdulkareem2018, Xu2021, Lin2022, Neffkeeaax3370, DelRio-Chanona2021, Frank2019, Moro2021}. Nevertheless, a granular understanding of workers' skill trajectories and their resulting impacts on individuals remains an ongoing area of exploration. Furthermore, these frameworks aim to capture complementarities or synergies between capabilities, knowledge, and skills \cite{AndersonKatharineA2017Snam, Alabdulkareem2018, Neffkeeaax3370, gomez2016explaining, hidalgo2018principle}. That is, jobs combine skills that complement one another. We contribute to these ongoing efforts by constructing a directed skill network that expresses how skills build or depend on one another, conceptualizing trajectories with conditional probability.

In this paper, we propose that the skill composition of jobs not only reflects complementarities but also the innate cognitive constraints of how individuals learn. That is, jobs not only combine synergistic skills but also skills that build on one another. This aligns with an understanding of skill acquisition as a cumulative, sequential trajectory that builds pyramidal skill structures where higher-level skills are nested in most basic layers of expertise \cite{WilkSteffanieL1995GtJC}. Students are taught calculus only after they have mastered the basics of algebra and geometry. We infer such dependencies by analyzing how skills co-occur in jobs and the construction of asymmetric skill networks in which the directed arrows describe skill dependencies.

These dependencies turn out to integrate one of the core concepts of traditional human capital theory into the network-based complexity approach: human capital specificity. Since its inception, the distinction between general and specific skills has been a hallmark of human capital theory, explaining why market economies typically underinvest in general skills \cite{becker2009human}, why acquiring specific skills creates hold-up problems \cite{williamson2007economic}, and why workers often face earning losses when they are displaced from their jobs \cite{jacobson1993earnings}. However, this distinction also matters because general skills constitute a foundational layer in an individual's human capital, on top of which more specific skills can be developed. Just like the way mastering calculus requires a prior understanding of algebra and geometry, these education and career paths are both sequential and cumulative, building on each other, and thus create a high-dimensional space of possibilities for job opportunities \cite{WilkSteffanieL1995GtJC, Jovanovic1997}.

The sequential nature of skill trajectories has important implications for professional development and, therewith, socio-economic outcomes because they mean that certain career paths are only feasible after prior investment in foundational skills \cite{heckman2011economics, Autor2013, SchwabeHenrik2020Awsa, NelsonDylan2022, Goldin2008, Azoulay2021}. 
As a consequence, specialization entails not just an increase in the volume of learning and investments in education and training \cite{BenJ2009} but also the existence of structured, sequential, and nested cumulative paths that can either enable or restrict specific career trajectories. These structured pathways systematically shape professional development and thus the socio-economic landscape at large, leading not only to differential rewards but also differential accessibility and feasibility of career options based on earlier choices \cite{Autor2013, SchwabeHenrik2020Awsa, NelsonDylan2022, Goldin2008, Azoulay2021}. Thus, to succeed in this complex environment, individuals must acquire the right set of skills, knowledge, and abilities \cite{Mincer1974, Becker1962, Lucas1988, Neffke2013, Neffkeeaax3370, Stephany2024, Jovanovic1997}. Yet, the most sought-after skills in today's economic and social sphere are often not readily accessible but are instead nested within specific domains, requiring a progressive accumulation of knowledge and expertise to unlock.

\begin{figure*}[!h]
    \centering
    \includegraphics[width=\textwidth]{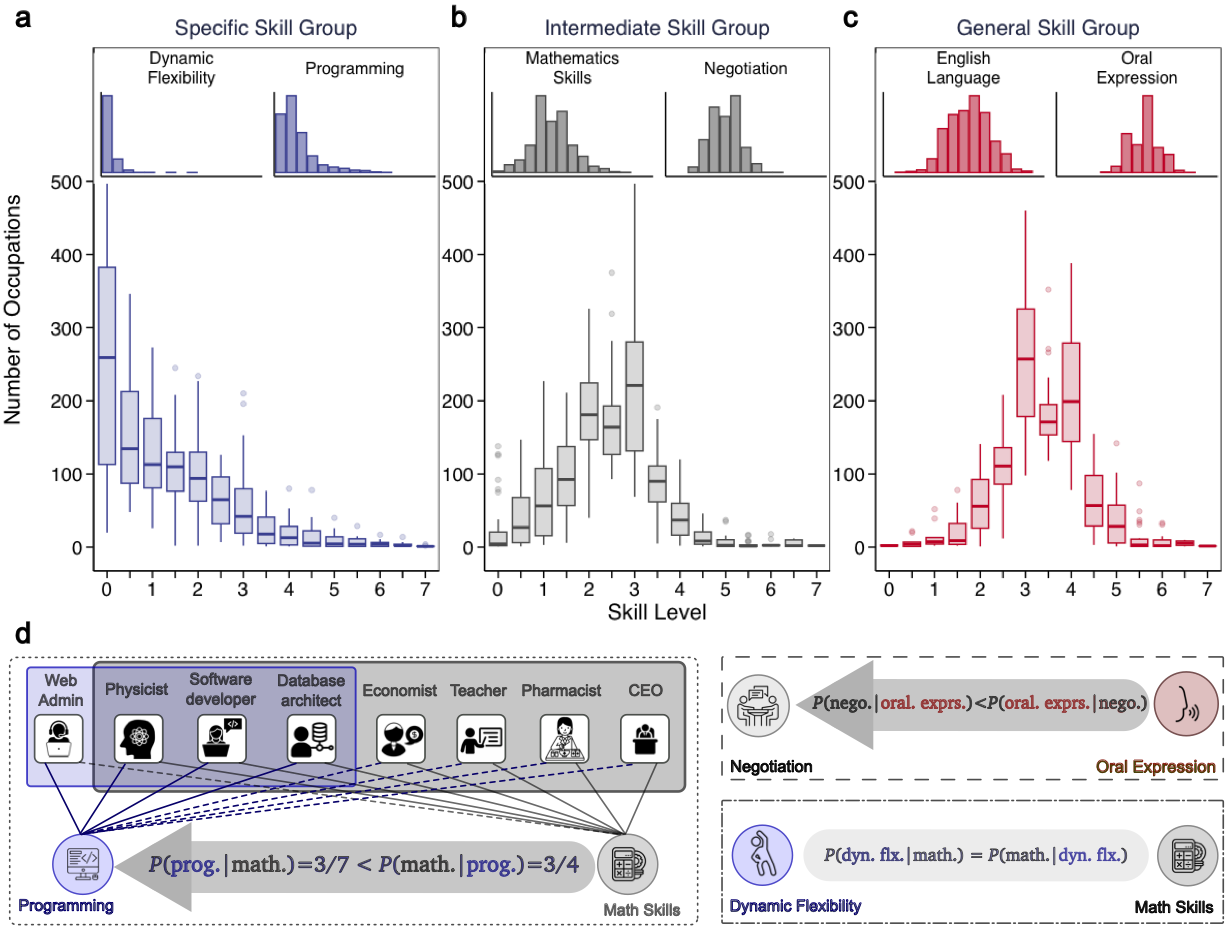}
    \caption{\textbf{Skill Level Distributions and Dependencies.}  
    \textbf{(a-c)} Average number of occupations requiring each skill level for the three groups (see SI Section \ref{supsec:skill clustering} for details.) 
    Skills are grouped based on their characteristic skill level distribution shapes, exemplified by the insets, and labeled as \textit{General} (31 skills), \textit{Intermediate} (43 skills), and \textit{Specific} (46 skills). The shapes indicate that Specific skills (blue) are needed only in a few jobs, while most jobs require high proficiency in General skills (red).  
    \textbf{(d)} Schematic illustrating our inference method for dependency between skill pairs using the asymmetric conditional probability of one skill being required given another. 
    For example, if requiring Math skills is more probable given the presence of Programming (compared to the reverse), we infer a directional dependency: Math $\rightarrow$ Programming, weighted by the level of asymmetry (see Methods). Similarly, Oral Expression $\rightarrow$ Negotiation, but Math $\not\rightarrow$ Dynamic Flexibility, as their presences are independent events, that is, $P(\text{Math}|\text{Dyn. Flex.}) = P(\text{Dyn. Flex.}|\text{Math})$.
    }
    \label{fig:Figure 1}
\end{figure*}

In this paper, we show that this hierarchical network yields a description of human capital that not only recovers broad, well-established job categories but also helps predict career transitions and wage curves. To do that, we analyze skill portfolios and their underlying structures using publicly accessible national surveys complemented by a proprietary dataset. We differentiate specialized skills, those required by select occupations, from general skills, those widely required across occupations (Fig.~\ref{fig:Figure 1}). We then construct a nested hierarchical structure of skill dependencies, employing conditional probabilities of the presence of one skill given the existence of another in occupations \cite{Jo2020}. Our method reveals that not every skill is embedded in a nested structure, resulting in a partially nested hierarchical structure among skills (Fig.~\ref{fig:Figure 2}). Therefore, we quantify each skill's contribution to the overall nested architecture of the network and find that skills contributing significantly to the nested architecture are rewarded most (Fig.~\ref{fig:nestedness}), echoing the nesting nature of economic complexity \cite{Hidalgo2009, Saavedra2011}.
 
By examining three different datasets (median occupational ages, synthetic birth cohorts of individuals, and 70 million job transitions in resumes), we uncover that nested branches are evidence of specialization and career advancement. That is, as individuals progress up the career ladder, they need to acquire and apply skills on nested specialization branches (Fig.~\ref{fig:age}).  
Moreover, we find most of the wage premiums for these nested specializations are conditional on foundational, general prerequisite skills they are nested in, unlike unnested specializations without prerequisite skills (Fig.~\ref{fig:wage curves}). 
This pattern suggests deeply rooted structural disparities in race/ethnicity and gender (Fig.~\ref{fig:Figure 7}). 
Finally, we examine structural changes in the skill network over time and find a wider gap between nested and unnested branches, suggesting potential barriers to upward mobility (Fig.~\ref{fig:historical skill change}). 

Structural properties of skill nestedness in human capital can provide actionable insights. The methodologies we employ introduce a scalable metric for skill categorization, enabling our analysis to extend to more granular levels. The nestedness metric effectively captures shifts in dependency intensity, providing a nuanced view of labor market polarization. As data on workplace skills, knowledge, capabilities, and tasks become increasingly granular, our approach extends to analyzing skills at finer resolutions, evaluating their diverse contributions to nestedness. This capability to identify changes in skill requirements across occupations complements the traditional context-informed categories, which may not adjust as readily to these changes; the flexibility and adaptability of our framework are useful for understanding the evolving landscape of skills and its impact on career development and socio-economic disparities. As the labor market continues to evolve, with new skills emerging and older ones becoming obsolete, our model acts as a comprehensive and dynamic tool for tracking these shifts and their wider implications.

\section*{Results}

\subsection*{Skill Generality (Individual Occurrences)} \label{sec: skill hierarchy}

The distinction between general and specialized skills is widely acknowledged, but a systematic quantification of this divide has been lacking \cite{Becker1962, Poletaev2008, Gathmann2010, FergusonJohn-Paul2013, Leung2014, Merluzzi2016, Byun2018, Fini2022, Byun2023, RotundoMaria2004Svgs}. 
Therefore, our study starts with examining, quantifying and classifying the generality of skills based on their breadth of application across occupations, using publicly available survey data from the U.S. Bureau of Labor Statistics (BLS).
These surveys provide detailed observations on the job requirements for thousands of occupational titles, including the importance and required level of each skill, knowledge, or ability necessary for workers to perform their occupational tasks.

Figure~\ref{fig:Figure 1} illustrates the existence of skills with varying degrees of occupational demand, characterized by their level distribution shapes across occupations with broad versus narrow applications. 
Here, demand denotes the number of occupations requiring the skill at a given level, ranging from 0 to 7. Specialized skills, such as Fine Arts and Programming (blue), are required only by select occupations, often at high levels (6 or 7), but not across a broad range of occupations.  
This leads to a distribution shape that primarily peaks at the 0-1 levels with a long tail.
In contrast, skills considered general (red), such as Oral Expression and Critical Thinking, are widely needed at elevated levels, with distributions that peak at levels 3-4, indicating their general applicability across most jobs.

To systematically classify skills, we group them based on similar level distribution shapes, which we interpret as indicators of broad versus narrow utility of skills (See Methods).
Figure~\ref{fig:Figure 1}~(a-c) show distribution shapes for the resulting skill groups, calculated by averaging the number of occupations that require the given skill levels within each group, which sketches the distinct level profile curve of that skill group.
The inset examples demonstrate that some skills are specialized, meaning they are not widely required across occupations but are critically needed at high levels in specific job contexts. These skills are identified and grouped into the \textit{specific} skill set. In contrast, skills relevant to a wide spectrum of roles are labeled as the \textit{general} skill set.

These classifications, detailed in SI-Table \ref{tab:skill_groups}, align with our common understanding of general and specialized skill categories. 
Nevertheless, we ensure the robustness of our findings by testing our results against different group sizes and clustering algorithms (see SI Sec.~\ref{supsec:skill clustering}). 
In addition to the distribution-based approach, skill generality can also be measured by the median skill levels required across occupations. For example, the median level for general skills is 3.34, for intermediate skills, it is 2.37, and for specific skills, it is 0.87, reflecting the skewed shape of niche skills.
In the following, we additionally show that these generality measures are consistent with network-based measures of generality \cite{Mones2012}.
Throughout the paper, our results are color-coded for consistency: general (red), intermediary (gray), and specific skills (blue).

\begin{figure*}[!h]
    \centering
    \includegraphics[width=\textwidth]{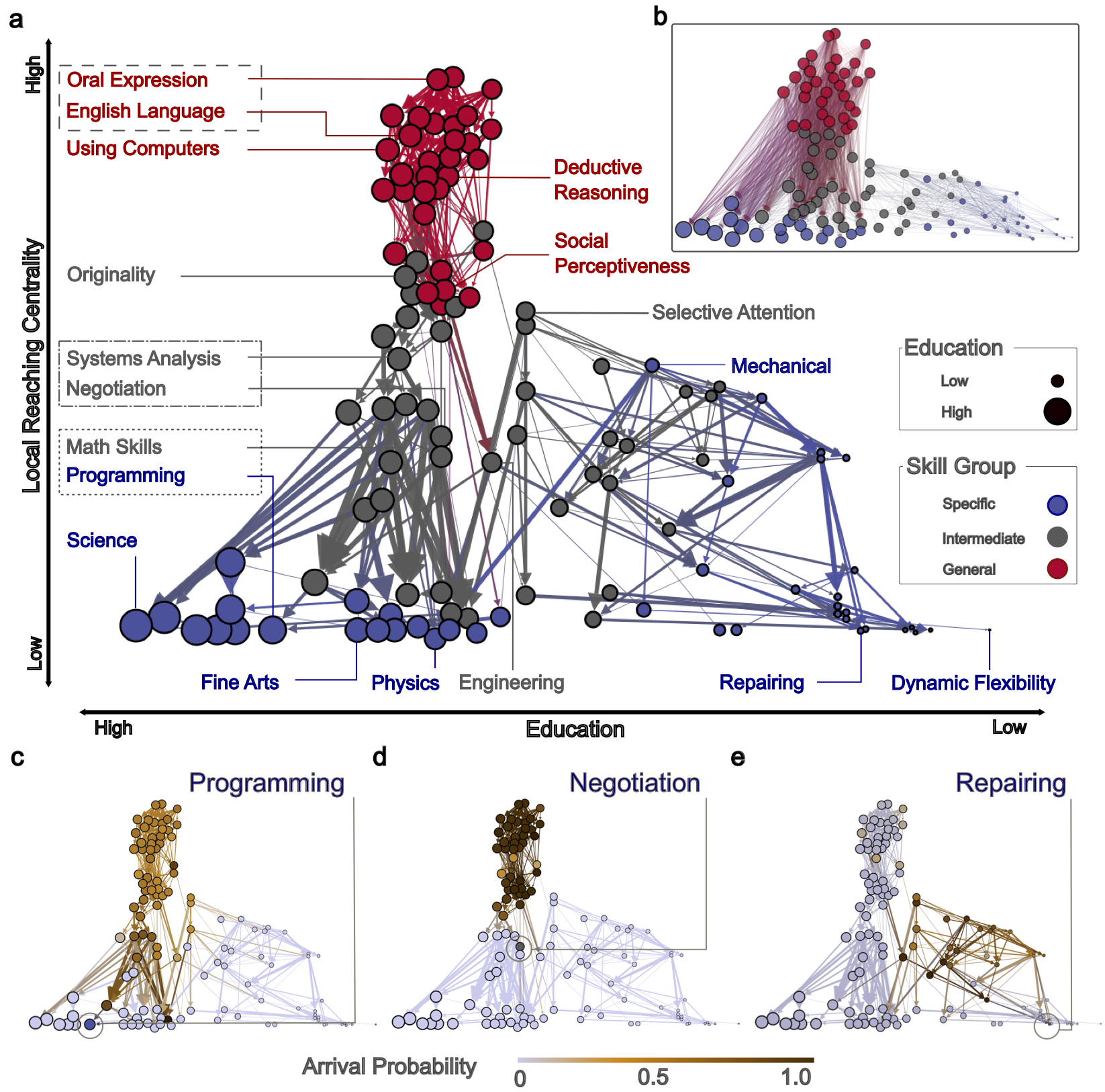}
    \caption{\textbf{Skill Dependency Hierarchy.} 
    \textbf{(a-b)} Dependency hierarchy is constructed from the aggregated weighted directions of all skill pairs. Node sizes are proportional to education levels and colored by the groups in Fig.~\ref{fig:Figure 1}.  
    A node's horizontal and vertical positions are, respectively, its educational attainment and local reaching centrality. 
    Defined as the proportion of the skills reachable from each node or the number of interdependent skills, the centrality is a reasonable indicator for skill generality \cite{Mones2012}. b shows the backbone of the network for better local visualization, while c shows the full network with normalized weights.
    \textbf{(c-d)} Reachability (arrival probability) from each skill to Programming, Negotiation, and Repairing (highlighted) \cite{norris1998markov}. Dark hues indicate a higher likelihood of arriving at the focal skill (see Methods). Contrary to the well-nested Programming and Negotiation, Repairing does not predominantly rely on general skills, indicating its unnested nature. 
    }
    \label{fig:Figure 2}
\end{figure*}

\subsection*{Skill Hierarchy (interdependency)}

The disparate skill level profiles captured by our empirical generality skill groups suggest a hierarchical structure among skills, with some serving as prerequisites for others. This hierarchy has been a longstanding topic of interest in fields such as labor economics, sociology, and management, but it has not been systematically analyzed   \cite{Becker1962, Neal1995, Parent2000, Poletaev2008, Gathmann2010, FergusonJohn-Paul2013, Leung2014, Merluzzi2016, Byun2018, Fini2022, Byun2023, Leahey2007, Teodoridis2018, Heiberger2021}. 
As such, we propose a method to quantify these relations by calculating how often occupations that require niche skills also require general skills and compare this to the inverse---how often needing general skills predicts the need for certain niche skills. If general skills are indeed prerequisites for niche skills, much like how most college curricula have fundamental courses preceding specialized ones, we should expect to find an asymmetry in these probabilities.

We operationalize the pairwise dependencies between skills using the information asymmetry in occupational skill requirements, following \cite{Jo2020}. The approach involves calculating the conditional probability of requiring one skill ($skill_A$) given the presence of another skill ($skill_B$), denoted as $p(skill_A | skill_B)$, and comparing it to the reverse probability, $p(skill_B | skill_A)$. This comparison allows us to assign directionality to the skill dependencies. If skill $A$ is contingent on skill $B$, meaning that the application or acquisition of skill $A$ is dependent on that of skill $B$, then $p(skill_A | skill_B)$ will be greater than $p(skill_B | skill_A)$.

In cases where skill $A$ and skill $B$ are independent events across occupations, the direction disappears as the conditional probabilities will be equal. This is because when two events are independent, $p(skill_A | skill_B)$ is expressed as $p(skill_A)p(skill_B)$, which is then the same as $p(skill_B | skill_A)$. Similarly, if two skills are rarely applied together within occupations, both base probabilities will be close to zero, $p(skill_A, skill_B) \simeq 0$, indicating no statistical dependency between them.
In both cases, co-occurrences are purely a result of their random independent events of either occupational need or individual workers' properties and, thus, not influenced by any underlying relationship. Therefore, an asymmetry in the conditional probabilities reveals how skill $A$ relies on skill $B$ for its application or acquisition, indicating the importance of the order in which skills are acquired or applied. 

It is important to acknowledge that this directionality does not provide a detailed understanding of the underlying process. The directionality could arise from the acquisition sequence, such as the learning process, or the requirement sequence through job seniority in organizations. What's happening at the individual worker level is inferred rather than directly measured in the current study because our empirical evidence is based on occupational attributes. 
Disentangling these factors would require more micro-level analyses, yet it is a promising avenue for future research. In this study, we focus on providing a phenomenological understanding of the structure of skill dependencies and their consequences for individuals.

Figure~\ref{fig:Figure 1}~(d) illustrates our inference method using select examples. Given the skill level distributions, the conditional probability of math skills given programming skills, $p(skill_{math} | skill_{prog})$, is higher than $p(skill_{prog} | skill_{math})$, resulting in the directional dependency $math \rightarrow programming$.
This direction is consistent with our common understanding and educational curriculum; to understand the complexity of a program, we need to have a minimum knowledge of math. The same holds true for Negotiation skills being conditional on Oral Expression. Moreover, developing and applying Math skills depends on advancements in Deductive and Inductive Reasoning, which are in the general group (red) of Fig~\ref{fig:Figure 1}~(c).
These create dependency branches, suggesting we will expect more than one depth to the hierarchical network.

These cross-group dependencies resemble biological mutualistic interactions where specialist species (i.e., niche skills) preferentially interact with generalists (i.e., general skills), suggesting a nested hierarchical skill integration \cite{Bascompte2003, Saavedra2011, Saavedra2009, Staniczenko2023}. However, the result is not always obvious; not every skill exhibits such dependency chains. Some specialized skills, like Dynamic Flexibility, may not be contingent on more general skills like mathematical prowess, which is again consistent with our common understanding. This can be calculated as $p(skill_{dyn.flx} | skill_{math})$ and $p(skill_{math} | skill_{dyn.flx})$. We find these two are independent events in which both expressions equal $p(skill_{dyn.flx}) p(skill_{math})$, resulting in no directional dependency in our methodological framework.

Figure \ref{fig:Figure 2}~(a) shows the backbone of the resulting hierarchical network obtained by aggregating the empirically derived dependencies across all skill pairs. The network extends from general to specialized skills, incorporating their directional dependencies (the full network is shown in Fig.~\ref{fig:Figure 2}~b).
Nodes are colored by generality group as in Fig.~\ref{fig:Figure 1} and positioned based on educational requirements (x-axis) and Local Reaching Centrality (y-axis), a measure of skill generality denoting the number of other skills reachable from the focal skill \cite{Mones2012}. 
The network reveals distinct specialization paths and a partially nested architecture. Methods and SI Sec.~\ref{supsec:conditional dependencies} provide detailed parameters for statistical filterings and the threshold for directionality and backbone structure for Fig.~\ref{fig:Figure 2}.

Constructing a network structure from these conditional directions provides a methodologically consistent definition of general and specific skills using reaching centrality \cite{Mones2012} as an alternative measure for generality, as this can reflect the mass of interdependent nodes on the focal node (0.71 correlated). 
Chains of dependencies for select examples are also well embedded as expected, such as Deductive Reasoning to Math skills to Programming, exemplifying the nesting of skills in the skill hierarchy. Negotiation has a different set of dependencies compared to Programming, including Systems Analysis.
Supplementary Information Secs.~\ref{supsec:RN vs. NP} and \ref{supsec:hispanic skill entrapment} offer brief case studies highlighting the role of dependency chains in career progress and specialization.
Finally, we include the fully labeled visualizations of Fig.~\ref{fig:Figure 2}~(a-b) in SI Figs.~\ref{fig:figure_2b_labeled} and \ref{fig:full_figure_2b_labeled} for further examinations.

\subsection*{Skill Nestedness Contributions} 

\begin{figure*}[!h]
    \centering
    \includegraphics[width=0.8\textwidth]{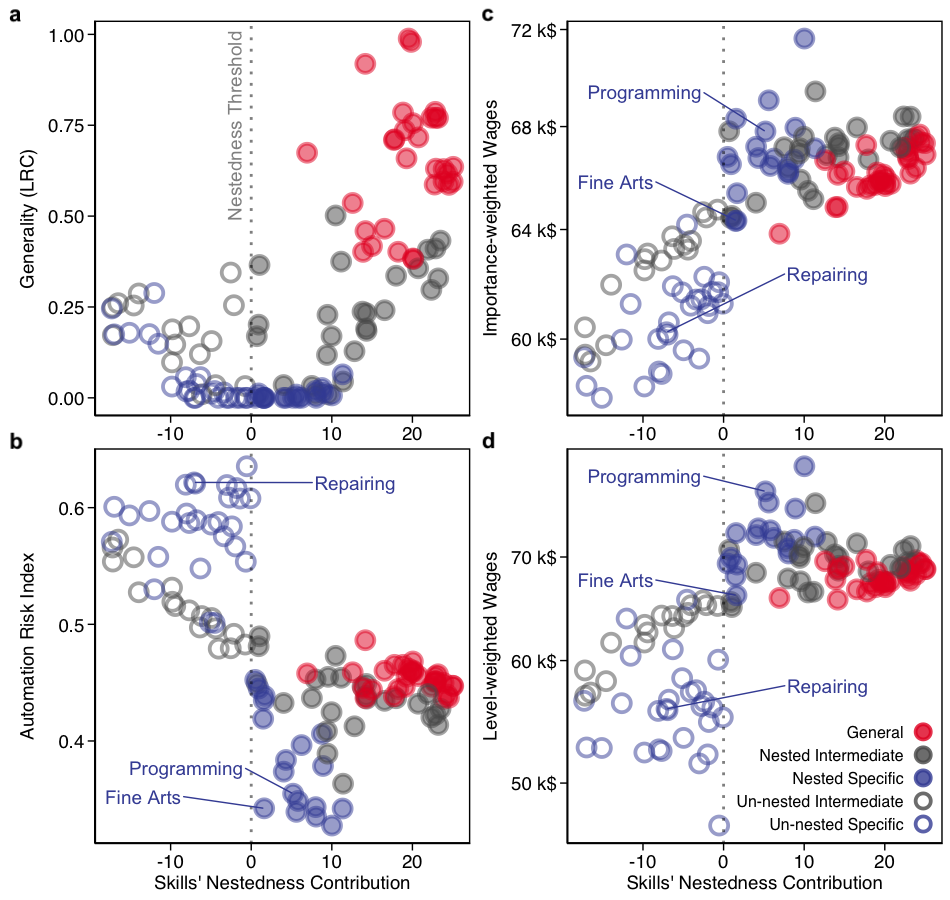}
    \caption{
    \textbf{Skill Nestedness Contributions Score.} Skills' nestedness score is highly indicative of their generality (a), risk of automation  (b), and their value (c-d). Skill Nestedness Contributions are measured following \cite{Saavedra2011}. Generality is measured by Local Reaching Centrality, as in Fig.~\ref{fig:Figure 2},
    Automation risk Index and Value for each skill is calculated, following \cite{Frey2017, Frank2019, Frank2022}.
    We divide skills into \textit{nested}, positive contributions, and \textit{un-nested}, negative contributions toward the nested skill structure.
    }
    \label{fig:nestedness}
\end{figure*}

Figure \ref{fig:Figure 2} also illustrates that the alignment of skills within a nested structure is not uniform. While some skills, such as Programming and Negotiation, seamlessly integrate with general skills in a nested pattern, others break from this arrangement, creating an uneven, tree-like hierarchy.  
This reveals a \textit{partially} nested architecture in human capital, indicating that specific skills don't consistently subordinate to general skills \cite{Saavedra2011, Baldwin2014}. 

To systematically quantify and differentiate these observations on skills, we introduce the ecological measure of nestedness and individual contribution scores where specialist species engage preferentially with generalists \cite{Bascompte2003}.  This analogy extends to the skill ecosystem, where general human capital forms the bedrock for the acquisition and application of more specialized skills \cite{Saavedra2011, Saavedra2009}. 
Therefore, we first measure an overarching nested structure in human capital $N$. There are a number of different ways to measure nested structures. We employ several measures commonly used in ecology, such as the overlap index ($N_c$), checkerboard score, Temperature, and NODF, to ensure the analysis withstands the test of different nestedness measurements \cite{Stone1990, Almeida-neto2008, write1992, write1992, Saavedra2011} (See SI Sec.~\ref{suppsec:nestedness} for the full analyses and robustness tests).

Next, we calculate a skill's nestedness contribution score $c_s$ to assess its alignment with the overarching nested structure $N$ \cite{Saavedra2011}. This score is derived by comparing the actual nestedness ($N$) with a null expectation where a focal node $s$ is randomly distributed across occupations without any underlying dependencies such as $p(A|B)$, which is expressed as $c_s = (N - <N_s^{\ast}> ) / {\sigma{N_s^{\ast}}}$. 
Here, $N$ denotes the empirically observed nestedness in our survey dataset, while $<N^{\ast}s>$ and $\sigma{N^{\ast}_s}$ are the average and expected standard deviation of the nestedness of the random condition, respectively \cite{Saavedra2011}.
We conduct 5,000 simulations for $<N^{\ast}s>$ and $\sigma{N^{\ast}_s}$. In each simulation, occupations using the focal skill $s$ are randomly selected, keeping the skill degree constant. This method allows us to maintain consistency with actually observed patterns of niche and general skills but destroy the dependencies such that we identify how dependencies positively/negatively contribute to the overarching nestedness structure.

Skills with a high nestedness contribution ($c_s$) are foundational to a hierarchical framework of human capital, suggesting a systematic progression from general to specialized skills toward layered learning paths that demand lengthy mastery effort \cite{Saavedra2011, Hausmann2011}. 
Such a pattern suggests a complex process of human capital formation characterized by interdependent skill acquisition pathways. These pathways are possibly essential for the emergence of specialized skills. In addition, they have profound implications for wages and education and contribute to disparities in demographics and opportunities \cite{Autor2014}.

Figure \ref{fig:nestedness}~(a) shows that highly specialized skills (blue) do not contribute equally to the overall nested structure and are thus divided into those with negative and positive contributions. 
As expected from Fig.~\ref{fig:Figure 2}, skills like Programming exhibit a positive impact on nestedness, indicating a strong reliance on vertical dependencies within their application domains.
In contrast, skills like Repairing, which also belong to the group blue in Fig.~\ref{fig:Figure 1}, are not heavily dependent on such structured dependencies and are quantified as having a negative contribution to nestedness.

We corroborate these findings with simulations of arrival probability from each focal skill. 
Figure \ref{fig:Figure 2}~(d-f) highlights the distinct interaction patterns among two types of specific skills: those that are primarily nested under general skills, such as Programming or Negotiation, and those that primarily interact with other niche skills, such as Repairing. We calculate arrival probabilities to the focal skill nodes and color other nodes according to their arrival probabilities to the focal node (see Methods) \cite{norris1998markov}. 
Unlike the well-nested Programming and Negotiation skills, only a handful of other skills are relatively more easily reachable from Repairing than other skills, which are mostly in the same parts of the skill tree.

Figures \ref{fig:nestedness}~(b-d) demonstrate that the nestedness score, a structural attribute, can translated into socio-economic properties. These findings suggest that skills with high nestedness contributions are more likely to be associated with lower risks of automation and higher wages, as they are integral to a deeply interconnected structure that demands considerable investment for mastery \cite{Davidson1898book, Autor2003, Frey2017}. Such skills play a crucial role in creating a distinctively hierarchical human capital with vertically intricate dependencies, fostering specialized niches that potentially affect wages, education, and demographics. In contrast, skills with negative nestedness contributions, such as Repairing, do not exhibit the same level of dependence on structured hierarchies and may be more susceptible to automation and lower wages. This highlights the importance of considering not only the generality of skills but also their position within the skill hierarchy when assessing their socio-economic implications.

The relationship between nestedness contributions and socio-economic outcomes underscores the significance of the skill hierarchy in shaping the labor market. By understanding the structural properties of skills and their interdependencies, we can better predict the impact of technological change on different skill domains and inform policies aimed at promoting skill development and mitigating the risks of job displacement.

For the remainder of this paper, we simplify the exposition by defining skills according to their skill group and the sign of their nestedness score $c_s$. Skills with $c_s > 0$ are indexed as \textit{nested}, while those with $c_s < 0$ are considered \textit{un-nested} skills.
We continue to refer to general skills as such since all skills in that group have positive nested scores.

\begin{figure*}[!h]
    \centering
    \includegraphics[width=\textwidth]{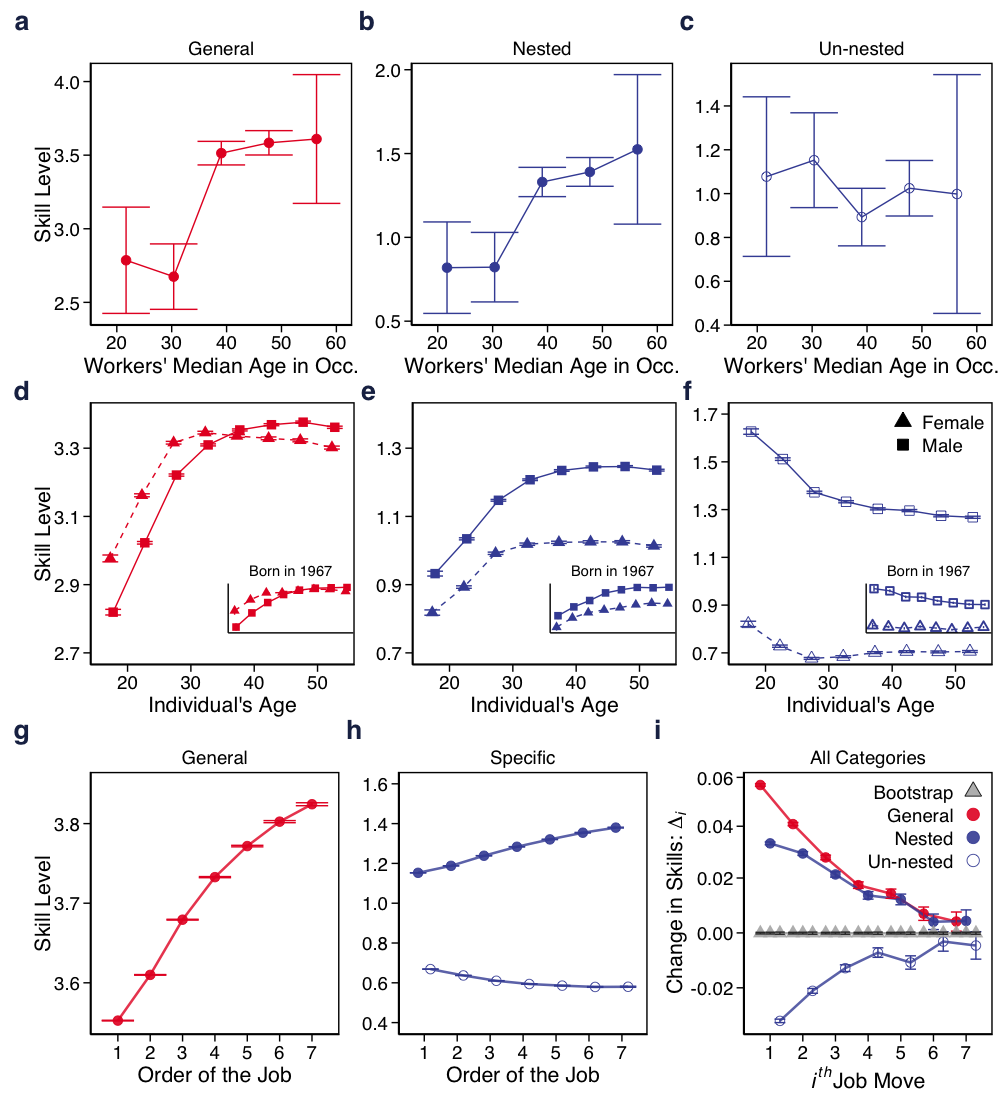}
    \caption{
    \textbf{Skill Compositions with Occupational Ages and Career Trajectory.}
    \textbf{(a-c)} Average skill levels of occupations (and 95\% confidence intervals), segmented by occupations' employees' median ages. Levels of general and nested skills rise with an occupation's median age, while unnested skills do not vary across median-age groups.
    \textbf{(d-f)} Average skill levels (and 95\% confidence intervals) against age in synthetic birth cohorts. The insets isolate cohorts born in 1967, whereas the main figures average across all cohorts. Notably, general and nested skills rise markedly until around age 30, with declining unnested skills. Moreover, gender gaps also become more pronounced around 30. 
    \textbf{(g-h)} Average skill levels (and 95\% confidence intervals) over identified job sequences as documented in resumes for general, nested, and unnested skills. 
    \textbf{(i)} Changes in skill levels in consecutive job transitions. Skill profiles are typically stabilized within the initial five jobs.  The grey triangles indicate bootstrapped results where the sequences of jobs are randomized.
    }
    \label{fig:age}
\end{figure*}
\subsection*{Skill Categories in Career Trajectories}

In this section, we examine how the derived skill structure uncovers individual career trajectories through three empirical observations: median ages for occupations, synthesized birth cohorts from individual surveys, and job transitions in resumes. Each data source provides unique strengths and weaknesses, which, when combined, complement each other and sketch a coherent picture of career paths.  

We begin our analysis with occupational ages, as it is reasonable to expect progression and skill development to correlate closely with age due to the substantial investment of time and the dense set of prerequisites they demand \cite{Argote1990,Jovanovic1997,Nedelkoska2015}. 
Figure \ref{fig:age}~(a-c) shows the levels of general, nested, and unnested skills in occupations, segmented by their median ages, computed using the Current Population Survey (CPS) (see Methods).  
The outcomes align consistently with our predictions \cite{Jovanovic1997}. Occupations with median ages over 30 demand high levels of both general and nested skills, while unnested skills, supposedly lacking interdependencies, do not demonstrate any significant correlations with ages.

To examine if our results hold across career trajectories, we construct synthetic birth cohorts using the CPS microdata, which provides yearly repeated cross-sectional surveys but does not allow longitudinal tracing of respondents long enough for us to trace a few decades. Therefore, we connect snapshots of surveys through their birth years to mimic career trajectories   \cite{Acemoglu2011,Hermo2022}. 
For example, we construct a 1967 cohort for Fig.~\ref{fig:age}~(d-f), excluding observations of non-full-time respondents and those below age 17 or above 55. We then repeat this for different birth cohorts. 

Figures \ref{fig:age}~(d-f) show the skill composition of synthetic birth cohorts from 1980 to 2022, with insets for the 1967 cohort. Consistent with the findings in Fig.\ref{fig:age}~(a-c), age 30 emerges as a significant transition point. General and nested skills concurrently increase sharply until around 30, when unnested skills experience a moderate decrease. After the age of 30, the rise in overall skill levels stabilizes.

The advantage of the second dataset is the information on both the age and demographics of individuals, allowing us to decompose our findings by gender. 
Differentiating skill trends by gender uncovers a gap in specializations that emerges around 30. Men continue to grow their general and nested skills until their 50s, whereas for women, the increase in these skills hits a plateau in their early 30s, the typical age range for first-time mothers in the US.
Supplementary Information Secs.~\ref{supsec:Parenthood_Male_vs_Female} and \ref{supsec:female job sorting} further investigate the influence of parenthood on male and female workers by slicing data by those with and without children as well as the influence of sorting into jobs based on schedule and working hours, respectively. 
These findings are robust to conditioning out yearly economic conditions (SI Fig.~\ref{fig:individuals' age and skill - year effects}).
In the following sections and in Fig.~\ref{fig:Skill Age Gender Race Trends - year effects}, we offer more detailed breakdowns of these gender disparity trends with respect to race and ethnicity.
Notably, education does not fully account for the growth in skill documented by our analysis.
As SI Fig.~\ref{fig:individuals' age and skill and education} shows, the share of educational attendance is negligible after the age of 30, while skill growth continues. Similar patterns, in more modest magnitudes, emerge for workers with no more than high school diplomas (SI Fig.~\ref{fig:individuals' age and skill - no college}.)

Lastly, we complement our findings using resume datasets that record individual job transitions, encompassing over 70 million job transitions documented in over 20 million resumes. While these data provide a direct record of individual workers' job sequences, they are not publicly accessible, do not include age or gender information for detailed analyses, and are known for biased sampling, favoring more nested job roles. Hence, while valuable for corroborating previous findings, they cannot replace the previous datasets.

Figures \ref{fig:age}~(g-h) show the average skill levels required in job sequences held across career paths, and Fig.~\ref{fig:age}~(i)  displays changes in skill requirements for the $i$th job transition, $\Delta_i$, excluding job transitions within the same occupation ($\Delta_{i}= 0$).  
On aggregate, career journeys unfold with increasing stocks of both general and nested skills ($c_s > 0$), suggesting that nested specialization paths require simultaneous increases in nested specific skills along with their dependency skills.  In addition, we find skill portfolios typically stabilize within the first five job transitions ($\Delta_{i > 5} \approx 0$), and in the first three jobs ($i < 3$), nested skills require more general skills than later ($\Delta^{general}_{i < 3} \gg \Delta^{nested}_{i < 3}$), after which they become comparable ($\Delta^{general}_{i > 3} \approx \Delta^{nested}_{i > 3}$). 
The continued growth in general skills across career paths suggests that these skills need to be continuously enhanced regardless of career stage. As a benchmark, we create bootstrapped job sequences (gray marks around zero) that randomize the order of jobs as if there were no career development, confirming that the observed trends are indeed attributed to career developments (see SI Sec.~\ref{supsec: bootstrapping BG} for details).

To explore nested specialization, we choose registered nurses (RNs) and nurse practitioners (NPs) by analyzing resume data to understand how skill and wage differences manifest in career trajectories. 
Supplementary Information Fig.~\ref{fig:RN vs. NP} shows the additional skills (necessary to prescribe medicine and diagnostic tests) in higher-paying NP positions appear in nested paths with growth in both general and dependent niche skills such as medicine, therapy, biology, science, and chemistry (see SI-Sec.~\ref{supsec:RN vs. NP} for the detailed analysis).  
In addition, SI Sec.~\ref{supsec:hispanic skill entrapment} makes a case wherein insufficient levels of certain general skills preclude the development of the dependent niche skills, once again highlighting how our framework teases out pathways for developing human capital.

All three empirical observations consistently depict nested specializations (i.e., growth in both general and nested skills) throughout career trajectories, while unnested skills are left relatively underdeveloped. The resume analysis offers direct evidence of a recurring yet counterintuitive pattern: valuable specialization is not just about developing niche skills; it is conditional on advancing the required more general skills. 
This suggests that the conventional model, where basic general skills precede advanced specialized skills, is not entirely accurate. Instead, career paths tend to unfold with increasing emphasis on general skills and their dependent, nested skills. While research has emphasized the role of education, Fig.~\ref{fig:age} (and SI Figs.~\ref{fig:individuals' age and skill and education} and \ref{fig:individuals' age and skill - no college}) reveal that skill advancement continues long after the age of schooling, suggesting nested specialization pathways operate through but also beyond education \cite{Jovanovic1997, Mincer1974, Arrow1962, Lucas1988, Hermo2022}, challenging the commonly held role of education in developing human capital. 

One might argue that our findings are driven by management/administration jobs, which are typically undertaken later in careers with higher wages. 
To ensure they do not drive our findings, we repeated the entire analysis without these factors and found consistent results (see SI Sec.~\ref{sec:robustness check: no managers}). Also, we repeated the entire analysis, excluding social skills, and again
our results remained robust, suggesting that our findings are attributed to the intrinsic structure of skills rather than the influence of particular social skills or managerial jobs (see SI Sec.~\ref{sec:social skills} for the full analyses).
\subsection*{Skill Categories and Wage Premiums} 

\begin{figure*}[!h]
    \centering
    \includegraphics[width=\textwidth]{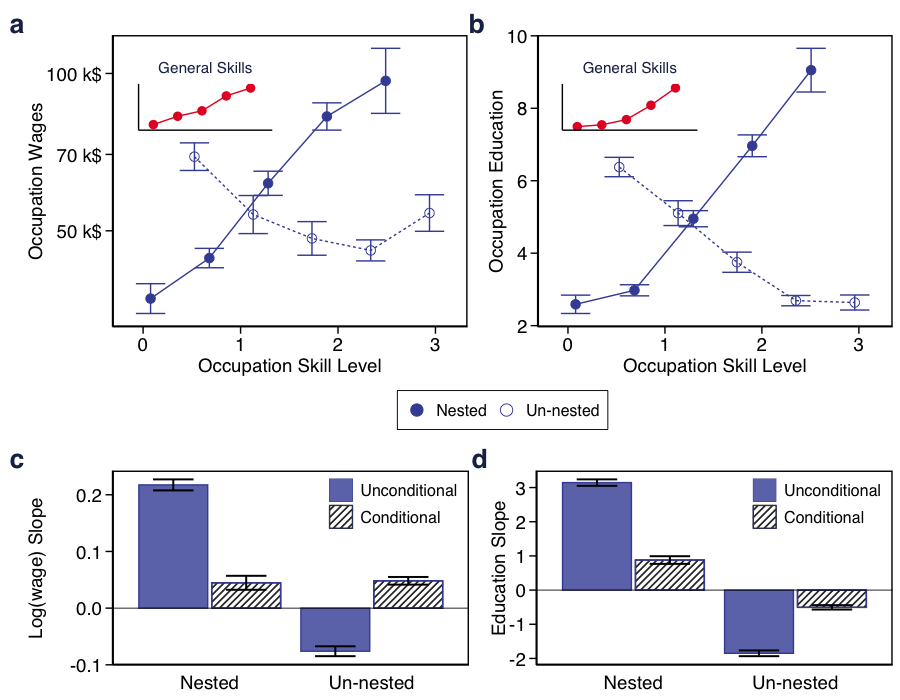}
    \caption{\textbf{Skill Wage Premiums and Educational Requirements}. \textbf{(a)} Occupations' average annual wage and \textbf{(b)} required education levels plotted against skill levels (with 95\% confidence intervals), 
    and their respective slopes (blue bars) in \textbf{(c-d)}, and standard errors. 
    The substantial wage premiums and higher educational requirements associated with nested specializations much reduced (shaded bars) after controlling for general skill levels (insets), implying that the bulk of investments in and returns to specialization are conditional on the accumulation of general skills. The initial wage penalty for unnested specializations turns into a wage premium once general skill levels are controlled for.  
    }
    \label{fig:Wage}
\end{figure*}

Figure~\ref{fig:Wage}~(a-b) supports our premise that nested specialization patterns are associated with wage premiums. In particular, we find that educational requirements and average annual wages tend to rise with rising requirements of nested skills in an occupation. However, a closer examination of the observed wage premiums for nested skills (blue bar) in Fig.~\ref{fig:Wage}~(c) reveals that such premiums almost fully disappear when we control for the occupation's general skill requirements (shaded bar). This suggests that general skills are integral to the deployment of nested skills. In contrast, unnested skills ($c_s < 0$) seem to be associated with wage penalties. However, controlling for general skill requirements now turns this penalty into a wage premium that is comparable in magnitude to the nested skill premium. This shows that unnested skills are also valued in the labor market. However, their wage premium is not immediately apparent because unnested skills tend to correlate with an \emph{absence} of general skills.

Further analyses in SI Sec.~\ref{supsec: add - returns to skill} demonstrate that these results are robust to controlling for education, training, and workplace experience and hold across subsamples of major occupational groups. 
Again, the results are not driven by managerial occupations or social skills, usual suspect factors in wage premium (see the results in SI-Table \ref{tab:wage reg on skill endowment}, and SI Figs.~\ref{fig:SI_education_skill_level}-\ref{fig:SI_wage_skill_level},
\ref{fig:Figure 3 full | major occupation groups}, \ref{fig:returns_to_skills_hierachy_gen_dependence_cor_no_manager}, and \ref{fig:social skills}).  

\subsection*{Disparity in Demographic Groups}

\begin{figure*}[!h]
    \centering
    \includegraphics[width=\textwidth]{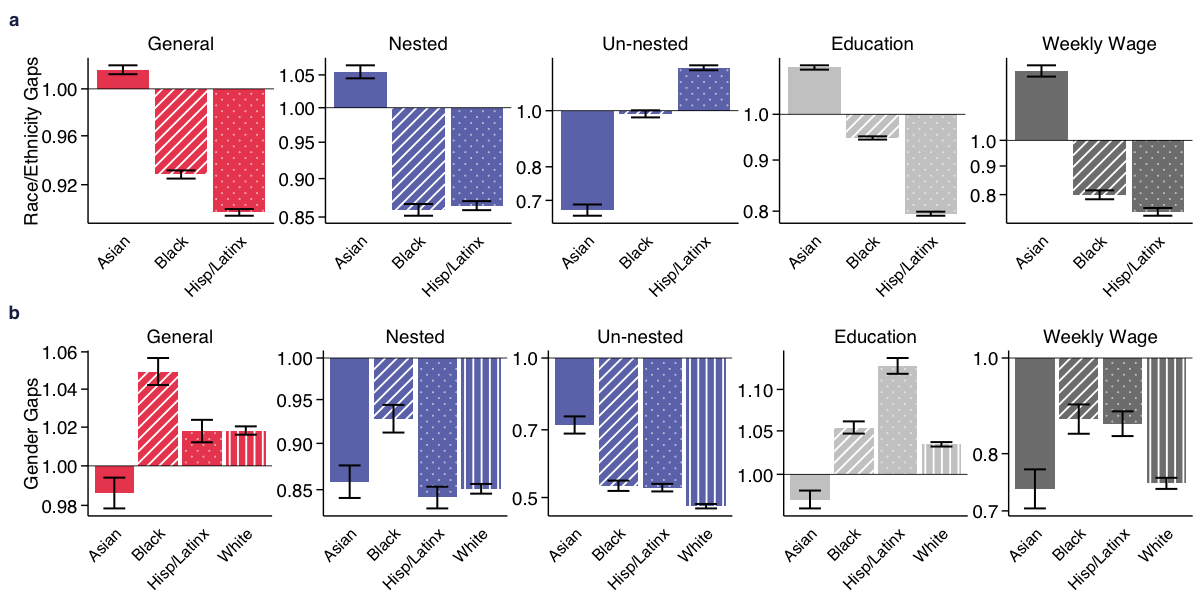}
    \caption{\textbf{Skill Disparity in Demographic Distribution of race/ethnicity and gender}
    \textbf{(a)} The relative average skill level, education level, and weekly wages for Asian, Black, and Hispanic/Latinx workers compared to White workers (expressed as a ratio).
    \textbf{(b)} The relative average skill level, education level, and weekly wages for female workers compared to male workers. 95\% confidence intervals for each estimated ratio are calculated by bootstrapping subsamples (see Methods). These differentials are robust to measurement (SI Fig.~\ref{fig:Tong et al race gender skill distribution}), follow similar age trends seen in Fig.~\ref{fig:age}, and are robust to time-variant economic factors (Fig.~\ref{fig:Skill Age Gender Race Trends - year effects}.) SI Figs. \ref{fig:Temporal Race Gaps - Skills, Education, Wages} and \ref{fig:Temporal Gender Gaps - Skills, Education, Wages}, further show the gaps have narrowed over time.  
    } \label{fig:Figure 7} 
\end{figure*}

To gain a better understanding of the role that skill differences may play in labor market inequalities, we examine how skills vary across demographic groups. Figure \ref{fig:Figure 7}~(a) compares skill, education, and wage differences across race/ethnic groups against their White peers. The results, first of all, show large wage gaps between Black and Hispanic workers on the one hand and  Asian workers and the baseline of White workers on the other hand. These wage gaps are accompanied by employment in jobs with lower requirements of nested skills for Black and Hispanic workers. However, for Hispanic workers, there is another potentially important factor: elevated unnested skill requirements. 

We explore this further in a brief case study of how language-skill requirements may keep workers out of jobs that require certain nested skills (see Supplementary Information Sec.~\ref{supsec:hispanic skill entrapment}). To do so, we leverage the hierarchical nature of our skill network. This allows us to distinguish between nested skills that depend on (general) language skills and nested skills that don't. Within the group of Hispanic workers, we find particularly large gaps in language-dependent nested skills compared to other nested skills for workers who have recently moved to the US. Such workers may instead develop un-nested skills, leading to ``skill traps'' that are associated with long-run wage penalties (SI Fig.~\ref{fig:wage curves}). Taken together, these findings indicate that closing wage gaps for Black workers may require different solutions than for Hispanic workers. 

Figure \ref{fig:Figure 7}~(b) focuses on skill gaps between men and women across social groups. The most pronounced disparities exist in nested and unnested specializations. Except for in the Asian subsample, women tend to work in occupations that require higher levels of education and general skills than men. However, this does not translate into higher levels of nested skills, where women tend to fall behind men. These disparities are likely to contribute to the well-known gender wage gap we observe in the right-most panel. Encouragingly, this gap has narrowed over time, as demonstrated in SI Fig. \ref{fig:Temporal Gender Gaps - Skills, Education, Wages}. However, the disconnect between education and general skills on the one hand and wages and nested skills on the other is puzzling. 
Supplementary Information Sections \ref{supsec:Parenthood_Male_vs_Female} and \ref{supsec:female job sorting} probe deeper into these gender gaps. This analysis suggests that parenthood, as well as the fact that women often work in jobs with more regular and predictable work schedules, impact both wages and skill development \cite{Bertrand2009, Goldin2015, Canon2016}. In fact, whereas having children is associated with reduced general and nested skills for women, men with children tend to have higher levels of general and nested skills than men without children in the household. When it comes to work schedules, Similarly, we find that the association between gender and nested skill requirements at work is reduced by over a third when we control for irregular hours and overtime in an occupation. 

Finally, Section \ref{section: add - geographical distribution of skills} of the Supplementary Information studies the geographic distribution of skills, showing that general skills concentrate in densely populated urban areas. This finding is in line with prior work that highlights the diverse and complex economic activities that are found in large urban economies \cite{Youn2016, gomez2016explaining, Hong2020, Balland2020, Bettencourt2014, Gomez-Lievano2021}. Moreover, this greater concentration of general skills in large cities can account for about one-third of the well-established urban wage premium \cite{Glaeser2001}.

In summary, the analysis of skill categories across demographic groups reveals a complex interplay between skills, education, and wages that leave an imprint on macro-level labor market disparities between societal groups. Although a deeper analysis of the causes and consequences of these disparities is beyond the scope of the current paper, our results highlight that analyzing skill gaps solely through the lens of educational attainment overlooks aspects of human capital that have an important impact on a variety of labor market disparities. Moreover, the complex interaction between wages and skill types suggests that considering such aspects may provide valuable insights for labor market policies: addressing long-lived disparities in the labor market may require targeted interventions that go beyond traditional educational programs and instead consider how different skill categories shape labor market outcomes.

\subsection*{Widening Gap in the Skill Structures}

\begin{figure*}[!h]
    \centering
    \includegraphics[width=\textwidth]{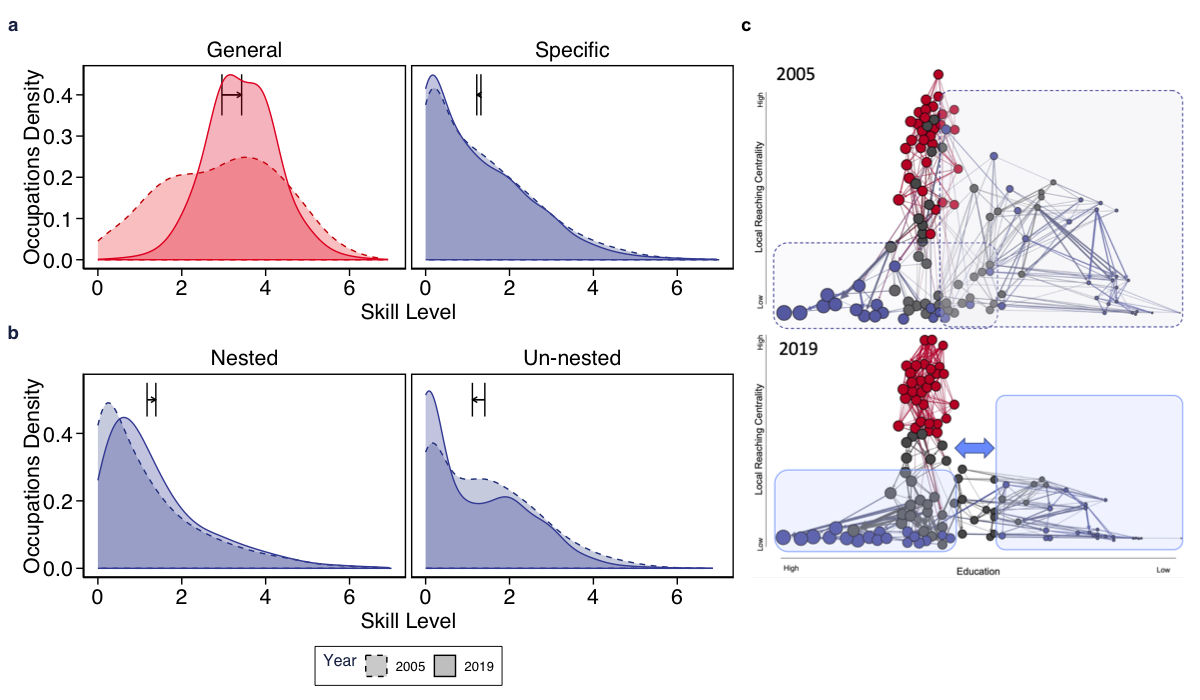}
    \caption{\textbf{Historical Changes in the Skill Structure}
    \textbf{(a)} Distribution of skill levels for different skill groups in 2005 and 2019. The arrow indicates the shift in average skill levels from 2005 to 2019. Unlike the positive shifts in general skills, the shift in specific skills is not as noticeable.
    \textbf{(b)} Distribution of skill levels for nested and unnested skills in 2005 and 2019. The arrow shows the shift in average skill levels from 2005 to 2019. While nested skills follow the shift in general skills, the demand for unnested skills has decreased.
    \textbf{(c)} Comparison of skill hierarchy structures between 2005 and 2019. The changes in the structure of skill hierarchies over time highlight an increasing divide in the dependencies of nested and unnested skills and the widening gap between them.} 
    \label{fig:historical skill change}
\end{figure*}

The historical changes in the skill structure, as shown in Figure \ref{fig:historical skill change}, raise concerns given the important roles that nested and unnested specializations play in career progression and demographic and regional disparities. These changes reignite the debate over the widening job polarization \cite{Autor2013, Alabdulkareem2018, Althobaiti2022}. 

Figure \ref{fig:historical skill change}~(a) indicates an increase in the demand for general skills, as evidenced by the shift from the dotted to the solid distribution. This increase in demand for general skills corresponds to higher wage premiums over the recent decade, suggesting that the economy has been rewarding workers with a broad set of skills (see SI Fig.~\ref{fig:Wage and education 2003}). However, the seemingly static distribution of specific skills masks underlying changes in the application of nested and unnested skills. As shown in Figure \ref{fig:historical skill change}(b), there has been a rise in the application of nested skills and a decline in the use of unnested skills between 2005 and 2019, reiterating the importance of considering skill interdependencies when analyzing changes in the skill structure. 

These changes have led to a more nested skill structure, as indicated by the decreased checkerboard score (from 438.67 to 356.4) and temperature (from 40.07 to 31.89), and increased NODF (from 39.06 to 41.72) and $N_c$ (from 573,873 to 651,030) between 2005 and 2019 \cite{Stone1990, Almeida-neto2008}. A lower checkerboard score and temperature, along with a higher NODF, signify a more nested structure.

However, the shift towards greater overall nestedness has been uneven across different skill sets. Figure~\ref{fig:historical skill change}~(c) demonstrates this as a widening gap between the nested and unnested branches over the decades. Supplementary Fig.~\ref{fig:Figure_7_supp}~(b) shows a strengthening in the connections among nested skills, indicating their growing complexity and mutual dependence, while SI Fig.~\ref{fig:Figure_7_supp}~(d) reveals weakened dependency chains of unnested skills. All in all, this trend is observed as a broadening and deepening of nested skill branches within the hierarchy in Fig. ~\ref{fig:historical skill change}~(c), reflecting an increase in the complexity and interdependence of these specialized skill areas \cite{DemingDavidJ2020EDCJ, Tong2021}.

Indeed, the chasm between the two types of specializations has alarmingly broadened within the educational domain over the last two decades \cite{Xu2021, Lin2022, Althobaiti2022}. In response to this, the structural changes in the hierarchical tree network are concerning, given the significance of these specializations for future career developments and wage premiums.  They reveal an economy wherein the structure of valuable human capital has grown more nested, reinforcing barriers to workers without the necessary fundamental skills, who are often entrapped in unnested specialization pathways (see SI Figs.~\ref{fig:wage curves} and \ref{fig:hispanics language skills}). The widening gap between the nested and unnested specialization paths could indicate strongly rooted chronic disparity. 

The increased demand for general skills and the shifting balance between nested and unnested skills have important implications for workers and policymakers. While the rising wage premiums associated with general skills suggest that workers who invest in developing a broad skill set may be better positioned to succeed, the growing importance of nested skills and the declining use of unnested skills may exacerbate existing inequalities and create new barriers to entry for certain occupations.

As the skill structure becomes more complex and interdependent, policymakers and educators must develop strategies to ensure that individuals from all demographic groups and regions have access to fundamental skill development opportunities. Failing to do so may exacerbate disparities and hinder economic mobility for certain segments of the population. To mitigate the potential negative consequences of the changing skill structure, it is crucial to invest in education and training programs that foster both general skills and unlock valuable specialized pathways. Our results suggest that providing individuals with a strong foundation in general skills and the opportunity to develop nested specializations is essential for navigating the increasingly complex labor market and achieving better career outcomes.



\section*{Discussion} 

Human capital has traditionally been quantified in terms of years of schooling or work experience, yielding important insights about wage curves and returns to education \cite{Mincer1974, Becker1962}. With the arrival of detailed data on tasks that people perform at work and the skills they require to do so, a more granular assessment of human capital became feasible, juxtaposing, for instance, cognitive and manual skills, routine and non-routine skills, or STEM and social skills \cite{Autor2003, goos2007lousy, Deming2017}. However, these dichotomies are often ad hoc, tailored to test specific assumptions about trends in the labor market, such as mechanization, computerization, and the rising importance of soft skills.

In contrast, the complexity approach to human capital analyzes labor markets through network analysis, providing a more comprehensive and data-driven perspective. Understanding the network architecture in complex economic systems—spanning technology, input-output, supply-chain, trade, products, and skills—has yielded insights into socioeconomic phenomena \cite{Neffke2013, Baldwin2014, Alabdulkareem2018, McNerney2022, Mariani2019, DelRio-Chanona2021, hidalgo2018principle, Elliott2014, Acemoglu2012, Elliott2022}. These insights both corroborate and contest established theoretical frameworks, including the underlying causes of economic disparities between countries and their potential developmental trajectories by analyzing trade networks \cite{hidalgo2018principle, mcnerney2021bridging}; the pace of technological innovation and economic growth through technology networks \cite{mcnerney2011, McNerney2022}; differences in labor productivity and resilience through the lens of skill and occupation networks \cite{Neffke2013, Neffkeeaax3370}; and economic network resistance and persistence using business networks \cite{Saavedra2011, Moro2021, Frank2018, DelRio-Chanona2021}. These models and methods translate a range of structural properties into quantifiable and actionable insights.

Our empirical study aims to add a new layer to these structural properties by illustrating how connections within these networks are conditional and how structures become increasingly nested as complexity and specialization grow. This method provides structural insights into prior empirical findings that cognitive skills are clustered themselves and valued more highly than physical skills, based on patterns of co-occurrence \cite{AndersonKatharineA2017Snam, Alabdulkareem2018, Xu2021, Lin2022, Neffkeeaax3370, DelRio-Chanona2021}. Within this networked framework, we observe a system where more central nodes, or skills, are rewarded more substantially along the network's nested branches, suggesting that the value attributed to cognitive skills in previous findings is interdependent with the increasingly nested structure of the skill network. Consequently, this finding leads us to move beyond the traditional dichotomy of cognitive versus physical skills towards a structural classification of skills as either nested or unnested.

Our research thus bridges economic theories that recognize hierarchical structures to explain progression and wage premiums \cite{Mincer1974, Becker1962} and the economic complexity model for understanding economic development \cite{hidalgo2018principle}, where the hierarchical organization of skills and their societal implications have been taken for granted rather than empirically verified \cite{Neffkeeaax3370}. Our work aims to offer an empirical framework as a network of skills in which ties capture skills' directional interdependencies, distinguishing pathways to specialization. Notably, we show that relying solely on the information embedded in the network of skills and occupations allows for a quantification of skills based on the concept of nestedness, independent of economic and social variables and without any presupposed or context-informed labeling of skills. Our analysis of wage and demographic disparities shows the predictive capability of this minimal approach for various socioeconomic factors.

The hierarchical structure and its inherent directionalities add a new dimension to the rising field of economic complexity, providing a deeper understanding of how knowledge is accumulated within a population and how its precedence relations between activities are expressed in the economic activities of a firm, city, region, or country \cite{Hidalgo2015, OClery2021, Balland2020, Hidalgo2021, Harmand2015, Hausmann2011, Hidalgo2007, Hidalgo2009,tacchella2012new, Park2019}. The directional dependencies that we propose break the symmetry in traditional co-occurrence networks for a better understanding of structural changes in economic complexity \cite{mcnerney2021bridging, OClery2021, yang2022scaling, Hong2020, hidalgo2018principle}.

In increasingly complex, large teams, social skills become crucial when specialization requires workers to coordinate with team members possessing different specialized skills \cite{Carlile2002, Wuchty2007, yoon2023, Neffkeeaax3370, Wu2019Science, Borner2018}. Our framework identifies and locates social skills embedded in the skill structure along with general and nested skills (SI Fig. \ref{fig:social skills}-a), explaining their recent growth and significant role in wage premiums (see SI Fig. \ref{fig:social skills}-b) \cite{Liu2013, Deming2017, Borghans2014, Weinberger2014, Lindqvist2011, Kuhn2005, Deming2018, Kogan2021, VanderWouden2023, Evans2024}. Nevertheless, our results go beyond the contributions of social skills and managerial occupations to wage premiums, as the results are robust to their absence in analyses (see SI Sections \ref{sec:social skills} and \ref{sec:robustness check: no managers}). Therefore, social skills are valuable not just because of their role in sociality but because of their structural properties, serving as foundational building blocks of human capital to enable further valuable specialization and more complex organizations.

The structural implications of our findings extend beyond individual careers and their associated rewards, suggesting potential consequences for not only intra-generation career mobility but also perhaps inter-generation career mobility. In this context, Figure~\ref{fig:historical skill change} presents a disconcerting trend, illustrating the widening gaps across nested branches over the span of a decade. We speculate that these growing disparities may be attributed to the increasing complexity of the economy and the deepening of individual specializations. As the skill structure becomes more intricate and the dependencies between skills more pronounced, individuals who successfully navigate these nested pathways may reap significant benefits, while those who struggle to acquire the necessary skills may face limited opportunities for advancement, potentially leading to entrenched inequalities that persist across generations. However, we acknowledge that the current study does not fully underpin these implications due to the lack of detailed datasets and a comprehensive analytical framework. Therefore, we recognize the need for further research to examine these critical questions and unravel the long-term consequences of the evolving skill structure on intra-generation and inter-generation mobility in order to inform policies and interventions aimed at promoting equitable access to skill acquisition, fostering inclusive economic growth and mitigating the potential for widening disparities within and across generations.

There are limitations in inferring the dynamics. First, our current empirical findings do not establish a causal relationship between semantic categories and structural manifestations, presenting an important question for future research using theoretical frameworks and computational models. 
Second, our analysis leverages datasets of occupational ``requirements'' of skills, that is, skills that are applied, which is not a direct measure of skill acquisition.  In essence, the manner in which skills are learned remains outside our observational scope. We assume that skills applied in the workplace have been acquired beforehand but not long before. This suggests that an individual may have competencies in arithmetic, linear algebra, and programming, which might not be fully exploited until they progress in their career. Although possible, such instances are presumed to be rare and not economically sensible, as individuals typically do not seek to acquire skills that are not immediately necessary, which probably pay less. This presumption rests on the belief that individuals strive to optimize their earnings and learning opportunities within their limited time and resources, making the phenomenon of being overqualified for job requirements relatively uncommon. Fundamentally, we suggest that there is a reluctance to engage in learning and skill development without direct application or compensation; thus, they occur relatively together. Related, our unit of analysis is jobs rather than individual employees, limiting our ability to discern the co-occurrence of learning and skill application. Future research could benefit from surveys targeting employees to gather nuanced data on individual skill portfolios as opposed to relying solely on job surveys. Finally,  our data primarily describe the U.S. labor market, which has idiosyncrasies in its education system, industrial composition, and urban structure. 
How well these findings generalize to other work settings, such as entrepreneurship \cite{Murray2023}, and economies, especially those at different stages of development \cite{Autor2022}, remains a task for future research.

Essentially, the underlying assumption is that people are less inclined to learn and develop skills unless these are directly applied or rewarded in their roles. In addition to the implicit mechanism of learning, our unit of analysis is not the employee but the job. The lack of granularity in our empirics makes it hard to identify whether learning goes together with the application of skills. In the future, conducting surveys of employees for detailed observations of individuals' skill endowments, rather than job surveys.  Finally, our data primarily describe the U.S. labor market, which has idiosyncrasies in its education system, industrial composition, and urban structure. 
How well these findings generalize to other work settings, such as entrepreneurship \cite{Murray2023}, and economies, especially those at different stages of development \cite{Autor2022}, remains a task for future research. 

In conclusion, our study introduces a novel approach to understanding the structure of skills in the labor market, shedding light on the pathways to specialization and the mechanisms driving skill value and resilience. While our study has limitations, it lays the groundwork for future research to explore the generalizability of our findings and investigate the relationships between skills, education, and socioeconomic outcomes.

\section*{Data and Methods} \label{sec: method}

\subsection*{Datasets}
\textbf{Occupational Information Network (O*NET) }
includes survey records of job-oriented attributes and worker-oriented descriptors conducted by The Bureau of Labor Statistics (BLS) \cite{Peterson1999ONET}. Job-oriented attributes include educational requirements, workplace experience, and training. Worker-oriented descriptors include 120 work-relevant knowledge, abilities, and skills (labeled \textit{skills} throughout the text for brevity).
Each occupation includes a list of skills with their sophistication levels (or intensity) and the importance of those requirements, each resulting in an occupation-skill matrix. 
Our main analysis uses the level, but the other variable is highly corrected (0.94), and therefore, our findings are robust to the choice of measurements.
We have obtained two versions: 2019, to avoid concerns over contaminating data with signals from the COVID-19 pandemic, and 2005, the first version with a consistent skill topology and available education covering a significant number of occupations.

\noindent \textbf{Occupational Employment and Wage Statistics (OEWS)}
offers wages and employment information at different granularity levels (nation-wide, region-specific, and industry-specific).
We have used nationwide, region-specific data for 2005 and 2019 and combined them with their respective year from O*NET.
Note that including and aggregating data from several years before and after 2005 and 2019 does not change our results.
In the resulting combined data, occupational units were aggregated at the 6-digit SOC codes (OEWS is available at the 6-digit level, while O*NET is available at the 8-digit SOC level).

\noindent \textbf{Current Population Survey (CPS)}
is a monthly survey of households conducted by the Bureau of Census for the Bureau of Labor Statistics.
It offers a representative sample of the population obtained in each round that offers statistics on various aspects of the labor force \cite{Flood2022}.
From the Labor Force Statistics component of CPS, we obtain the median age of workers in occupations for 2019.
From the CPS microdata, we acquire employment and demographic information on households between 1980 and 2020, including occupation, wage, hours worked, gender, and race/ethnicity information.
Matching with SOC occupational units requires a crosswalk described in the corresponding section.

\noindent \textbf{Burning Glass Resume Data }
includes 70 million job sequences (8-digit SOC) documented in 20 million individuals' resumes between 2007 and 2020 from Burning Glass (also known as Lightcast).
Burning Glass applied AI tools to submitted resumes, digitizing their text and mapping them to occupational titles consistent with BLS SOC codes, allowing for easy integration with O*NET data.




\subsubsection*{Skill Generality Groups} 

For each skill, O*NET reports the required levels needed for workers of each occupation to perform their tasks. We call the distribution of the number of occupations that require skill at varying levels the \textit{level distribution}.
The shape of a skill's level distribution illustrates its generality across occupations, shown in Fig.~\ref{fig:Figure 1} (a). 
As such, we group skills by their similar distribution shapes by $k$-mean clustering algorithms with correlation metrics. Figure \ref{fig:Figure 1} (b) shows the characteristic shapes of each skill group.  
We provide three statistical tests for optimal $k$ and show the findings are qualitatively robust to some variations (see SI Sec.~\ref{supsec:skill clustering}). Throughout analyses, we mainly analyze the effects of general and specific skills to filter possible noises.

This group is consistent with the local reaching centrality measure, which was used to embed nodes vertically in Fig.~\ref{fig:Figure 2} (b). 
The local reaching centrality is defined as the proportion of the skill hierarchy structure that is reachable from a skill via outgoing edges \cite{Mones2012}. The higher reaching centrality in the hierarchy structure is, therefore, the more interdependent skills. As such, this measure offers additional indicators of skill generality. 

\subsubsection*{Conditional Probabilities for Skill Hierarchy Structure}

The conditional probability that infers the directionality operates on binary values, but skill levels are recorded in continuous variables [0,7], which makes it hard to apply the conditional probability method. We use the disparity filter to extract a statistically significant presence/absence in an occupation-skill matrix \cite{Serrano6483}.
Parameters are chosen such that i) the rank of skill terms in the strength (from the weighted network) and degree (in the binary network) is preserved, ii) the rank of occupations' skills of each category in the weighted network is preserved in the binary network.
Supplementary Information Section \ref{supsec:skill-occ} discusses details and compares the state of data before and after the transformation.

We then calculate conditional probabilities of every pair of skills in the transformed (binary) matrix to infer dependence and directions between two skills, following \cite{Jo2020}.
We first account for the significance of conditional appearances, subject to a threshold, $z_{th}$.
Here, $z_{th}$ is a threshold for the extent to which we eliminate chance from two skills being used in the same occupation.
Given the significant skill pair conditional appearances, we estimate conditional probabilities $P(u|v)$ and $P(v|u)$. 
The direction of dependence $v \rightarrow u$ is set when $P(u|v)$ is \textit{substantially} greater than $P(v|u)$, subject to a parameter $\alpha_{th}$, which is differentially weighted for each pair of skills so that it accounts for heterogeneous skill node degrees (see Eq.~\ref{eq: a_th} in SI Section \ref{supsec:conditional dependencies}).
The magnitude of the dependence is a parametric function of the difference between the conditional probabilities of observing $u$ and $v$, and the null model that accounts for the estimated number of shared occupations between them, given the degrees of $u$ and $v$.as shown by Eq.~\ref{eq: dependency weight} in SI Section \ref{supsec:conditional dependencies}.
Figure \ref{fig:Figure 1} (d) broadly illustrates the intuition behind this methodology.
Figure \ref{fig:Figure 2} (a) presents a backbone structure of the aggregated all skill pairs, where the edge weights follow the magnitudes of pairwise dependencies, as described above.
Figure \ref{fig:Figure 2} (b) offers the full network.
Please see SI Section \ref{supsec:conditional dependencies} and \cite{Jo2020} for the detailed procedures and choices of parameters and thresholds. 


\subsubsection*{Reachability with Arrival Probability} 
To quantify what are the chances of getting to the focal skill $j$ given the pre-requisite skill $i$, we calculate reachability from one skill to a focal skill. It is basically arrival probability, or a version of hitting probability, of a random walk \textit{arriving} at $j$ from node $i$ given 
the weighted skill dependency network \cite{norris1998markov}.
For source and target skills $i \neq j$, this is numerically equivalent to first deriving the probability of random walks of length $l$ by raising the weighted-directed adjacency matrix (skill dependency network in Fig.~\ref{fig:Figure 2}), $M$, to power $l$, and then calculating  $R_{i,j} = \Sigma_l M^l_{i,j}$.
We obtain the final arrival probability by summing over a sufficient number of path lengths until reaching saturation points. To compute arrival probabilities for focal skills (such as Programming, Negotiation, and Repairing) in Fig.~\ref{fig:Figure 2} (b-f), we apply the R package \textit{markovchain} \cite{MarkovchainRPackage}.


\subsubsection*{Nested and Unnested Skill Categories} 
Nestedness is a structural characteristic that describes interactions in an ecological system, where specialist species often interact with a subset of generalists. 
Unlike ecological systems, however, SI-Fig.~\ref{fig:occ_skill_nestedness_mat} shows the skill-occupation matrix is a noisy nested structure far from the perfect upper-left triangle when sorted by marginal totals (fills).
This imperfect nested structure may account for the constraints on occupations (limited carrying capacity), introducing severe competition between skill species. Indeed, SI-Fig.~\ref{fig:occ_vs_skill_importance_avg_cos} shows, unlike broad skill generality, the occupation's scope is narrowly distributed, indicating 
that the total amount of skill levels embodied in an occupation is not much different from each other, regardless of how much they are paid and how advanced education is needed (see SI Sec.~\ref{suppsec:nestedness}).

We attribute occupations' limited scope of skills to the limited attention and cognition/physiological capacity that individual workers can offer. There is only so much a single person can equip and do for a single job \cite{BenJ2009, DUNBAR1992}. Thus, individuals' capacity restricts how many skills an occupation can bundle. This constraint explains the process of specializations needed for a complex job. The structure now includes not only nested structure but also mutually exclusive presences, possibly due to competition between skills within an occupation. 
In contrast to occupations, skills do not have such constraints. Therefore, for limited occupation scope, we only consider the skills' contribution to nested structure.

This constraint distinguishes the nestedness of extensive economies such as nations, regions, and urban areas from the nestedness of occupations in that specializations dominate the evolution of the labor market while others are dominated by diversification. 
As a result, the skill-occupation matrix is expected to be modular as well as nested with mutually exclusive modules. \textit{Nested-modular matrix} is a complicated structure and will be beyond our current scope \cite{Fortuna2010, VanDam2021}. Here, we will focus on individual skills' contributions to the nested structure and differentiate skills that contribute to the nested structure from those that do not. 

Therefore, we quantify a skill's contribution to the nested structure, i.e., nested score, $c_s$, defined as a deviation from a null model where the edges of a focal node $s$ to occupations are randomly reassigned, that is, $c_s = (N - <N^{\ast}_s> ) / {\sigma_{N^{\ast}_s}}$.
$N$ is a nestedness score, and $<N^{\ast}_s>$ and $\sigma_{N^{\ast}_s}$ are the means and standard deviation derived from the null model \cite{Saavedra2011}. For each focal skill $s$, we run 5,000 iterations \cite{SergeiMaslov2002}. We employ the overlap index, checkerboard score, Temperature, and NODF, nestedness scores commonly used in ecology, to quantify nestedness $N$ \cite{write1992, Stone1990, Atmar1993, Almeida-neto2008}.
In addition, we only consider skill's contribution and do not occupation's contribution.
To obtain discrete categorizes, any non-general skill with $c_s>0$ is called ''nested'' skills, and ''un-nested'' otherwise. The resulting skill categories are shown in Fig.~\ref{fig:nestedness}.
The detailed allocation of skills to these categories are outlined in SI Table \ref{tab:skill_split_C_nestedness}, and SI Sec.~\ref{suppsec:nestedness} offers details and robustness checks.



\subsubsection*{Educations}
Education variables in O*NET are categorized into twelve discrete grades, ranging from below high school (1) to post-doctorate (12). 
Each occupation includes the proportion to which corresponding sampled employees had to have a given educational level to be hired. 
With this information, we calculated an occupation's associated education variable as a weighted average of the employees. 
For instance, Chief Executives' expected education variable $<edu>_o$ is calculated as $\Sigma_e f_e \cdot edu_e $ where $f_e$ is a fraction of CEO whose education is $e$, and $edu_e$ is a corresponding value of education category, ranging 1 for below high school to 12 for post-doctorate.
For an educational requirement to a skill $s$, $<edu>_s$, we average the skill's education levels of occupations, $<edu>_o$, weighted by the level of skill, $\text{Level}$, that is $\frac{\Sigma_o \; <edu>_o \cdot \: \text{Level}_{o,s} }{\Sigma_o \text{Level}_{e,o}}$.

\subsubsection*{Demographic Distribution of Skills}

Median ages of workers in each occupation are derived from the Current Population Survey (CPS) of the year 2019, and synthetic birth cohorts from individuals born in each year are created from the individuals' survey conducted jointly by the U.S. Census Bureau and the Bureau of Labor Statistics \cite{Flood2022}. Different occupational taxonomies between the two datasets are mapped by the BLS crosswalk.

\noindent\textbf{Synthetic birth cohorts}:
The Current Population Survey (CPS) conducts monthly surveys to obtain a representative sample of the population in each round \cite{Flood2022}.  However, this longitudinal survey does not span over a long period of time, which presents a challenge when attempting to analyze long-term trends. To address this issue, we employ the concept of synthetic cohorts. Synthetic cohorts are constructed by stitching together snapshots of individuals born in the same year across different survey rounds. For example, to create a synthetic cohort for those born in 1970, we first identify people whose birth year was 1970 in the CPS surveys conducted in 1995, 1996, 1997, and so on, up to 2015. We then plot the data for this cohort as if we have been following the individuals born in 1970 throughout their ages, as shown in the inset of Figure 4.

It is important to note that this cohort is referred to as a ``synthetic birth cohort'' because it is not a real cohort in the traditional sense. The individuals surveyed by CPS in each round are different, even though they were all born in the same year. By following individuals born in the same year across multiple survey rounds, we can track changes in the behaviors or characteristics of interest as people age, albeit with different individuals representing the cohort at each point in time.

While synthetic cohorts do not provide the same level of individual-level consistency as true longitudinal studies, they offer a valuable tool for analyzing long-term trends and changes within a specific age group when long-running longitudinal data is not available. This approach allows researchers to leverage the representative nature of the CPS surveys to gain insights into the evolution of various social, economic, and demographic characteristics over time, and thus a common practice across various literature \cite{Acemoglu2011, KambourovGueorgui2013ACNO, Hermo2022, Aeppli2022}.

\noindent\textbf{Demographic analysis}: CPS microdata also include gender and race/ethnicity demographic information. 
We chose four categories, Whites, Blacks, Asians, and Hispanic, as they are the bulk of the sample, and any individuals of Hispanic background are included in that category for Fig.~\ref{fig:Figure 7}. 
To avoid attrition and early retirement, we include only full-time workers employed at the time of the survey, earning at least \$10,000 annually, and between 18 and 55. 
For each demographic category, the average skill level is calculated for their occupational composition.
The microdata records individuals' wages and the number of hours worked. 
We adjust wages for inflation and account for the number of hours worked to compute an adjusted weekly wage, which is readily comparable across the population. 
The race/ethnic disparities in Fig.~\ref{fig:Figure 7} are a ratio of each demographic quantity (general level, nested level, unnested levels, education, and weekly wages) to those of White workers, following \cite{Tong2021} identifying a dominant social group, a social group if it is at least 1.5 times more likely to be employed in the focal occupation. 
Likewise, the gender gap within each race/ethnicity is measured as a ratio of those quantities to those of male workers.
Because we do not have a matched sample, we obtain 95\% confidence intervals by random sub-sampling. In each iteration, we take 10\% of the subpopulation of interest, for instance, Asian male and Asian female workers, and estimate all corresponding measures. 
Repeating this sampling and estimation process in 10,000 iterations, we obtain the distribution for each estimation and derive the 95\% confidential intervals.
The skill, education, and wage estimations of Fig.~\ref{fig:Figure 7} average over the years. Supplementary Figs.~\ref{fig:Temporal Race Gaps - Skills, Education, Wages} and \ref{fig:Temporal Gender Gaps - Skills, Education, Wages} capture temporal patterns of these factors, exhibiting the gaps have narrowed over time.
In addition, SI Figs.~\ref{fig:Skill Age Gender Race Trends} and \ref{fig:Skill Age Gender Race Trends - year effects} show the skill differentials between male and female workers that start around the age of 30 (main Fig.~\ref{fig:age}), manifest across racial and ethnic groups.


\subsubsection*{Skill Compositions in Career Trajectories}
The expected skill levels of each category in the career sequences. 
We studied over 70 million job sequences (8-digit SOC) in 20 million individual resumes from Burning Glass Institute between 2007 and 2020.  
We then calculate the expected skill levels in $i$th job by averaging the skill levels of those occupations appearing in $i$th sequences, shown in Fig \ref{fig:age} (g-h). 
From these sequences of averaged skill levels, we calculate skill level changes in $i$th job transition levels, $\Delta_i$, shown in Fig. \ref{fig:age} (i).  

We exclude job transitions shorter than one year or within an occupation (i.e., moving from one company to another without changing the occupation) for our primary analyses. 
The decision to remove such occupations arises from the oddity we observed in most such jobs. For instance, various janitors or models became CEOs immediately or with overlapping periods.
Nevertheless, our findings are robust to this decision (see SI Sec \ref{supsec: skill dependencies and age} for details). 

To see if the observed trends are truly attributed to career trajectories, we shuffle job history in resumes, bootstrapping the job sequences to produce a benchmark and compare it with the skill changes we empirically observed in career moves in Fig. \ref{fig:age} (i), confirming that the empirically observed trends are unique to the career trajectories.

\subsubsection*{Temporal Evolution of Skill Structure}

We utilize this evolution of skill structure to demonstrate the implication of our constructed nestedness skill structure. 
We choose two sufficiently apart datasets to capture the structural difference, that is, 
version 9.0 in 2005 because it is the first version comparable to the most recent version while offering satisfactory coverage of occupational information (such as education and wage), and version 24.1 in 2019 because it is the most recent version without the potential contamination of irregular patterns due to the pandemic. 
The empirical challenge is that the classification system is continuously updated in response to technological progress, economic transformation, and social reconfiguration \cite{Park2020}.

We created a crosswalk between occupation classifications in 2005 and 2019 that is not immediately available but only between two consecutive years.  
Occupation codes in 2005 are matched to those in 2006, and then those in 2006 to 2009, ... to 2019. Our crosswalk automatically matches 968 occupations in 2019 skill data and 941 unique occupations in 2005 skill data, and the rest are manually matched.
Using these occupations and their skill levels in 2005, we construct the skill structure of 2005 in Fig. ~\ref{fig:historical skill change} (c), using comparable parameters and layouts for both years to make the networks most comparable (see SI). 


\section*{Acknowledgement}
H. Y. and M. H. acknowledge the support of the National Science Foundation Grant Award Number EF-2133863.
The authors are grateful to Yong-Yeol Ahn, Inho Hong, Hyunuk Kim, Balazs Lengyel, Muhammed Yildirim, James McNerney, Morgan Frank, Ljubica Nedelkoska, Christopher Esposito, Ulrich Schetter, Serguei Saavedra, James Evans, and Brian Uzzi for their valuable discussions and feedback.
F.N. gratefully acknowledges financial support from the Austrian Research Agency (FFG), project \#873927 (ESSENCSE).


\printbibliography

\clearpage
\beginsupplement
\fontsize{11}{12}\selectfont
\section*{Supplementary Information: Nested Skills in Labor Ecosystems: A Hidden Dimension of Human Capital}
\thispagestyle{empty}


Section \ref{supsec:skill clustering} offers details on the statistical derivations and robustness checks corresponding to the results on Generality in the main text (Figs.~\ref{fig:determining_k_level}-\ref{fig:skill_level_dist_cor_k=4}), and the resulting skill groups (Tab.~\ref{tab:skill_groups}) introduced in the main Fig.~\ref{fig:Figure 1}~(a-c) and used across the paper.

Section \ref{suppsec:nestedness} expands on the nestedness of occupation-skill networks in part shown in Fig.~\ref{fig:nestedness} and used throughout the paper.
It describes the rationale for (Fig.~\ref{fig:occ_vs_skill_importance_avg_cos} and \ref{fig:occ_skill_nestedness_mat}), the methodology of measuring skill-level contribution to nestedness, alternative measurement of skills' contributions, the results based on different measures (Fig.~\ref{fig:Skill Nestedness Contribution - C}-\ref{fig:Skill Nestedness Contribution - NODF}), and the resulting split of skills based on nestedness we used throughout the paper (Tab.~\ref{tab:skill_split_C_nestedness}).
It also includes an alternative approach to splitting skills based on correlation, which yields consistent results(Tab.~\ref{tab:skill_split_alt}).

Section \ref{supsec:conditional dependencies} articulates the construction of the skill hierarchy of the main Fig.~\ref{fig:Figure 2}~(a and b).
It describes how we derive conditional probabilities between pairs of skills (which is briefly introduced in Fig.~\ref{fig:Figure 1}~(d), the choice of parameters (Figs.~\ref{fig:comparison_skill_deg_before_after_binarization}-\ref{fig:YY z-score distribution}), and sensitivity analysis (Figs.~\ref{fig:a-threshold and z-threshold sensitivity analysis}).
Figs.~\ref{fig:full_figure_2b_labeled} and \ref{fig:figure_2b_labeled} show the full and backboned skill hierarchy network with all skill labels attached.
The section highlights the linkage between our skill hierarchy and a skill co-occurrence network (Fig.~\ref{fig:Network_Skill_Complementarity_RCA_modularity_colored}), and offers two cases based on comparing registered nurses with nurse practitioners (Fig.~\ref{fig:RN vs. NP}), and the skill entrapment of some immigrants (Figs.~\ref{fig:skills_of_different_Hispanics}-\ref{fig:hispanics language skills}), to showcase how the skill hierarchy captures career progress.

Section \ref{supsec: skill dependencies and age} expands on the temporal analyses reported in the main Fig.~\ref{fig:age}. We have explicated the preparation process (Figs.~\ref{fig:BG_skill_change_fig_full}-\ref{fig:BG_fullfig_cleaned} and Tab.~\ref{tab:odd job sequences}) Bootstrapping of the job sequences in resume data (Fig.~\ref{fig:BG_skill_change_single_bootstrap}-\ref{fig:BG_skill_change_bootstraps}), and included the result for all skill cateogires (Fig.~\ref{fig:BG_fullfig_dynamic}).
We also include details about the analysis of median age of workers (Fig.~\ref{fig:occupations' median age and skill - full fig}), and the analysis of synthetic birth cohorts based on CPS (Fig.~\ref{fig:individuals' age and skill - year effects}).
As robustness checks, we also show that the skill development observed in the main Fig.~\ref{fig:age} continues long after education (Fig.~\ref{fig:individuals' age and skill and education}) and also emerges for individuals without college education (Fig.~\ref{fig:individuals' age and skill - no college}).

Section \ref{supsec: add - returns to skill} expands the analyses of the main Fig.~\ref{fig:Wage}, capturing the correlation of occupational wages, educational requirement, and experience with their average levels of each skill category (Fig.~\ref{fig:SI_education_skill_level}-\ref{fig:SI_wage_skill_level}), supplements these results by robustness checks, using alternative measures of skill levels (Fig.~\ref{fig:SI_education_skill_level_top5}-\ref{fig:SI_wage_skill_level_top5}), and regression analyses (Tab.~\ref{tab:wage reg on skill endowment}).
We show the robustness of the main wage findings across major occupational groups (Fig.~\ref{fig:Figure 3 full | major occupation groups}), replicate the main Fig.~\ref{fig:Wage} based on the data of year 2005 (Fig.~\ref{fig:Wage and education 2003}), and finally show the correlation between levels of each skill categories and occupational automation risk \cite{Frey2017} (Fig.~\ref{fig:occ_FOautomation_skill}).

Section \ref{section: add - geographical distribution of skills} offers a descriptive geographic analysis of skill distribution.
We offer evidence that part of the urban wage premiums is explained by the distribution of general and nested skills (Tab.~\ref{tab:urban wage premium}, and Figs.~\ref{fig:geo_dist_level_general_skills_employment_weighted}-\ref{fig:skill_and_manufacturing_full}), but leave an in-depth study of the topic for future work.

Section \ref{section: add - demographic distribution of skills} extends the demographic skill analysis of the main Fig.~\ref{fig:Figure 7}, in Fig.~\ref{fig:racial and gender skill endowments}.
Figs.~\ref{fig:Temporal Race Gaps - Skills, Education, Wages} and \ref{fig:Temporal Gender Gaps - Skills, Education, Wages} capture temporal patterns of these factors, exhibiting the gaps have narrowed over time.
Fig.~\ref{fig:Skill Age Gender Race Trends} depicts that racial/ethnic and gender differentials in skills follow similar age trends observed in the main Fig.~\ref{fig:age}, and are robust to time-variant economic factors (Fig.~\ref{fig:Skill Age Gender Race Trends - year effects}).
In Fig.~\ref{fig:Parenthood_Male_vs_Female}, we highlight the differential influence of parenthood on male and female workers, observed in the diverging growth of general and nested skills in the main Fig.~\ref{fig:age}.

Section \ref{supsec: historical skill change} expands on the changes in occupational skill requirements between 2005 and 2019 (the main Fig. \ref{fig:historical skill change}), and the resulting changes in the skill hierarchy (Figs.~\ref{fig:occupation groups historical changes to skill cluster and type levels}-\ref{fig:Figure_7_supp}). 
The section also provides a brief discussion of the changes in the occupational taxonomy (Fig.~\ref{fig:soc match coverage}).

Section \ref{supsec:robustness checks} offers a battery of robustness checks on whether administrative and managerial occupations (Figs.~\ref{fig:determining_k_kmeans_70bins_correlation_no_manager_start}-\ref{fig:skill_and_age_no_manager} and Tab.~\ref{tab:list of manager occupations}) or social skills (Figs.~\ref{fig:social skills}) derive the increasingly important role of general skills.

\clearpage
\tableofcontents

\clearpage
\pagenumbering{arabic} 
\section{Skill Groups} \label{supsec:skill clustering}

We obtain data-driven categories of skill generality by grouping skills based on their Level Distributions. We employ a $k$-means clustering algorithm (see Fig.~\ref{fig:Figure 1} in the main) and supplement the results with two more measures of skill generality (the average skill level and occupation counts), explained later in this and the next sections. Here, we discuss clustering skills based on their distribution shapes, as Fig. 1 shows in the main text.

\begin{figure*}[!h]
    \centering
    \includegraphics[width=\textwidth]{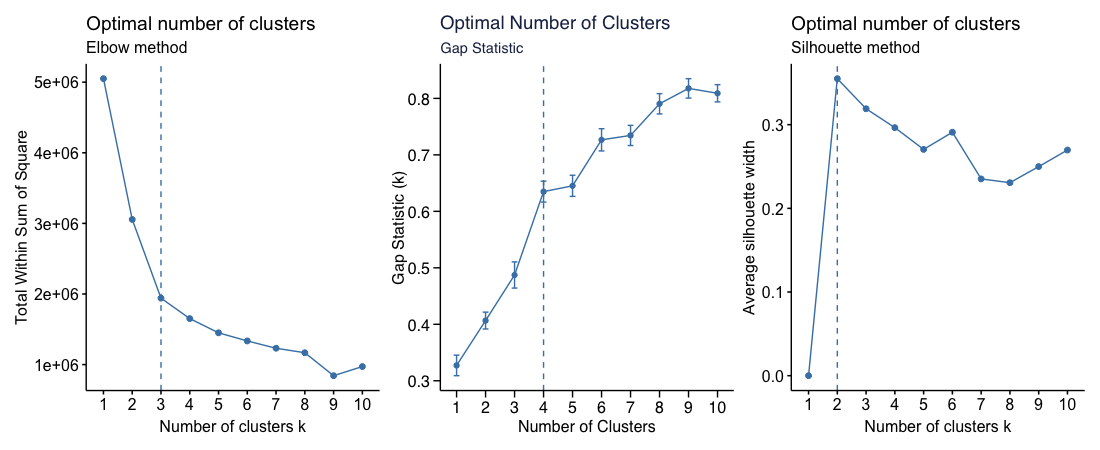}
    \caption{\textbf{Statistical tests to determine the optimal $k$ for $k$-mean clustering algorithms}. The figure shows the results of Elbow method, Gap statistic, and Silhouette analysis}
    \label{fig:determining_k_level}
\end{figure*}

The $k$-means clustering algorithm requires two inputs, a distance metric and the number of clusters, $k$. We choose the correlation distance (as in equation 1) for the former and $k=3$ for the latter. We did not use Euclidean distance because it does not differentiate the shape distributions as inputs compared to correlation. 
Second, we choose $k=3$ because it seems to be in the range of optimal numbers (2-4) from various statistical tests shown in Fig.~\ref{fig:determining_k_level}. Finally, we provide two alternative categories of skill generality, which are consistent with the results of the $k$-means clustering.

To measure the correlation similarity among the distribution shapes, we binned the distribution with intervals of 0.1. For instance, the skill level ranges from 0 to 7, resulting in a vector of 35 entries, each corresponding to bins of [0,0.10), [0.10, 0.20),... 
Correlation similarities are measured across these vectors. 
Table S1 shows the assignment of skills resulting from $k$-means clustering (based on correlation similarity and $k=3$) used in the main text. 

\begin{equation}
    d = 1 - \frac{\Sigma_i x_i y_i - \frac{1}{n}\Sigma_i x_i \Sigma_i y_i}
    {\sqrt{\Sigma_i x_i^2 - \frac{1}{n}(\Sigma_i x_i})^2 \sqrt{\Sigma_i y_i^2 - \frac{1}{n}(\Sigma_i y_i})^2}
\end{equation}

\begin{figure*}[!h]
    \centering
    \includegraphics[width=\textwidth]{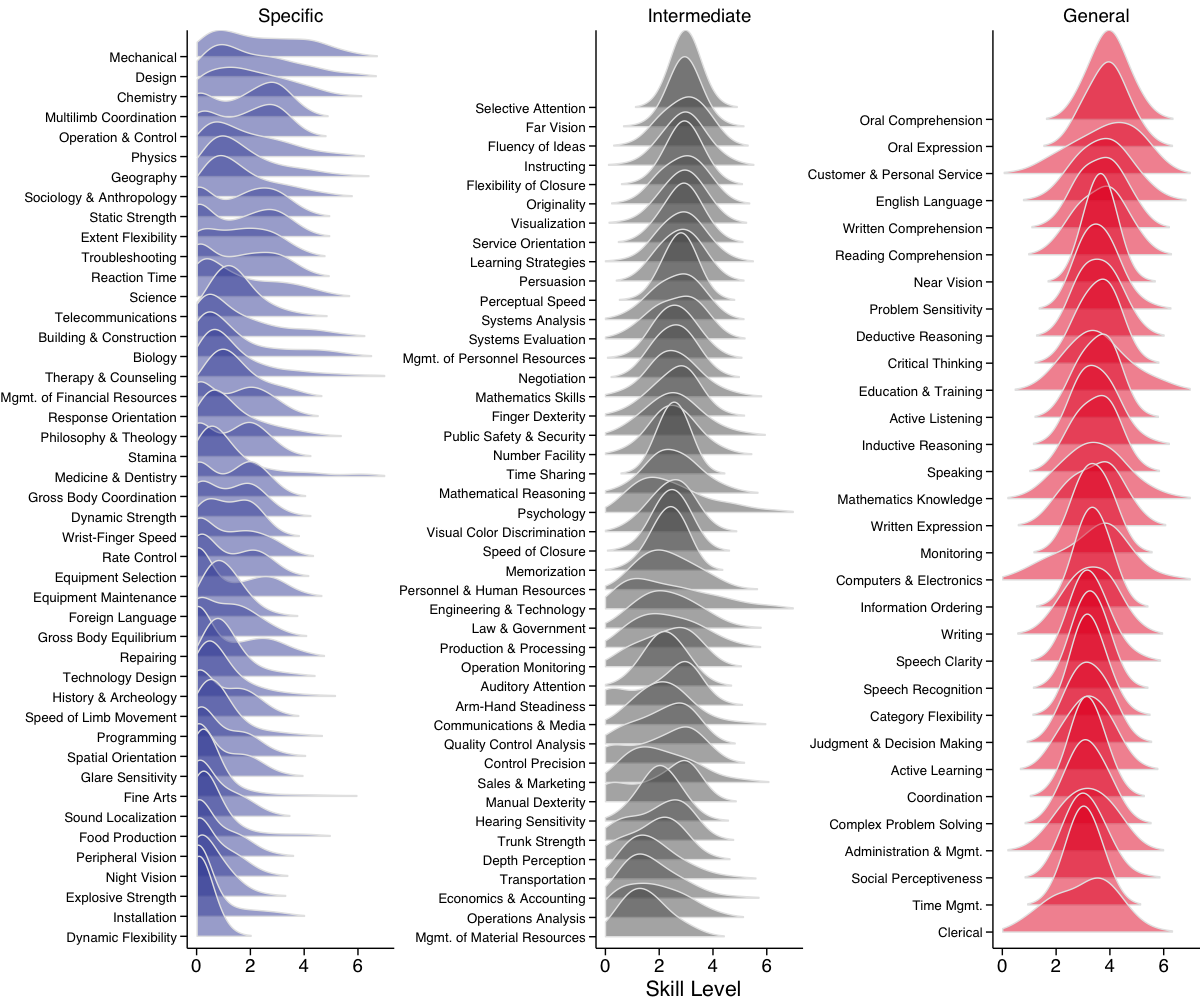}
    \caption{\textbf{Skill Level Distribution with $k = 3$.} Skills are in descending order of generality. The depicted distribution of skills is used in the main text.}
    \label{fig:skill_level_dist_cor_k=3}
\end{figure*}

\begin{figure*}[!h]
    \centering
    \includegraphics[width=.9\textwidth]{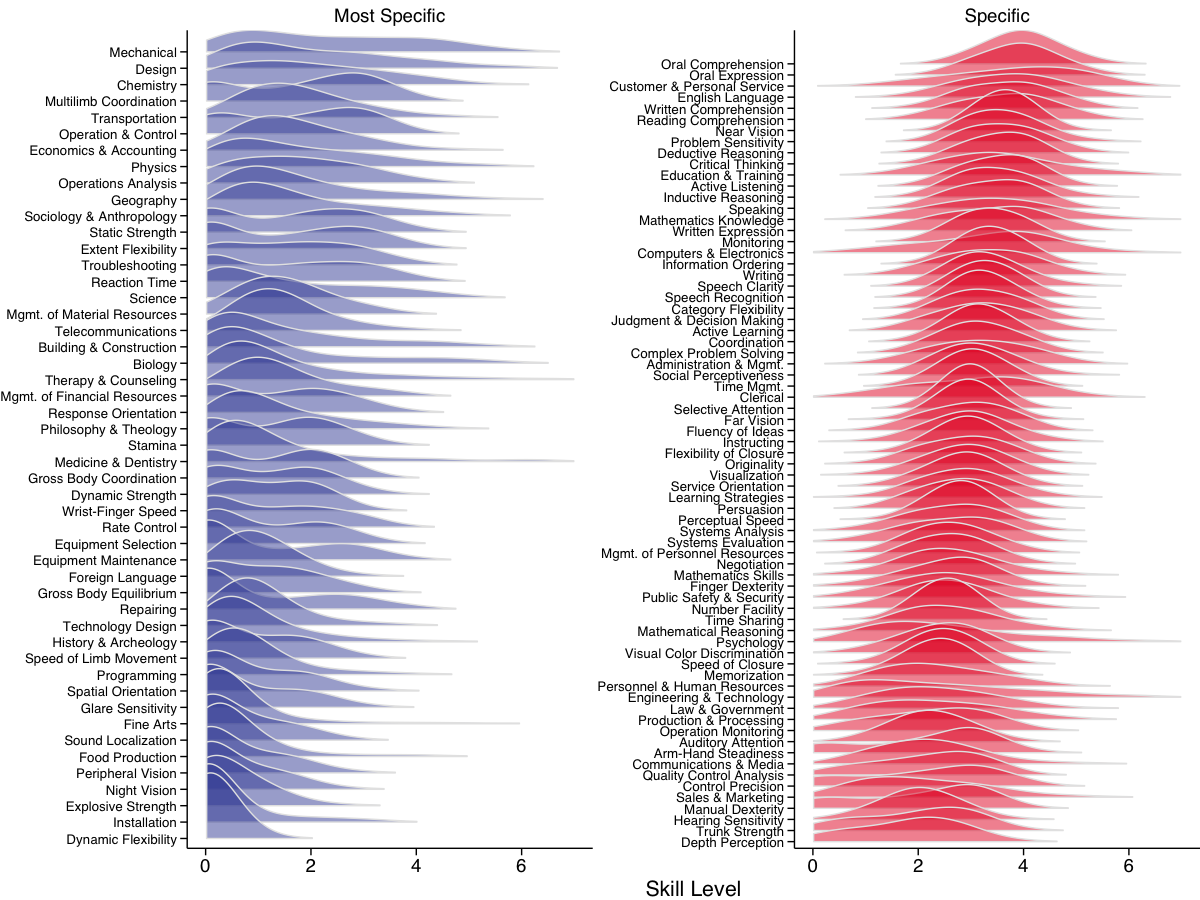}
    \caption{\textbf{Skill Level Distribution with $k = 2$.} Skills are in descending order of generality.}
    \label{fig:skill_level_dist_cor_k=2}
\end{figure*}

\begin{figure*}[!h]
    \centering
    \includegraphics[width=.9\textwidth]{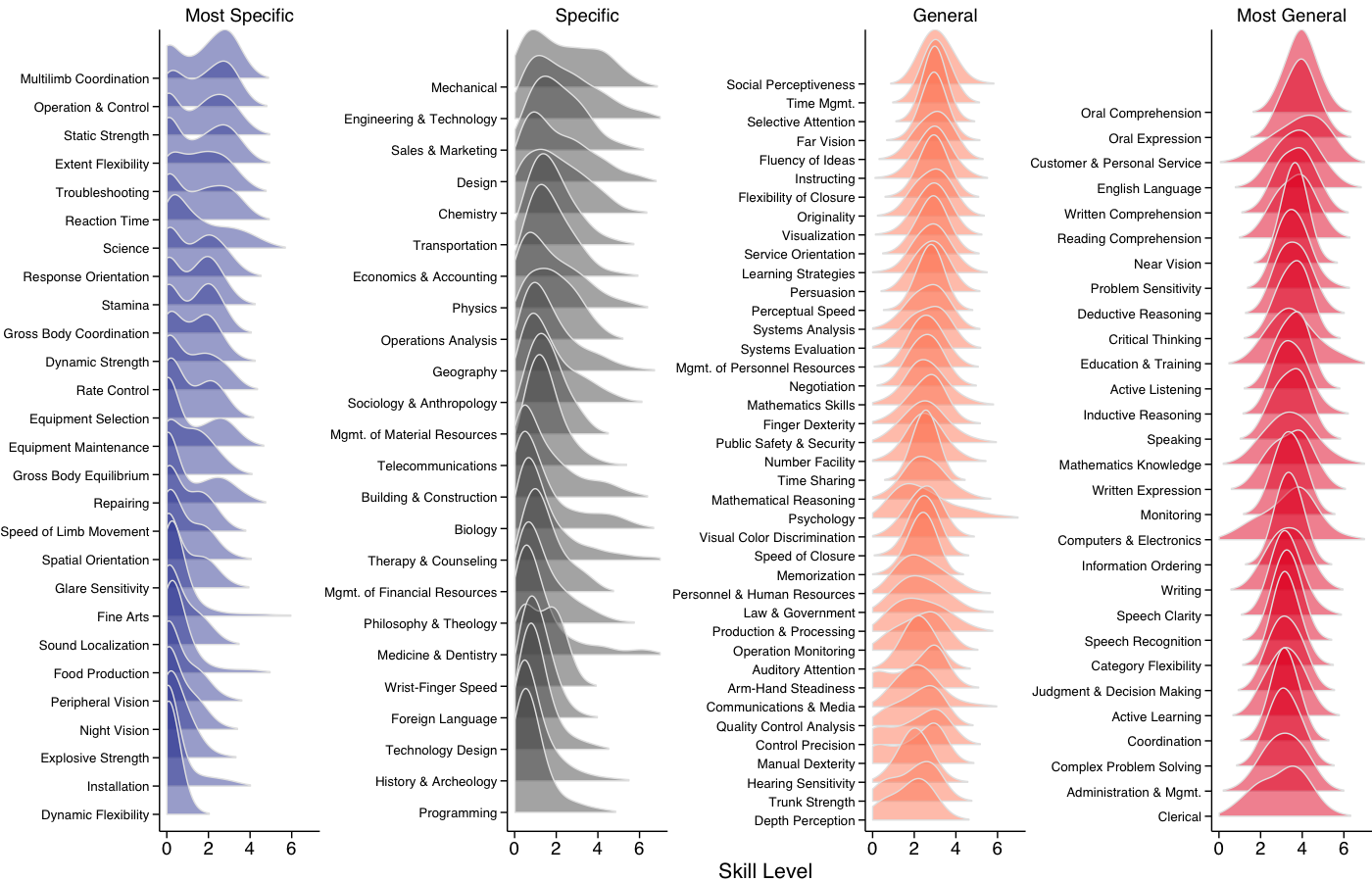}
    \caption{\textbf{Skill Level Distribution with $k = 4$.} Skills are in descending order of generality.}
    \label{fig:skill_level_dist_cor_k=4}
\end{figure*}

We use three statistical tests to determine the optimal $k$. These include the elbow method, silhouette analysis, and gap statistics, as shown in Fig.~\ref{fig:determining_k_level}. These results suggest optimal numbers from 2 to 4. We provide the clusters resulting from each choice of $k$ in Figs.~\ref{fig:skill_level_dist_cor_k=3}-\ref{fig:skill_level_dist_cor_k=4}.

The conventional \textit{Elbow method} calculates the within-cluster sum of squares for different numbers of clusters $k$ in order to find a sharp decline from one $k$ to another followed by a more gradual decrease in slope, where we find $k=3$ is the best. 
\textit{Silhouette analysis} \cite{ROUSSEEUW198753} measures the similarity of each observation with the cluster to which it is assigned, producing a metric that ranges from -1 (dissimilar) to 1 (similar). In determining the optimal $k$, one looks for the value at which the average (silhouette width) is maximum, providing $k=2$ for the optimal number. 
The \textit{Gap statistic} \cite{Tibshirani2001} compares the total intracluster variation for different $k$ with their expected values under a null model (i.e., a distribution with no obvious clustering, generated using (1,000 iterations of) Monte Carlo simulations of the sampling process,) wherein maximal intracluster variation is desired, providing $k=4$ for the optimal number.
To determine the optimal number of clusters, $k$, based on Gap Statistic, we used the criterion proposed by \cite{Tibshirani2001}, wherein the smallest $k$ such that the \textit{change} in intracluster variation, $f$, is smaller than an (error-adjusted) standard deviation, $s$ of the null model ($f(k+1) - f(k) \geq s_{k+1}$).

Figures \ref{fig:skill_level_dist_cor_k=3}-\ref{fig:skill_level_dist_cor_k=4} show individual skills within categories.
The number of groups does not change the order of generality of skills, which is central to our analysis. The context of our study encourages a focus on the most and the least general skills because those epitomize two skill categories of broad theoretical interest: general skills and specialized skills. Therefore, it is a practical choice for us to start with three clusters, focus primarily on the two extremes, and subject the skills in the remaining cluster to secondary examination.
Given the visual shapes of distributions and the semantic benefit of differentiating the most general and moderately general skills (so-called intermediate skills), we continue using $k=3$ in the main text.
To reduce any inherent noise due to the skills between general and specifics, we choose $k=3$ and focus on general and specifics.
Table \ref{tab:skill_groups} shows the resulting split and offers some supporting statistics.

\clearpage
\footnotesize
\input{Nature_HB_2023/tabsNHB/Jul_19_2023_-_Skill_Group_Assignments.tex}
\normalsize

\normalsize
\clearpage
\section{Skill Nestedness} \label{suppsec:nestedness}

\subsection{Nested Modular structure in Skills and Occupations}

\begin{figure*}[!h]
    \centering
    \includegraphics[width=0.7\textwidth]{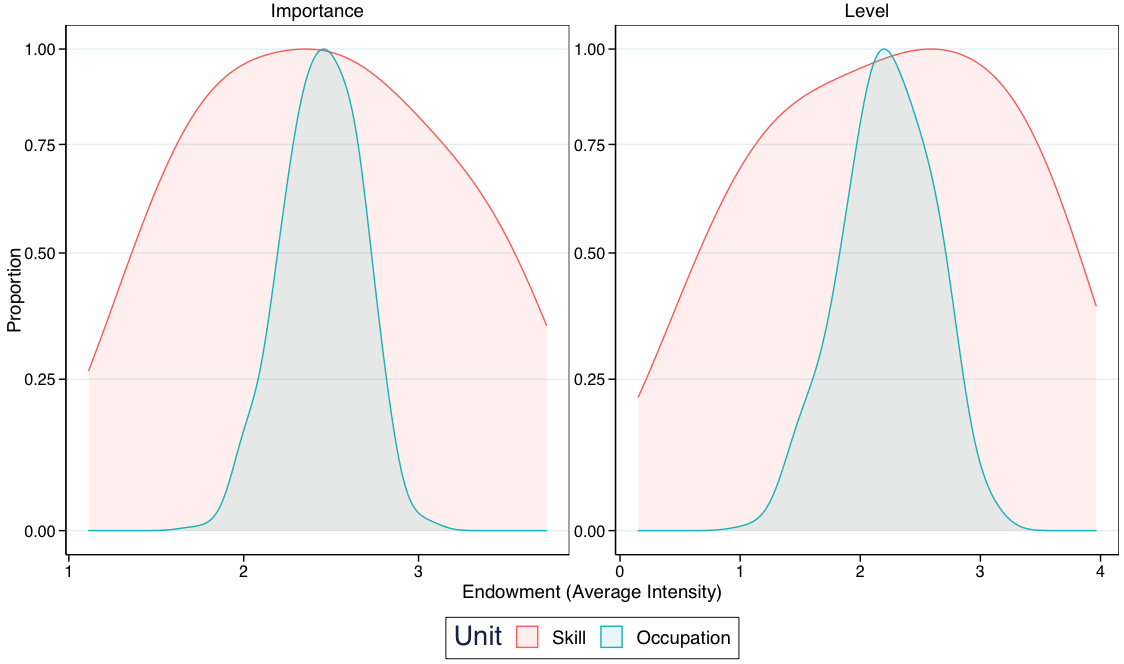}
    \caption{\textbf{Scaled Density Distribution of Skill and Occupation Endowments.} Endowment for skills and occupations is calculated by averaging the intensity values for each column and row, respectively. We compute endowment using both intensity measures of Importance and Level. The results contrast for endowment distribution of skills from occupations. Unlike skills, occupations show closer average Importance values. This finding implies occupation's attention is constrained. Hence, they must allocate their limited attention to skills.}
    \label{fig:occ_vs_skill_importance_avg_cos}
\end{figure*}

\begin{figure*}[!h]
    \centering
    \includegraphics[width=.5\textwidth]{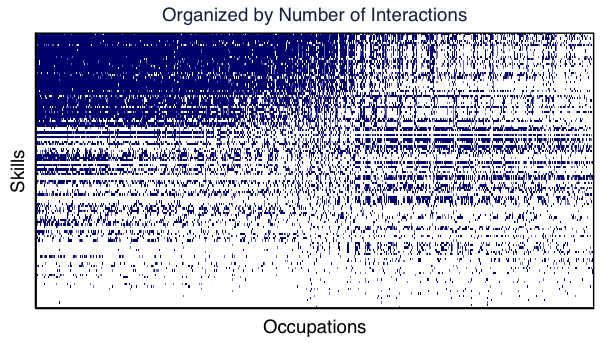}
    \caption{ \textbf{The skill-occupation matrix} The occupations and skills of the matrix are arranged in descending order of their marginal fills (along the x and y axes, respectively.)}
    \label{fig:occ_skill_nestedness_mat}
\end{figure*}

As the scope of knowledge expands, the need for specialization grows.
Unlike findings of the economic complexity about the nested landscape of national, regional, and urban capabilities \cite{Hidalgo2009, Bustos2022}, occupations often bundle few skills and therefore encompass much narrower knowledge domains.
While the main focus of our paper is revealing the underlying structure of workplace skills, noting the distinction is vital.
Here, we empirically offer evidence of the difference between occupation and skill scopes in two ways before discussing nestedness in the skill space.
We find higher variation among skills (in the number of occupations that demand a skill) than among occupations (in the number of skills an occupation demands), as seen in Fig.~\ref{fig:occ_vs_skill_importance_avg_cos}), suggesting a non-trivial nested structure (Fig.~\ref{fig:occ_skill_nestedness_mat}). 

Figure \ref{fig:occ_vs_skill_importance_avg_cos} shows the Level Distribution of skills (red) and the distribution of the total skill amounts in occupations (blue). To obtain the skill Level Distribution, one measures the demand for each skill and makes a distribution. For example, how much English skills are needed for the entire labor market or how much Physics skills are needed across occupations. The former is more broadly used (i.e., general) and therefore has a higher demand than the latter. A skill's demand is calculated by summing skill levels/importances in the occupations (red). 
Similarly, by adding the total levels/importance of each occupation, one obtains occupations' skill endowments, the total level of skills needed to undertake the job's tasks.

Figure \ref{fig:occ_vs_skill_importance_avg_cos} shows, unlike broad skill generality, occupation's endowment is narrowly distributed. This narrow distribution indicates that the total amount of skills needed for an occupation is not much different from each other, regardless of how much they are paid and how advanced education is needed. We attribute occupations' limited scope of skills to the limited scope or attention that individual workers can offer. There is only so much a single person can equip and do. Thus, individuals' capacity restricts how many skills occupations can bundle. This constraint explains the process of specializations needed for a complex job. In contrast to occupations, skills do not have such constraints. While some skills are niche, general skills epitomize expertise of widespread demand, as they are needed in most occupations.

This stark difference in the scope of occupations and skills requires quantifying the nestedness structure of the skill-occupation matrix differently. In ecological terms, there is no site/area/biome (occupation) that is large enough to nest other sites (occupations), whereas there are species (skills) that can nest other species (skills) as they can appear anywhere. 
This explains why Fig.~\ref{fig:occ_skill_nestedness_mat} shows the noisy nested structure in the skill-occupation matrix, far from the perfect nested triangle. The skill-occupation structure allows mutually exclusive presences, possibly due to competition between skills within an occupation. 

We construct and measure a nested structure of a skill-occupation matrix in Fig.~\ref{fig:occ_skill_nestedness_mat}. The original skill-occupation matrix's entry is a continuous variable (indication of the degree or point along a continuum to which a particular descriptor is required or needed to perform the occupation). But most conventional nestedness analyses require binary entries, and thus, we employ a disparity filter to make the matrix binary entries of statistically significant presences (see Sec.~\ref{supsec:skill-occ}). 
We then sort the matrix entries in descending orders of marginal totals \cite{Bascompte2003}. 
As Fig.~\ref{fig:occ_skill_nestedness_mat} shows, the result deviates from the perfect nested structure as an upper-left triangle. Nevertheless, the upper left is highly populated, indicating a nested structure. This imperfect nested structure may account for the constraints on occupations (limited carrying capacity), introducing severe competition between skill species. This constraint distinguishes the nestedness of extensive economies of nations, regions, and urban areas from occupation's nestedness, for which specializations dominate the evolution more than diversification. 
As a result, the skill-occupation matrix is expected to be modular as well as nested with mutually exclusive modules. \textit{Nested-modular matrix} is a complicated structure and will be beyond our current scope \cite{Fortuna2010}. Here, we will focus on individual skills' contributions to the nested structure and differentiates skills that contribute to the nested structure from those that do not. 

\subsection{Skill's contribution to Nestedness}

The above evidence reveals that the landscape of human capital is partially nested due to matching increasing complexity with specialization (perhaps, niche constructions), resulting in a nested-modular structure. 
We speculate that increasing complexity mainly generates nested structure, and specialization mainly generates modular structure. We think mathematical modeling of a labor ecosystem according to this insight can be extremely interesting, and we leave this for future work while we focus on empirical observations and quantifications for now.

Due to the structural complications, described above, conventional approaches for quantifying nested structure (sorting the matrix to observe an upper triangle or calculating presences/absences accounting for a well-defined nested structure) are likely imperfect.
Instead, we look for skills' individual contributions to the current nested structure compared to their counterfactual contributions under a null model. 
For instance, \cite{Saavedra2011} proposes such an approach based on the idea of randomizing edges for a focal skill and comparing the nestedness in the simulated network with the observed value in the system. 
In our case, we create counterfactual worlds as if a focal skill can appear equally likely in any occupation. This equally probable null hypothesis randomly chooses a focal skill's occupations (edges) without considering education, domain knowledge, industrial requirements, or historical contingency, imposing the current socio-economic structure. 
Then, we measure an increase/decrease in nestedness by destroying the current imposition.  
For simplicity, this method is only available for a presence/absence bipartite network \cite{Bastolla2009, Almeida-neto2008}. We use the disparity filter \cite{Serrano6483} because the method preserves degree heterogeneity, which is crucial to distinguishing general from niche skills.
We explain this method in more detail in the supplementary section \ref{supsec:conditional dependencies}.

We use three commonly used metrics of nestedness (checkerboard score, Temperature, and NODF) to quantify nestedness $N$ at the level of the skill-occupation matrix.
Checkerboard score measures the deviation from nestedness as checkerboard appearance of fills \cite{Stone1990}. This score is consistent with the well-known nestedness index, $N_c$, counting the number of times that a species' presence at a site correctly predicts its presence at richer sites and sums these counts across species and sites \cite{write1992}. 
The presence of a checkerboard, an empty site when the nested site predicts the fill, decreases the nestedness. 
Temperature measures as the total number of ``surprises'' on the assumption of a perfectly nested matrix as temperature increases thermal noises to destroy perfect structure \cite{Atmar1993}. Although this is a great measure, this index has its underlying assumption that the system is actually following the mechanism for a perfect nested structure if there is no temperature. 
NODF quantifies nested overlaps, the notion that all species in a poor habitat are present in richer habitats, and decreasing fill (marginal totals of interactions between habitats and species) \cite{Almeida-neto2008}.

Now that we identify the null hypothesis to generate a focal skill's counterfactuals and nestedness indexes let's calculate skills' contributions to nestedness, $c_s$. 
For each skill, we run at least 1,000 simulations, wherein, the focal skills' ties to occupations are randomly shuffled, keeping the number of ties constant. Therefore, all ties of the focal skill (meaning the skill's generality) are preserved. Then, we measured a nestedness index of the generated matrix mentioned above, as $N^{\ast}$.
We quantify a skill $s$'s contribution as:

\begin{equation} \label{eq:nestedness contribution}
    c_s = \frac{N - <N^{\ast}_s>}{\sigma_{N^{\ast}_s}}
\end{equation}

$<N^{\ast}_s>$ and $\sigma_{N^{\ast}_s}$ denote the mean and standard deviation of the nestedness of the simulated matrix, in which skill $s$'s edges were randomized.

\begin{figure*}[!h]
    \centering
    \includegraphics[width=\textwidth]{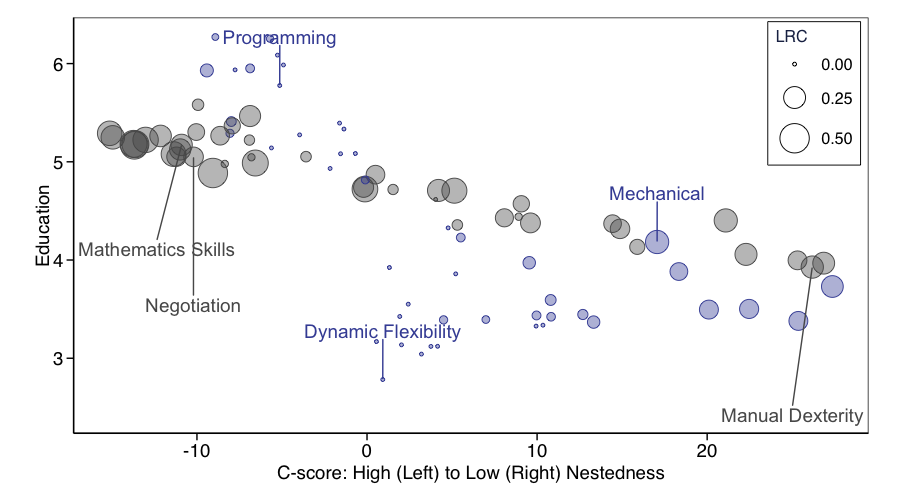}
    \caption{\textbf{Nestedness Contribution of Skills based on checkerboard} \cite{Stone1990}.}
    \label{fig:Skill Nestedness Contribution - C}
\end{figure*}

\begin{figure*}[!h]
    \centering
    \includegraphics[width=\textwidth]{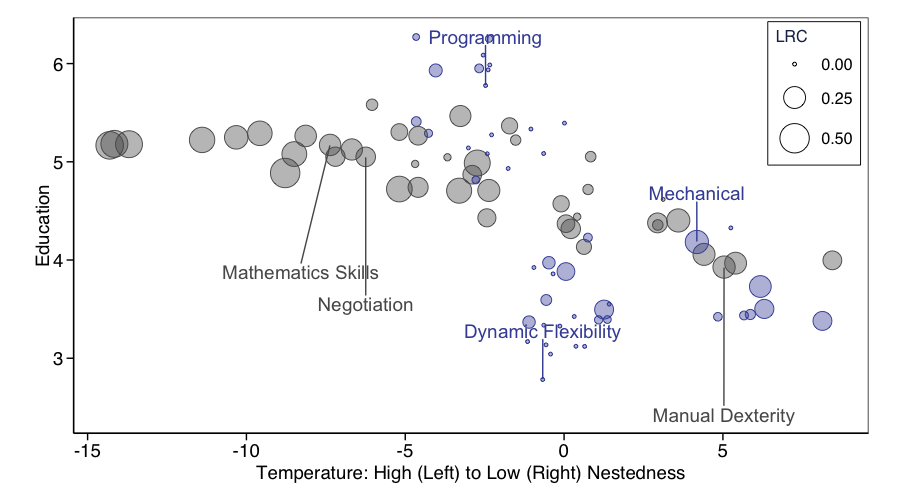}
    \caption{\textbf{Nestedness Contribution of Skills based on Temperature} \cite{Atmar1993}.}
    \label{fig:Skill Nestedness Contribution - BINMATNEST}
\end{figure*}

\begin{figure*}[!h]
    \centering
    \includegraphics[width=\textwidth]{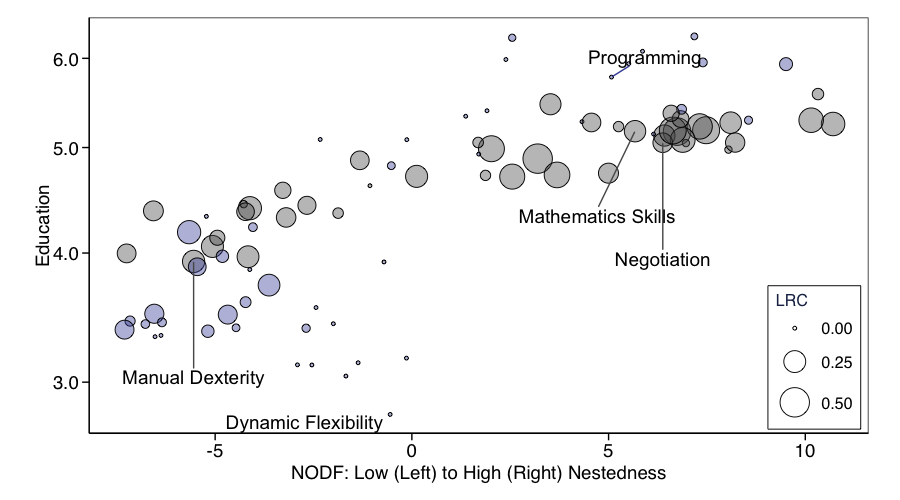}
    \caption{\textbf{Nestedness Contribution of Skills based on NODF}\cite{Almeida-neto2008}. Extreme values arise from large denominators for some of the skills.}
    \label{fig:Skill Nestedness Contribution - NODF}
\end{figure*}

\subsection{Nested and Un-nested Skills}
Figs \ref{fig:Skill Nestedness Contribution - C}, \ref{fig:Skill Nestedness Contribution - BINMATNEST}, and \ref{fig:Skill Nestedness Contribution - NODF} show the nestedness contribution of skills using checkerboard score, Temperature, and NODF, respectively.
We are particularly interested in examining the relationship between nestedness contribution and the position of skills in our hierarchy whose vertical position is a local reaching centrality and the horizontal position is education attainment. 
In addition, we would like to compare skills at the same generality level to avoid comparing apples to oranges. 
For example, it is not fair to compare general skills to specific skills as they have more edges. 
Given that general skills mass at the root of those dense dependency webs, we show more closely examine the nestedness contribution of intermediate and specific skills, and use the measurement to split them into categories of nested and un-nested. 

Table \ref{tab:skill_split_C_nestedness} shows the resulting split of skills into categories or subtypes based on the checkerboard score \cite{Stone1990} also shown in Fig.~\ref{fig:Skill Nestedness Contribution - C}.
To be clear, we refer to the result of our skill clustering based on generality \textit{skill clusters} (general, intermediate, and specific) and refer to the further split made based on nestedness \textit{skill categories} or \textit{skill subtypes} (general, nested intermediate, nested specific, un-nested intermediate, un-nested specific.). 

\footnotesize
\input{Nature_HB_2023/tabsNHB/Jul_15_2023_Skill_Split_C-score_Nestedness_Simul_Jan_2023.tex}
\normalsize

\subsection{Alternative Approach for Deriving Skill Categories} \label{supsec:alt splitting skills}
We split the skills of each cluster (general, intermediate, and specific) based on their correspondence with general skills.
We measure such correspondence $C$ by calculating the correlation between the importance of given skill $i$, and the importance of each of the general skills $j$:

\begin{equation}
    C^{<Level>}_{i,j\in <\text{general}>} = \frac{\Sigma_o(Level_{i,o}- \mu_{Level_{i}}) (Level_{j,o}- \mu_{Level_{j}})}{\sqrt{\Sigma_o(Level_{i,o}- \mu_{Level_{i}})^2 \Sigma_o(Level_{j,o}- \mu_{Level_{j}})^2}}
\end{equation}

Aggregating values of $C_{i,j}$ over general skills $j$, we obtain a measure of correspondence between skill $i$ and the set of general skills, $C_{i,<general>}$.
Then, we compare skill $i$ to other skills $l$ in the same cluster $k$ to which $i$ belongs— given our assignment from supplementary section \ref{supsec:skill clustering}.
To do so, for skills $l$ of cluster $k \in \{\text{intermediate}, \text{specific} \}$, we calculate the mean correlation to general skills:

\begin{equation}
    C^{<mean>}_{k} = mean_{l \in k} {C_{i,<general>}}
\end{equation}

Finally, we suggest a skill $i$ of cluster $k \in {Specific, Intermediate}$ is nested if it depends on general skills above the mean level and call it '\textit{nested}' if $C_{i,<general>} \geq C^{<mean>}_{k}$, and suggest it is independent of general skills and call it '\textit{un-nested}', otherwise.
Table \ref{tab:skill_split_alt} shows the resulting assignment of skills based on this approach.
One obtains a similar split of skills if the Importance measure instead of Level is used.

\clearpage
\footnotesize
\input{Nature_HB_2023/tabsNHB/Jul_15_2023-_Alternative_Skill_Split-_Avg_Correlation_with_General_Skill.tex}
\normalsize

\clearpage
\section{Conditional Skill Dependencies} \label{supsec:conditional dependencies}

To obtain the skill structure, as seen in main Fig.~\ref{fig:Figure 2}, we extract conditional probabilities of the appearance of a skill $u$, given the appearance of another, $v$, in the skill-occupation matrix, which was used for nested structure in the previous section \cite{Jo2020}.

\subsection{Skills-Occupation Matrix} \label{supsec:skill-occ}
The original skill-occupation matrix's entry is a continuous variable (indication of the degree, or point along a continuum, to which a particular descriptor is required or needed to perform the occupation). But most conventional nestedness analyses, used in section \ref{suppsec:nestedness}, and conditional probability measures for main Fig.~\ref{fig:Figure 2}, require binary entries.
Thus, we employ a disparity filter to make the matrix binary entries of statistically significant presences \cite{Serrano6483}. 

\begin{figure*}[!h]
    \centering
    \includegraphics[width=\textwidth]{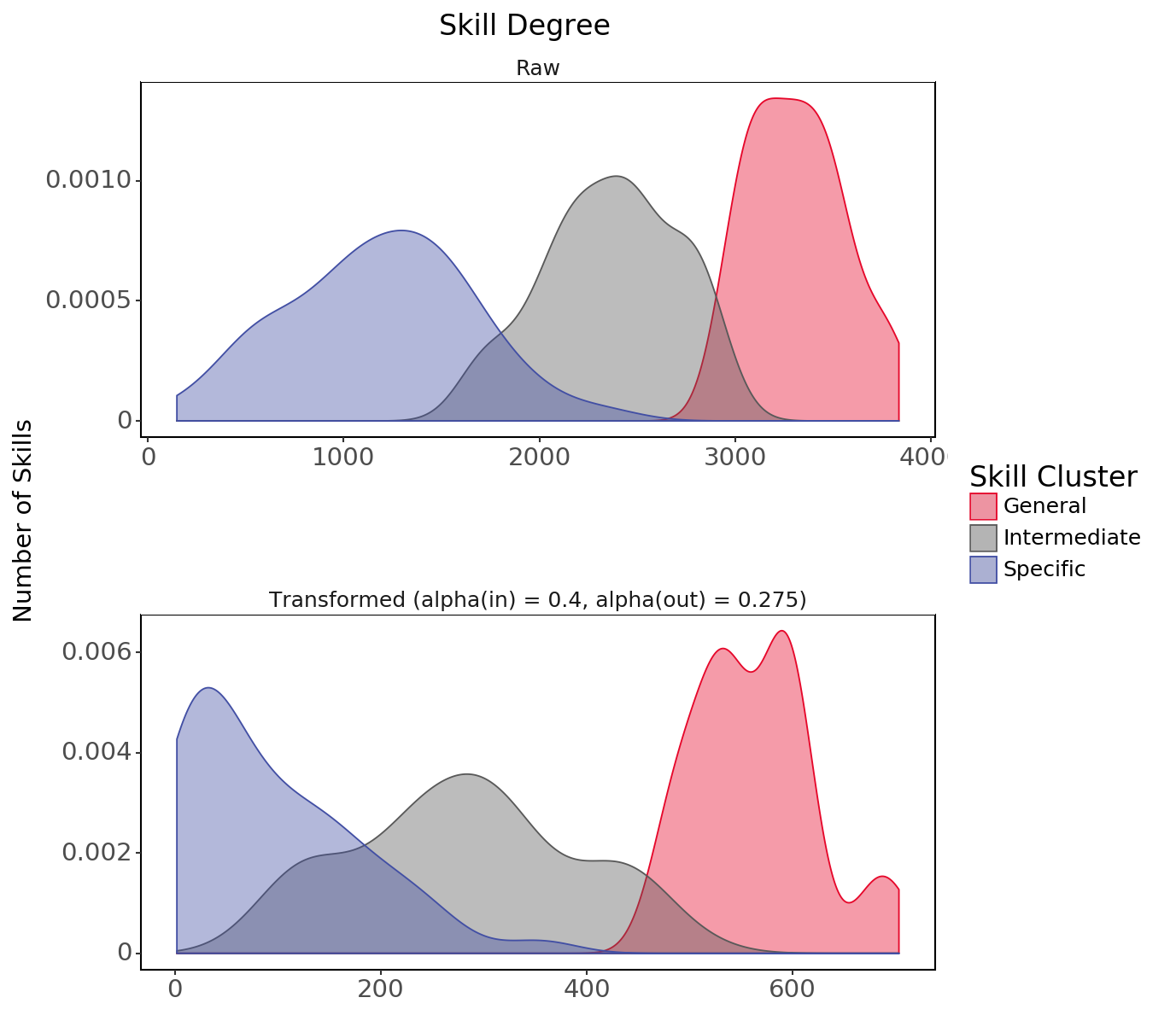}
    \caption{\textbf{Skill Degrees as Validity Check on Choosing Parameters of Obtaining Skill-occupation Network Backbone}. The figure compares the degree distribution of skills in each skill group before and after the transformation. Our emphasis is on the distinction between the distribution of three types of skills (their overlap) and their relative position to the raw data. Indeed, the Pearson correlation between skills' strengths (sum of edge weights) and their transformed degree is 0.95. The ranking of skills across these two measures is also preserved (correlation is 0.97).}
    \label{fig:comparison_skill_deg_before_after_binarization}
\end{figure*}

\begin{figure*}[!h]
    \centering
    \includegraphics[width=\textwidth]{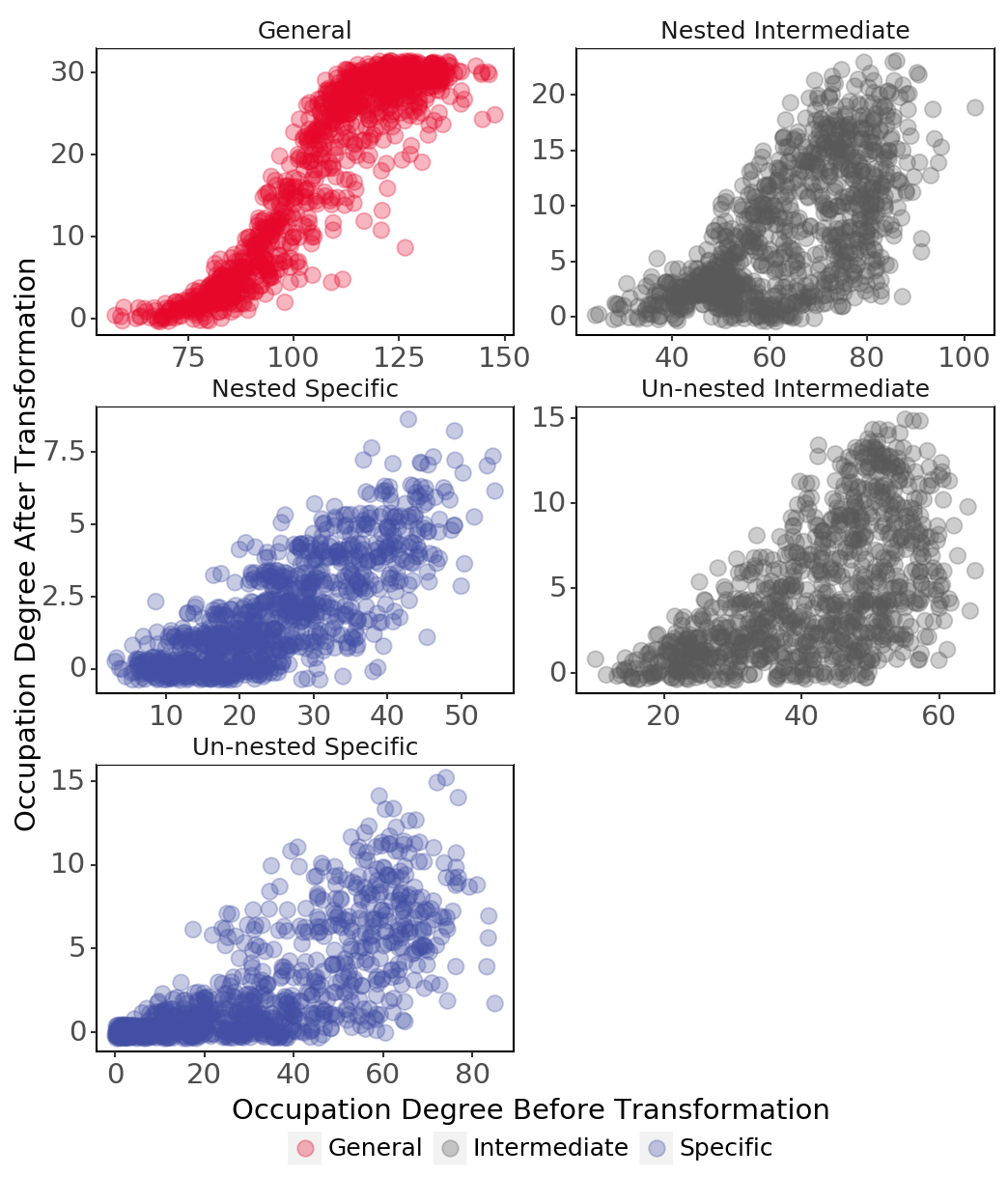}
    \caption{\textbf{Occupational Degrees as Validity Check on Choosing Parameters of Obtaining Skill-occupation Network Backbone}. The figures compare the degree distribution of occupations before and after the transformation. Our emphasis is faithfulness across each of the five skill subtypes.}
    \label{fig:comparison_occ_deg_before_after_binarization}
\end{figure*}

We chose this method for two reasons.
First, it allows the user to set different restrictions on the skill and occupation sides of the bipartite network. This feature is desirable given the differences in the strength and degree distributions of occupations and skills.
Second, it accommodates heterogeneous degree distribution, which we know is a key characteristic of our skill side.
In choosing the parameters, we ensured the resulting binary network satisfied the following conditions. First, the filtered network has to remain faithful to the skill and occupation degree distributions (macro-level features). We show that the filter indeed kept the distribution shapes in Fig.~\ref{fig:comparison_skill_deg_before_after_binarization} for the skills strength distribution and Fig.~\ref{fig:comparison_occ_deg_before_after_binarization} for occupations' strength correlations.
The Pearson correlation between skills' strengths (sum of edge weights) and their transformed degree is 0.95. The ranking of skills across these two measures is also preserved (correlation is 0.97).
Note that we used comparisons between node strengths and node degrees because of our idiosyncratic empirical data structure. Each occupation includes a survey for every skill, resulting in every occupation having every skill entry with numbers ranging from 0 to 7. 
The surviving skills preserve the distribution and ranking of occupations (correlations between occupations' strength and ranking before and after transformation are 0.79, and 0.79, respectively).

In the end, the parameter pair ($\alpha_{in} = 0.4$, $\alpha_{out} = 0.275$) results in 33,865 (29\%) edges. 
We also conducted validity checks on the choice of parameters by examining the sampled results (5\% of occupations).
The test compares the survived and eliminated skills to common sense. For example, have the links between ``Surgeon" and the skill ``Medicine and Dentistry", and ``Programmer" and skill ``Programming" survived? Conversely, has the link between ``Mathematician" and "Explosive Strength", defined as \textit{The ability to use short bursts of muscle force to propel oneself (as in jumping or sprinting), or to throw an object}, been eliminated?
The goal of this exercise is to ensure the parameters are not set too strictly or too lenient, and that the retained information in ties conforms to expectations.

\subsection{Skill Dependency from Conditional Probabilities} 
We first account for the significant conditional appearances, and discount for noise from independent co-appearances (of two skills in occupation) by random chance with a z-score threshold, $z_{th}$ \cite{Jo2020}. 
That is, we account only for those skills that appear together more than randomly expected by $z_{th}$ magnitude. 
Here, $z_{th}$ is a threshold for the extent to which we eliminate chance from two skills appearing in the same occupation.
\begin{equation} \label{eq: YY z-score}
    z_{u,v} = \frac{N(u,v) - \mu}{\sigma} > z_{th}
\end{equation}

Where $\sigma^2 = \frac{N(u).N(v)}{\|O\|} \frac{{\|O\|}- N(v)}{\|O\|} \frac{{\|O\|}- N(u)}{\|O\| - 1 }$ and $\mu = \frac{N(u).N(v)}{\|O\|}$, are the standard deviation and mean of a hypergeometric distribution for the expected co-occurrence of skills (that arise the under the null model of a bipartite configuration model that preserves skill degrees \cite{Jo2020}.) $N(u)$ and $N(v)$ denote the number of occupations that demand skill $u$ and $v$, respectively and $\|O\|$ denotes the total number of occupations.
 
We now estimate conditional probabilities $P(u|v)$ and $P(v|u)$ and assign a direction to them.
The direction, $u \rightarrow v$, is determined when $P(u|v)$ is \textit{substantially} greater than $P(v|u)$. 
Once again, we wouldn't consider every $P(u|v)$ that is insignificantly greater (smaller) than $P(v|u)$, but only those that are sufficiently greater (smaller) to be considered as a \textit{dependent} structure. $\alpha_{th}$ sets the minimum difference between two conditional probabilities so that they are considered to have directional dependence. 
This threshold has to be differentially applied to each skill pair due to the heterogeneous skill node degrees.  
Therefore, the threshold $\alpha_{th}$ is weighted by $(\frac{k_{max}}{min(k_u, k_v)})$ to be applied to filter $[P(u|v)-P(v|u)] \neq 0 $ 

\begin{equation} \label{eq: a_th}
     |P(u|v)-P(v|u)| \: > \: (\frac{k_{max}}{min(k_u, k_v)}) \times \alpha_{th}
\end{equation}
Where $k$ denotes the number of other skills with ties to the focal skill, and $k_{max}$ denotes the biggest degree observed among skills.

The magnitude of the dependence between $u$ and $v$, $w_{u \rightarrow v}$, follows the parametric function introduced by \cite{Jo2020}:

\begin{equation} \label{eq: dependency weight}
    w_{u \rightarrow v} = \frac{min(k_u, k_v)}{k_{max}}\Bigl( \frac{N(u,v)}{N(v)} - \frac{N(u,v)}{N(u)} \Bigr)
\end{equation}

In simple terms, the direction of arrows shows whether by observing skill $v$ in an occupation, it is (more) likely also to observe skill $u$ (than the other way around).
The magnitude of dependence, used as weights in the main Fig.~\ref{fig:Figure 2}, is a parameteric function of the difference between the conditional probabilities of observing $u$ and $v$, and the null model that corresponds to the estimated number of shared occupations between them, given the degrees of $u$ and $v$.
The final network is shown in Fig.~\ref{fig:full_figure_2b_labeled} and used across all analysis, but for the main Fig.~\ref{fig:Figure 2} and \ref{fig:historical skill change} that depict the parsimonious versions, from a directed acyclic graph (DAG) \cite{Jo2020}. 

\subsubsection*{Choice of Parameters}
\begin{figure*}[!h]
    \centering
    \includegraphics[width=0.5\textwidth]{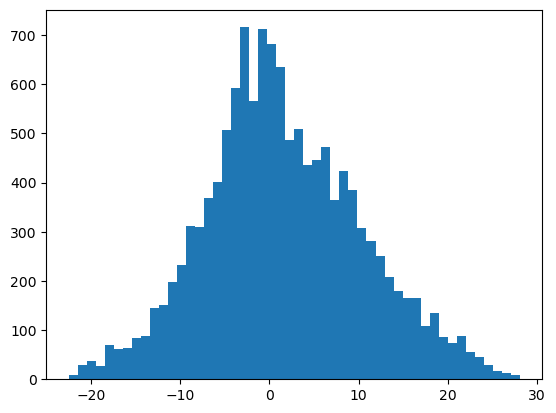}
    \caption{\textbf{Histogram of Z-scores Resulting from Equation \ref{eq: YY z-score} for Skill Co-appearances.} The x-axis shows the $z_{th}$ values derived from equation \ref{eq: YY z-score} on our data, and the y-axis shows the number of co-appearance links falling into a given range of $z_{th}$ values. We focus on $z_{th}$ values between 4 and 6.}
    \label{fig:YY z-score distribution}
\end{figure*}

There are two parameter choices  $z_{th}$ and $\alpha_{th}$. Here, we present results across different parameters to ensure the robustness of our findings. 
We choose the first parameter, $z_{th}$, in such a way that we remove about two-thirds of the edges.
Fig.~\ref{fig:YY z-score distribution} shows the distribution of z-scores for all skill co-appearance edges.

\begin{figure*}[!h]
    \centering
    \includegraphics[width=.95\textwidth]{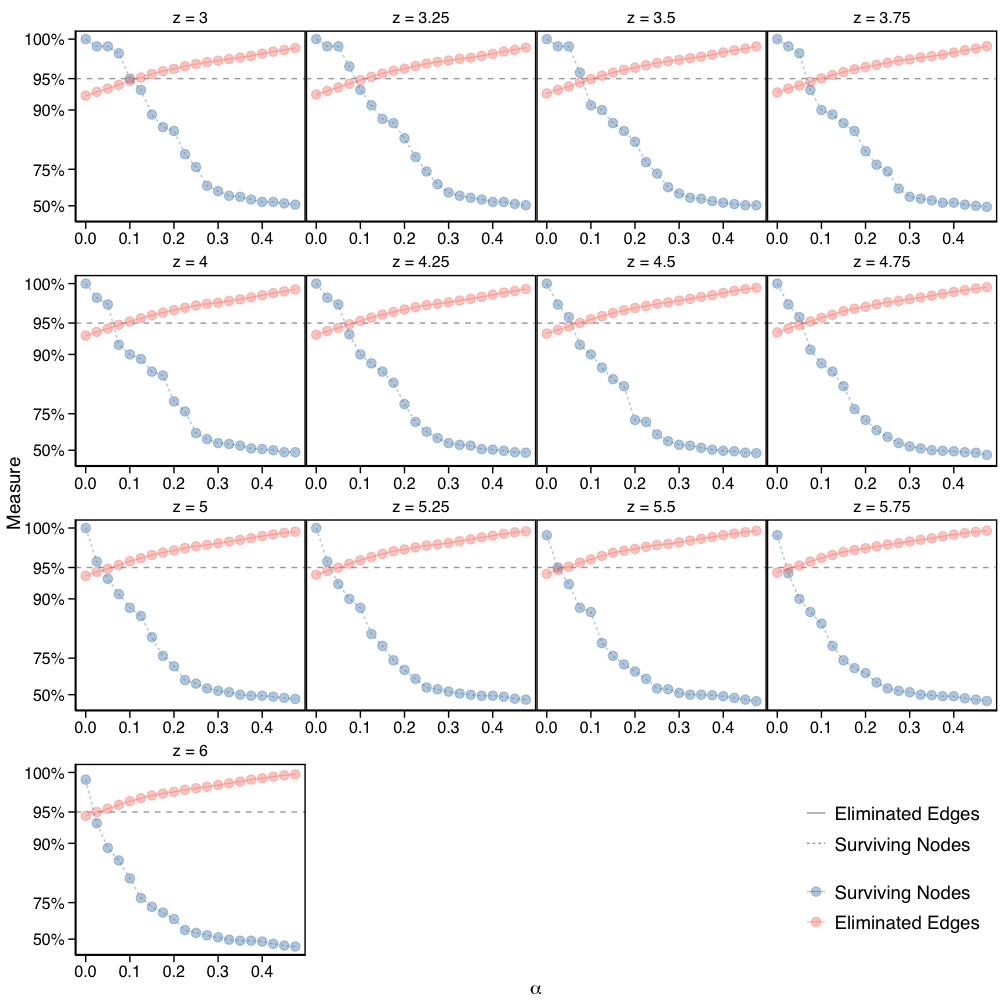}
    \caption{\textbf{Sensitivity Analysis on Parameters Used for Obtaining Significant Skill Dependencies.} The x-axis shows the $\alpha_{threshold}$ values. Each panel shows a certain $z_{threshold}$ as in equation \ref{eq: YY z-score}, and the y-axis shows the rate of node survival or edge elimination. Ideally, no more than 5\% of skills are eliminated, while about 95\% of ties between skills were removed as statistically insignificant. $z_{threshold} = 4.75$ and $\alpha_{threshold} = 0.05$ is a possible solution used in the main text. The resulting backbone from a number of combinations is offered in the following.}
    \label{fig:a-threshold and z-threshold sensitivity analysis}
\end{figure*}

Ideally, one chooses $\alpha_{th}$ as strictly as possible to remove insignificant links without the loss of skills. If all edges associated with a given skill are removed as a result of the two steps of the algorithm, the skill is eliminated from the resulting dependency network.
However, as our sensitivity analysis shows, retaining too many statistically insignificant links weakens our ability to extract conditional dependencies robustly.
Therefore, retaining more statistically significant edges inevitably impose the cost of losing several skills.
Fig.~\ref{fig:a-threshold and z-threshold sensitivity analysis} offers a sensitivity analysis on the interaction of $z_{th}$ and $\alpha_{th}$.
Ideally, no more than 5\% of skills are eliminated, while about 95\% of ties between skills were removed as statistically insignificant.
The combination $z_{th} = 4.75$ and $\alpha_{th} = 0.05$ is a possible solution used in the main text.
At this level, only five skills are eliminated from the network.

A consideration is whether the shape of the skill dependency in Fig.~\ref{fig:Figure 2}~(b and c) is influenced by the choice of parameters. 
We conduct a robustness check wherein we visualize the resulting networks from the combination of values of $z_{th}$ between 4 and 5.5 and values of $\alpha_{th}$ between 0.01 and 0.1.

Throughout, a disjointed structure emerges, wherein a set of specialized skills (blue and gray) have closer connections with the general skills (red), than other specialized skills.
Even in networks obtained from a lenient $\alpha_{th}$, the hierarchical structure is visible, and in most, one can distinguish between a more closely knit web of skills that manifest stronger dependence on generals skills (manifest higher connection to red nodes), and a second set of skills, decoupled from the first, which manifest a comparatively shallow dependency web.
We withdrew the visualized network for the sake of brevity. These visuals are sharable upon request.

\clearpage
\subsubsection*{Visualizing the Skill Hierarchy}
\begin{figure*}[!h]
    \centering
    \includegraphics[width=\textwidth]{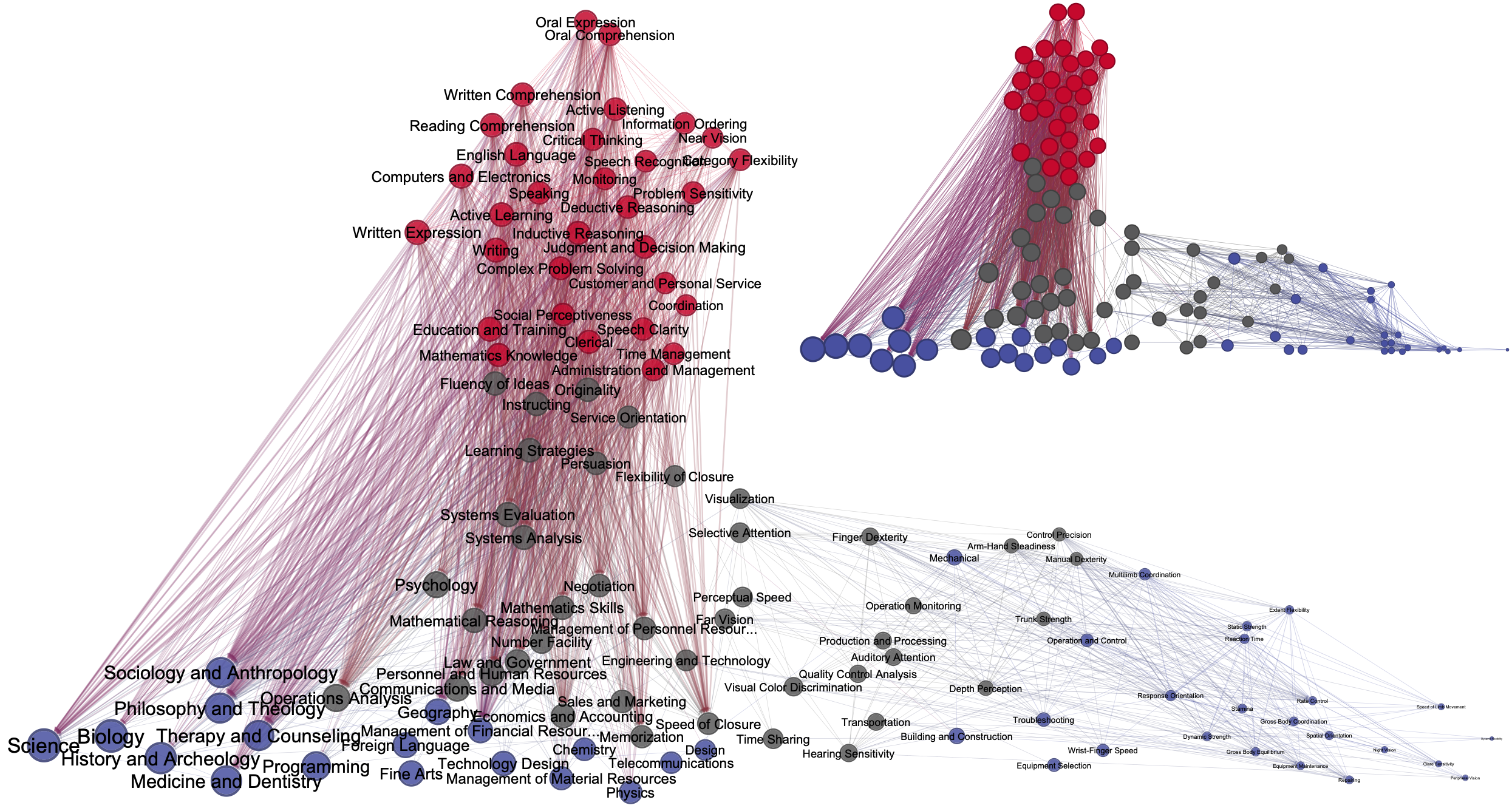}
    \caption{\textbf{Skill Dependency Network.} The layout is adjusted to accommodate the labels. The original layout (without overlap) is shown in the top right corner.}
    \label{fig:full_figure_2b_labeled}
\end{figure*}

The main Fig.~\ref{fig:Figure 2} is the backbone of a network with parameters $z_{th} = 4.75$ and $\alpha_{th} = 0.05$.
Fig.~\ref{fig:full_figure_2b_labeled} shows the full skill network, which contains 115 nodes and 1,796 dependency relationships.
The following skills are eliminated from the graph because neither of their dependency relationships passed the statistical significance test: \textit{Installation, Explosive Strength, Sound Localization, Food Production, Public Safety, and Security}.
In the backbone network, any direct path is eliminated where there exists an indirect path through dependencies. As a result, the backbone contains only 395 edges, accommodating visualization.
Nonetheless, we perform all calculations on the skill network and not its backbone.

\begin{figure*}[!h]
    \centering
    \includegraphics[width=\textwidth]{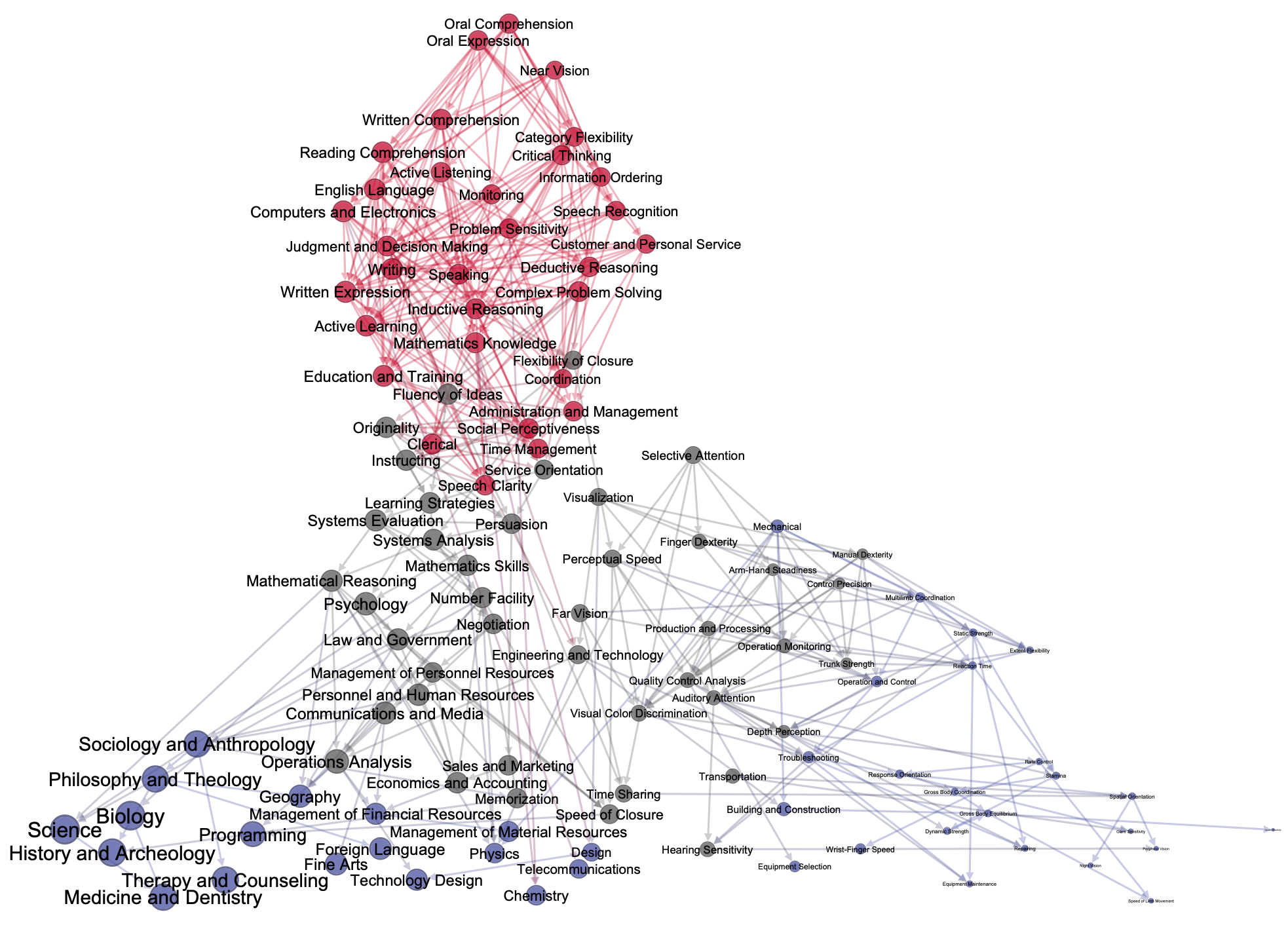}
    \caption{\textbf{Labeled Skill Dependency Backbone.} Node positions are adjusted to accommodate labels.}
    \label{fig:figure_2b_labeled}
\end{figure*}

For both the skill network and its backbone, we use a layout technique inspired by \cite{Mones2012, Kosack2018}.
We determine the vertical placement of a skill based on its local reaching centrality \cite{Mones2012}, defined as the number of nodes achievable from the focal node. This highly correlates with a skills' demand, defined as the total level values, across occupations (Pearson correlation 0.89). 
The horizontal position is proportional to the skill's association with education, calculated as the weighted education of occupations (using levels as weights).
We pass normalized values for both the vertical and horizontal axes through a Lambert cylindrical projection to Gephi for visualization.
 Fig.~\ref{fig:figure_2b_labeled} shows the network in the main Fig.~\ref{fig:Figure 2}~(b) with all nodes labeled and the position of nodes adjusted to accommodate labels.

\newpage


\subsection{Linkage to Skill Co-occurrence Networks} \label{supsec:skill co-occurrence}

Our work builds on a vast literature that conceptualizes the landscape of skills as a co-occurrence network \cite{Neffkeeaax3370, AndersonKatharineA2017Snam, Alabdulkareem2018, Frank2019}.
Indeed, without directionally, the hierarchical network is in excellent agreement with such previously constructed skill networks.
Here, we follow the approach used in \cite{Alabdulkareem2018} that identifies communities of cognitive versus physical from a pairwise co-occurrence network. We obtain such a network in two steps (using O*NET skill data from 2019):

\begin{enumerate}
    \item Measuring the "effective use of skill" by occupation based on RCA as follows:
    \begin{equation}
        RCA(s,j)=\frac{Importance(s,j)/\Sigma_{s'\in S}Importance(s,j')}{\Sigma_{j'\in J}Importance(s',j')/\Sigma_{s'\in S,j' \in J}Importance(s',j')}
    \end{equation}
    where $s$ denotes a given skill, and $j$ a given occupation. $S$ and $J$ denote the population of skills and occupations respectively. An skill-occupation is 'effective'— i.e., $e(i,j)=1$ if $RCA(s,j)$— and is not— i.e., $e(i,j)=0$, otherwise.
    
    \item Using $e(i,j)$ values, authors derive pairwise skill "complementarity" proportional to the number of times skills $s$ and $s'$ co-appeared in an occupation as follows:
    \begin{equation}
        \theta(s,s')=\frac{\Sigma_{j\in J}e(s,j).e(s',j)}{max\big(\Sigma_{j\in J}e(s,j), \Sigma_{j\in J}e(s',j)\big)}
    \end{equation}
\end{enumerate}


There are two key messages.
First, we explain that the cluster of \textit{General} skills resides at the center of such a skill co-occurrence network— in fact, the ordering of skill specificity based on our skill clusters is predictive of how far the skills lie towards the fringes of the skill co-occurrence network.
Second, the dichotomy of cognitive versus non-cognitive skills has tight connections with the disjointed structures we found and called nested and un-nested skills, respectively.



\begin{figure*}[!h]
    \centering
    \includegraphics[width=.7\textwidth, trim={0 5.5cm 0 5.5cm},clip]{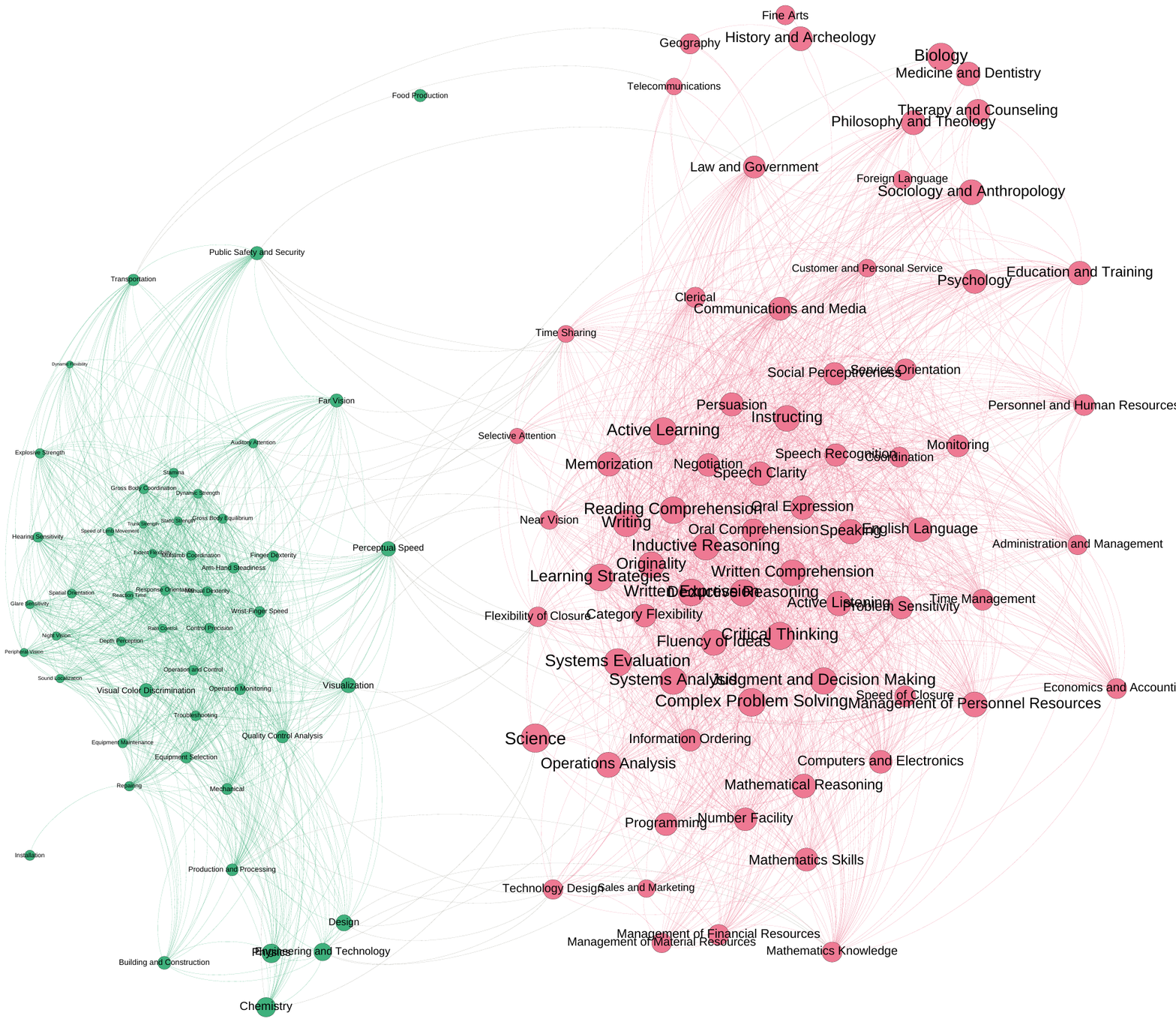}
    \caption{\textbf{Network of Pairwise Skill "Complementarity."} In this figure, edges denote skill complementarity relationship if they passed the threshold of $\theta(s,s')>0.5$— authors used a threshold of $\theta(s,s')>0.6$ which leads to several isolated nodes in 2019 data. We assign an average education value to each skill based on the educational requirements of occupations. A skill's associated education level is only impacted by occupations that use the skill 'effectively'— i.e., $\forall j:  RCA(s,j)$, nodes are colored based on their modularity communities.}
    \label{fig:Network_Skill_Complementarity_RCA_modularity_colored}
\end{figure*}

Fig.~\ref{fig:Network_Skill_Complementarity_RCA_modularity_colored} shows a network representation
of skills based on the pairwise "complementarity" values manifests the bi-modal structure reported by \cite{Alabdulkareem2018}.
There are several departure points, however.
First, we restrict our workplace skills to the so-called \textit{knowledge, abilities, and skills}, disregarding \textit{work activities}, while the latter is commonly used in co-occurrence networks constructed using O*NET.
Our rationale for not including work activities is that they are job descriptions (i.e., generalized forms of job tasks that are specific to jobs).
In contrast, abilities, knowledge, and skills are characteristics of workers' expertise, which are our primary concern.

Second, the community on the right is an ensemble of "cognitive" skills, while the left group corresponds to mostly "physical" skills, which are consistent with our nested and un-nested skills, respectively.
However, several skills, such as \textit{Physics, Design, and Chemistry} seem out of place at the bottom end of the left community, and are classified by our approach as nested skills— which appear consistent with their wage and educational associations.
The key advantage of our method is that we can predict numerous implications of skills based solely on the informationed embedded in the occupation-skill networks, without the need for knowing the content of the skill, as labeled by cognitive or physical skills.
Our skill hierarchy, in effect, offers an explanation for why certain skills, known to be cognitive, are more valuable, based on the investments necessary to satify their dense and nested web of dependencies.

\subsection{Skill Hierarchy Captures Career Progress (Specialization)} \label{supsec:RN vs. NP}

A key advantage of integrating the conceptual distinction between general and niche skills with a structural network approach to studying skills is that the aggregation of pairwise skill interdependencies reveals pathways of progress (what has come to be known as “specialization”). However, the structure of our skill hierarchy implies that progress entails co-development in certain niche skills and the prerequisites, often more general skills.

\begin{figure*}[!h]
    \centering
    \includegraphics[width=.85\textwidth]{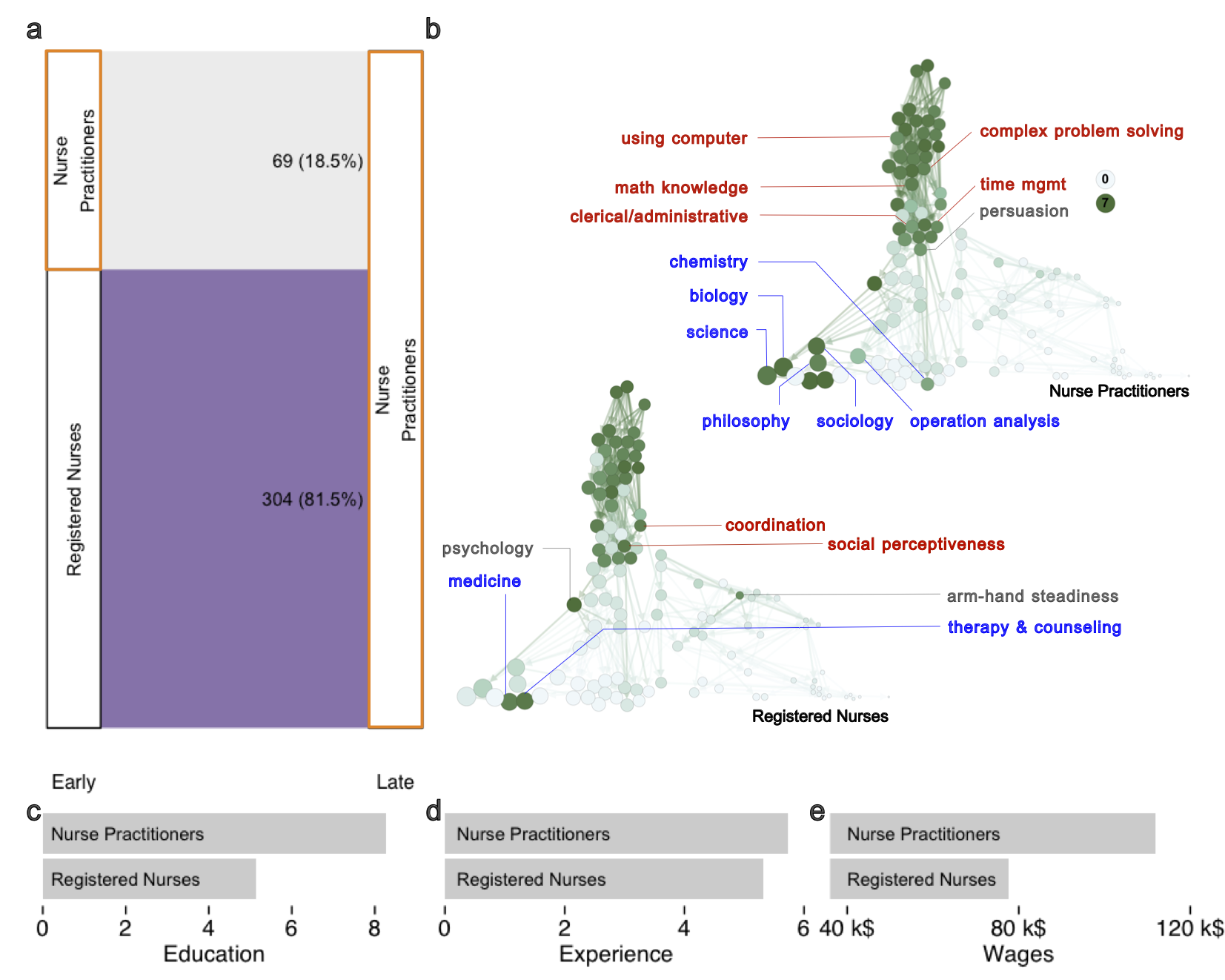}
    \caption{\textbf{Transition between Registered Nurses (RNs) and Nurse Practitioners (NPs).}
    \textbf{(a)} uses resume data from Burning Glass Technology to capture the transition statistics between RNs and NPs. We restrict the analysis to individuals with at least five listed occupations in their resume and define their early career occupations as the most appeared occupation in the first three jobs, similarly late career occupations as the most appeared in the fourth jobs and onward. We disregard individuals whose early and late careers are neither RN nor NP. Including these individuals would not change the result but significantly complicate the exposition. One expects that higher wages for NPs would attract RNs \textbf{(e)}. Indeed, most NPs were RNs early on. However, only a subset of RNs progresses to NP jobs, suggesting barriers to entry, summarized in higher experience and educational requirements \textbf{(c-d)}.
    \textbf{(b)} captures the skill requirements of RNs and NPs, highlighting the advantage of integrating the conceptual distinction between general and niche skills with a structural network approach to studying skills in revealing pathways of progress (also known as “specialization”). The structure of our skill hierarchy also implies that progress entails co-development in certain niche skills and the prerequisite, often more general skills.}
    \label{fig:RN vs. NP}
\end{figure*}

Here, we explore a case study of such progress based on the skill requirement differentials of registered nurses (RNs) versus nurse practitioners (NPs). Compared to RNs, NPs prescribe medicine and diagnostic tests and command higher wages (Fig.~\ref{fig:RN vs. NP} e). Without any cost, someone equipped with the skills of an RN would ideally prefer to work as a nurse practitioner to benefit from higher payoffs.
However, as Fig.~\ref{fig:RN vs. NP}~(a) shows, only a subset of individuals who are RNs early on in their careers (i.e., for whom RN appears most in their first three jobs listed in their Burning Glass Technology resume) manage the switch to the better-paid NP jobs later in their careers (i.e., NP appears most after their third jobs listed in their Burning Glass Technology resume). The fact that most NPs were initially RNs (81.5\%) corroborates our interpretation of the path from RN to NP as one that entails career progress. The transition statistics captured in Fig.~\ref{fig:RN vs. NP}~(a) are also consistent with the higher experience and the more extended training needed for nurse practitioners to develop the necessary skills (Fig.~\ref{fig:RN vs. NP} c-d).

The correlation between education and wages observed at the cross-section of RNs and NPs agrees with the economic theory narrative. However, only a structural approach can reveal the skill development involved in such a transition, highlighting skill growth pathways, seen in Fig.~\ref{fig:RN vs. NP}~(b).
While RNs require high levels of medicine and therapy (niche skills), psychology (intermediate skill), coordination, and social perceptiveness (among many other general skills), the transition into an NP requires further levels of those skills as well as significant development of science, biology, chemistry (among other niche skills), persuasion (intermediate skill), as well as higher knowledge of math, time management, complex problems solving, administrative and computer skills (among other general skills). In contrast, arm-hand steadiness used at high levels by RNs is not as intensely utilized by NPs. 

Comparing RNs’ with NPs’ skills showcases that our approach teases out meaningful progression (or specialization) pathways embedded in the skill requirement of occupations. The co-development of niche and the relevant general skills underpin what we call a nested specialization path. In the following, we offer evidence that the pattern observed in RNs’ and NPs’ careers emerges across individuals in other occupations.

\subsection{Skill Hierarchy Captures Skill Entrapment} \label{supsec:hispanic skill entrapment}

\begin{figure*}[!h]
    \centering
    \includegraphics[width=.75\textwidth]{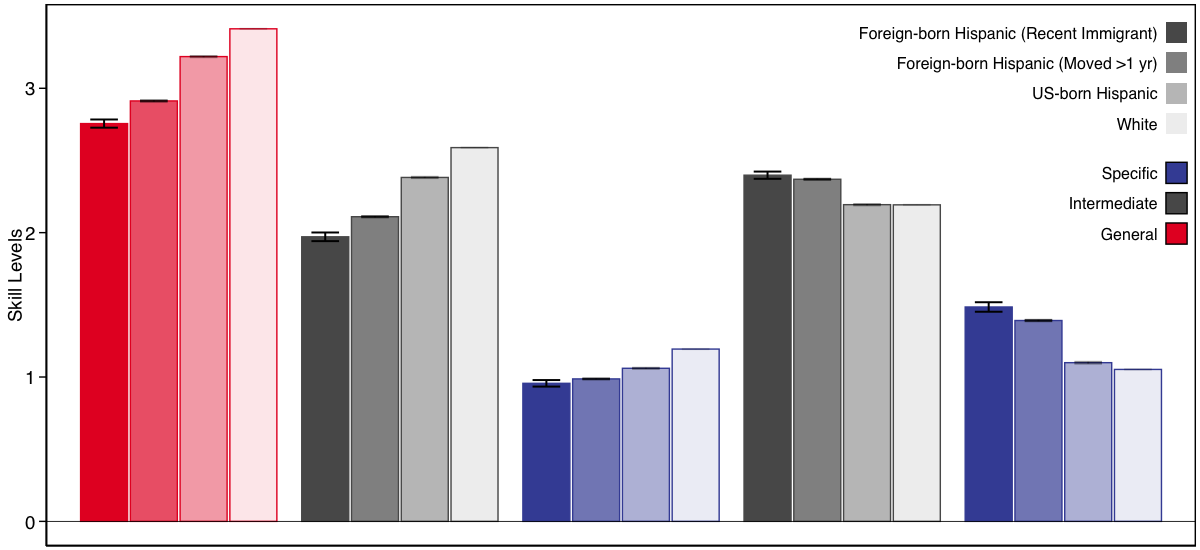}
    \caption{\textbf{Comparison of the Skill Levels of Hispanic Immigrants and White Workers} We distinguish between four groups of workers (i. foreign-born Hispanics who have migrated less than a year to the US from the time of survey, ii. foreign-born Hispanics who have been in the US for more than a year, iii. US-born Hispanics, and iv. the White workers) and map their average skill levels for each skill category. Recently migrated foreign-born Hispanics have the least levels of general and nested skills and most un-nested skills.
    }
    \label{fig:skills_of_different_Hispanics}
\end{figure*}

As the main Fig.~\ref{fig:Figure 7} and SI Sec.~\ref{section: add - demographic distribution of skills} show, Hispanics tend to possess relatively high levels of un-nested skills but are underprivileged in gender nested skills.
This unbalance leads to skill entrapments with possibly early rewards and long-term wage penalties, as SI Fig.~\ref{fig:wage curves}.
Our skill hierarchy allows us to explore one possible driver of this skill unbalance for Hispanics.

\begin{figure*}[!h]
    \centering
    \includegraphics[width=.78\textwidth]{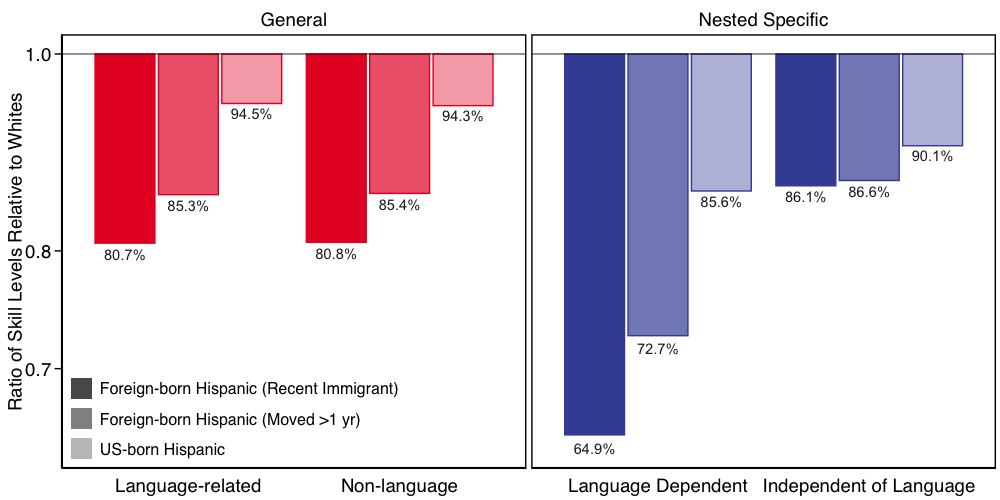}
    \caption{\textbf{Language Barriers Manifest in Lower Levels of Language-related Nested Skills for Hispanics}. The figure depicts, for the Language-related and non-language general skills and Language-dependent (defined as the skills in the top 25\% arrival probability to the mentioned language skills) and Language-independent nested skills, the ratios of skills levels for the different Hispanic subgroups relative to White workers. The results depict that the language-dependent nested skills vary significantly more across the Language-dependent subset, supporting our suspicion that Hispanic workers, at least in part, suffer from their language skills, which prevents them from acquiring/applying downstream skills.
    }
    \label{fig:hispanics language skills}
\end{figure*}

We suspect language skills are barriers to some hispanic workers, particularly early on in their careers in the US, hampering the acquisition of (language-related) general and (the downstream) nested specific skills, but less so the acquisition of un-nested skills.
To test this, we split the sample of individuals from the CPS into four subgroups, ordered based on their likely level of English proficiency: \textit{Hispanics born outside of the US who immigrated less than a year before the survey, Hispanics born outside the US who have been in the US for more than a year, Hispanics born inside of the US, and White workers}.
We map the average skill levels of each of the above subgroups for each skill category in Fig.~\ref{fig:skills_of_different_Hispanics}, below. As hypothesized,the foreign-born Hispanics who recently migrated to the US have the lowest levels of general and nested skills and have the highest unnested skills. The suspected ranking of English proficiency of each subgroup is consistent with their ranking in terms of general, nested and unnested skills.
Next, we investigate the role of language skills directly Fig.~\ref{fig:hispanics language skills}.

Our network allows us to directly identify which nested skills more closely depend on language general skills. To do so, we first identify six general skills as “language-related”: i.~English Language, ii.~Oral Expression, iii.~Oral Comprehension, iv.~Written Expression, v.~Written Comprehension, and vi.~Speaking. One can quantify the dependence of each nested skill, i, on each of the mentioned language skills, j, by deriving the arrival probability of a random walk starting from the mentioned language general skills, $P_{i,j}^{<arrival>}$. Aggregating these probabilities over the language general skills, we obtain $P_i^{<arrival>}=\Sigma_j P_{i,j}^{<arrival>}$.
We flag nested specific skills at the top 25\% of skills in terms of their average arrival probability, $P_{i,j}^{<arrival>}$, obtaining the following skills: i.~History \& Archeology, ii.~Management of Material Resources, iii.~Management of Financial Resources, iv.~Programming, v.~Philosophy \& Theology. Splitting general and nested skills by their language associations (general skills into Language-related and Non-language skills, and nested skills into Language dependent and Language independent), we obtain the average skill levels of individuals for the previously defined subgroups of workers (Hispanic and White based on their place of birth and time since immigration). In Fig.~\ref{fig:hispanics language skills}, we show the ratios of skills levels for the different Hispanic subpopulation groups relative to White workers for the Language-related and non-language general skills and Language-dependent.

The results show that the skill gaps between Hispanic subpopulations and White workers mimic the implied language gradient: the less proficient in English a subgroup will be, the larger the gap is to White workers in language-dependent nested skills but not in language-independent specific skills. This supports our hypothesis that the skill gaps for Hispanic workers as a whole are, at least in part, due to language barriers.


\clearpage
\section{Skill Categories in Career Trajectories} \label{supsec: skill dependencies and age}
Main Fig.~\ref{fig:age} supplements our inference of the skill structure from O*NET, which relies on cross-sectional data, with longitudinal evidence in line with the notion that one actually acquires or advances general skills when they progress in their career and acquire more specific skills.
Here, we provide additional evidence and robustness checks on the analysis of main Fig.~\ref{fig:age}, based on resume (Burning Glass) data, occupational median age, and skill acquisition reflected in synthetic birth cohorts we created using CPS microdata.

\subsection{Resume Data} \label{section: add - burning glass}

Unlike O*NET, Burning Glass resume data offers longitudinal observation of skill acquisition and will allow us to conduct a more strict test of our skill structure.
We keep track of one's occupations in the resume data, from which longitudinal skill acquisition is inferred. 

\subsubsection*{Preparing Burning Glass Data}

\begin{figure*}[!h]
    \centering
    \includegraphics[width=\textwidth]{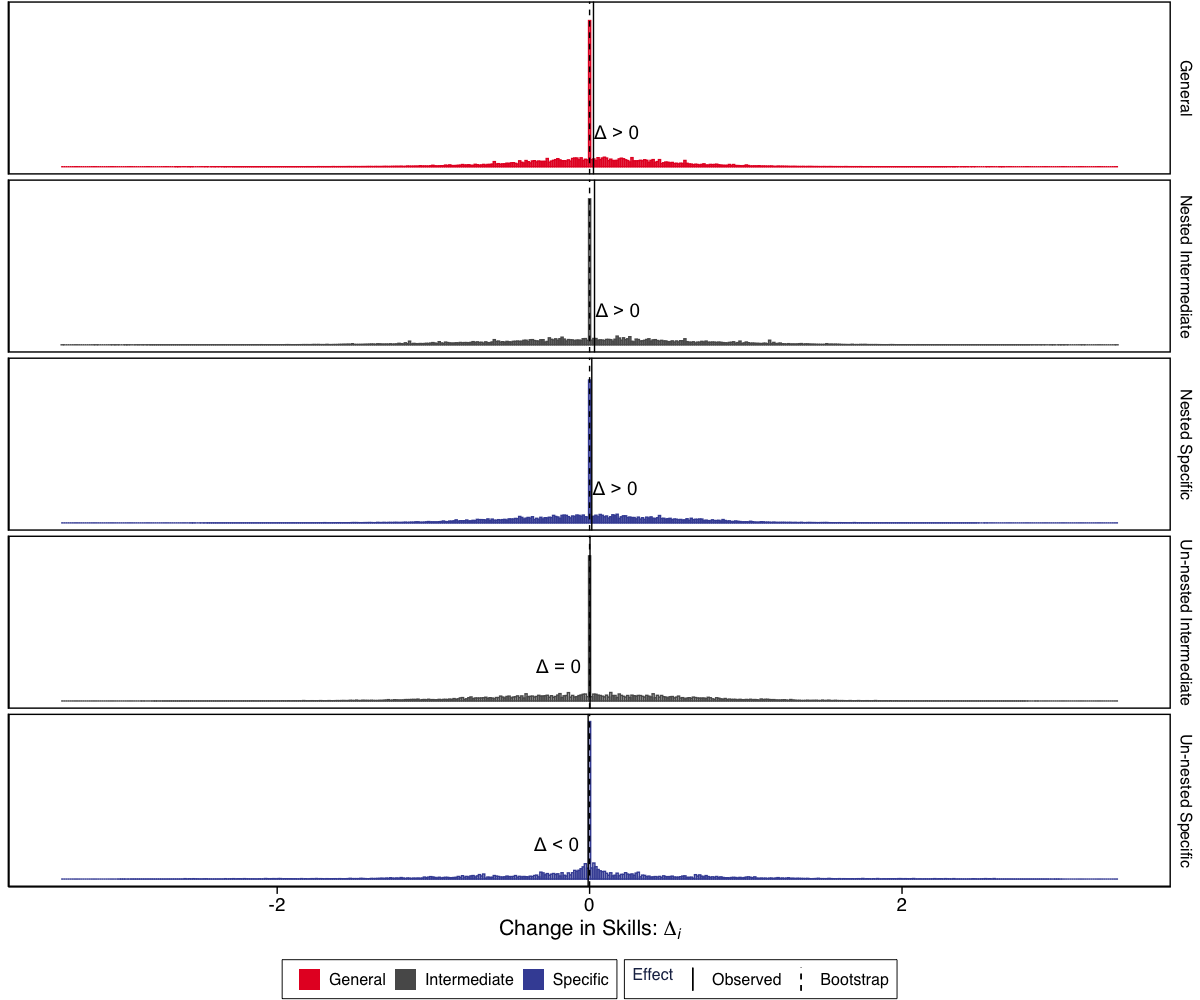}
    \caption{\textbf{Changes in Skill Levels in Individuals' Career Moves.} The distribution shows the Burning Glass resume data. A minority of career moves produce extreme values, stretching the skill change distributions' tails due to imperfect data.
    }
    \label{fig:BG_skill_change_fig_full}
\end{figure*}

The following discussion describes choices made in cleaning the data, revealing robustness to such choices in terms of the direction of the results, although the magnitude may vary slightly.
We studied over 20 million resumes from the Burning Glass data, which amounts to over 70 million job moves.
For each move, we link the source and destination occupations to skills from O*NET in 2019.
Excluding all within-occupation moves— which amount to no skill change— we calculate a skill level change across our skill categories and show the result. Fig.~\ref{fig:BG_skill_change_fig_full} as the distribution of career moves for resumes in the Burning Glass sample— after removing within-occupational career moves.

As can be seen in Fig.~\ref{fig:BG_skill_change_fig_full},
A minority of career moves produce extreme values, stretching the skill change distributions' tails.
Table \ref{tab:odd job sequences} shows a few such cases from the data.
For instance, the resume with the ID \textit{652855}, serves as a janitor for a short period (4 months) before seemingly claiming a Chief Executive role. Resume with ID \textit{1723696} held overlapping jobs as a Medical Health Technician and a Middle School Teacher.
Studying the career moves that correspond to such skill changes, we noticed a significant proportion arise from short job stints and coinciding jobs— some seemingly voluntary part-time commitments.

\begin{figure*}[!h]
    \centering
    \includegraphics[width=\textwidth]{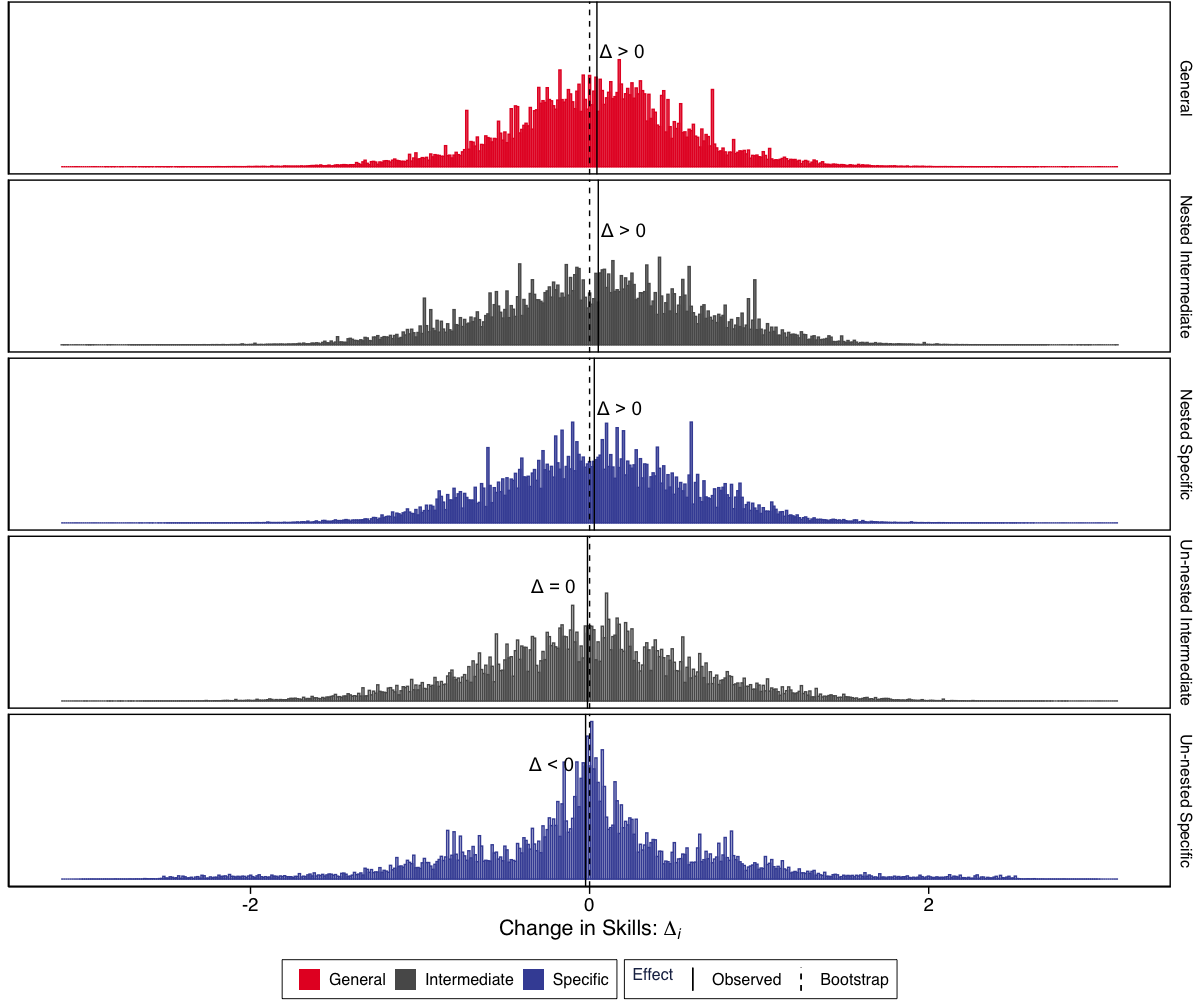}
    \caption{\textbf{Changes in Skill Levels in Individuals' Career Moves.}
    Most career moves amount to small changes in skills. On aggregate, general, nested skills experience increases on aggregate, while un-nested skills record non-positive changes.
    Nested skill changes closely correlate with changes in general skills. In contrast, there is almost no noticeable relationship between changes in general and un-nested skills.
    Importantly, randomizing the sequence of job transitions (bootstrap) eradicats the direction of skill acquisition in the observed data.
    }
    \label{fig:BG_skill_change_fig_clean}
\end{figure*}

We removed such jobs from our resume sample.
Particularly, we kept jobs if they lasted at least 12 months— we arrived at the threshold after studying the career moves that correspond to the thousand largest absolute skill changes.
Furthermore, sorting jobs for each resume based on starting date and end date, we removed any job that had a shorter length and overlapped with another— that is, we remove a job $j_r$ from a resume $r$, if it had a later or equal start date with another job $j'_r$, but did not have a later end date.
We also removed jobs for which we could not extract the start and end date— we used \textit{Python's \textit{dateparser} version 1.1.1.} for the extraction.
The resulting sample was 9,382,602 career moves and 5,361,751 resumes.
Fig.~\ref{fig:BG_skill_change_fig_clean} shows the resulting skill change distributions.

\begin{figure*}[!h]
    \centering
    \includegraphics[width=\textwidth]{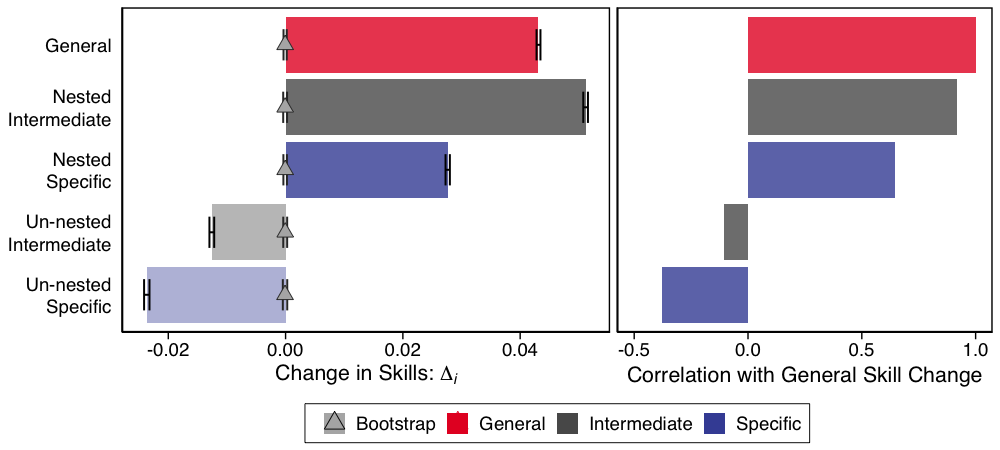}
    \caption{\textbf{Expected Changes of Skill Levels for Each Career Moves.}
    For each career move, we linked the source and destination occupations to skills from O*NET in 2019. We calculate a skill level change across our five skill subtypes.
    For each skill sub-type, we measure changes in skill levels, $\Delta_s$, corresponding to each career move as the average of differences between the skill levels of the target and source occupations.
    \textbf{(a)} shows average changes in skill levels for skill subtypes. On aggregate, general, nested skills experience increases on aggregate, while un-nested skills record non-positive changes.
    \textbf{(b)} shows the correlation between general skills and each skill category resulting from individuals' career moves. Nested skill changes are closely related to changes in general skills. In contrast, there is almost no noticeable relationship between changes in general and un-nested skills.
    }
    \label{fig:BG_fullfig_cleaned}
\end{figure*}

\begin{figure*}[!h]
    \centering
    \includegraphics[width=.7\textwidth]{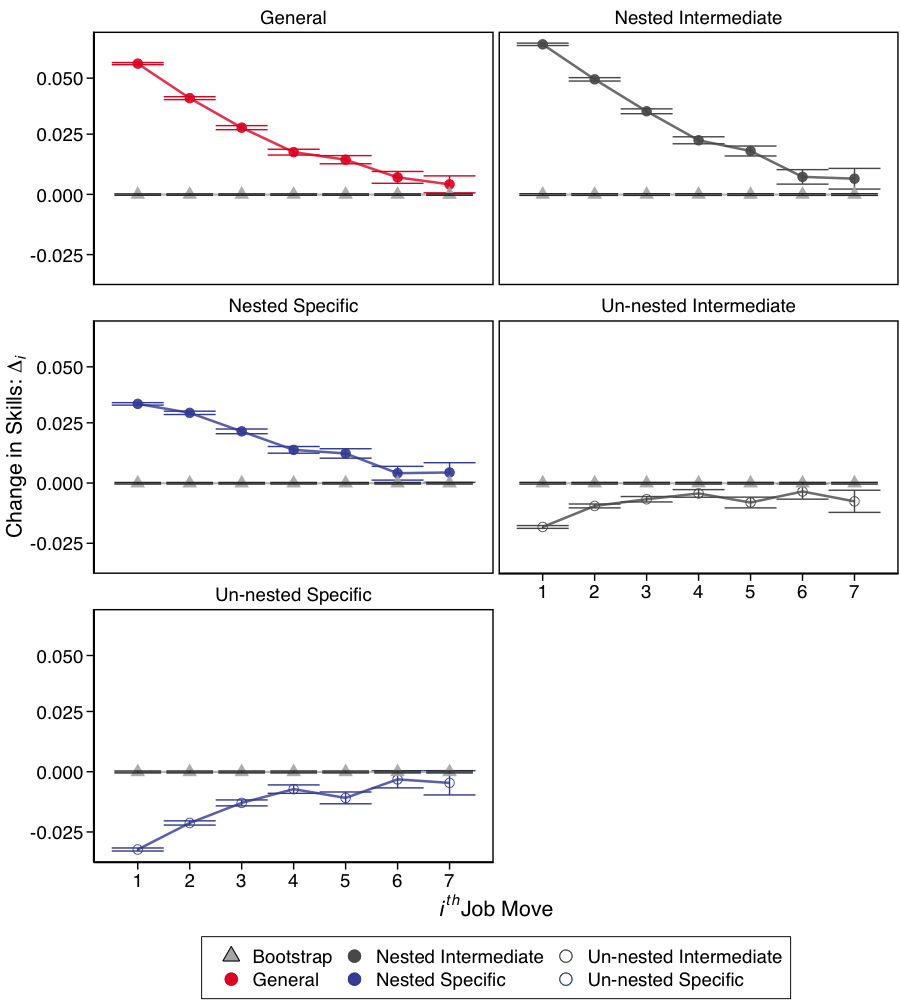}
    \caption{\textbf{Changes in all skill category levels in consecutive job transitions.}
    }
    \label{fig:BG_fullfig_dynamic}
\end{figure*}

The main text analyzes the levels and patterns of skill change across general and (nested and un-nested) specific skills.
Fig.~\ref{fig:BG_fullfig_cleaned} shows the net effects, i.e., the average change in levels resulting from job moves across all skill categories, and the correlation between the change in the levels of general skills and changes in the level of other skill categories resulting from job transitions.
Fig.~\ref{fig:BG_fullfig_dynamic} supplements main Fig.~\ref{fig:age} (i) by providing the changes in all skill category levels resulting from consecutive job transitions.

\footnotesize
\input{Nature_HB_2023/tabsNHB/Jul_15_2023__Weird_Job_Sequences.tex}

\normalsize

\newpage

\newpage
\subsubsection*{Expected Skill Change from Random Job Transition} \label{supsec: bootstrapping BG}
\begin{figure*}[!h]
    \centering
    \includegraphics[width=.7\textwidth]{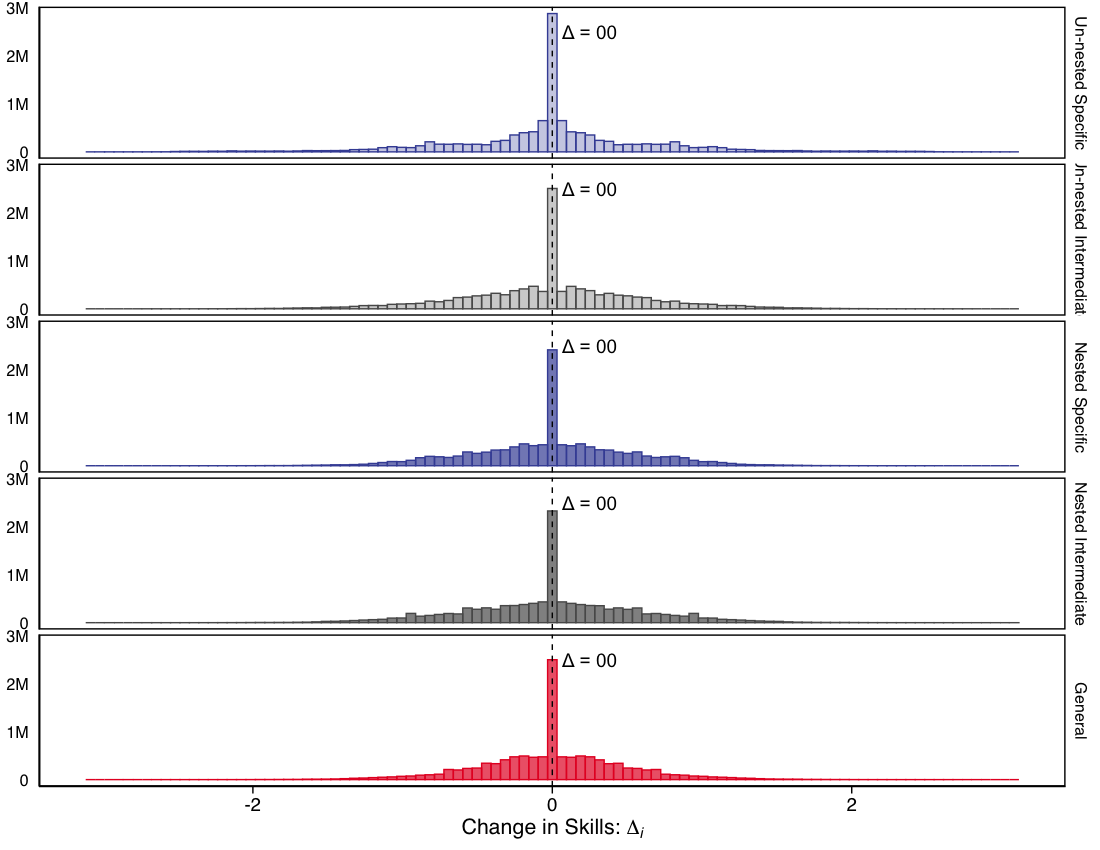}
    \caption{\textbf{Changes in Skill Levels in \textit{Bootstrapped} Individuals' Career Moves.}
    The distribution of changes in skill levels visibly differs from what we obtain from the observed career moves.
    }
    \label{fig:BG_skill_change_single_bootstrap}
\end{figure*}

Furthermore, we bootstrapped our resume sample to produce a benchmark and compare it with the skill changes we obtained from observed career moves.
For each resume in our sample, we randomly permuted the order of career moves and measured the skill changes again.
 Fig.~\ref{fig:BG_skill_change_single_bootstrap} shows one such bootstrap.
It is visible in Fig.~\ref{fig:BG_skill_change_single_bootstrap} that the randomization eradicated the direction of skill changes we had obtained from the observed career moves— in Fig.~\ref{fig:BG_skill_change_fig_clean}.

\begin{figure*}[!h]
    \centering
    \includegraphics[width=.7\textwidth]{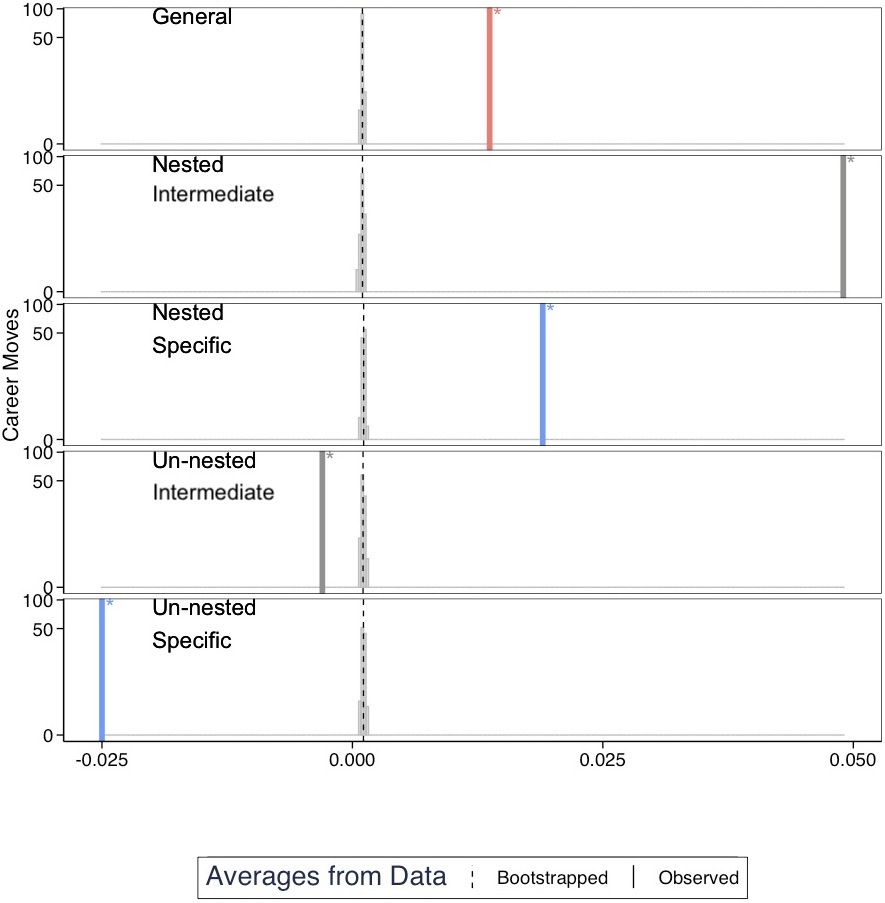}
    \caption{\textbf{Distribution Skill Changes from Bootstrapped Career Moves.}
    }
    \label{fig:BG_skill_change_bootstraps}
\end{figure*}

Fig.~\ref{fig:BG_skill_change_bootstraps} further shows the distribution of average skill changes for 100 bootstraps.
The fact that resulting skill changes from a null model differ significantly from our observed results ensures our results are meaningful signals of individuals' career moves, pointing to the dependencies between (general and nested) skills.

\clearpage

\subsection{Occupational Median Age}
\begin{figure*}[!h]
    \centering
    \includegraphics[width=.95\textwidth]{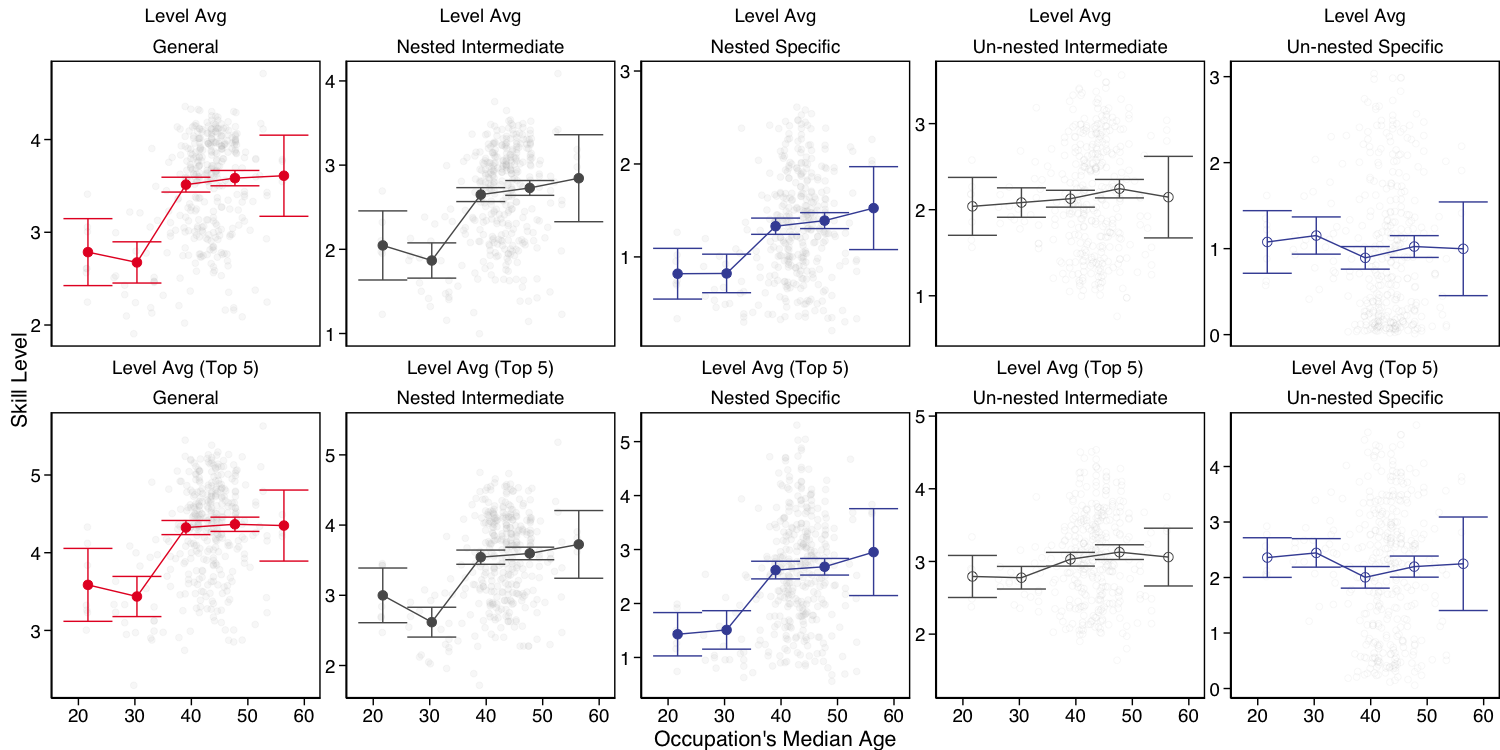}
    \caption{
    \textbf{Median Age of Workers in Occupation and Changes in Skill Categories.}
    }
    \label{fig:occupations' median age and skill - full fig}
\end{figure*}

Fig.~\ref{fig:occupations' median age and skill - full fig} shows the trends of average skill levels and the average levels of the top 5 skills in each category against occupations' median age.
This analysis supplements the main Fig.~\ref{fig:age} (a-c).

\subsection{Individuals' Age and Skills}
\begin{figure*}[!h]
    \centering
    \includegraphics[width=.95\textwidth]{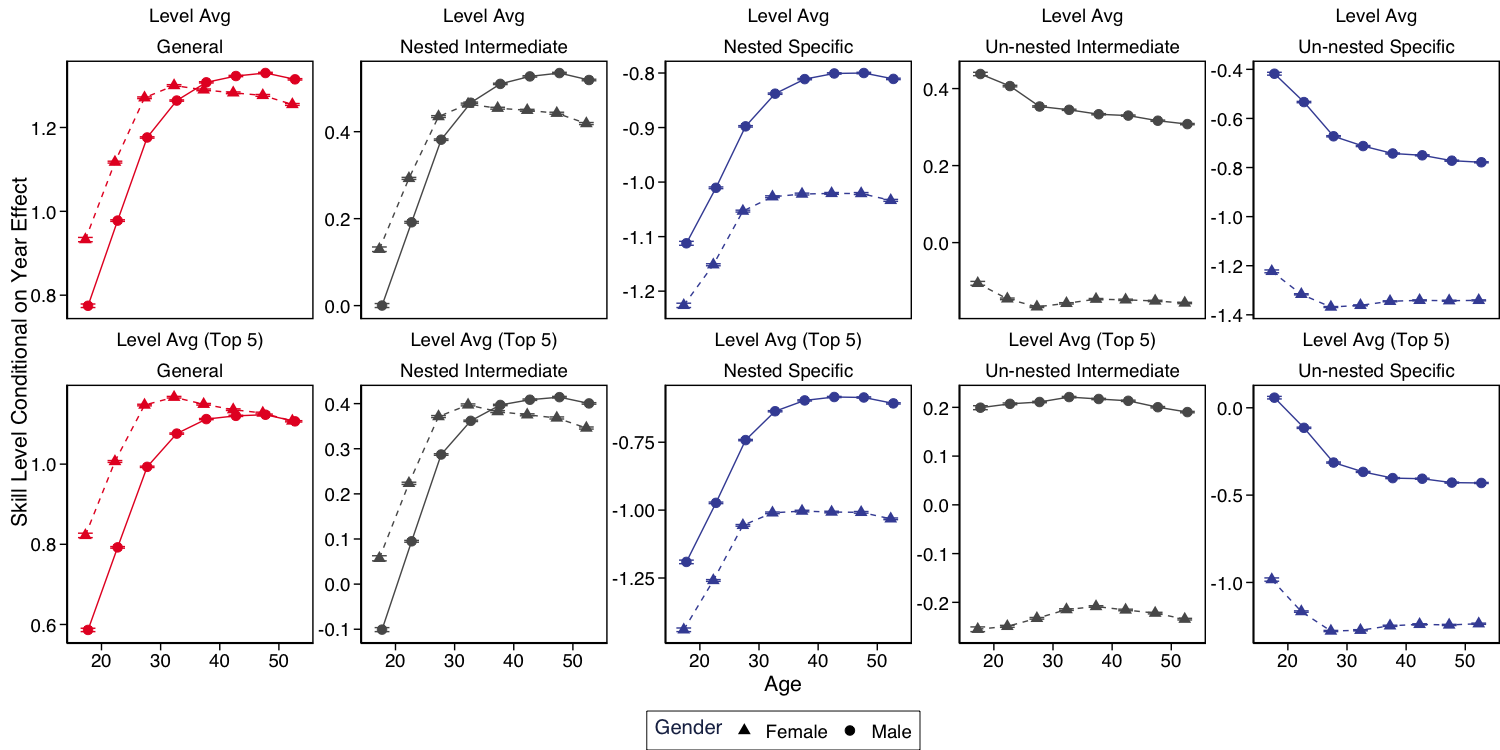}
    \caption{
    \textbf{Individuals' Skill Acquisition and Age.}
    }
    \label{fig:individuals' age and skill - year effects}
\end{figure*}

Fig.~\ref{fig:individuals' age and skill - year effects} shows the trends of average skill levels and the average levels of the top 5 skills in each category as individuals age, accounting for the year effect.
This analysis supplements the main Fig.~\ref{fig:age} (d-f) by controlling for varying annual economic situations.
The top 5 skills are determined based on the highest levels of skills in each category and are inferred for the individual based on their occupation.
The results are consistent with the main figure.

\begin{figure*}[!h]
    \centering
    \includegraphics[width=\textwidth]{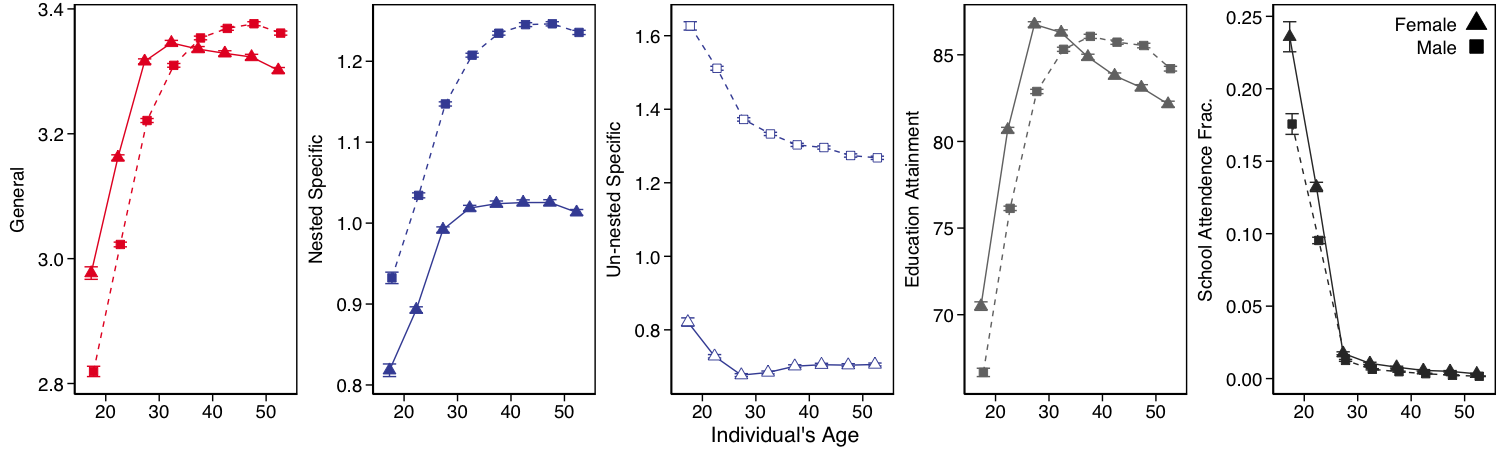}
    \caption{
    \textbf{Evolution of skill, age and education.} To measure education, we have used educational attainment and the fraction of individuals who attend school as functions of age, both taken from the Current Population Survey (CPS). The education attainment variable ranges from 2 (i.e., no schooling) to 125 (i.e., doctorate degree). To obtain the fraction of the sample attending school, we utilized the information in the CPS variable SCHOOLCOL that documents attending high school (1 or 2) or college/university (3 or 4) or not attending school (5). We transformed the information so that if an individual attends school (1,2,3 or 4), it receives a value of 1, and if not attending, it has a value of 0. Even though by the age of 30, education plateaus and school attendance drops significantly, skill growth continues, manifesting the presence of other mechanisms for skill accumulation apart from education.
    }
    \label{fig:individuals' age and skill and education}
\end{figure*}

\begin{figure*}[!h]
    \centering
    \includegraphics[width=.95\textwidth]{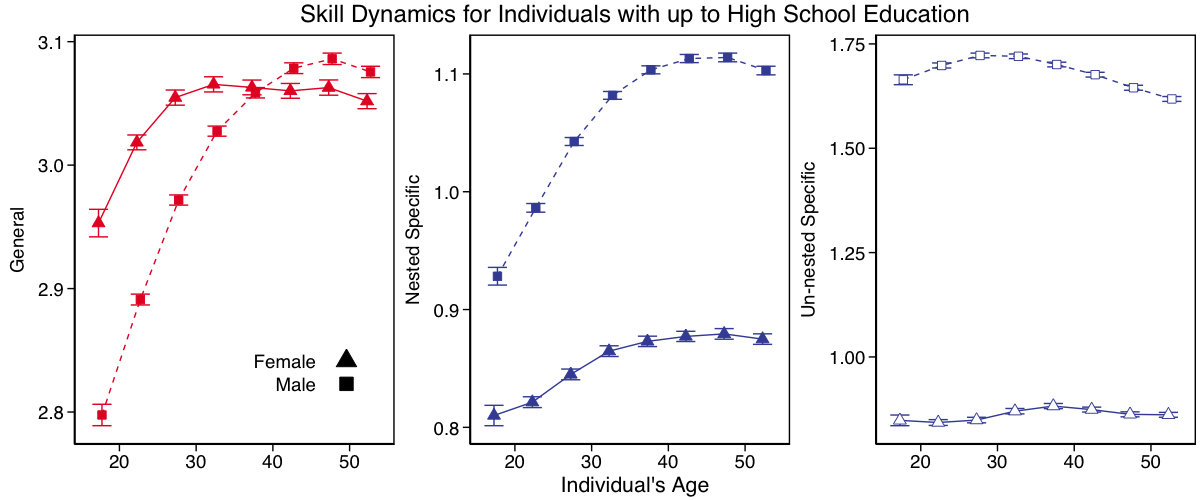}
    \caption{
    \textbf{Skill Acquisition and Age for Individuals with no College Education.} The figure replicated the skill-age analysis (Fig. \ref{fig:age}) for the subset of individuals who have obtained no more than a high school diploma (values of less than or equal to 073 on the CPS education attainment variable.) The patterns resemble the skill accumulation across the population, even though the levels of general and nested skills are lower compared to the population-level estimates.
    }
    \label{fig:individuals' age and skill - no college}
\end{figure*}

As robustness checks, we also show that the skill development observed in the main Fig.~\ref{fig:age} continues long after education (Fig.~\ref{fig:individuals' age and skill and education}) and also emerges for individuals without a college education (Fig.~\ref{fig:individuals' age and skill - no college}).

\clearpage
\section{Skill Investment and Payoffs} \label{supsec: add - returns to skill} 

\subsection{Investment and Payoffs of Skill Subtypes}

\begin{figure*}[!h]
    \centering
    \includegraphics[width=.7\textwidth]{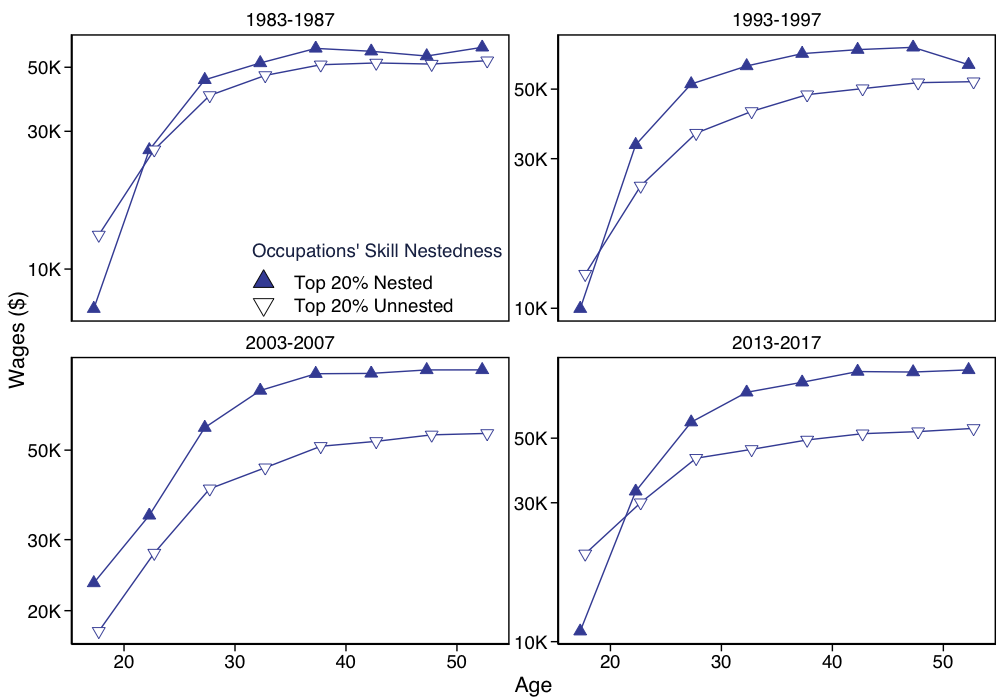}
    \caption{\textbf{Wage Curves for Occupations with Primarily Nested vs. Primarily Un-nested Skills.}
    We average over the levels of nested and un-nested skills of each occupation in our sample and pick occupations at the top 20\% of the nested skills as the most nested and occupations at the top 20\% of the un-nested skills as the most un-nested. Matching these occupations to the individuals in the CPS, we can obtain estimates of wages for individuals in these occupations at different ages. To avoid conflating long-run economic factors, we show the wage-age curves for four 5-year periods: 1983-1987, 1993-1997, 2003-2007, 2013-2017. Un-nested jobs have an early wage lead which quickly evaporates with age.}
    \label{fig:wage curves}
\end{figure*}

Figure \ref{fig:wage curves} shows “wage curves” that depict wages as a function of age for individuals in the most nested and the most un-nested occupations.
The figures capture entrapment due to un-nested skills.
To obtain wage curves, we averaged over the levels of nested and un-nested skills of each occupation in our sample. We picked occupations at the top 20\% of the nested skills as the most nested, and occupations at the top 20\% of the un-nested skills as the most un-nested. Matching these occupations to the individuals in the CPS, we can obtain estimates of wages for individuals in these occupations at different ages.
To avoid conflating long-run economic factors, we show the wage-age curves for four distinct periods of 5-years: 1983-1987, 1993-1997, 2003-2007, 2013-2017.
In three of the four periods, un-nested jobs have an early wage lead, which quickly evaporates with age. The pattern is consistent with the notion that learning is steeper in occupations with more complex tasks \cite{Jovanovic1997,Nedelkoska2015}. To arrive at a complete picture, one would need to account for the higher cost of education associated with nested occupations. Hence, the wage offsets observed in the figure may occur later in individuals’ lives in terms of real earnings once the cost of education is accounted for.

\begin{figure*}[!h]
    \centering
    \includegraphics[width=\textwidth]{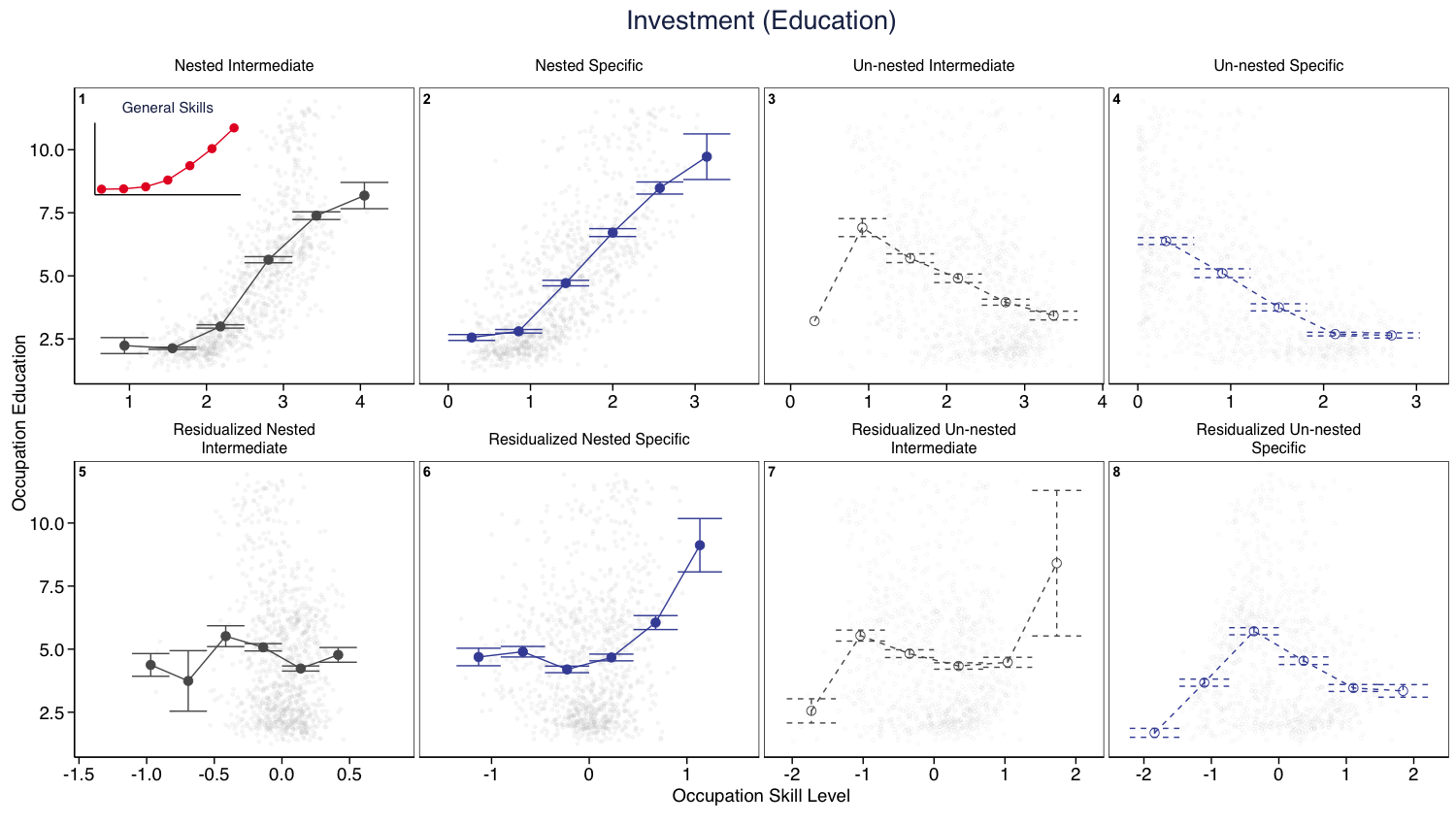}
    \caption{\textbf{Relationship between Occupations' Educational Requirement and Skill Subtypes.}}
    \label{fig:SI_education_skill_level}
\end{figure*}

\begin{figure*}[!h]
    \centering
    \includegraphics[width=\textwidth]{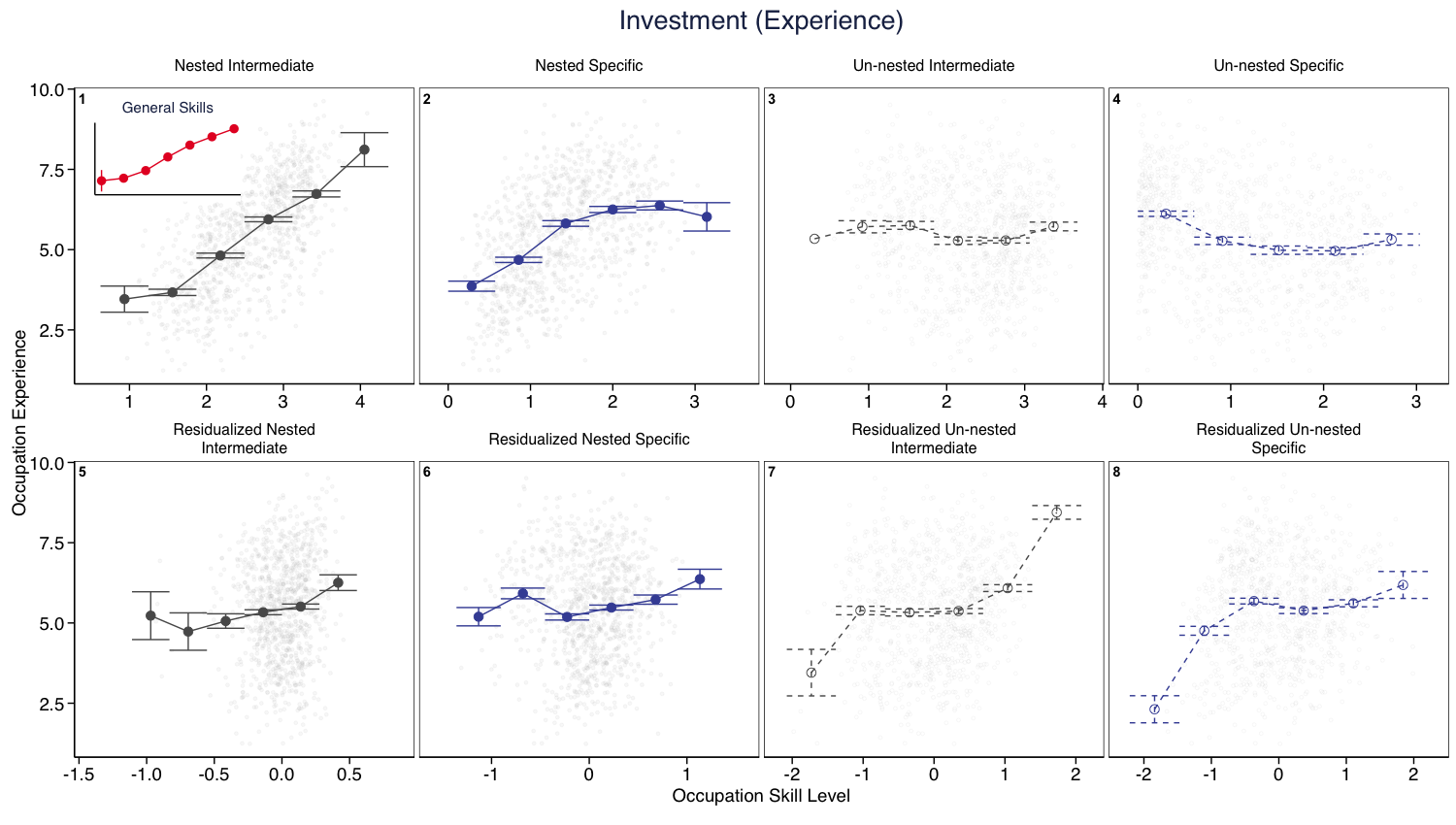}
    \caption{\textbf{Relationship between Occupations' Workplace Experience and Skill Subtypes.}}
    \label{fig:SI_experience_skill_level}
\end{figure*}

\begin{figure*}[!h]
    \centering
    \includegraphics[width=\textwidth]{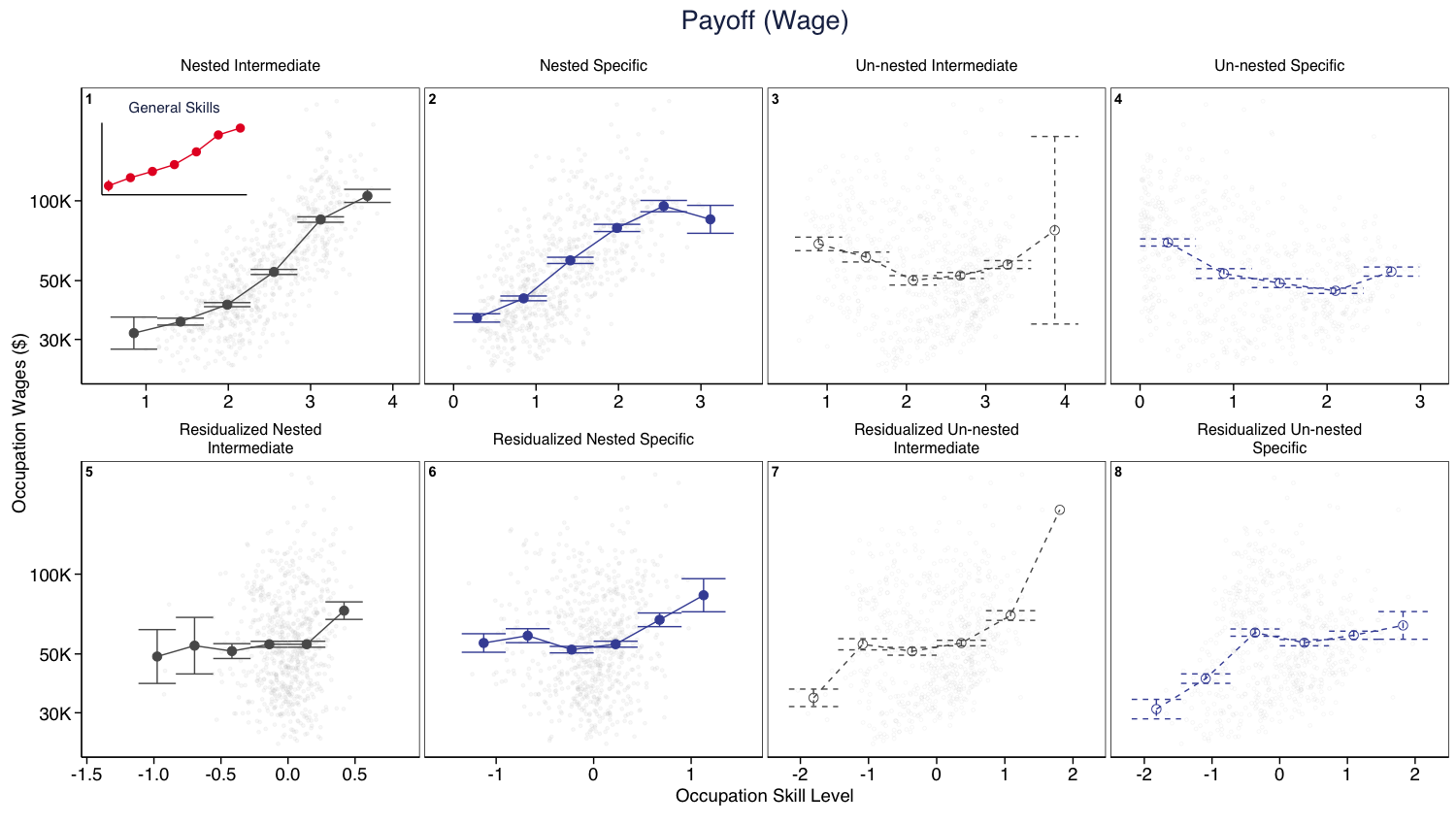}
    \caption{\textbf{Relationship between Occupational Wages and Skill Subtypes.}}
    \label{fig:SI_wage_skill_level}
\end{figure*}

Figures \ref{fig:SI_education_skill_level}-\ref{fig:SI_wage_skill_level} capture a similar analysis to the main Fig.~\ref{fig:Wage} for all skill subtypes, separating the relationship between skills and occupation educational requirement, occupation workplace experience, and wages, respectively.
In each figure, the upper panel depicts the bivariate relationship between each nested or un-nested and intermediate or specific skill subset and a corresponding work measure (educational requirement, workplace experience, and wages).
The inset shows the relationship between general skills.
The lower panels control for general skills when regressing the work measure on the corresponding skill subset. The residualized form shows the partial association between the skill subtype and work measure.
For nested skills, the relationship with education weakens but for experience and wages it almost disappears, consistent with the main text and our intuition that general skills derive a large part of the signal.
For un-nested skills, the predominantly negative relationships reverse to modest positive, consistent with the intuition that cetris paribus, un-nested skills behave as if human capital.
Hence, they require training, accumulate experience, and contribute to wages, albeit modestly.

\begin{figure*}[!h]
    \centering
    \includegraphics[width=\textwidth]{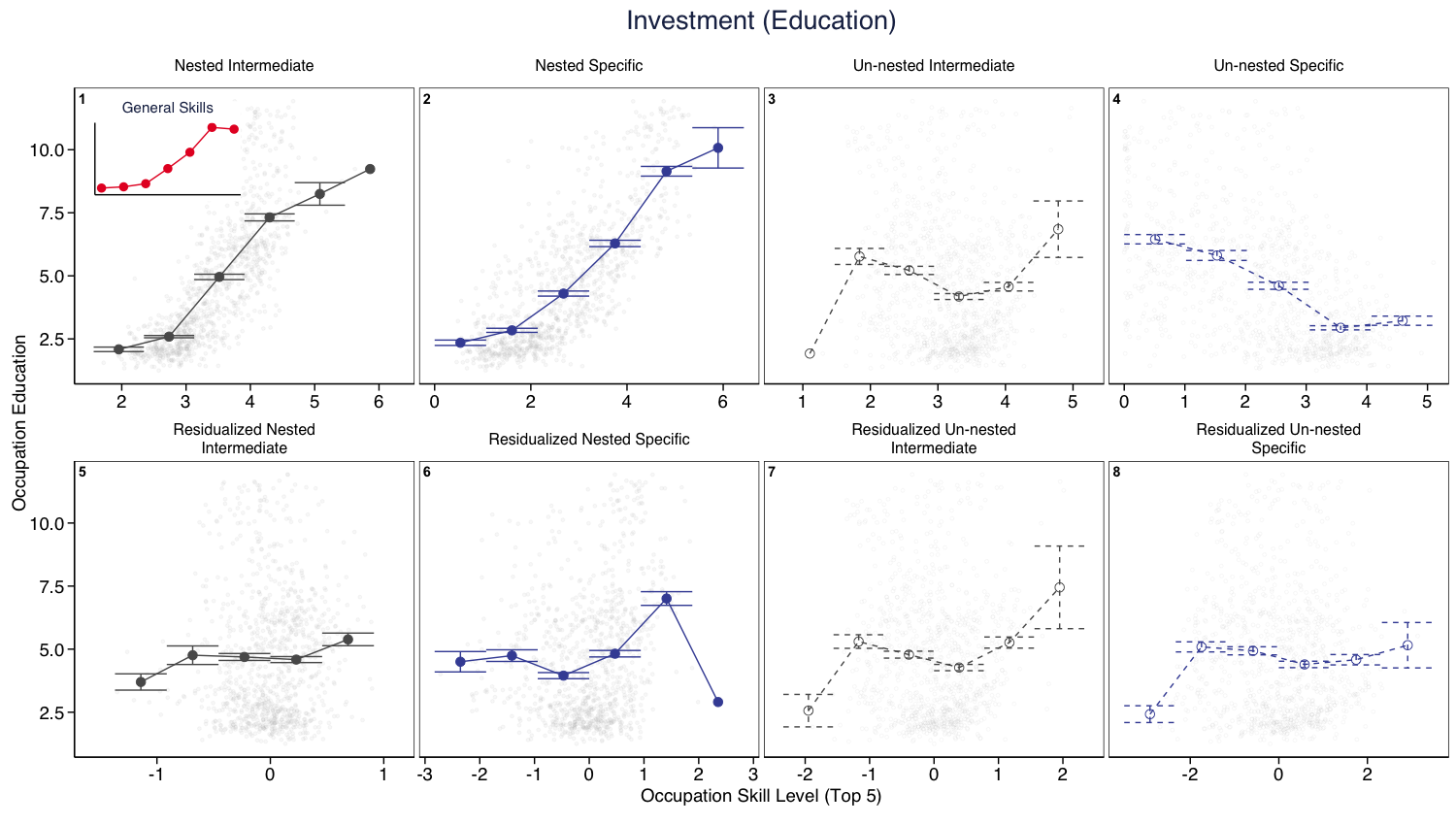}
    \caption{\textbf{Relationship between Educational Requirement and Occupation's Top 5 Skills in Subtypes.}}
    \label{fig:SI_education_skill_level_top5}
\end{figure*}

\begin{figure*}[!h]
    \centering
    \includegraphics[width=\textwidth]{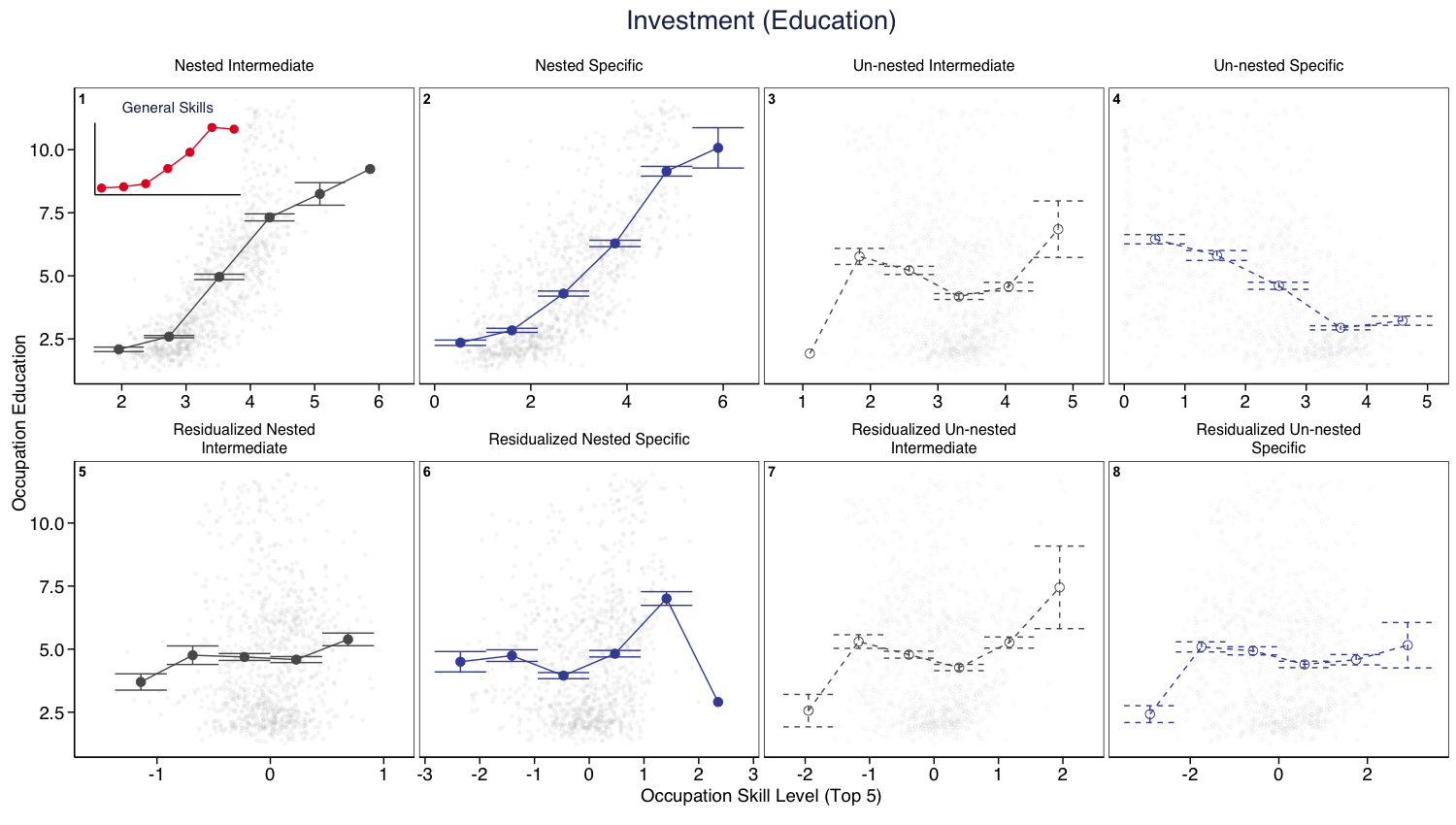}
    \caption{\textbf{Relationship between Workplace Experience and Occupation's Top 5 Skills in Subtypes.}}
    \label{fig:SI_experience_skill_level_top5}
\end{figure*}

\begin{figure*}[!h]
    \centering
    \includegraphics[width=\textwidth]{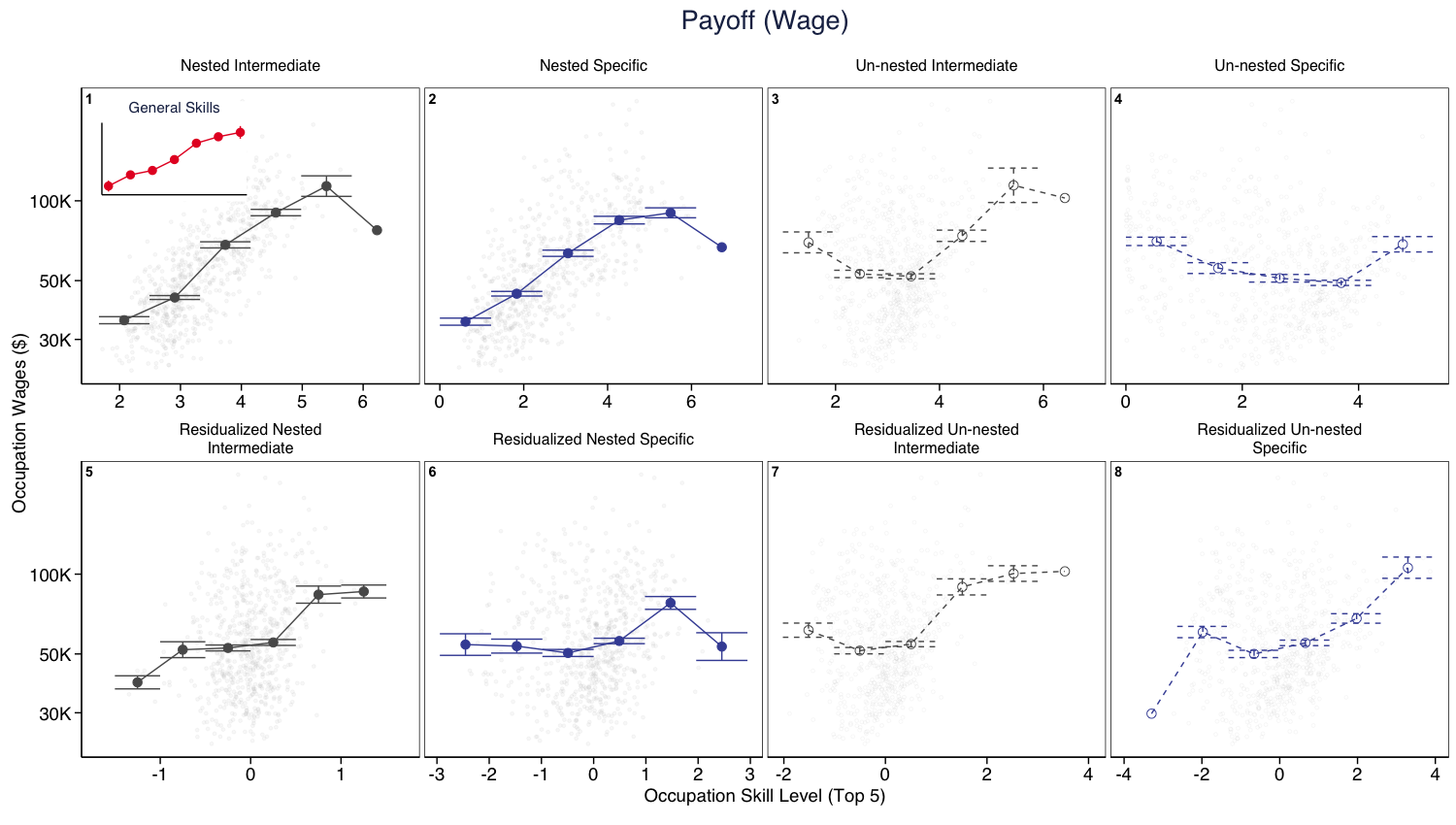}
    \caption{\textbf{Relationship between Wages and Occupation's Top 5 Skills in Subtypes.}}
    \label{fig:SI_wage_skill_level_top5}
\end{figure*}

Figs. \ref{fig:SI_education_skill_level_top5}-\ref{fig:SI_wage_skill_level_top5} repeat the above analyses with the minor difference that the skill level is calculated not as the average of all skills that belong to a subtype, but as the average level of each occupations' top 5 skills in each skill category.
The nature of the relationships is robust to this change— while slopes vary modestly.

Table \ref{tab:wage reg on skill endowment} supplements previous figures by comparing the partial effect of nested and un-nested categories for each skill group on wages. It also introduces conventional control variables of human capital, such as education, experience, and training.
The slopes are consistent with previous results and are robust (both statistically and in magnitude) to adding human capital controls.
Note that we do not run a regression including all subtypes because of the biases introduced by adding pre-treatment variables— general skills are prerequisites to nested skills. 

\begin{table}[!h]
    \centering
    \caption{\textbf{Wage Regression on Skill Endowment.}}
    \resizebox{\columnwidth}{!}{
    \input{Nature_HB_2023/tabsNHB/Jul_15_2023__Wage_Regression_on_Skill_Endowments.tex}}
    \label{tab:wage reg on skill endowment}
\end{table}

\newpage
\subsection{Automation Risk and Skills}

\begin{figure*}[!h]
    \centering
    \includegraphics[width=\textwidth]{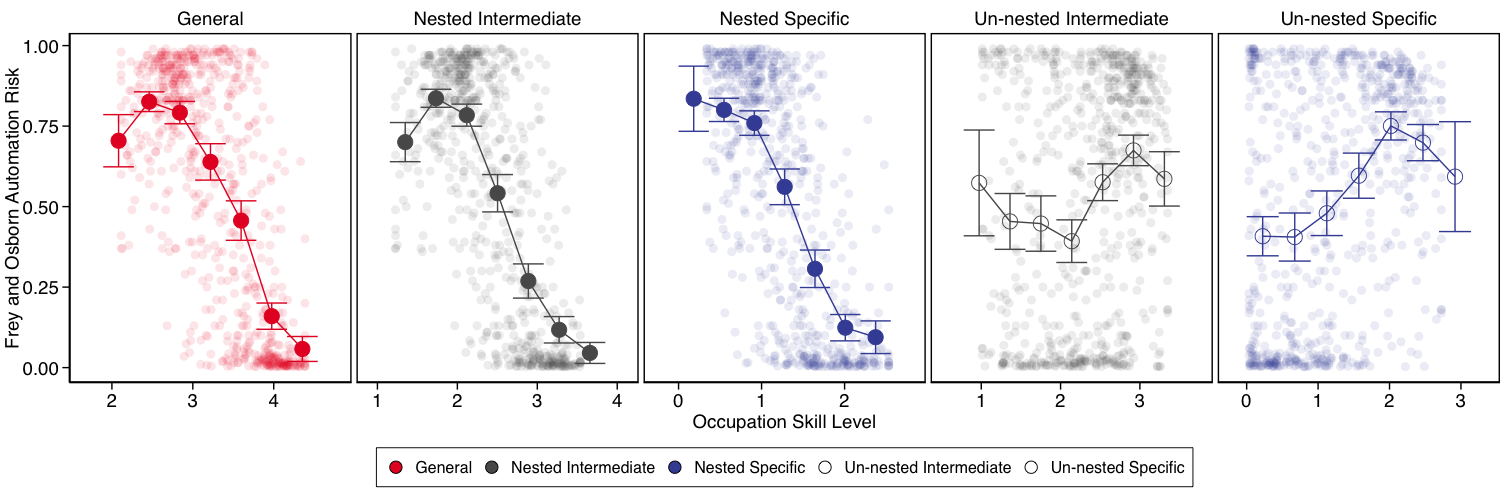}
    \caption{\textbf{Occupations' Automation Risk Index Against their Average Levels in each Skill Category.}}
    \label{fig:occ_FOautomation_skill}
\end{figure*}

Given the broad interest in understanding human capital and automation risk, we plot occupations' automation risk index \cite{Frey2017} against their average levels in each skill category in Fig.~\ref{fig:occ_FOautomation_skill}.

\subsection{Skill Payoffs for Different Occupations}

\begin{figure*}[!h]
    \centering
    \includegraphics[width=\textwidth]{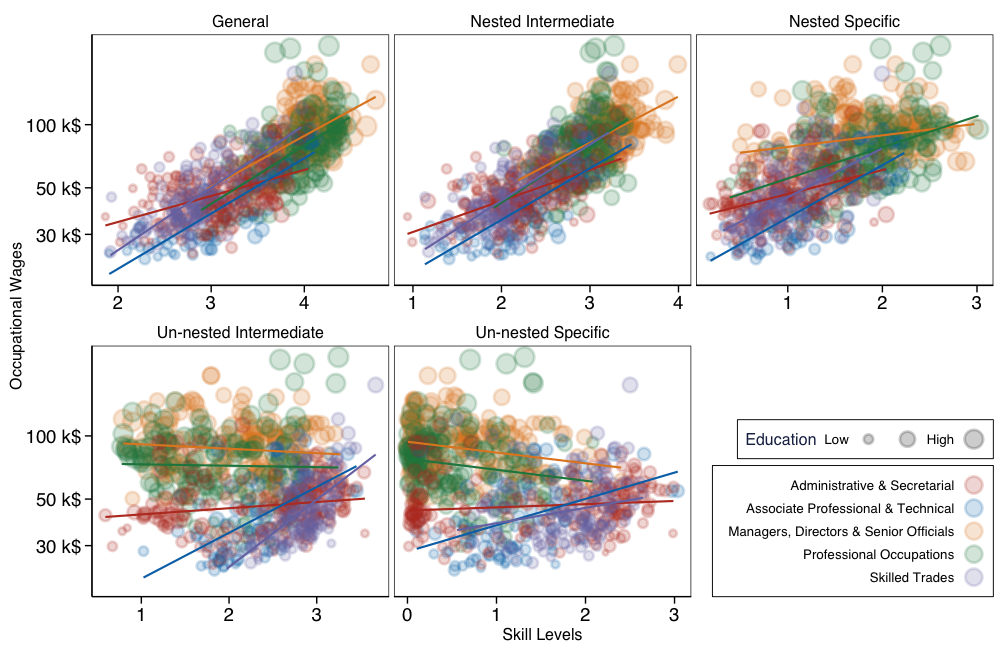}
    \caption{\textbf{Wage Returns to Different Types of Skill Endowment for each Major Occupational Group}. Each point corresponds to an occupation. The setup supplements the main Fig.~\ref{fig:Wage}, highlighting the benefits of higher levels of nested skills. However, un-nested skills only improve wages of Professional occupations and Skilled traders. This in itself underpins multi-dimensionality skills.}
    \label{fig:Figure 3 full | major occupation groups}
\end{figure*}

Fig.~\ref{fig:Figure 3 full | major occupation groups} relates returns to skills for each major occupational group— 1-digit SOC. The key pattern is that all occupational groups, despite varying in their skill endowments, benefit from higher levels of nested skills. 
However, un-nested skills only improve wages of Professional occupations and Skilled traders. 

Interestingly, managerial occupations command high general skills.
Section \ref{sec:robustness check: no managers} of the supplementary document examines (and finds evidence against) the possibility that the returns to general skills are largely a managerial phenomenon.

\subsection{Skill Investment and Payoffs in 2005}

\begin{figure*}[!h]
    \centering
    \includegraphics[width=\textwidth]{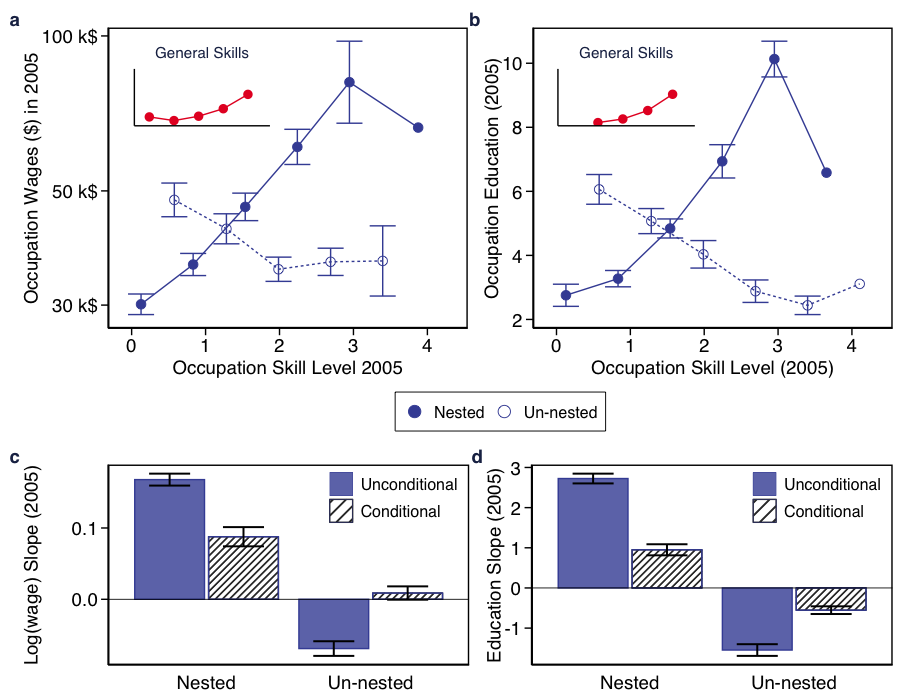}
    \caption{\textbf{Investment and Payoffs of Different Specific Skills in 2005.}}
    \label{fig:Wage and education 2003}
\end{figure*}

In Fig.~\ref{fig:Wage and education 2003}, we repeat our analysis of investment and payoffs to skills (main Fig.~\ref{fig:Wage}) for 2005, finding results consistent with the growing importance of general skills.
The figures show lower associated education and payoffs to general skills than the main figure (\ref{fig:Wage}).

\clearpage
\section{Skills' Geographic Distribution} \label{section: add - geographical distribution of skills}

An in-depth analysis of how skills interface with urban growth is beyond the scope of this work, we provide a brief descriptive analysis, here. Overall, urban areas are more endowed with general skills.
In contrast, rural areas are less likely to carry general skills.
This is consistent with the concentration of more innovative and complex economic activity \cite{Hong2020, Balland2020} and the concentration of managerial and administrative occupations in larger cities.
We test and find support for the hypothesis that skills, in particular concentration of general skills, explain away part of the urban wage premiums.
Upon grouping cities by manufacturing employment relative to the national average, we find that cities highly specialized in manufacturing tend to exhibit lower levels of nested specialization but higher levels of unnested specializations (Fig.~\ref{fig:skill_and_manufacturing_full}).
This shows that cities indeed specialize in distinct directions. Interestingly, skill patterns shift in a non-linear fashion across cities with increasing concentrations of manufacturing employment.
Both a strong dependence on and a complete absence of manufacturing correlate with adverse skill bases, i.e., skill bases dominated by unnested skills and a lower prevalence of general and nested skills. Conversely, skills that typically command high wage premiums are overrepresented in cities with intermediate levels of manufacturing activity.

\subsection{Counties' Skill Endowments}
Using the occupational employment for Metropolitan and nonmetropolitan areas\footnote{\tiny\url{https://www.bls.gov/oes/}} published by the Bureau of Labor Statistics (BLS), one can map the geographical distribution of skills.
BLS uses Core-based Statistical Areas (CSAs) as geographic units, which are more coarse than the county level.
US counties follow the Federal Information Processing System (FIPS) taxonomy.
To obtain employment at the level of FIPS and map skill information onto US counties, we used a crosswalk also provided by BLS \footnote{\tiny\url{https://www.bls.gov/oes/current/msa_def.htm}}.
We aggregate occupation skills at the level of \textit{general, nested intermediate and specific, and un-nested intermediate and specific}.
Taking an average for each US county using the county employment of occupations as weights, we derive a regional measure of skill endowment for each skill sub-type.

\begin{figure}[ht]
    \centering
    \includegraphics[width=\textwidth, trim={5cm 0 2cm 2cm},clip]{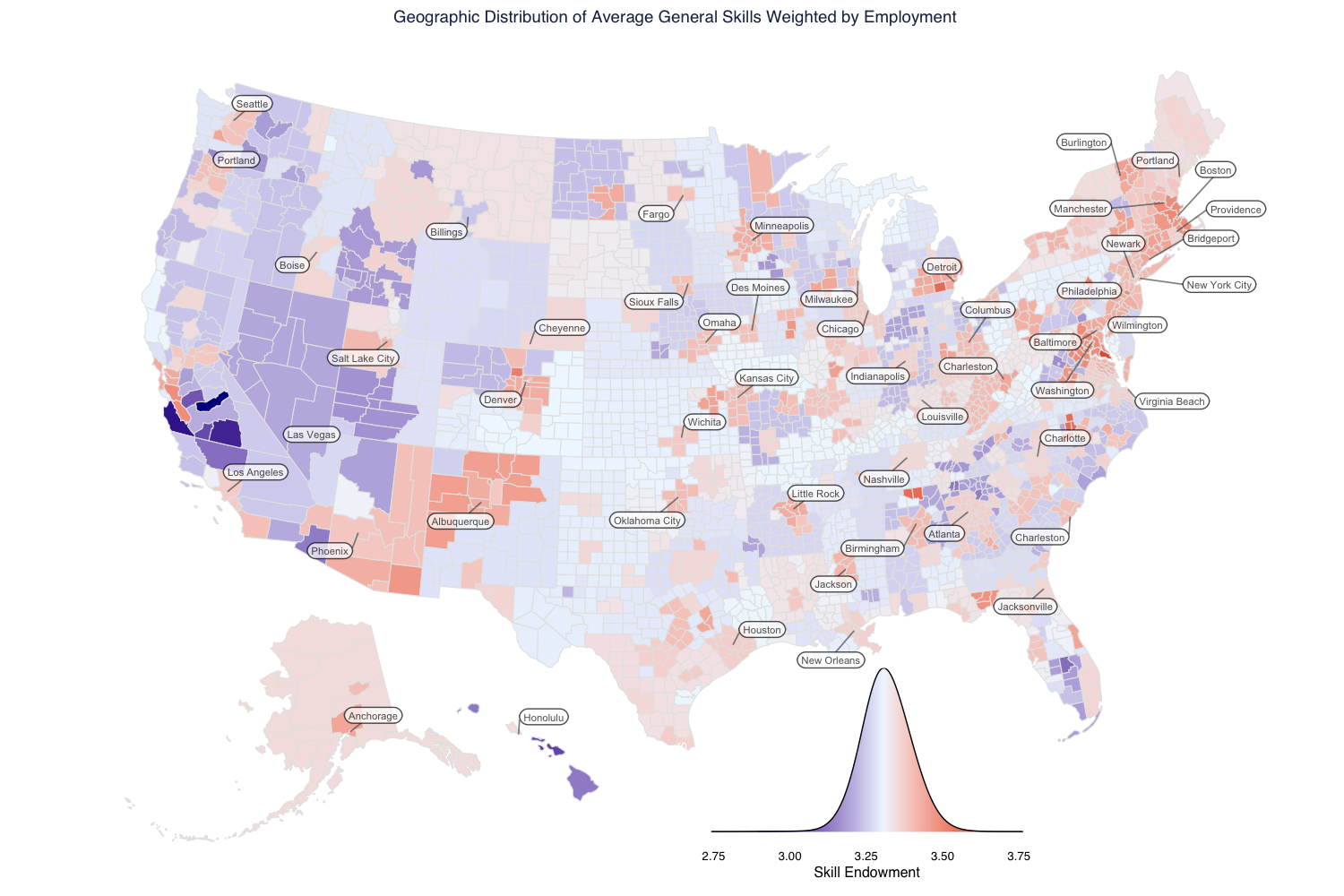}
    \caption{\textbf{Geographic Distribution of Average General Skills Weighted by Employment.} The bottom histogram shows the distribution of the corresponding stock of skills across the US economy. The y-axis shows the number of unique FIPS with the respective skill level.
    Overall, urban areas are more endowed with general skills— seen in red. In contrast, rural areas are less likely to carry general skills— seen in blue. This is consistent with the concentration of more innovative and complex economic activity \cite{Hong2020, Balland2020} and the concentration of managerial and administrative occupations in cities.}
    \label{fig:geo_dist_level_general_skills_employment_weighted}
\end{figure}

\begin{figure}[!h]
    \centering
    \includegraphics[width=.875\textwidth, trim={5cm 0 2cm 2cm},clip]{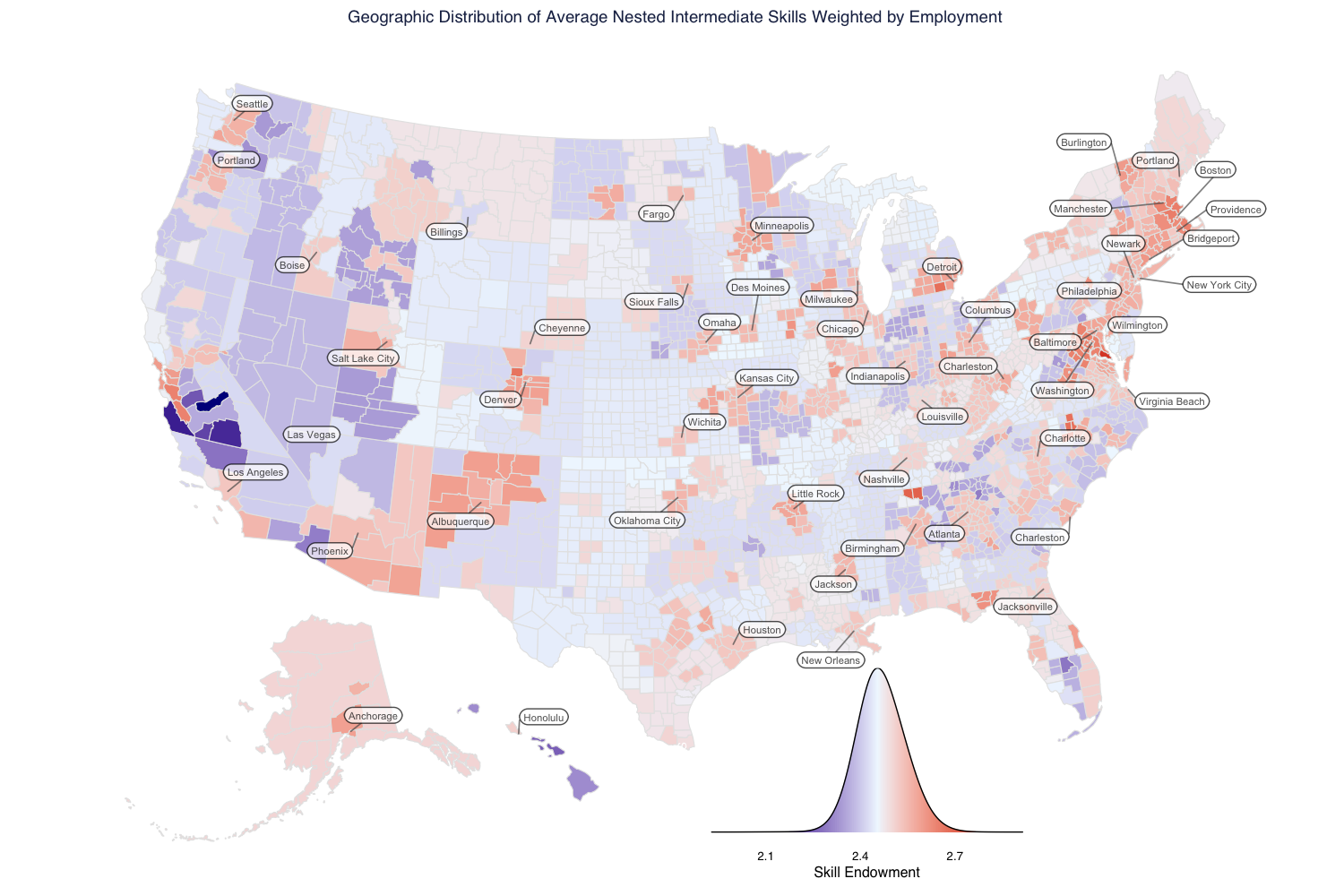}
    \caption{\textbf{Geographic Distribution of Average Nested Intermediate Skills Weighted by Employment.}
    The bottom histogram shows the distribution of the corresponding stock of skills across the US economy. The y-axis shows the number of unique FIPS with the respective skill level.}
    \label{fig:geo_dist_level_nested_common_skills_employment_weighted}
\end{figure}

\begin{figure}[!h]
    \centering
    \includegraphics[width=.875\textwidth, trim={5cm 0 2cm 2cm},clip]{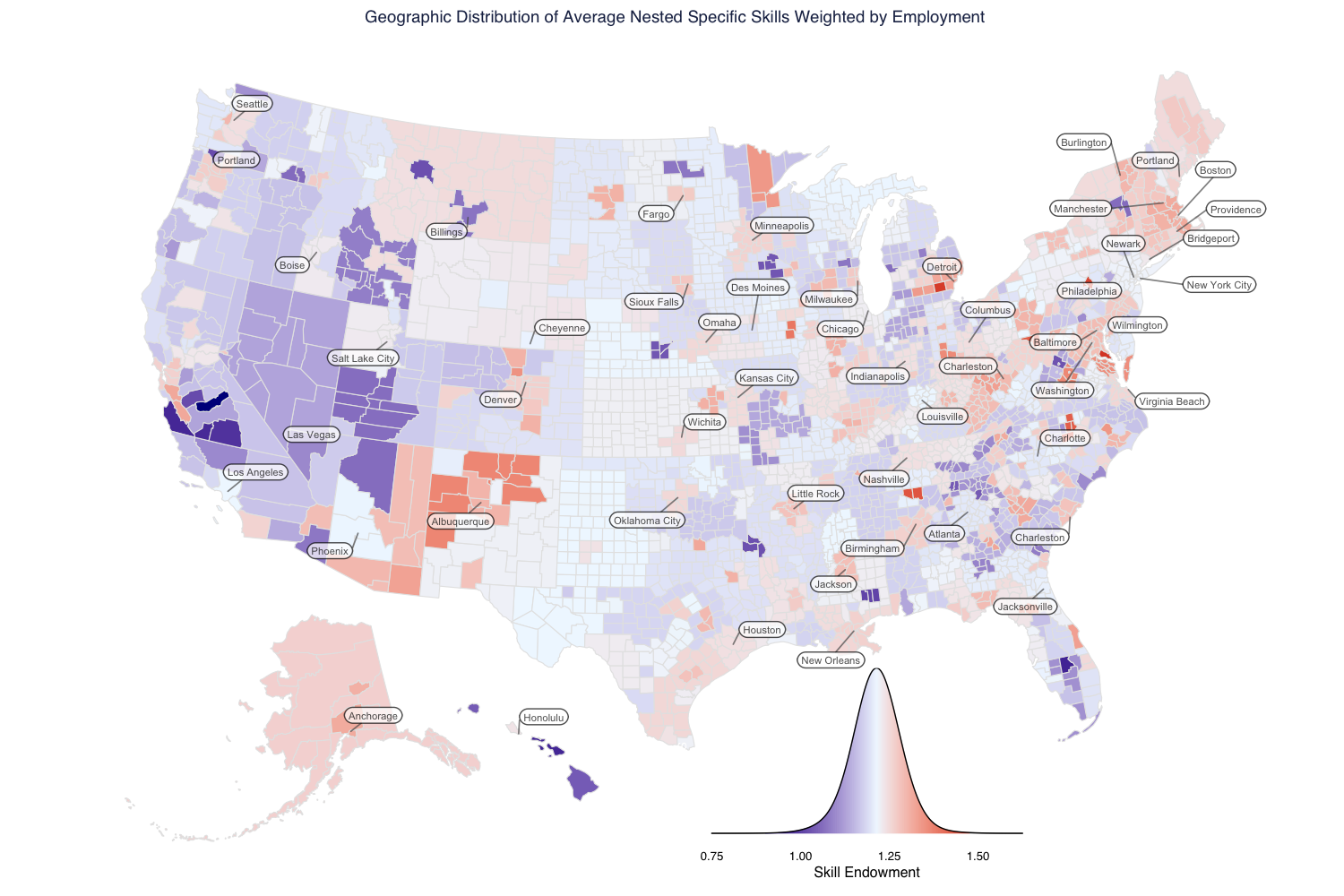}
    \caption{\textbf{Geographic Distribution of Average Nested Specific Skills Weighted by Employment.}The bottom histogram shows the distribution of the corresponding stock of skills across the US economy. The y-axis shows the number of unique FIPS with the respective skill level.}
    \label{fig:geo_dist_level_nested_specific_skills_employment_weighted}
\end{figure}

\begin{figure}[!h]
    \centering
    \includegraphics[width=.875\textwidth, trim={5cm 0 2cm 2cm},clip]{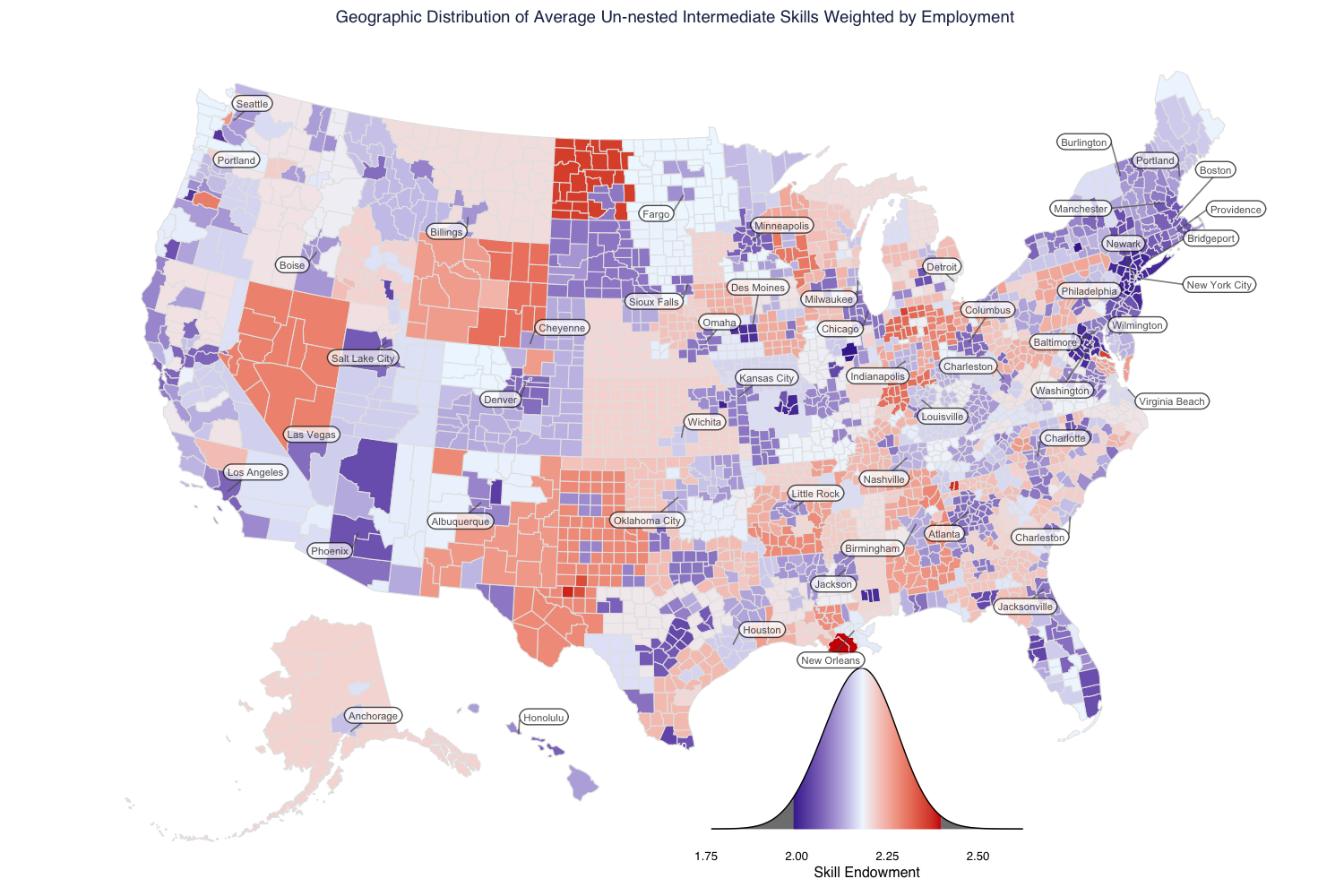}
    \caption{\textbf{Geographic Distribution of Average Un-nested Intermediate Skills Weighted by Employment.}The bottom histogram shows the distribution of the corresponding stock of skills across the US economy. The y-axis shows the number of unique FIPS with the respective skill level.}
    \label{fig:geo_dist_level_common_specific_skills_employment_weighted}
\end{figure}

\begin{figure}[!h]
    \centering
    \includegraphics[width=.875\textwidth, trim={5cm 0 2cm 2cm},clip]{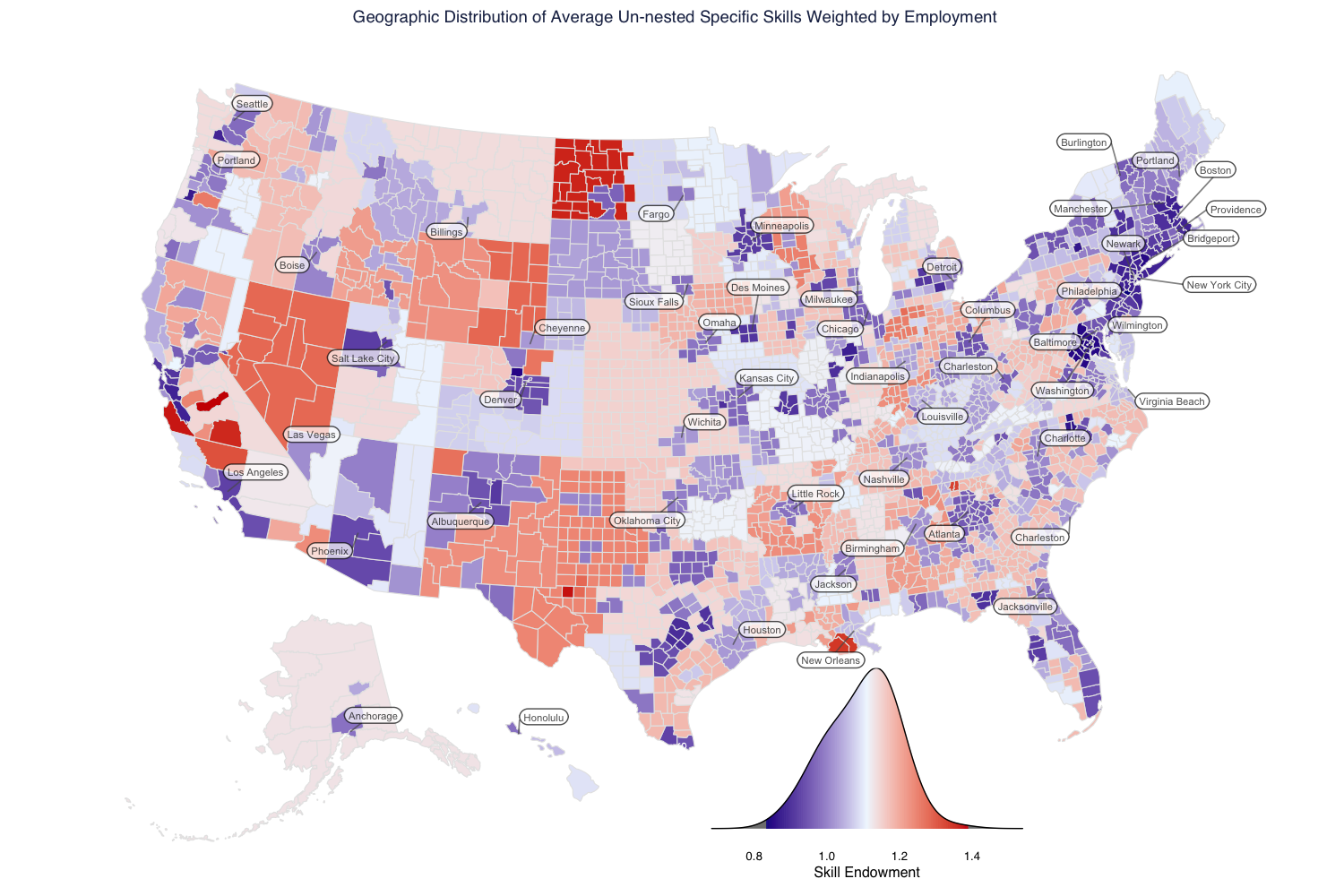}
    \caption{\textbf{Geographic Distribution of Average Un-nested Specific Skills Weighted by Employment.}The bottom histogram shows the distribution of the corresponding stock of skills across the US economy. The y-axis shows the number of unique FIPS with the respective skill level.}
    \label{fig:geo_dist_level_un-nested_specific_skills_employment_weighted}
\end{figure}

Overall, our analysis (Figs.~\ref{fig:geo_dist_level_general_skills_employment_weighted} through \ref{fig:geo_dist_level_un-nested_specific_skills_employment_weighted}) show a clear concentration of general skills in densely populated urban areas, reflecting the diverse and complex economic activities found in these locales \cite{Glaeser1999, Wheeler2001, Youn2016, gomez2016explaining, Hong2020, Balland2020, Bettencourt2014, Gomez-Lievano2021}. Large cities tend to have higher levels of general and nested skills (also seen in Fig.~\ref{fig:skill_and_population_full}).
For instance, New York and Washington D.C. harbor significant financial and state employment.
Moreover, even in states with comparatively rural structures, such as Indiana, Iowa, Nebraska, and Kansans, state capitals, where the local state is likely to reside, command a higher level of general skills— than their neighboring counties.
A secondary driver of the abundance of general skills in urban areas is the specialization needed for accomplishing complex economic tasks.
For instance, Boston, Seattle, and San Francisco (the latter not shown on the map) are tech hubs and command a strong stock of general (, and as seen in Fig.~\ref{fig:geo_dist_level_nested_specific_skills_employment_weighted}, specific) skills\footnote{
Finer-grained insights can also be obtained from these maps.
For instance, the most extreme concentration of general skills (or lack thereof) is observed in less populated cities that are specialized in a certain industry.
The significant proportion of the focal industries' workers relative to the total employment highlights the skills used by those workers.
The most extreme concentrations of general skills (or lack thereof) are observed in less populated cities that are specialized in a certain industry.
The significant proportion of the focal industries' workers relative to the total employment highlights the skills used by those workers.
The five most and least endowed counties with general skills are shown on the map— as italicized text.
For instance, St. Mary County (Maryland) is an air force and aerospace hub with companies such as Lockheed Martin and Boeing, and military naval air station Patuxent River among the top employers.
Another example is Chatham and its neighboring counties, Durham (hosting Duke University), Orange (hosting the University of Carolina at Chapel-Hill), and Person, which have fostered one of the fastest growing tech sectors in the US, earning the nickname of \textit{Research Triangle}.
Other notable concentration points of general skills are Limestone and Madison (Alabama), hosting numerous aerospace and automobile manufacturing facilities, and Washtenaw county (Michigan) hosting the University of Michigan Ann Arbor and its off-sprung businesses.
In contrast, Madera (California), and its neighboring counties, Tulare, Kings, and Monterey)Highlands (Florida), Yuma (Arizona), Hall (Georgia), and Kalawao (Hawaii) are primarily designated agricultural areas, accruing unnested skills.
}.
However, the starkest disparities between smaller and larger cities are seen in the prevalence of unnested skills, which are significantly less common in cities with over a million inhabitants, a known threshold for cities transitioning towards innovative economic specializations \cite{Hong2020}.
While most workers with nested skills need general s kills, the concentration of managerial and other supporting roles also needs high levels of general skills.
Hence, examine and find evidence consistent with the hypothesis that the accumulation of general skills indeed explains part of the value generated in large cities (Tab.~\ref{tab:urban wage premium}).

\begin{table}
    \centering
    \caption{\textbf{General Skills Explain Urban Wage Premiums.}}
    \resizebox{\columnwidth}{!}{
    \input{Nature_HB_2023/tabsNHB/Mar_25_2024__urban_wage_premiums.tex}}
    \label{tab:urban wage premium}
\end{table}

\subsection{Skills and Population}
\begin{figure*}[!h]
    \centering
    \includegraphics[width=\textwidth]{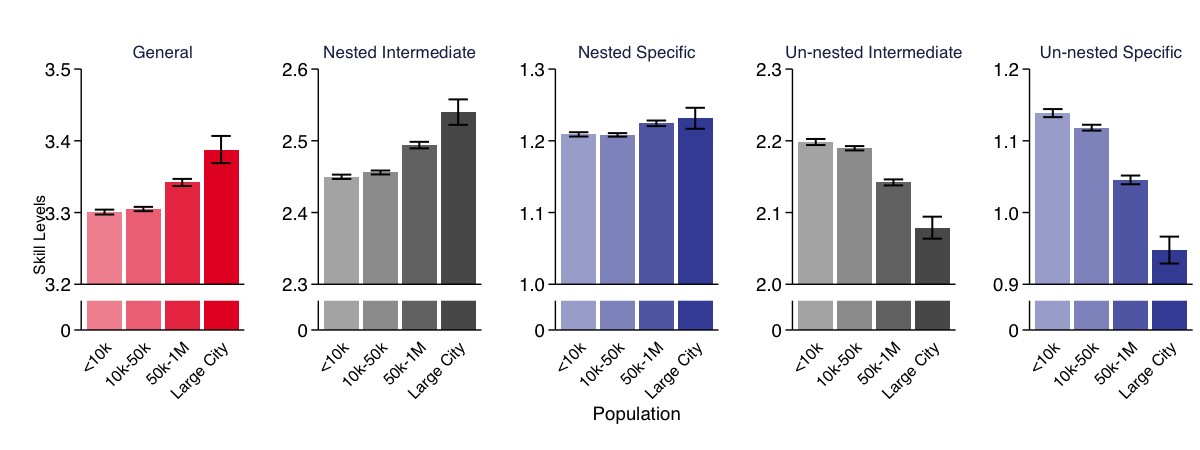}
    \caption{\textbf{Population Size and Skills}
    The figure shows for cities of different size the levels of all our skill categories, highlighting the statements for nested and un-nested specific skills also hold for the corresponding intermediate skills.}
    \label{fig:skill_and_population_full}
\end{figure*}

We divide cities into four mutually exclusive groups by population (below 10 thousand, below 50 thousand, below 1 million, and more than a million inhabitants) based on 2010 Census population estimates.
Skill endowment for each city group is taken as the average of counties, and 95\% confidence intervals are shown.
Fig.~\ref{fig:skill_and_population_full} shows for cities of different size the levels of
all skill categories.

We also test the hypothesis that the accumulation of general skills indeed explains part of the value generated in cities \cite{Lucas1988, Bacolod2009, gomez2016explaining}.
we test that hypothesis directly by utilizing the CBSA size (CBSASZ) variable from CPS microdata, which carries information about the size of the metropolitan area in which the surveyed individual resides (since 2004).
The values range from 0: areas of $<100,000$ inhabitants that do not meet the threshold of a metropolitan area to 6: over 5 million inhabitants. We transform these brackets to cities below and above 1M population \cite{Hong2020}.

In the model (1) of Tab.~\ref{tab:urban wage premium}, we first regressed the log wage reported by individuals to CPS on the size of the metropolitan area in which they reside, obtaining partial correlations that signify the urban wage premiums (the baseline is areas of $<1M$ inhabitants.) In the second model of the table, we add general skills of individuals (which we obtain from matching to O*NET the occupation associated with each individual in the CPS microdata). That means that large cities tend to have more people in occupations with general skills. This bias toward more general-skill intensive activities explains over one-third of the urban wage premiums \cite{Lucas1988, Glaeser2001, Bacolod2009, Gomez-Lievano2021}. Adding nested and un-nested specific skills first without and then with general skills in models 3 and 4, respectively, have similar effects.

\subsection{Skills and Manufacturing Industries}
\begin{figure*}[!h]
    \centering
    \includegraphics[width=\textwidth]{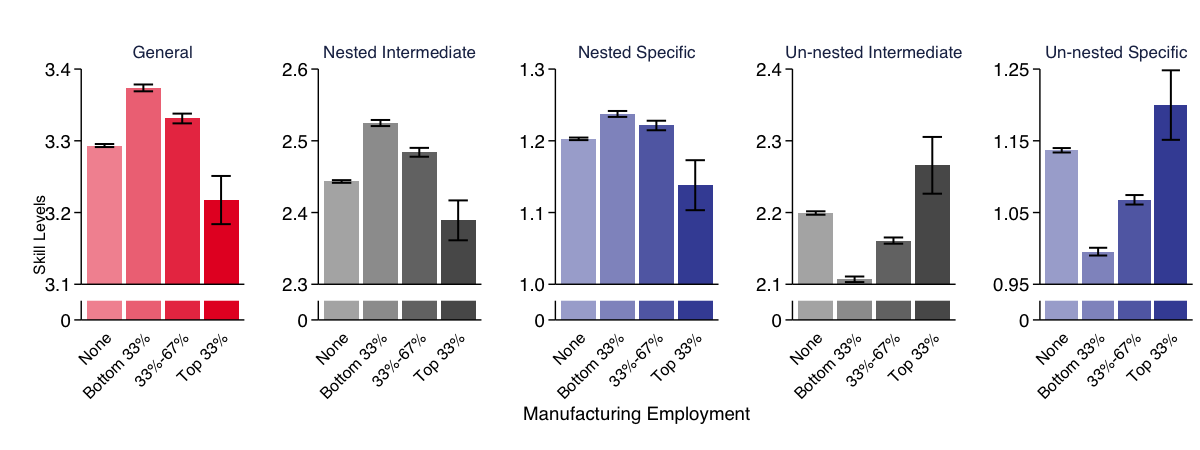}
    \caption{\textbf{Intensity of Manufacturing Industries and Skills}
    The figure shows for cities of different manufacturing concentrations the levels of all skill categories, highlighting the statements for nested and un-nested specific skills hold also for the corresponding intermediate skills.}
    \label{fig:skill_and_manufacturing_full}
\end{figure*}

We divide cities into four mutually exclusive groups based on the intensity of their manufacturing industries.
We use US Census County Business Patterns from 2019 that report industry employment for metropolitan areas to quantify manufacturing presence.
At the 2-digit naics codes, we take 31-33 as manufacturing industries and calculate the location quotient of manufacturing employment (the ratio of manufacturing employment from the metro area total employment over the nationwide ratio).
Matching metro areas to counties, we designate counties with no manufacturing employment to group "None", and group the rest based on quotient 33\% and 66\% quantiles of the measure into bottom, middle, and top.
Fig.~\ref{fig:skill_and_manufacturing_full} shows for cities of different manufacturing concentrations the levels of
all skill categories.

\clearpage
\section{Skills' Demographic Distribution} \label{section: add - demographic distribution of skills}

Using the CPS household data between 1980 and 2022, we derive the skill endowment across racial (White, Black, Hispanic/Latinx, and White) and gender (Female and Male) groups in each skill category.
Restricting to full-time workers employed at the time of the survey, who are between 18 and 55,
we apply the mentioned features to examine the prevalence of skills among individuals of different gender and racial groups.
Individuals' skills are infered based on their coded occupations in the CPS data by linking it to the occupational skill requirement in O*NET.
The two datasets, however, use different occupational taxonomies.
As a result, one needs to map CPS and O*NET occupations.
We use a crosswalk offered by BLS\footnote{\tiny\url{https://www.census.gov/topics/employment/industry-occupation/guidance/code-lists.html}}, which maps a CPS occupation to 542 out of 968 occupations in O*NET 8-digit SOC codes.
Note that CPS offers various racial categories. We use Whites, Blacks, and Asians, which constitute the bulk of the sample.
CPS data also contains a separate (from race) variable for identifying Hispanic individuals.
We create a fourth racial category for Hispanics and associate any individual of Hispanic background with that category.
Next, we calculate the endowment of each skill category for each of the resulting four demographic categories and (binary) gender groups.

\begin{figure}[!h]
    \centering
    \includegraphics[width=\textwidth]{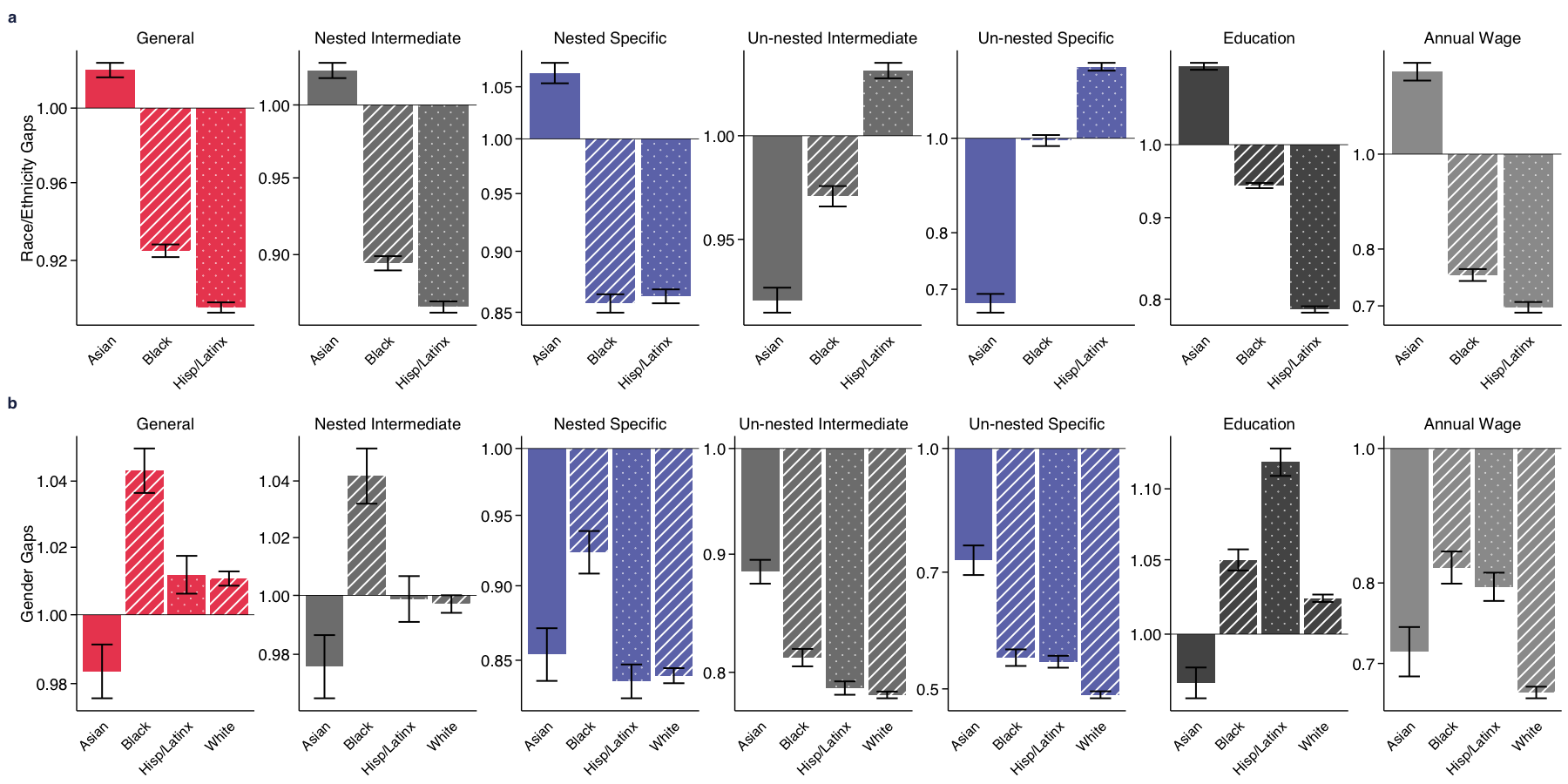}
    \caption{\textbf{Skill Disparity in Demographic Distribution} of race/ethnicity and gender adding the information on intermediate skills and annual wage.}
    \label{fig:racial and gender skill endowments}
\end{figure}

Figure~\ref{fig:racial and gender skill endowments} replicates the main Fig.~\ref{fig:Figure 7}, adding the information on intermediate skills and annual wage. We omit the weekly wage results for brevity.

\begin{figure}[!h]
    \centering
    \includegraphics[width=\textwidth]{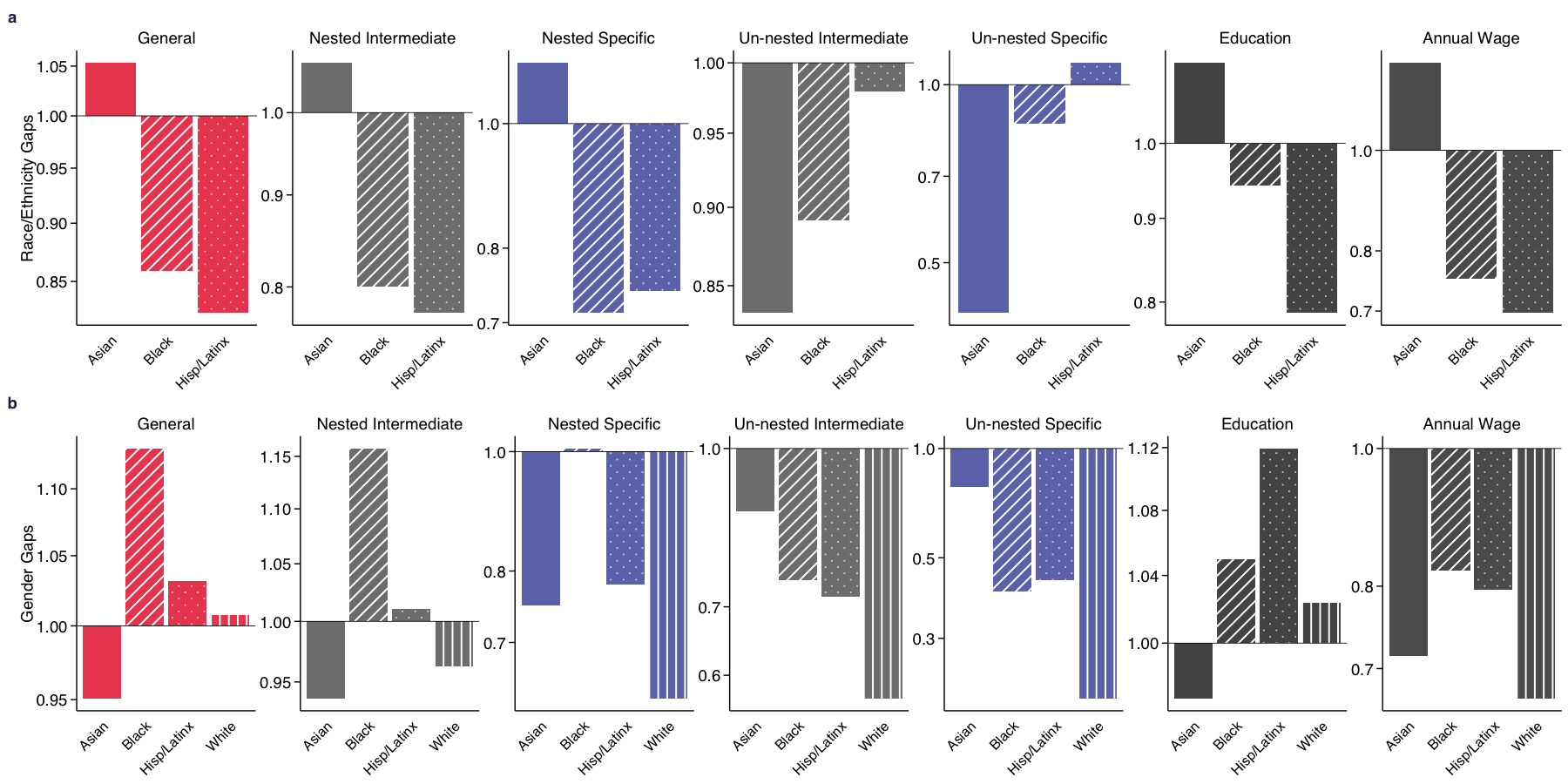}
    \caption{\textbf{Skill Disparity in Demographic Distribution} of race/ethnicity and gender with an alternative aggregation. Similar to our main Fig.~\ref{fig:Figure 7}, we use CPS micro data, however, follow the aggregation of \cite{Tong2021}. The results are consistent with our main figure.}
    \label{fig:Tong et al race gender skill distribution}
\end{figure}

As a robustness check, we used a different measurement of skills for demographics and found similar results, following Tong et al. \cite{Tong2021}.
They group occupations of different skill levels by corresponding workers’ dominant gender and race/ethnicity and calculate skill endowment across occupations from the same group.
In determining occupations’ “dominant” demographic characteristics, we link an occupation to a racial/gender group if it is 1.5 times or more likely to be employed in the focal occupation than its fraction in the sample.
We then aggregated skill endowments across racial and gender categories and show the results in
Fig.~\ref{fig:Tong et al race gender skill distribution}.
The results are consistent with our main Fig.~\ref{fig:Figure 7} and SI Fig.~\ref{fig:racial and gender skill endowments}.

\subsection{Parenthood and the Diverging Skills of Male and Female Workers} \label{supsec:Parenthood_Male_vs_Female}

\begin{figure*}[!h]
    \centering
    \includegraphics[width=\textwidth]{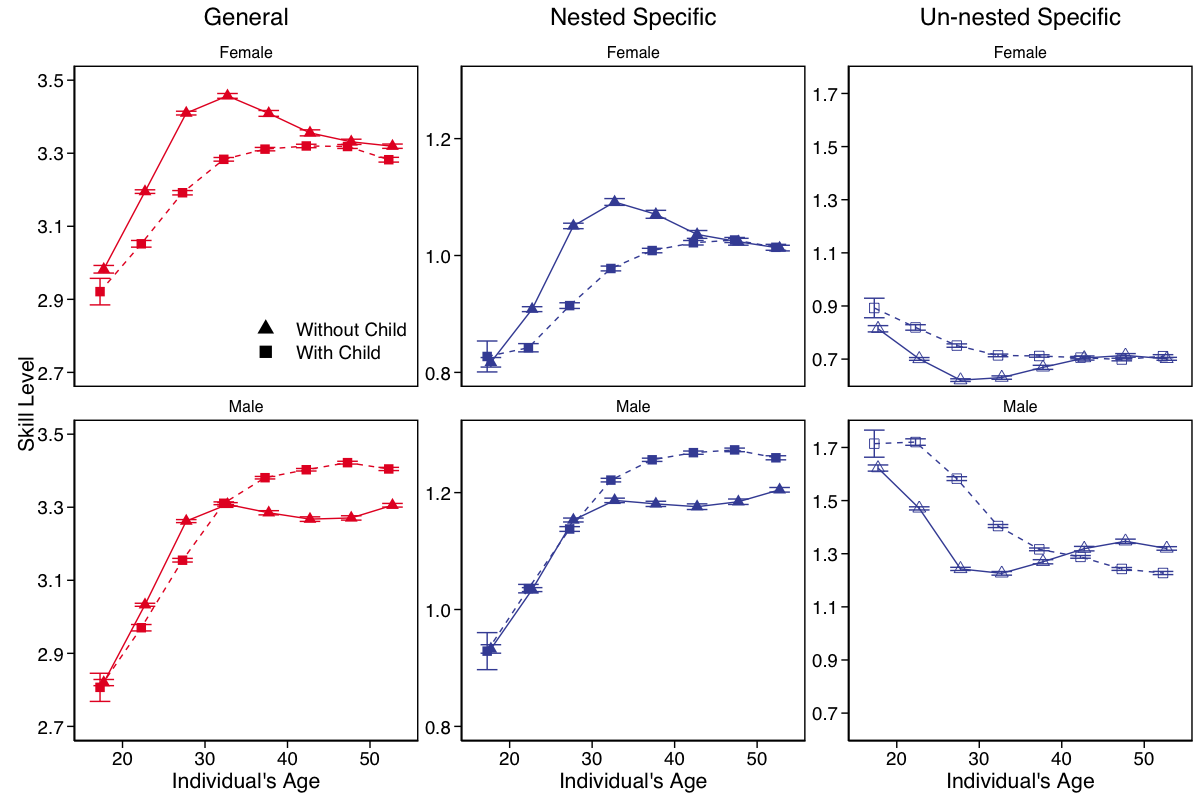}
    \caption{\textbf{Parenthood and the Diverging Skills of Male and Female Workers}.
    We track the skills manifested in the occupational compositions of birth cohorts as they age, splitting individuals based on their binary gender (Make/Female) and whether they lived with children at the time of the survey, obtaining the below figure. Each column shows the levels of a certain category of skills, while the rows show the results for men and women. The solid line (and triangles) show the pattern for people without children, while the dashed line (and squares) show the pattern for individuals with children, highlighting the drop in general and nested skills of mothers.
    }
    \label{fig:Parenthood_Male_vs_Female}
\end{figure*}

An intriguing pattern in the main Figs.~\ref{fig:age}~(d-e) is the diverging general and nested skills of men and women around the age of 30, when one expects some individuals to become parents.
Utilizing the number of children in the surveyed households recorded by CPS microdoa, we split our birth cohort sample into individuals \textit{with} and \textit{without} children. 
We replicate the analysis of the main Fig.~\ref{fig:age}~(d-f) by tracking the skills manifested in the occupational compositions of birth cohorts as they age, splitting individuals based on their binary gender (Male: lower panels; Female: upper panels of the below figure) and their binary parental status (with child: square; without child: triangle in the below figure) at the time of the survey.
Fig.~\ref{fig:Parenthood_Male_vs_Female} shows the result of aggregating skills for each subgroup. Each column shows the levels of a certain category of skills, while the rows show the results for a gender.
The solid line (and triangles) show the pattern for people without children, while the dashed line (and squares) show the pattern for individuals with children.

There is a pronounced gap in the general and nested skills between women with and without children. Please note that the later convergence is likely to arise from the fact that at higher ages, the “without" subgroup will mix families who never had any children with families whose children have already left the household.
In the latter families, caregivers may have been disadvantaged in their early careers, leading to lower skill levels at higher ages.
Contrary to the negative correlation with general and nested skills, women with children appear to sort into jobs that require higher un-nested skills (the SI Sec. \ref{supsec:female job sorting} offers partial evidence of female job sorting).
Interestingly, men with children tend to do better.
Especially men in jobs that require general and nested skills tend to be more intriguing and for longer periods compared to their counterparts without children.
The latter pattern for men may arise from sample selection effects or from the fact that the cost of raising children incentivizes acquiring skills that lead to better-paid careers.
Synthetic birth cohorts are not ideal data for this purpose, as they do not allow for tracking individuals over time. However, it is reasonable to believe this approach offers unbiased estimates of the population behavior.

\subsection{Gender and Jobs Sorting} \label{supsec:female job sorting}

Another intriguing pattern in the main Fig.~\ref{fig:age}~(d-f) is the diverging patterns of skill development between men and women, wherein women exhibit high levels of general skills, surpassing their male counterparts at certain ages but do not manifest the high levels of nested skills observed for male workers of the same age.
Lower levels of nested skills for women are also seen in the first column of the regression Tab.~\ref{tab:irregular hours} that predicts the gender of workers based on their general and nested skills in our CPS sample (Female = 1): general skills are associated with greater, but nested skills with smaller shares of women in an occupation.

One explanation for this pattern \cite{Bertrand2009, Goldin2015, Canon2016} is that women may avoid jobs with irregular or long working schedules.
This implies that despite their high levels of general skills and education, women may avoid jobs that require nested skills because of the working conditions of such jobs.
To examine that hypothesis, we examined whether adding descriptors of work schedule to the same regression diminishes the correlation between skills and the gender of the worker, as reported in column 1 of Tab.~\ref{tab:irregular hours}.

\begin{table}[!h]
    \centering
    \caption{\textbf{Regression analysis of the correlation between gender, skills, and irregular and long work schedule.}
    The first column offers a baseline model that predicts the gender (Female = 1) of the worker based on general and nested skills, showing a negative correlation with nested skills. Adding descriptors of irregular and long schedules in the second model explains away part of the predictive power of nested skills for workers’ gender. As such, part of the reason why women manifest high level s of general skills but comparatively low levels of nested skills is that jobs that require the latter categories of skills likely impose long and irregular work conditions, which have been found to deter female workers.}
    \resizebox{\columnwidth}{!}{
    \input{Nature_HB_2023/tabsNHB/Mar_25_2024__Irregular_Hours__Gender.tex}}
    \label{tab:irregular hours}
\end{table}

To implement this test, we matched individuals in the Current Population Survey (CPS) aged 18 to 55 who were in the workforce between 1980 and 2020 to the following information in the O*NET using their reported occupation code: skill information (namely, general and nested specific skills) and occupational work schedule (irregularity). Work schedule irregularity is collected as a part of the O*NET work context record as a categorical variable with three levels of
\textit{Regular} (established routine, set schedule),
\textit{Irregular} (changes with weather conditions, production demands, or contract duration), and
\textit{Seasonal} (only during certain times of the year).
This variable is reported for all occupations with weights associated with each category. For example, Chief Executive has the majority of weight in category 1, as it is primarily a job with a regular schedule. A surgeon has more weight, in comparison, on the irregular category. Using the weights, we obtained an aggregated “schedule irregularity” score for each occupation, wherein a value closer to 1 denotes a more regular schedule, and a value closer to 3 denotes a more irregular schedule.
Next, to proxy \textit{long working hours}, we follow Cha et al. \cite{Cha2014} to use the number of hours worked during the week in the CPS data and form a dummy variable that is one if the worker had worked more than 50 hours a week, and 0 otherwise.

Adding the descriptors of irregular schedules or long hours (in Tab.~\ref{tab:irregular hours} column 2) indeed diminishes the correlation estimated in the baseline, per the baseline model (column 1).
A unit increase in the nested specific skills required by a job, decreases the chances of the worker being female by 36\%. Adding schedule descriptors reduces that relation by more than one third, to about 26\%.

\subsection{Skills and Wage Gaps Have Narrowed Over Years}

Figures \ref{fig:Temporal Race Gaps - Skills, Education, Wages} and \ref{fig:Temporal Gender Gaps - Skills, Education, Wages} below show the temporal dynamics of skill, education, and wage gaps shown as averages in the main Fig.~\ref{fig:Figure 7}. These figures show the gaps have narrowed over years.

\begin{figure*}[!h]
    \centering
    \includegraphics[width=\textwidth]{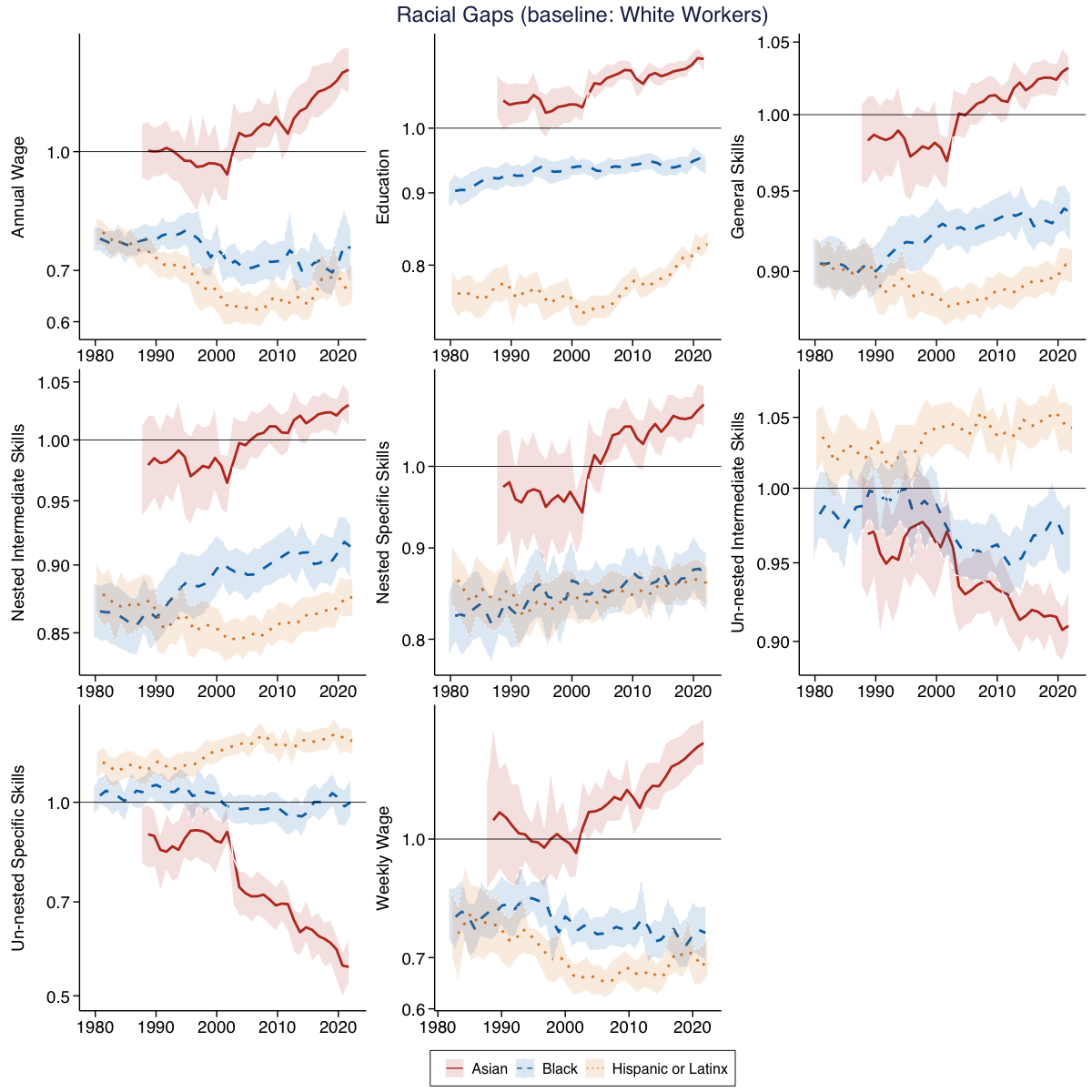}
    \caption{\textbf{Skill and Economic Race/Ethnicity Gaps over Time}.
    Following main Fig.~\ref{fig:Figure 7}, we use White workers as the baseline and show each measure of other demographics as a ratio over the values of White workers.
    In all cases, The 95\% confidence intervals are obtained by a random sub-sampling. In each iteration, we take 20\% of the subpopulation of interest at a certain year, for instance, Asian male and Asian female workers in 2020, and estimate all corresponding measures. Repeating this sampling and estimation process in 10,000 iterations, we obtain the distribution for each estimate (of the subpopulation of interest in that year) and derive the 95\% confidential intervals.
    }
    \label{fig:Temporal Race Gaps - Skills, Education, Wages}
\end{figure*}

\begin{figure*}[!h]
    \centering
    \includegraphics[width=\textwidth]{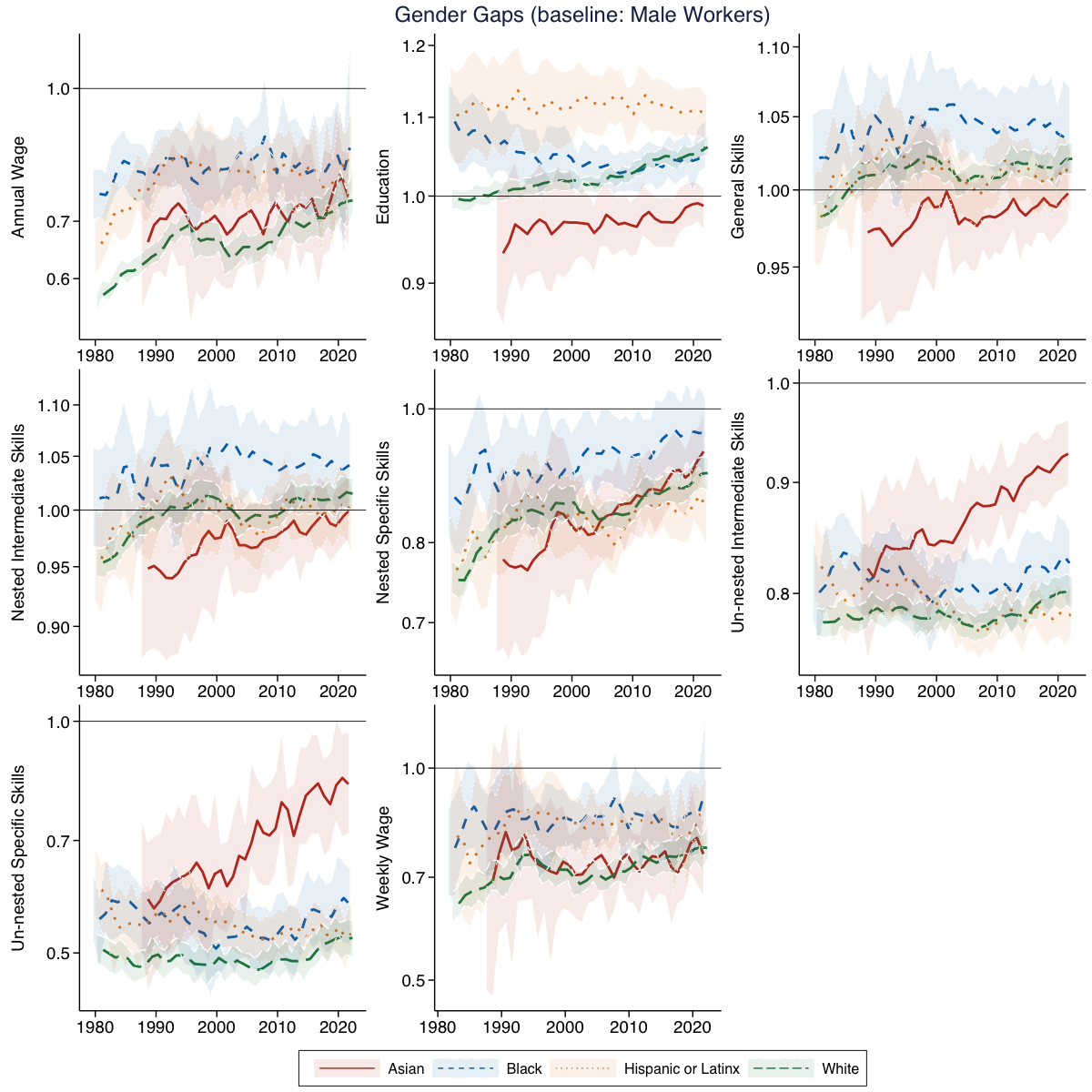}
    \caption{\textbf{Skill and Economic Inequality Across Genders over Time}.
    Following main Fig.~\ref{fig:Figure 7}, we use male workers in each racial group as the benchmark, showing the average value for women over men for each measure and each demographic.
    In all cases, 95\% confidence intervals are created as explained in Fig.~\ref{fig:Temporal Race Gaps - Skills, Education, Wages}.}
    \label{fig:Temporal Gender Gaps - Skills, Education, Wages}
\end{figure*}

\newpage
\subsection{Gender-Age Divergence of Skills across Demographic Groups}

\begin{figure*}[!h]
    \centering
    \includegraphics[width=\textwidth, trim={2cm 0 1cm 1cm},clip]{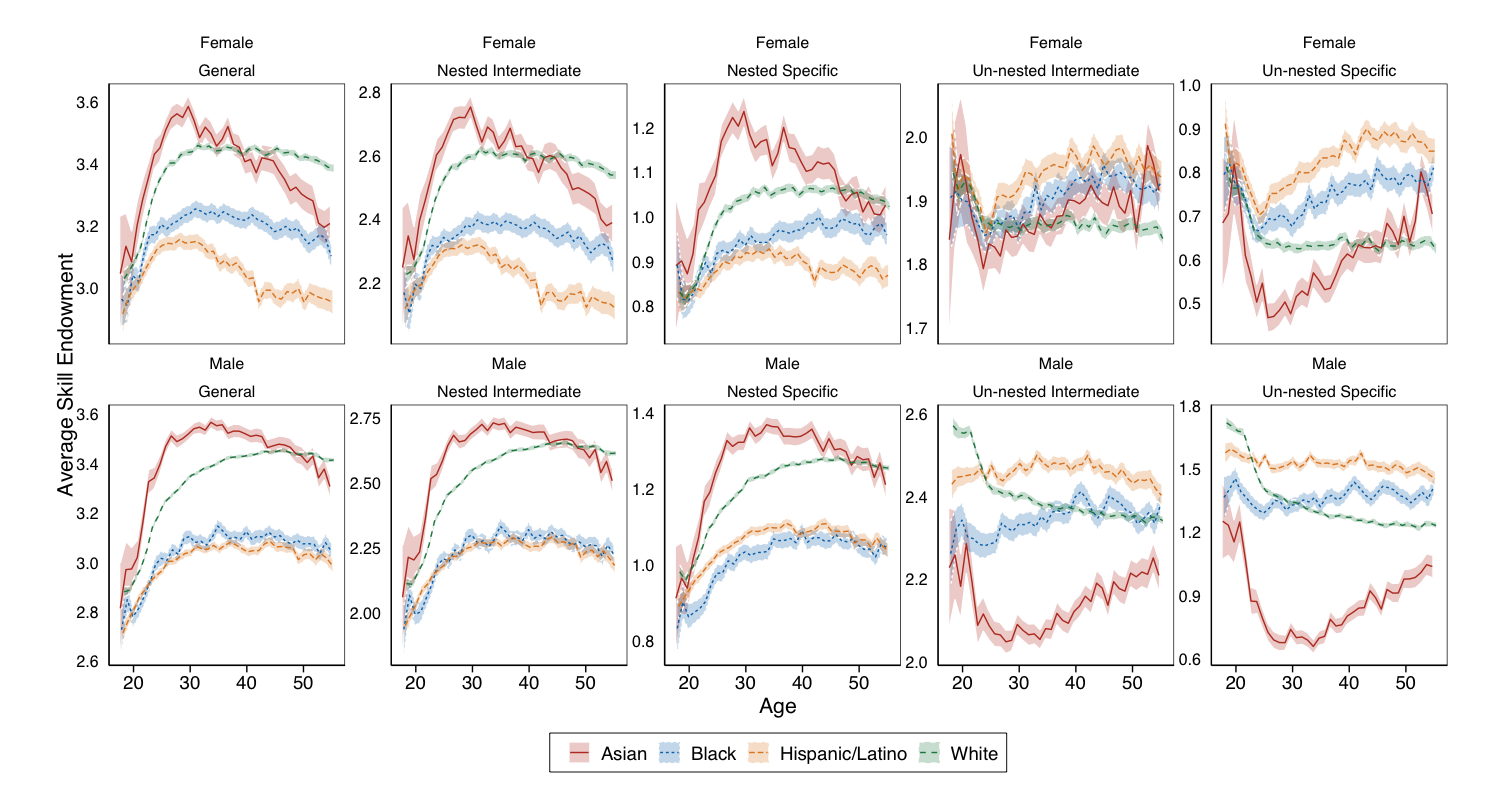}
    \caption{\textbf{Skill Acquisition Differences Across Gender and Race.}
    Using CPS household data between 1980 and 2022, we estimate the distribution of different skills over four racial categories, namely \textit{White, Asian, Black/African-American, Hispanic/Latino}, and for men and women.
    We infer individuals' skills from the skill requirement of their detailed occupation according to O*NET 2019 and calculate a skill endowment for a given race and gender in each skill subtype.
    We aggregate skill endowments for racial and gender groups over age, allowing us to estimate the usage of workplace skills for these subgroups as they age. Shaded areas show 95\% confidence intervals.}
    \label{fig:Skill Age Gender Race Trends}
\end{figure*}

\begin{figure*}[!h]
    \centering
    \includegraphics[width=\textwidth, trim={2cm 0 1cm 1cm},clip]{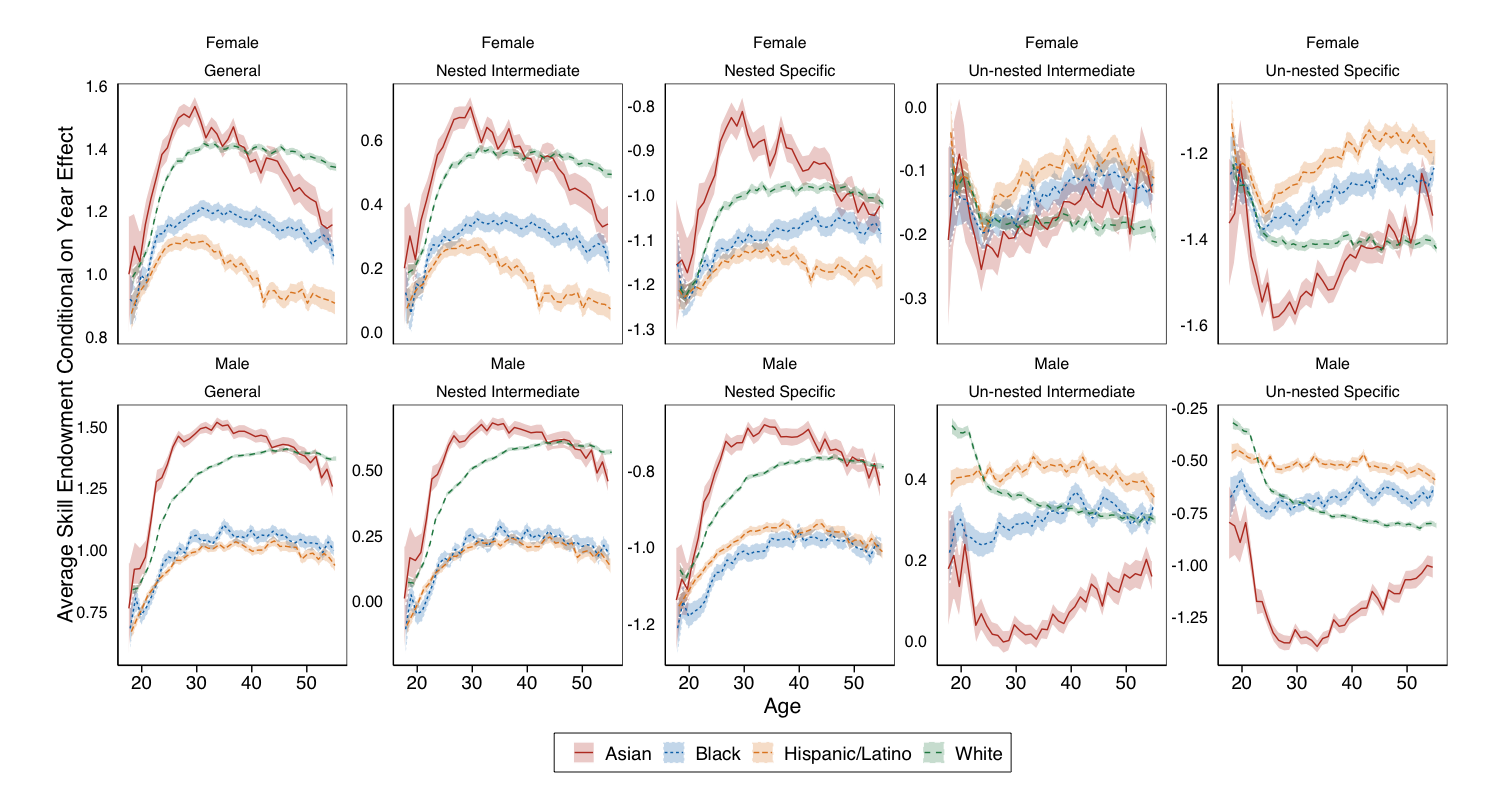}
    \caption{\textbf{Skill Acquisition Differences Across Gender and Race Conditional on Year Effects.}
    The setup follows Fig.~\ref{fig:Skill Age Gender Race Trends} with the minor difference that skill levels are first residualized by a year effect. The results are almost identical.
    }
    \label{fig:Skill Age Gender Race Trends - year effects}
\end{figure*}

Fig.~\ref{fig:Skill Age Gender Race Trends} replicates Fig.~\ref{fig:age}. However, it teases out time trends in skill acquisition for racial groups.
The skill differentials between male and female workers that start around the age 30s (main Fig.~\ref{fig:age}) manifest across racial and ethnic groups.
In most cases, female workers' (general and nested) skill accumulation plateaus in their mid-20s to 30s, while their male counterparts' skill stocks expand (even though slowly) up to their 40s and then plateaus.
Section \ref{supsec:Parenthood_Male_vs_Female} addresses the possible role of children in the divergence of skills.
Fig.~\ref{fig:Skill Age Gender Race Trends - year effects} replicate the exercise factoring in annual economic circumstances.

\clearpage
\section{Historical Patterns of Skill Change for Occupations} \label{supsec: historical skill change}
How have occupational skills (Level and Importance) changed over time? 
Do they manifest our theorized co-evolution of skills at different rungs of the skill hierarchy?
In other words, can we observe our inferred conditional dependence and independence on the level and importance of skills?

To answer these questions, we compared the level and importance of occupational skills reported by O*NET in 2019 and 2005.
When comparing levels and importance of skills across the two years, we use a crosswalk, explained in the supplementary section \ref{sec: onet taxonomy change}, to account for the changes in the taxonomy between 2005 and 2019 \cite{Hopson2021}.
We further capture changes in the skill structure by comparing the skill structure of 2019 to a past snapshot of O*NET from 2005.
We produced the mentioned 2005 skill structure using the same methodology and parameters as used for the 2019 skill dependency network.

\subsection{Changes in the Skill Levels}

\begin{figure*}[!h]
    \centering
    \includegraphics[width=.8\textwidth]{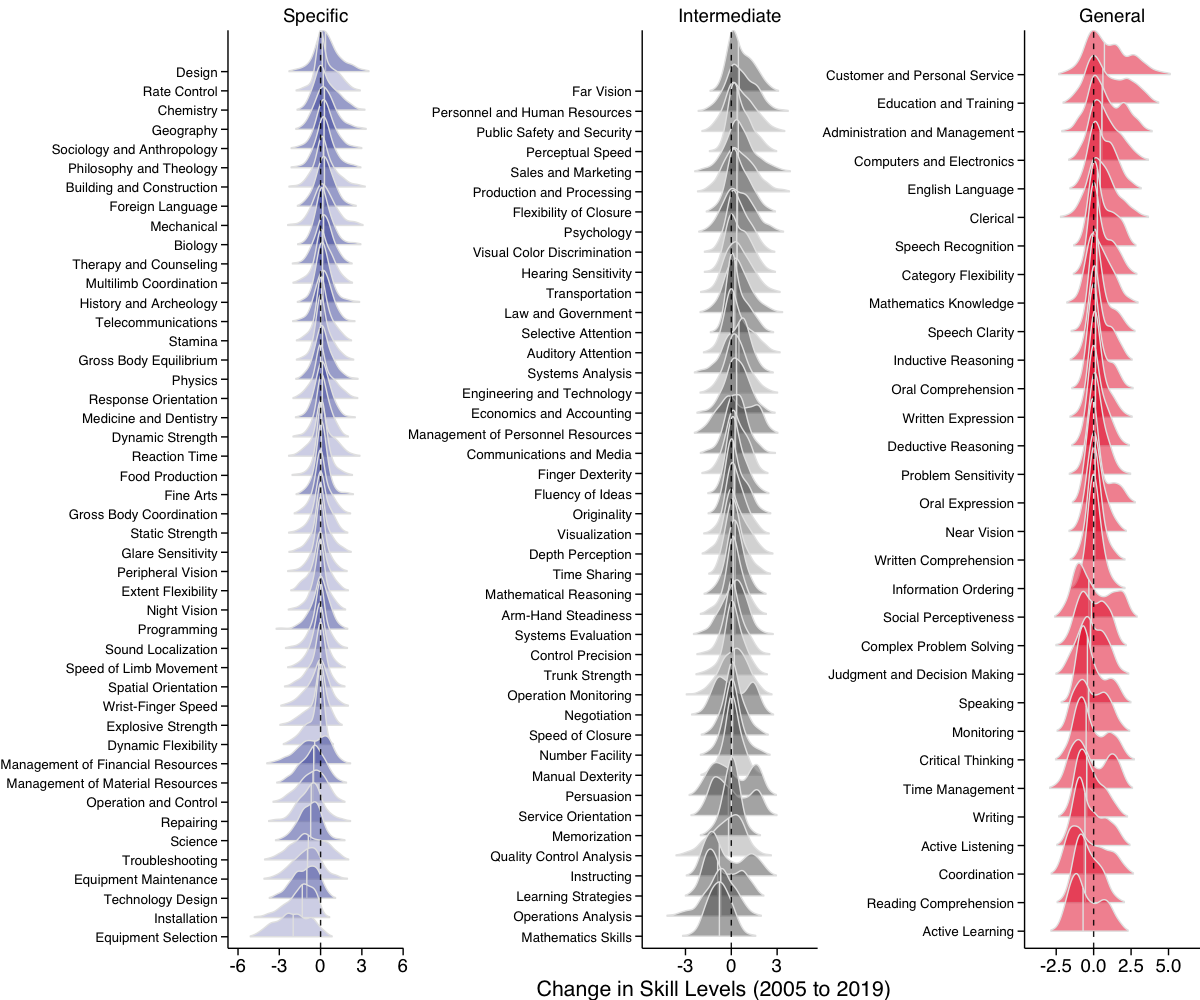}
    \caption{\textbf{Changes in the Level of Each Skill across Occupations between 2005 and 2019 as reported by O*NET.}}
    \label{fig:occupation groups historical changes to skill cluster and type levels}
\end{figure*}

Figure \ref{fig:occupation groups historical changes to skill cluster and type levels} shows the changes in the level of each skill across occupations between 2005 and 2019. 
For each occupation and skill, we subtracted the level O*NET reports in 2019 from the level in 2005. For each skill, we show the density plot of occupations based on their corresponding level change.
The white line for each skill denotes the median, and the dashed line corresponds to no change.

\subsection{Changes in the Skill Dependency Network}
\begin{figure*}[!h]
    \centering
    \includegraphics[width=\textwidth]{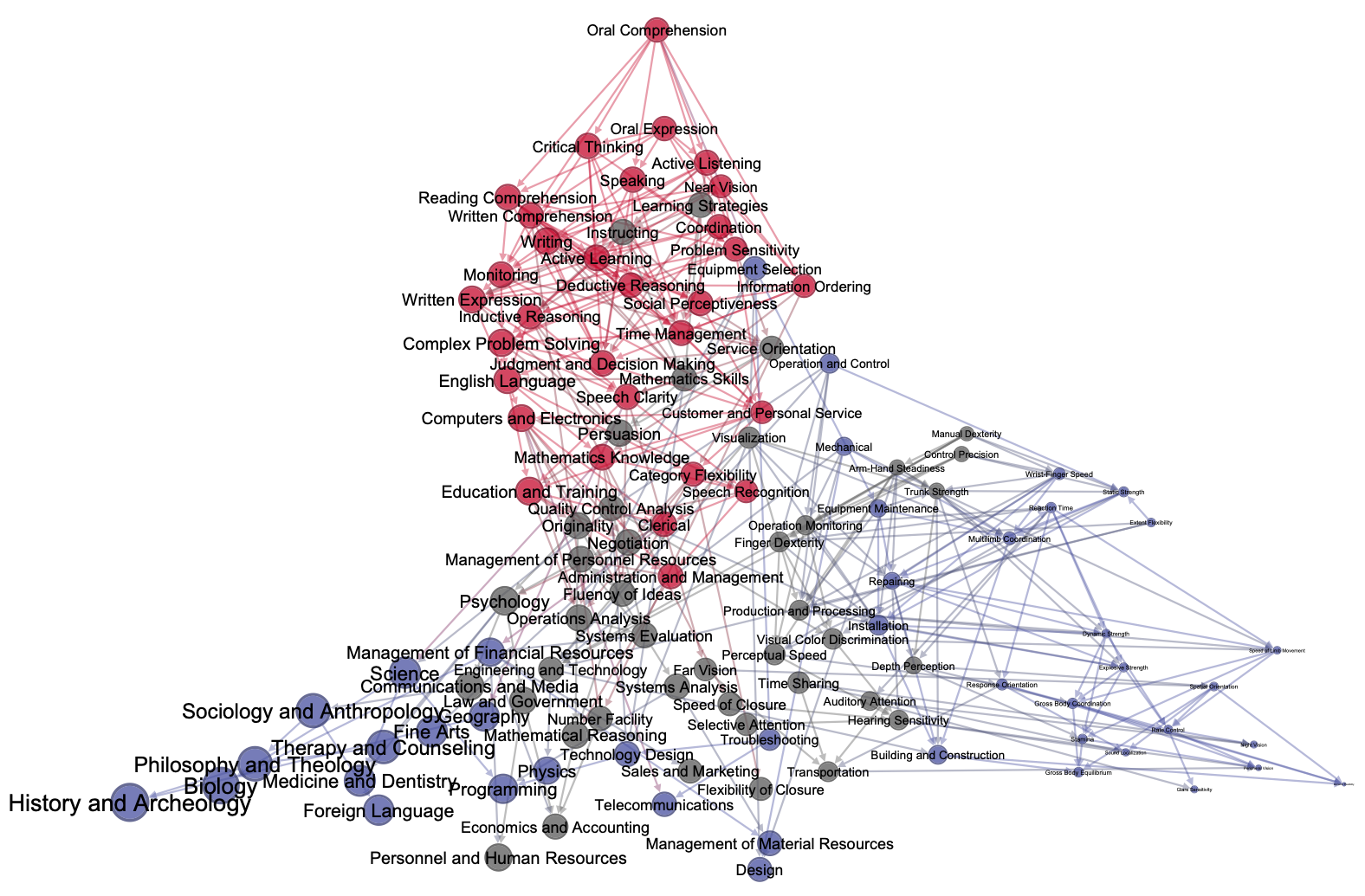}
    \caption{\textbf{Labeled Backbone of Skill Dependency for the year 2005}.
    }
    \label{fig:parsi_skill_dep_2005_labeled}
\end{figure*}

Fig.~\ref{fig:parsi_skill_dep_2005_labeled} shows the backbone of the skill dependency network based on 2005, as appears in Fig.~\ref{fig:Figure 7} with skill labels attached.

\begin{figure*}[!h]
    \centering
    \includegraphics[width=\textwidth]{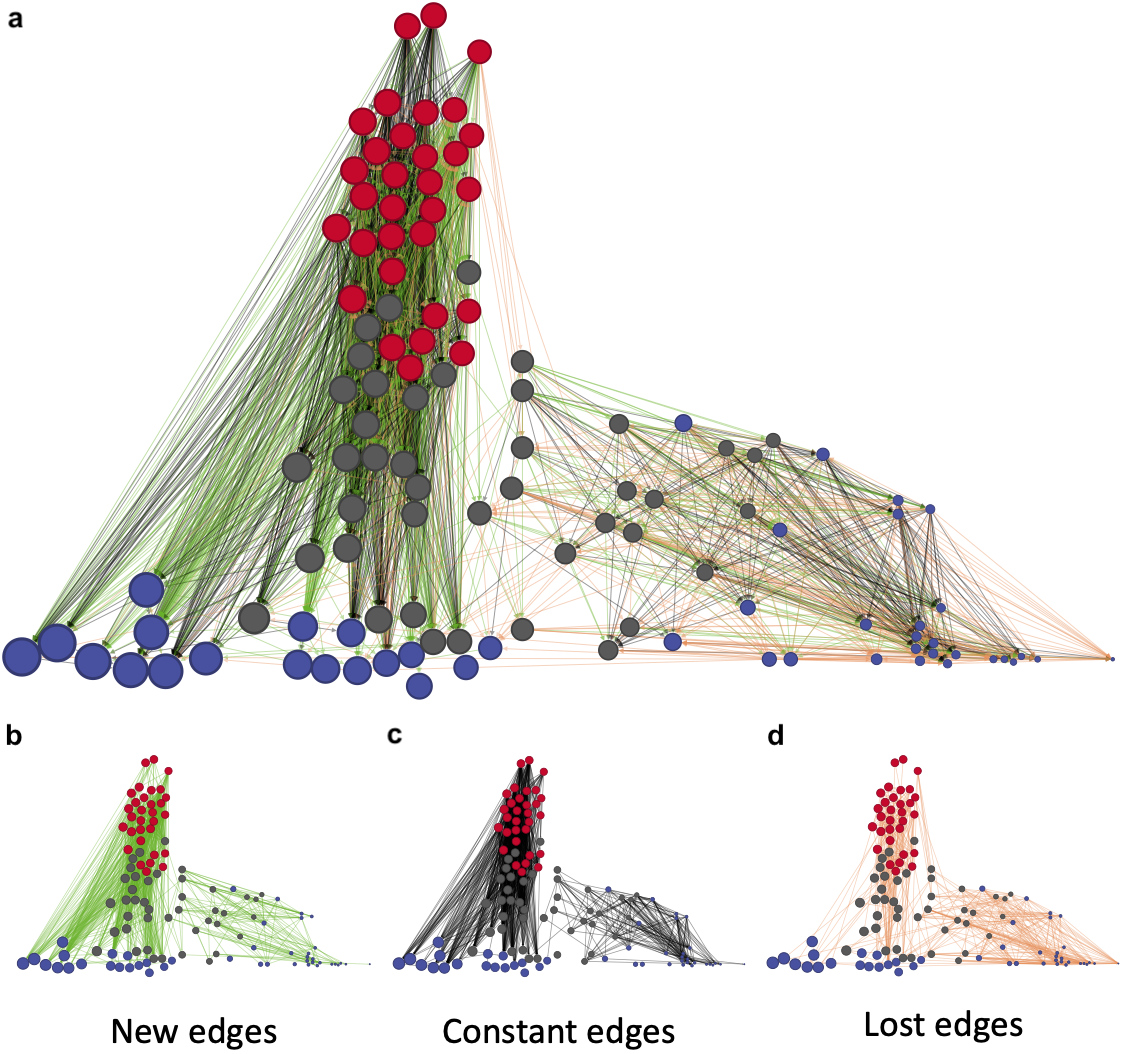}
    \caption{\textbf{Changes in the Skill Dependency Networks between 2005 and 2019}.
    We distinguish between three types of ties: \textbf{(b)} \textit{New edges}: dependency ties that were statistically insignificant in 2005 and became significant in 2019 (shown in green); \textbf{(c)} \textit{Constant edges}: dependency ties that were statistically significant in both 2005 and 2019 (shown in black); \textbf{(d)} \textit{Lost edges}: dependency ties that lost statistical significance in 2019 while being significant in 2005 (shown in orange). \textbf{(a)} The combined figure highlights the increasing intensity of dependencies in the nested part of the network, while a collapse of vertical dependencies in the un-nested section.}
    \label{fig:Figure_7_supp}
\end{figure*}

In Fig.~\ref{fig:Figure_7_supp}, we unpack the changes that manifest in distinct backbones of skill dependency networks in 2019 compared with 2005 (as shown in Fig.~\ref{fig:Figure 7}).
We compare the dependency ties between all skills that are present in both 2005 and 2019 networks (Installation, Explosive strength, Sound localization, Food production, Public safety, and security in 2019 and Memorization, Food production, Chemistry, and Public safety and security in 2005 are eliminated due to a lack of statistically significant ties.)
We distinguish between three types of ties: 1) \textit{New edges}: dependency ties that were statistically insignificant in 2005 and became significant in 2019 (shown in green); 2) \textit{Constant edges}: dependency ties that were statistically significant in both 2005 and 2019 (shown in black); 3) \textit{Lost edges}: dependency ties that lost statistical significance in 2019 while being significant in 2005 (shown in orange).
In Fig.~\ref{fig:Figure_7_supp}, we used the layout of our main Fig.~\ref{fig:Figure 2} (b), adjusting the distance between nodes slightly to visualize edges better. The edges are replaced with the three types described above.
The pattern of changes in dependency ties offers insights into the widening gap between the nested and un-nested parts of the skill structure over time.
Most new edges (green) are massed in the nested section. Particularly, a noticeable number of edges tie general skills to the most niche nested skills— these dependency ties are not visible in the backbone for better visualization. However, they are taken into account in all calculations.
There are virtually no new ties formed between the general and un-nested sections.
Few previously existing ties were no longer statistically significant in 2019 (orange).
Therefore, the figure highlights the increasing intensity of dependencies in the nested part of the network, while a collapse of dependencies in the un-nested section.
These patterns emphasize our findings about the changes in the nature of work.

\newpage

\subsection{Occupation Taxonomy}\label{sec: onet taxonomy change}

Taxonomy has changed over time \cite{Park2020}.
Our historical analysis comparing 2005 and 2019 data must consider such changes.
2005 O*NET complies to \textit{O*NET SOC 2000}, while 2019 O*NET uses \textit{O*NET SOC 2010}, with two other waves of taxonomy change between (2006 and 2009).
Therefore, identically encoded occupations may not be comparable across these two years, and matching them requires a crosswalk.

While O*NET reports\footnote{\url{https://www.onetcenter.org/taxonomy.html}} crosswalks between each consecutive taxonomy, there is no direct crosswalk between 2005 and 2019.
We created such a crosswalk to match occupations in 2005 and 2019 using the consecutive crosswalks mentioned above\footnote{
For instance, if occupation $X_{2000}$ in taxonomy 2000 is linked to $X_{2006}$ in taxonomy 2006, and $X_{2006}$ is matched to $X_{2009}$ in taxonomy 2009, and $X_{2009}$ is matched to $X_{2010}$ in taxonomy 2010, our crosswalk will link $X_{2000}$ to $X_{2010}$.}.

\begin{figure*}[!h]
    \centering
    \includegraphics[width=\textwidth]{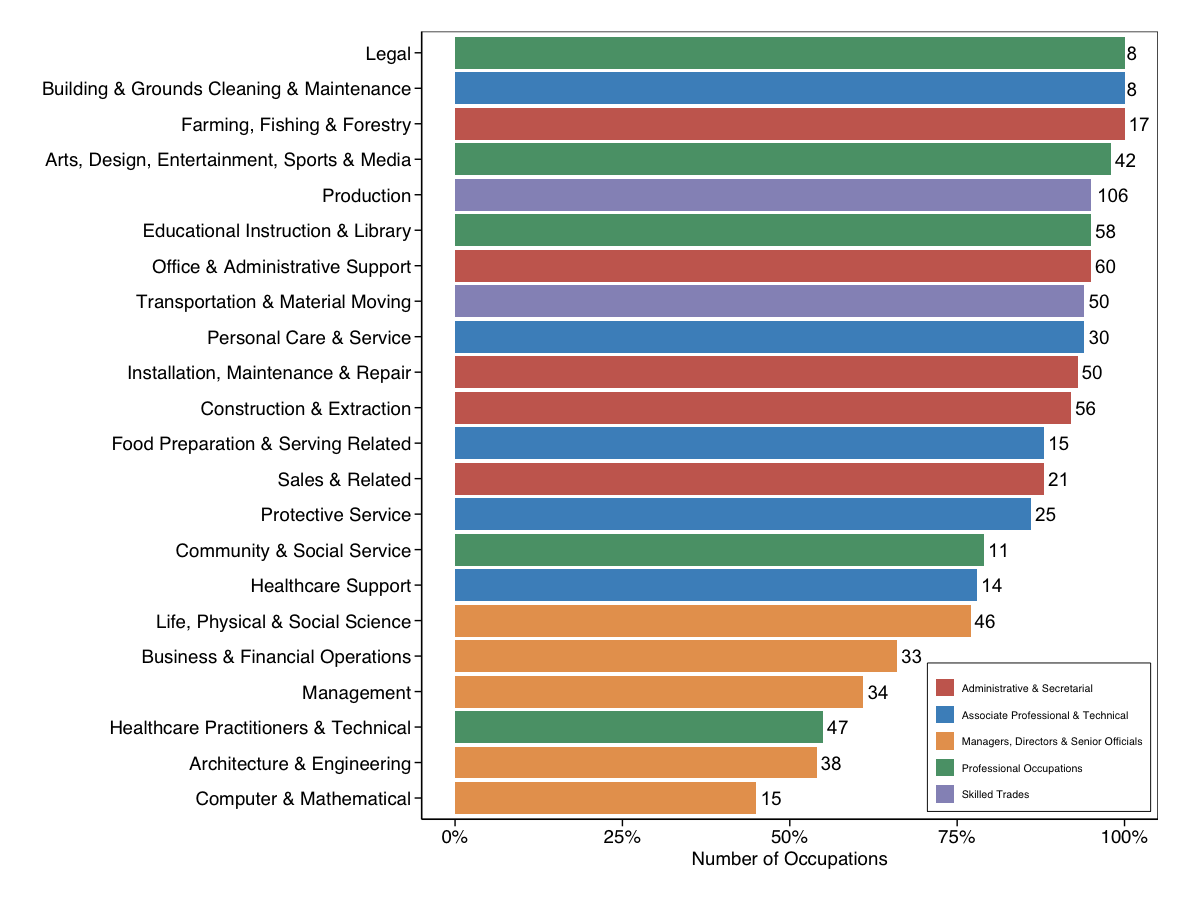}
    \caption{\textbf{Percentage and Number of Occupations from Occupational Groups Matched between 2005 and 2010.}}
    \label{fig:soc match coverage}
\end{figure*}

Our crosswalk matches 968 occupations in 2019 skill data and 941 unique occupations in 2005 skill data.
Out of 1,334 records in our crosswalk, 362 correspond to occupations whose SOC codes have changed.
Fig.~\ref{fig:soc match coverage} shows the number and percentage of occupations in the skill data we could match across both years. Groups such as Computer and Math, Engineering, Health, Management, and Business have the most number of occupations with skill information added between 2005 and 2019.
Therefore, it is important to consider the unbalanced nature of the data, when interpreting analysis on the changes of skills using O*NET.

\clearpage
\section{Robustness Checks with Management and Admin Occupations and Social Skills} \label{supsec:robustness checks}

In this section, we offer supplementary analyses to our main findings and produce pieces of evidence refuting several alternative explanations.
We begin by examining whether managerial occupations drive the importance of general skills.
Next, we discuss whether the general skills' effect is driven primarily by social skills, whose importance has been a topic of growing emphasis.
We continue by offering more detailed analyses of historical changes in skill requirements and skill acquisition with age.

\subsection{Role of Management and Administrative Occupations} \label{sec:robustness check: no managers}
Here, we test if the importance of general skills in the wage premium (Fig.~\ref{fig:Wage}) is driven by management or administrative occupations.
To do so, we identify such occupations, exclude them from our analyses at various stages, and examine the resulting changes. Table \ref{tab:list of manager occupations} lists these occupations with their annual wage and educational requirements.  
We identify those using the Standard Occupational Code (SOC) at the 2-digit level, wherein ``11'' denotes managerial occupations. 
In addition, we use descriptive terms for these occupation titles (manager, administrator, and director) to identify relevant occupations further using their titles. 
In total, we found and omitted a total of 75 occupations out of 968 occupations and collected them, sorted based on average annual wage and required education, in Table \ref{tab:list of manager occupations}.

\footnotesize
\input{Nature_HB_2023/tabsNHB/Apr_14__List_of_Managers.tex}
\normalsize

\begin{figure*}[!h]
    \centering
    \includegraphics[width=\textwidth]{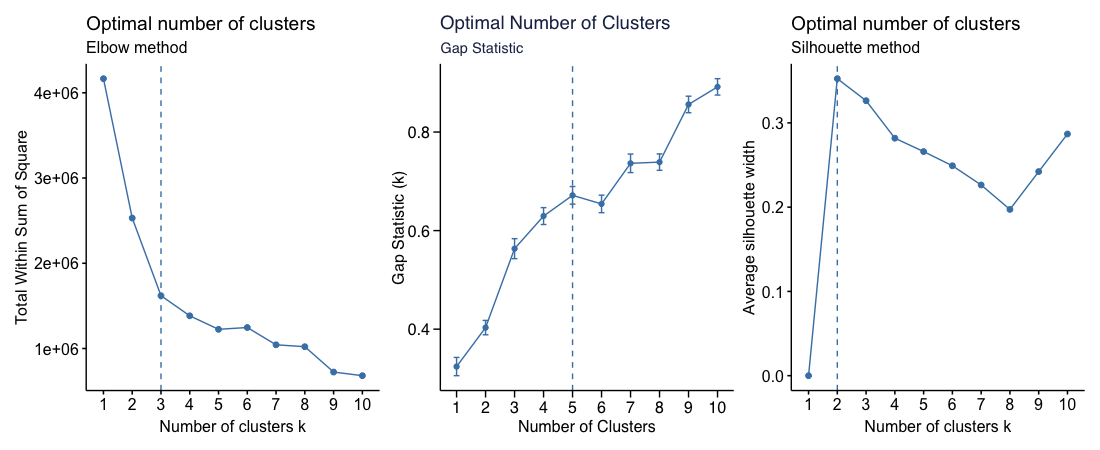}
    \caption{\textbf{Determining $k$ for $k$-mean Clustering at the Absence of Managerial Occupations.} Reproduction of Fig.~\ref{fig:determining_k_level} without management and administrative occupations. We use the Elbow method, Gap statistic, and Silhouette analysis to test the optimal $k$.}
\label{fig:determining_k_kmeans_70bins_correlation_no_manager_start}
\end{figure*}

\begin{figure*}[!h]
    \centering
    \includegraphics[width=\textwidth]{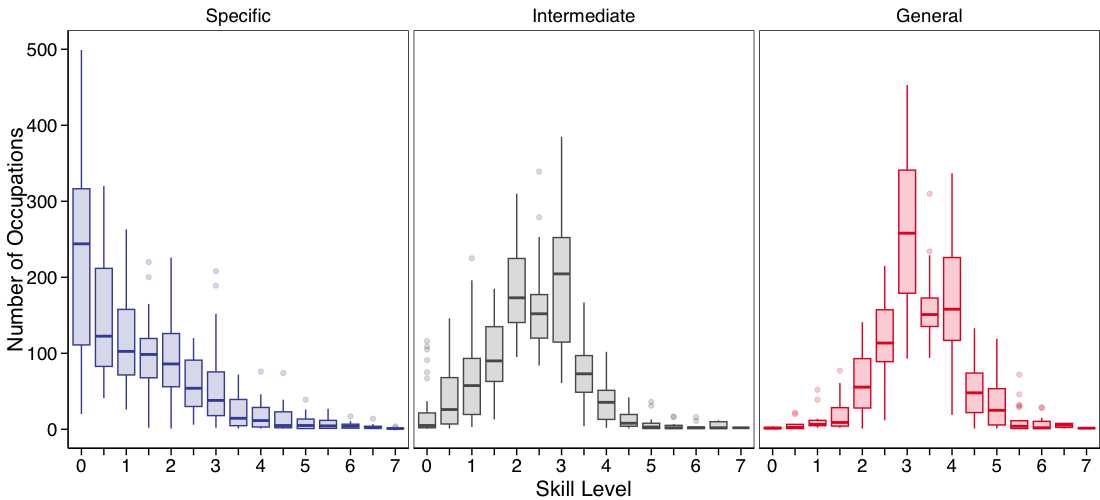}
    \caption{\textbf{Grouping Skills at the Absence of Managerial Occupations.} Reproduction of Fig.~\ref{fig:Figure 1} without management and administration occupations.}
\label{fig:profiles_k_kmeans_70bins_correlation_no_manager_start}
\end{figure*}

\begin{figure*}[!h]
    \centering
    \includegraphics[width=.8\textwidth]{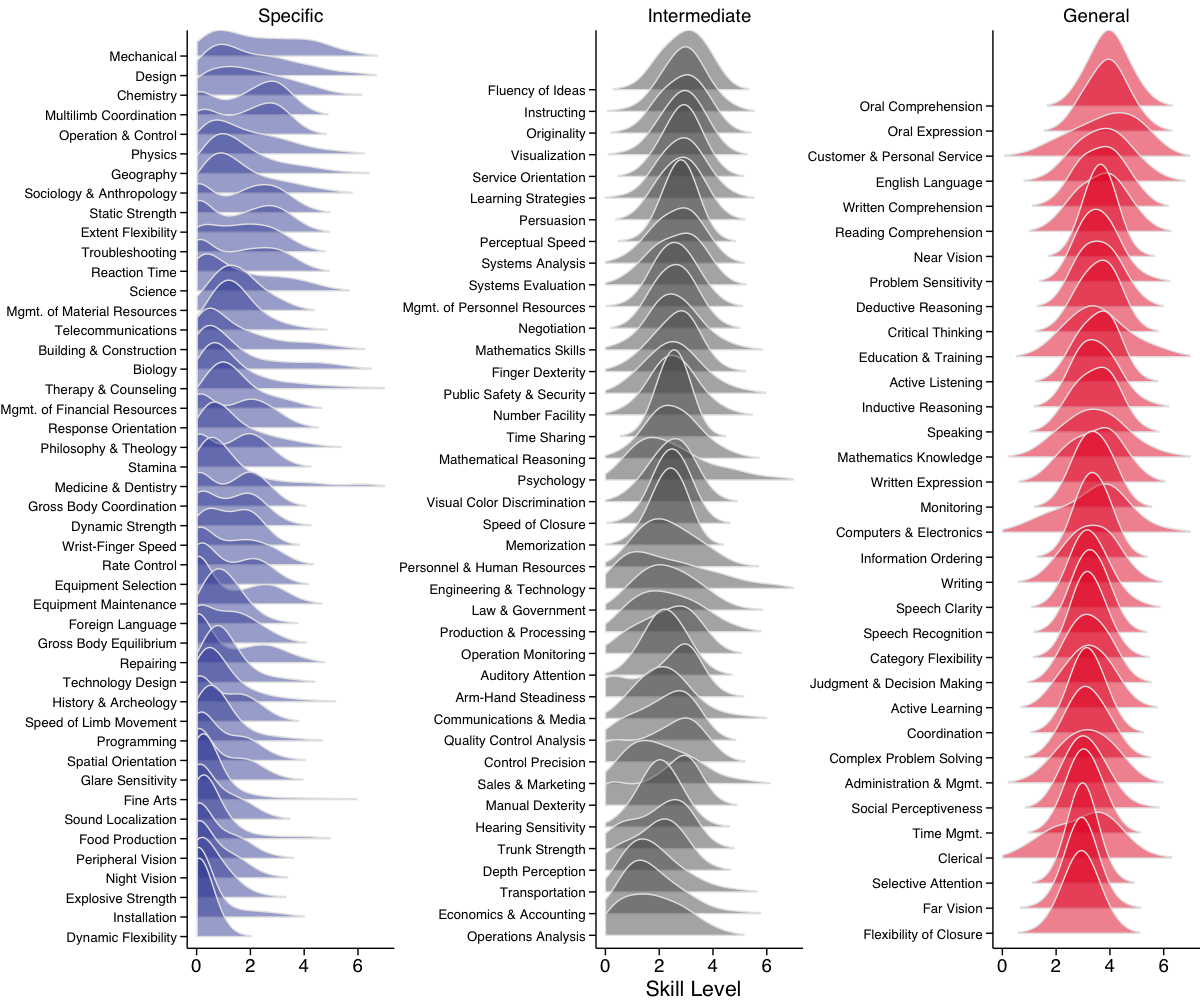}
    \caption{\textbf{Detailed Assignment of Skills among Clusters.} Reproduction of Fig.~\ref{fig:skill_level_dist_cor_k=3} without management and administration occupations.}
\label{fig:skill_clusters_k_kmeans_70bins_correlation_no_manager_start}
\end{figure*}

\begin{figure*}[!h]
    \centering
    \includegraphics[width=.8\textwidth]{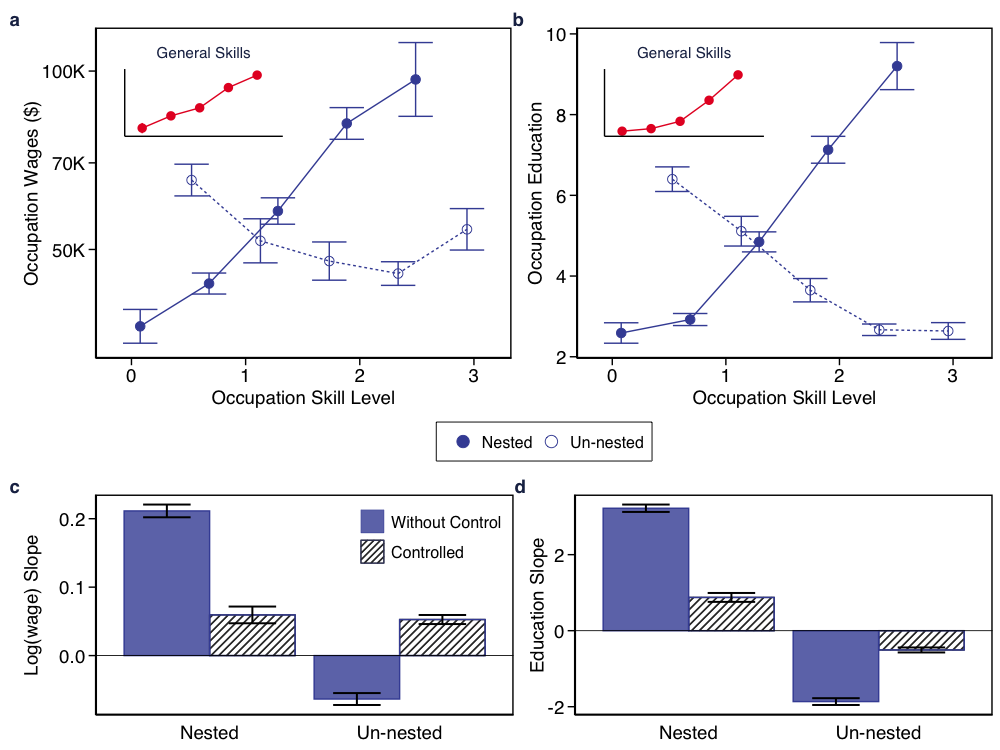}
    \caption{\textbf{Reproduction of Fig.~\ref{fig:Wage} in the main text} without management and administration occupations. Managerial occupations do not drive the wage premiums and the educational requirement.}
\label{fig:returns_to_skills_hierachy_gen_dependence_cor_no_manager}
\end{figure*}

\begin{figure*}[!h]
    \centering
    \includegraphics[width=.9\textwidth]{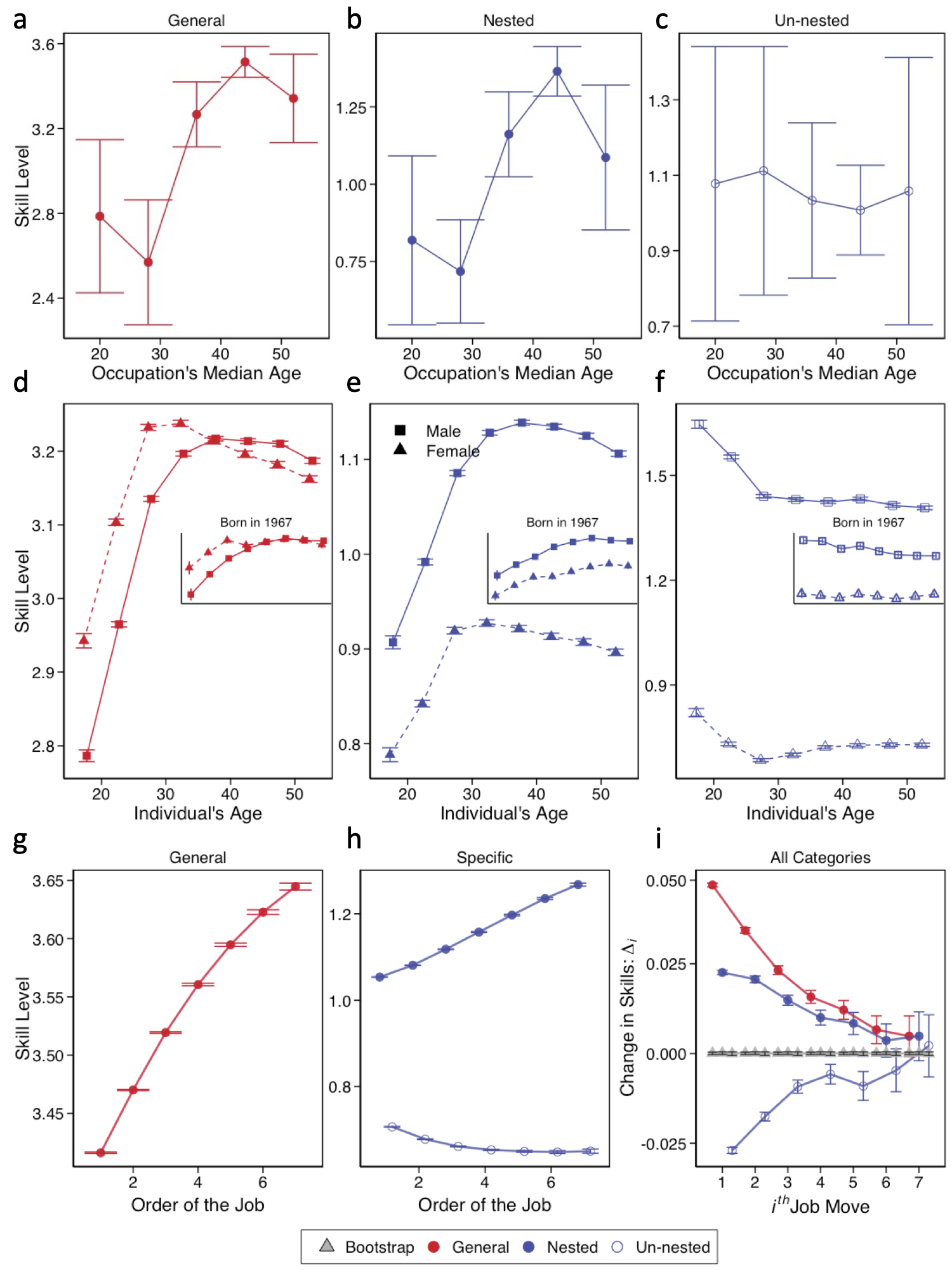}
    \caption{\textbf{Reproduction of Fig. 3 without Management and Administration occupations}. The findings are consistent, suggesting managerial occupations are not the primary drivers of increases in general or nested skills over time.
    \textbf{(a-c)} replicate the analysis of main Fig.~\ref{fig:age}~(a-c) on occupations' median age in the absence fo managerial occupations.
    \textbf{(d-f)} follows analysis Fig.~\ref{fig:age}~(d-f) of synthetic birth cohorts identified in CPS microdata except for excluding observations on individuals who held managerial occupations.
    \textbf{(g-i} follows the analysis of Fig.~\ref{fig:age}~(g-i) on resume data except for excluding observations on individuals who held managerial occupations.}
\label{fig:skill_and_age_no_manager}
\end{figure*}

In conclusion, our findings of skill clusters, skills wage premiums, and educational requirements are robust to the presence/absence of managerial occupations.
In Figs \ref{fig:determining_k_kmeans_70bins_correlation_no_manager_start} through \ref{fig:skill_clusters_k_kmeans_70bins_correlation_no_manager_start} we use $k$-mean clustering to group skills into profiles without considering managerial occupations. The results complement the supplementary section \ref{supsec:skill clustering}, establishing the robustness of our skill profiles.
Fig.~\ref{fig:skill_and_age_no_manager} shows that excluding managerial occupations does not diminish the acquisition of general and nested skills over time by analyzing occupational median age, skill composition of synthetic birth cohorts based on CPS microdata, and analysis of our resume sample. 
Fig.~\ref{fig:returns_to_skills_hierachy_gen_dependence_cor_no_manager} examines wage premiums and educational requirements of occupations in general and specific skills while excluding the information on managerial occupations in the analysis. We find these occupations do not drive the wage premiums, and educational requirements persist. These findings supplement the results in Fig.~\ref{fig:Wage}.  




\clearpage

\clearpage
\subsection{Role of Social Skills} \label{sec:social skills}
Given the growing importance of social skills \cite{Liu2013}, we examine the extent to which they shape the role we observed for general skills in our work.
There are six \textit{social skills} in O*NET taxonomy. 
They are \textit{Social perceptiveness}, the skill of being aware of other's reactions and understanding why they react as they do; \textit{Coordination}, the skill to adjust actions in relation to others' actions; \textit{Persuasion}, the skill to persuade others to change their minds or behavior; \textit{Negotiation}, bringing others together and trying to reconcile differences; \textit{Instructing}, the skill to teach others how to do something; 
and \textit{Service orientation}, actively looking for ways to help people.

\begin{figure*}[!h]
    \centering
    \includegraphics[width=\textwidth]{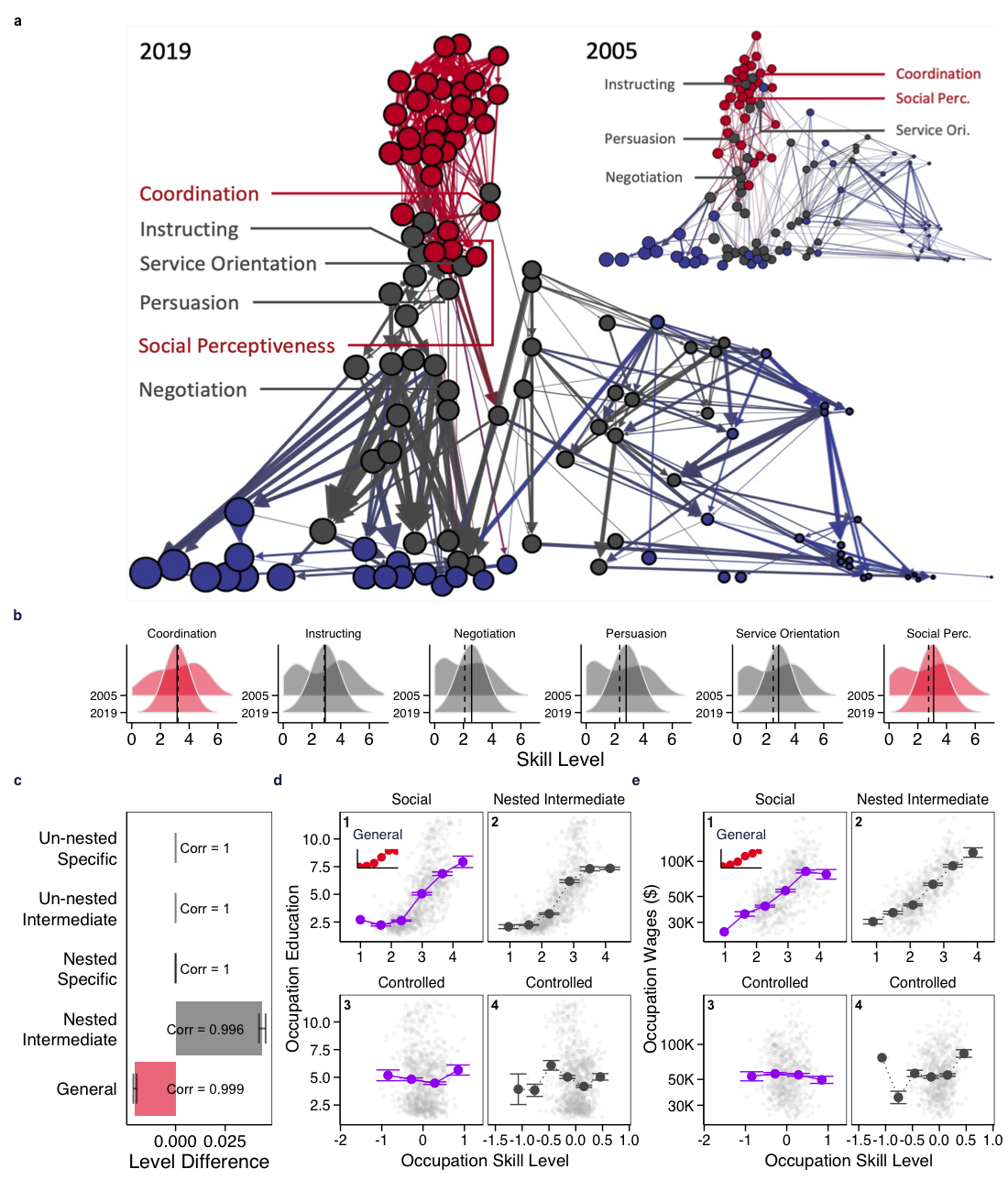}
    \caption{\textbf{Analysis of the Social Skills and their Relationship with our Work.}
    \textbf{(a)} Annotates the skill hierarchy of main text Fig.~\ref{fig:Figure 2} and that of the year 2005 with social skills included in O*NET.
    \textbf{(b)} Skill profiles of social skills in 2005 (average shown as dashed line) and in 2019 (average shown as a solid line).
    \textbf{(c)} The differences in skill levels when social skills are excluded.
    \textbf{(d-e)} The equivalent plots of Fig. 4 without social skills in analyses. 
    }
    \label{fig:social skills}
\end{figure*}

In Fig.~\ref{fig:social skills}-(a), we annotated these social skills in our skill hierarchy of the main text as well as the hierarchy of 2005 data (insets). 
We find these skills fall within our categories of general (Coordination and Social perceptiveness) and nested skill categories (Instructing, Service orientation, Persuasion, and Negotiation); and these skills are more demanded in 2019 than in 2005 as shown in Fig.~\ref{fig:social skills}-(b).

The average levels at which occupations in 2019 use negotiation, persuasion, social perceptiveness, and service orientation skills surpassed their levels in 2005.
However, social skills' position in the skill dependency network has moderately moved away from the most broadly used skills.
This means their comparative role has specialized moderately despite their absolute demand rise.
 Fig.~\ref{fig:social skills} (c) shows the changes in the level of each skill category resulting from omitting social skills. As these skills belong to general and nested intermediate categories, scores across other categories do not change. However, changes in the affected subgroups are also minimal, leading to a 0.998 correlation before and after social skills omission.
This offers evidence that social skills do not influence the significance of general skills.
We repeated analyses of wage premiums (Fig.~\ref{fig:Wage} in the main text) on categories excluding the aforementioned six social skills, ad our findings are intact as shown in Fig.  \ref{fig:social skills} (d-e).

\

\end{document}

%% file: Nature_HB_2023/tabsNHB/Jul_19_2023_-_Skill_Group_Assignments.tex
\begin{longtable}{@{\extracolsep{5pt}} 
    p{2 cm}|p{13cm}}
    \caption{\textbf{The Skill Group Assignment Resulting from Clustering Skills based on their Usage Distribution Shapes.}. Skills are ordered by their generality (their average levels demanded across occupations, and their marginal totals in the binary skill-occupation network).} 
    \label{tab:skill_groups}
\\[-1.8ex]\hline 

\hline \\[-1.8ex] 
Skill Group & Skill Titles (skill generality, occupation counts) \\ 
\hline \\[-1.8ex]
 
\endhead
\hline \\[-1.8ex] 
General \\(3.43, 563.61) & Oral Comprehension (3.96, 682), Oral Expression (3.9, 680), Customer and Personal Service (3.85, 604), English Language (3.76, 590), Written Comprehension (3.68, 598), Reading Comprehension (3.65, 593), Near Vision (3.64, 703), Problem Sensitivity (3.58, 571), Deductive Reasoning (3.58, 553), Critical Thinking (3.57, 594), Education and Training (3.55, 539), Active Listening (3.54, 593), Inductive Reasoning (3.48, 533), Speaking (3.45, 550), Mathematics Knowledge (3.43, 538), Written Expression (3.4, 532), Monitoring (3.38, 589), Computers and Electronics (3.38, 582), Information Ordering (3.37, 616), Writing (3.29, 514), Speech Clarity (3.28, 474), Speech Recognition (3.26, 589), Category Flexibility (3.23, 610), Judgment and Decision Making (3.21, 520), Active Learning (3.19, 531), Coordination (3.16, 534), Complex Problem Solving (3.15, 498), Administration and Management (3.15, 507), Social Perceptiveness (3.1, 486), Time Management (3.05, 477), Clerical (3.03, 492) \\


\hline \\[-1.8ex] 
Intermediate \\(2.44, \\281.42) & Selective Attention (2.99, 496), Fluency of Ideas (2.94, 451), Far Vision (2.94, 451), Flexibility of Closure (2.92, 420), Instructing (2.92, 435), Originality (2.88, 440), Visualization (2.86, 415), Service Orientation (2.85, 442), Learning Strategies (2.84, 387), Persuasion (2.79, 366), Perceptual Speed (2.76, 317), Systems Analysis (2.64, 321), Systems Evaluation (2.63, 337), Management of Personnel Resources (2.6, 245), Mathematics Skills (2.57, 284), Negotiation (2.57, 277), Finger Dexterity (2.56, 352), Public Safety and Security (2.55, 298), Number Facility (2.54, 264), Time Sharing (2.54, 164), Mathematical Reasoning (2.52, 267), Psychology (2.47, 305), Visual Color Discrimination (2.45, 207), Speed of Closure (2.44, 128), Memorization (2.34, 120), Personnel and Human Resources (2.33, 234), Engineering and Technology (2.33, 296), Law and Government (2.32, 241), Production and Processing (2.29, 302), Operation Monitoring (2.28, 250), Auditory Attention (2.23, 175), Communications and Media (2.21, 200), Arm-Hand Steadiness (2.21, 327), Quality Control Analysis (2.19, 216), Control Precision (2.17, 318), Sales and Marketing (2.09, 230), Manual Dexterity (2.08, 302), Hearing Sensitivity (2.04, 85), Trunk Strength (2.01, 238), Depth Perception (1.84, 122), Transportation (1.79, 119), Economics and Accounting (1.77, 129), Operations Analysis (1.75, 128) \\


\hline \\[-1.8ex] 
Specific \\(1.22, \\93.3) & Mechanical (2.4, 354), Design (2.08, 244), Chemistry (1.97, 204), Multilimb Coordination (1.88, 264), Operation and Control (1.79, 192), Physics (1.76, 147), Geography (1.71, 159), Sociology and Anthropology (1.65, 153), Static Strength (1.64, 211), Extent Flexibility (1.59, 234), Troubleshooting (1.58, 126), Reaction Time (1.53, 189), Science (1.52, 154), Management of Material Resources (1.5, 32), Telecommunications (1.48, 53), Building and Construction (1.47, 137), Biology (1.42, 152), Therapy and Counseling (1.42, 119), Management of Financial Resources (1.37, 69), Response Orientation (1.32, 72), Philosophy and Theology (1.31, 102), Medicine and Dentistry (1.26, 108), Stamina (1.26, 79), Gross Body Coordination (1.24, 55), Dynamic Strength (1.21, 56), Wrist-Finger Speed (1.2, 21), Rate Control (1.18, 93), Equipment Selection (1.09, 15), Equipment Maintenance (1.06, 95), Foreign Language (1.03, 17), Gross Body Equilibrium (1.03, 33), Technology Design (1.02, 19), Repairing (1.02, 91), History and Archeology (0.93, 52), Speed of Limb Movement (0.86, 11), Programming (0.84, 29), Spatial Orientation (0.83, 21), Glare Sensitivity (0.74, 13), Fine Arts (0.6, 43), Sound Localization (0.6, 3), Peripheral Vision (0.59, 8), Food Production (0.59, 44), Night Vision (0.53, 2), Explosive Strength (0.48, 4), Installation (0.37, 11), Dynamic Flexibility (0.15, 2) \\

\hline \\[-1.8ex] 
\end{longtable} 

%% file: Nature_HB_2023/tabsNHB/Jul_15_2023_Skill_Split_C-score_Nestedness_Simul_Jan_2023.tex
\begin{longtable}{@{\extracolsep{5pt}} 
    p{3cm}|p{12cm}}
    \caption{\textbf{The Skill Split Resulting from Nested Contribution of Skills.}} 
    \label{tab:skill_split_C_nestedness}
\\[-1.8ex]\hline 

\hline \\[-1.8ex] 
 Skill Category & Skill Titles \\ 
\hline \\[-1.8ex]
 \vspace{0.1 cm} \\
\endhead
Nested Specific (18 skills) & Biology, Chemistry, Design, Fine Arts, Foreign Language, Geography, History and Archeology, Management of Financial Resources, Management of Material Resources, Medicine and Dentistry, Philosophy and Theology, Physics, Programming, Science, Sociology and Anthropology, Technology Design, Telecommunications, Therapy and Counseling \\ 

\vspace{0.5 cm} \\
\hline \\[-1.8ex]

Un-nested Specific (28 skills) & Building and Construction, Dynamic Flexibility, Dynamic Strength, Equipment Maintenance, Equipment Selection, Explosive Strength, Extent Flexibility, Food Production, Glare Sensitivity, Gross Body Coordination, Gross Body Equilibrium, Installation, Mechanical, Multilimb Coordination, Night Vision, Operation and Control, Peripheral Vision, Rate Control, Reaction Time, Repairing, Response Orientation, Sound Localization, Spatial Orientation, Speed of Limb Movement, Stamina, Static Strength, Troubleshooting, Wrist-Finger Speed \\ 

\vspace{0.5 cm} \\
\hline \\[-1.8ex]

Nested Intermediate (25 skills) & Communications and Media, Economics and Accounting, Far Vision, Flexibility of Closure, Fluency of Ideas, Instructing, Law and Government, Learning Strategies, Management of Personnel Resources, Mathematical Reasoning, Mathematics Skills, Memorization, Negotiation, Number Facility, Operations Analysis, Originality, Personnel and Human Resources, Persuasion, Psychology, Sales and Marketing, Selective Attention, Service Orientation, Speed of Closure, Systems Analysis, Systems Evaluation \\  

\vspace{0.5 cm} \\
\hline \\[-1.8ex]

Un-nested Intermediate (18 skills) & Arm-Hand Steadiness, Auditory Attention, Control Precision, Depth Perception, Engineering and Technology, Finger Dexterity, Hearing Sensitivity, Manual Dexterity, Operation Monitoring, Perceptual Speed, Production and Processing, Public Safety and Security, Quality Control Analysis, Time Sharing, Transportation, Trunk Strength, Visual Color Discrimination, Visualization \\

\vspace{0.5 cm} \\
\hline \\[-1.8ex]

General (31 skills) & Active Learning, Active Listening, Administration and Management, Category Flexibility, Clerical, Complex Problem Solving, Computers and Electronics, Coordination, Critical Thinking, Customer and Personal Service, Deductive Reasoning, Education and Training, English Language, Inductive Reasoning, Information Ordering, Judgment and Decision Making, Mathematics Knowledge, Monitoring, Near Vision, Oral Comprehension, Oral Expression, Problem Sensitivity, Reading Comprehension, Social Perceptiveness, Speaking, Speech Clarity, Speech Recognition, Time Management, Writing, Written Comprehension, Written Expression \\ 
\vspace{0.5 cm} \\
\hline \\[-1.8ex] 
\end{longtable}

%% file: Nature_HB_2023/tabsNHB/Jul_15_2023-_Alternative_Skill_Split-_Avg_Correlation_with_General_Skill.tex
\begin{longtable}{@{\extracolsep{5 pt}} 
    p{3cm}|p{12cm}}
    \caption{\textbf{The Skill Split Resulting from Correlation Dependence Analysis described above.}} 
    \label{tab:skill_split_alt}
\\[-1.8ex]\hline 

\hline \\[-1.8ex] 
 Skill Category & Skill Titles \\ 
\hline \\[-1.8ex]
 \vspace{0.1 cm} \\
\endhead
Nested Specific (20 skills) & Biology, Building and Construction, Chemistry, Design, Fine Arts, Food Production, Foreign Language, Geography, History and Archeology, Management of Financial Resources, Management of Material Resources, Medicine and Dentistry, Philosophy and Theology, Physics, Programming, Science, Sociology and Anthropology, Technology Design, Telecommunications, Therapy and Counseling \\ 

\vspace{0.5 cm} \\
\hline \\[-1.8ex]

Un-nested Specific (26 skills) & Dynamic Flexibility, Dynamic Strength, Equipment Maintenance, Equipment Selection, Explosive Strength, Extent Flexibility, Glare Sensitivity, Gross Body Coordination, Gross Body Equilibrium, Installation, Mechanical, Multilimb Coordination, Night Vision, Operation and Control, Peripheral Vision, Rate Control, Reaction Time, Repairing, Response Orientation, Sound Localization, Spatial Orientation, Speed of Limb Movement, Stamina, Static Strength, Troubleshooting, Wrist-Finger Speed \\ 

\vspace{0.5 cm} \\
\hline \\[-1.8ex]

Nested Intermediate (23 skills) & Communications and Media, Economics and Accounting, Flexibility of Closure, Fluency of Ideas, Instructing, Law and Government, Learning Strategies, Management of Personnel Resources, Mathematical Reasoning, Mathematics Skills, Memorization, Negotiation, Number Facility, Operations Analysis, Originality, Personnel and Human Resources, Persuasion, Psychology, Sales and Marketing, Service Orientation, Speed of Closure, Systems Analysis, Systems Evaluation \\ 

\vspace{0.5 cm} \\
\hline \\[-1.8ex]

Un-nested Intermediate (20 skills) & Arm-Hand Steadiness, Auditory Attention, Control Precision, Depth Perception, Engineering and Technology, Far Vision, Finger Dexterity, Hearing Sensitivity, Manual Dexterity, Operation Monitoring, Perceptual Speed, Production and Processing, Public Safety and Security, Quality Control Analysis, Selective Attention, Time Sharing, Transportation, Trunk Strength, Visual Color Discrimination, Visualization \\ 

\vspace{0.5 cm} \\
\hline \\[-1.8ex]

General (31 skills) & Active Learning, Active Listening, Administration and Management, Category Flexibility, Clerical, Complex Problem Solving, Computers and Electronics, Coordination, Critical Thinking, Customer and Personal Service, Deductive Reasoning, Education and Training, English Language, Inductive Reasoning, Information Ordering, Judgment and Decision Making, Mathematics Knowledge, Monitoring, Near Vision, Oral Comprehension, Oral Expression, Problem Sensitivity, Reading Comprehension, Social Perceptiveness, Speaking, Speech Clarity, Speech Recognition, Time Management, Writing, Written Comprehension, Written Expression \\ 
\vspace{0.5 cm} \\
\hline \\[-1.8ex] 
\end{longtable}

%% file: Nature_HB_2023/tabsNHB/Jul_15_2023__Weird_Job_Sequences.tex
\begin{longtable}{@{\extracolsep{5pt}} 
    p{0.2cm}p{1.2cm}p{1.45cm}p{1.45cm}p{8.7cm}}
    \caption{A Select List of Job Sequences in Burning Glass data that yield extreme Skill Changes.} 
  \label{tab:odd job sequences} 
\\[-1.8ex]\hline 

\hline \\[-1.8ex] 
 & ID & Start & End & Occupation \\ 
\hline \\[-1.8ex]

1 & 652855 & Oct 2011 & Feb 2012 & Janitors \& Cleaners, Except Maids \& Housekeeping Cleaners \\ 
2 &   & Jun 2012 & Aug 2012 & Chief Executives \\ 
3 & 1723696 & Sep 1981 & Sep 1991 & Janitors \& Cleaners, Except Maids \& Housekeeping Cleaners \\ 
4 &   & Sep 1991 & Sep 1994 & Medical Records \& Health Information Technicians \\ 
5 &   & Sep 1993 & Sep 1993 & Middle School Teachers, Except Special \& Career or Technical Education \\ 
6 &   & Sep 1994 & Sep 1999 & Lodging Managers \\ 
7 & 18075175 & Jun 2007 & Jun 2007 & Chief Executives \\ 
8 &   & Aug 2009 & Aug 2009 & Janitors \& Cleaners, Except Maids \& Housekeeping Cleaners \\ 
9 &   & Aug 2010 & Aug 2010 & Cashiers \\ 
10 &   & Aug 2011 & Aug 2011 & Retail Salespersons \\ 
11 &   & Dec 2012 & Dec 2012 & Retail Salespersons \\ 
12 &   & Feb 2013 & Feb 2013 & Cashiers \\ 
13 & 18325881 & Jun 2022 & Oct 2022 & Medical \& Health Services Managers \\ 
14 &   & Oct 2022 & Jan 2022 & Medical \& Health Services Managers \\ 
15 &   & Jan 2022 & May 2022 & Human Resources Specialists \\ 
16 &   & May 2022 & Sep 2022 & Models \\ 
\hline \\[-1.8ex] 
\end{longtable}

%% file: Nature_HB_2023/tabsNHB/Jul_15_2023__Wage_Regression_on_Skill_Endowments.tex
\begin{tabular}{@{\extracolsep{5pt}}lD{.}{.}{-3} D{.}{.}{-3} D{.}{.}{-3} D{.}{.}{-3} D{.}{.}{-3} D{.}{.}{-3} } 
\\[-1.8ex]\hline 
\hline \\[-1.8ex] 
\\[-1.8ex] & \multicolumn{6}{c}{$log(Wage_{2019})$} \\ 
\\[-1.8ex] & \multicolumn{1}{c}{(1)} & \multicolumn{1}{c}{(2)} & \multicolumn{1}{c}{(3)} & \multicolumn{1}{c}{(4)} & \multicolumn{1}{c}{(5)} & \multicolumn{1}{c}{(6)}\\ 
\hline \\[-1.8ex] 
 $General$ & 0.251^{***} &  &  & 0.135^{***} &  &  \\ 
  & \multicolumn{1}{c}{(0.237$, $0.264)} &  &  & \multicolumn{1}{c}{(0.109$, $0.161)} &  &  \\ 
  & & & & & & \\ 
 $Intermediate^{Nested}$ &  & 0.260^{***} &  &  & 0.120^{***} &  \\ 
  &  & \multicolumn{1}{c}{(0.245$, $0.274)} &  &  & \multicolumn{1}{c}{(0.098$, $0.141)} &  \\ 
  & & & & & & \\ 
 $Intermediate^{Unnested}$ &  & 0.015^{**} &  &  & 0.034^{***} &  \\ 
  &  & \multicolumn{1}{c}{(0.002$, $0.028)} &  &  & \multicolumn{1}{c}{(0.020$, $0.047)} &  \\ 
  & & & & & & \\ 
 $Specific^{Nested}$ &  &  & 0.199^{***} &  &  & 0.042^{***} \\ 
  &  &  & \multicolumn{1}{c}{(0.182$, $0.215)} &  &  & \multicolumn{1}{c}{(0.020$, $0.064)} \\ 
  & & & & & & \\ 
 $Specific^{Unnested}$ &  &  & -0.051^{***} &  &  & 0.003 \\ 
  &  &  & \multicolumn{1}{c}{(-0.063$, $-0.039)} &  &  & \multicolumn{1}{c}{(-0.011$, $0.017)} \\ 
  & & & & & & \\ 
 \textit{Education} &  &  &  & 0.024^{***} & 0.034^{***} & 0.042^{***} \\ 
  &  &  &  & \multicolumn{1}{c}{(0.018$, $0.030)} & \multicolumn{1}{c}{(0.029$, $0.039)} & \multicolumn{1}{c}{(0.036$, $0.048)} \\ 
  & & & & & & \\ 
 \textit{Experience} &  &  &  & 0.014^{***} & 0.014^{***} & 0.023^{***} \\ 
  &  &  &  & \multicolumn{1}{c}{(0.008$, $0.021)} & \multicolumn{1}{c}{(0.008$, $0.021)} & \multicolumn{1}{c}{(0.016$, $0.030)} \\ 
  & & & & & & \\ 
 \textit{Training} &  &  &  & 0.038^{***} & 0.027^{***} & 0.042^{***} \\ 
  &  &  &  & \multicolumn{1}{c}{(0.029$, $0.048)} & \multicolumn{1}{c}{(0.017$, $0.037)} & \multicolumn{1}{c}{(0.031$, $0.052)} \\ 
  & & & & & & \\ 
 Constant & 3.909^{***} & 4.069^{***} & 4.550^{***} & 3.969^{***} & 4.043^{***} & 4.223^{***} \\ 
  & \multicolumn{1}{c}{(3.861$, $3.956)} & \multicolumn{1}{c}{(4.017$, $4.121)} & \multicolumn{1}{c}{(4.520$, $4.581)} & \multicolumn{1}{c}{(3.908$, $4.031)} & \multicolumn{1}{c}{(3.997$, $4.090)} & \multicolumn{1}{c}{(4.182$, $4.264)} \\ 
  & & & & & & \\ 
\hline \\[-1.8ex] 
Observations & \multicolumn{1}{c}{789} & \multicolumn{1}{c}{789} & \multicolumn{1}{c}{789} & \multicolumn{1}{c}{789} & \multicolumn{1}{c}{789} & \multicolumn{1}{c}{789} \\ 
R$^{2}$ & \multicolumn{1}{c}{0.622} & \multicolumn{1}{c}{0.607} & \multicolumn{1}{c}{0.470} & \multicolumn{1}{c}{0.686} & \multicolumn{1}{c}{0.703} & \multicolumn{1}{c}{0.653} \\ 
Adjusted R$^{2}$ & \multicolumn{1}{c}{0.622} & \multicolumn{1}{c}{0.606} & \multicolumn{1}{c}{0.469} & \multicolumn{1}{c}{0.684} & \multicolumn{1}{c}{0.701} & \multicolumn{1}{c}{0.651} \\ 
Residual Std. Error & \multicolumn{1}{c}{0.119
} & \multicolumn{1}{c}{0.122
} & \multicolumn{1}{c}{0.142
} & \multicolumn{1}{c}{0.109
} & \multicolumn{1}{c}{0.106
} & \multicolumn{1}{c}{0.115
} \\
F Statistic & \multicolumn{1}{c}{1,297.040$^{***}$ 
} & \multicolumn{1}{c}{607.527$^{***}$
} & \multicolumn{1}{c}{348.362$^{***}$
} & \multicolumn{1}{c}{428.084$^{***}$
} & \multicolumn{1}{c}{370.741$^{***}$
} & \multicolumn{1}{c}{294.761$^{***}$
} \\ 

\hline 
\hline \\[-1.8ex] 
\multicolumn{7}{p{1.5\linewidth}}{\textit{Note: OLS regressions are shown, with 95-percentile confidence intervals in parentheses ($^{*}$p$<$0.1; $^{**}$p$<$0.05; $^{***}$p$<$0.01). $R^2$, coefficient of determination, and adjusted $R^2$ is normalized for the models' number of variables.}} \\
\end{tabular} 

%% file: Nature_HB_2023/tabsNHB/Mar_25_2024__urban_wage_premiums.tex
\begin{tabular}{@{\extracolsep{5pt}}lD{.}{.}{-3} D{.}{.}{-3} D{.}{.}{-3} D{.}{.}{-3} } 
\\[-1.8ex]\hline 
\hline \\[-1.8ex] 
 & \multicolumn{4}{c}{\textit{Dependent variable:}} \\ 
\cline{2-5} 
\\[-1.8ex] & \multicolumn{4}{c}{Log(Wage)} \\ 
\\[-1.8ex] & \multicolumn{4}{c}{\textit{OLS}} \\ 
\\[-1.8ex] & \multicolumn{1}{c}{(1)} & \multicolumn{1}{c}{(2)} & \multicolumn{1}{c}{(3)} & \multicolumn{1}{c}{(4)}\\ 
\hline \\[-1.8ex] 
 Population > 1M & 0.082^{***} & 0.054^{***} & 0.056^{***} & 0.059^{***} \\ 
  & \multicolumn{1}{c}{(0.080$, $0.084)} & \multicolumn{1}{c}{(0.053$, $0.056)} & \multicolumn{1}{c}{(0.054$, $0.058)} & \multicolumn{1}{c}{(0.057$, $0.060)} \\ 
  & & & & \\ 
 General Skills &  & 0.269^{***} &  & 0.281^{***} \\ 
  &  & \multicolumn{1}{c}{(0.268$, $0.270)} &  & \multicolumn{1}{c}{(0.278$, $0.283)} \\ 
  & & & & \\ 
 Nested Specific Skills &  &  & 0.248^{***} & 0.007^{***} \\ 
  &  &  & \multicolumn{1}{c}{(0.246$, $0.250)} & \multicolumn{1}{c}{(0.005$, $0.010)} \\ 
  & & & & \\ 
 Un-nested Specific Skills &  &  & -0.073^{***} & 0.026^{***} \\ 
  &  &  & \multicolumn{1}{c}{(-0.074$, $-0.072)} & \multicolumn{1}{c}{(0.025$, $0.027)} \\ 
  & & & & \\ 
 Constant & 4.671^{***} & 3.787^{***} & 4.471^{***} & 3.709^{***} \\ 
  & \multicolumn{1}{c}{(4.670$, $4.673)} & \multicolumn{1}{c}{(3.783$, $3.792)} & \multicolumn{1}{c}{(4.469$, $4.474)} & \multicolumn{1}{c}{(3.702$, $3.717)} \\ 
  & & & & \\ 
\hline \\[-1.8ex] 
Observations & \multicolumn{1}{c}{635,554} & \multicolumn{1}{c}{635,554} & \multicolumn{1}{c}{635,554} & \multicolumn{1}{c}{635,554} \\ 
R$^{2}$ & \multicolumn{1}{c}{0.012} & \multicolumn{1}{c}{0.200} & \multicolumn{1}{c}{0.141} & \multicolumn{1}{c}{0.203} \\ 
Adjusted R$^{2}$ & \multicolumn{1}{c}{0.012} & \multicolumn{1}{c}{0.200} & \multicolumn{1}{c}{0.141} & \multicolumn{1}{c}{0.203} \\ 
Residual Std. Error & \multicolumn{1}{c}{0.368 (df = 635552)} & \multicolumn{1}{c}{0.331 (df = 635551)} & \multicolumn{1}{c}{0.343 (df = 635550)} & \multicolumn{1}{c}{0.331 (df = 635549)} \\ 
F Statistic & \multicolumn{1}{c}{7,845.032$^{***}$ (df = 1; 635552)} & \multicolumn{1}{c}{79,439.180$^{***}$ (df = 2; 635551)} & \multicolumn{1}{c}{34,872.520$^{***}$ (df = 3; 635550)} & \multicolumn{1}{c}{40,451.410$^{***}$ (df = 4; 635549)} \\ 
\hline 
\hline \\[-1.8ex] 
\textit{Note:}  & \multicolumn{4}{r}{$^{*}$p$<$0.1; $^{**}$p$<$0.05; $^{***}$p$<$0.01} \\ 
\end{tabular}

%% file: Nature_HB_2023/tabsNHB/Mar_25_2024__Irregular_Hours__Gender.tex
\begin{tabular}{@{\extracolsep{5pt}}lD{.}{.}{-3} D{.}{.}{-3} } 
\\[-1.8ex]\hline 
\hline \\[-1.8ex] 
 & \multicolumn{2}{c}{\textit{Dependent variable:}} \\ 
\cline{2-3} 
\\[-1.8ex] & \multicolumn{2}{c}{Gender Dummy (Female = 1)} \\ 
\\[-1.8ex] & \multicolumn{2}{c}{\textit{OLS}} \\ 
\\[-1.8ex] & \multicolumn{1}{c}{(1)} & \multicolumn{1}{c}{(2)}\\ 
\hline \\[-1.8ex] 
 General Skills & 0.203^{***} & 0.150^{***} \\ 
  & \multicolumn{1}{c}{(0.201$, $0.204)} & \multicolumn{1}{c}{(0.148$, $0.153)} \\ 
  & & \\ 
 Nested Skills & -0.357^{***} & -0.258^{***} \\ 
  & \multicolumn{1}{c}{(-0.359$, $-0.355)} & \multicolumn{1}{c}{(-0.261$, $-0.256)} \\ 
  & & \\ 
 Irregular Schedule &  & -0.338^{***} \\ 
  &  & \multicolumn{1}{c}{(-0.342$, $-0.334)} \\ 
  & & \\ 
 Long Hours Dummy ($>50$ weekly) &  & -0.176^{***} \\ 
  &  & \multicolumn{1}{c}{(-0.178$, $-0.174)} \\ 
  & & \\ 
 Constant & 0.097^{***} & 0.629^{***} \\ 
  & \multicolumn{1}{c}{(0.092$, $0.101)} & \multicolumn{1}{c}{(0.620$, $0.638)} \\ 
  & & \\ 
\hline \\[-1.8ex] 
Observations & \multicolumn{1}{c}{1,493,142} & \multicolumn{1}{c}{1,096,362} \\ 
R$^{2}$ & \multicolumn{1}{c}{0.072} & \multicolumn{1}{c}{0.108} \\ 
Adjusted R$^{2}$ & \multicolumn{1}{c}{0.072} & \multicolumn{1}{c}{0.108} \\ 
Residual Std. Error & \multicolumn{1}{c}{0.463 (df = 1493139)} & \multicolumn{1}{c}{0.455 (df = 1096357)} \\ 
F Statistic & \multicolumn{1}{c}{57,942.160$^{***}$ (df = 2; 1493139)} & \multicolumn{1}{c}{33,058.290$^{***}$ (df = 4; 1096357)} \\ 
\hline 
\hline \\[-1.8ex] 
\textit{Note:}  & \multicolumn{2}{r}{$^{*}$p$<$0.1; $^{**}$p$<$0.05; $^{***}$p$<$0.01} \\ 
\end{tabular} 

%% file: Nature_HB_2023/tabsNHB/Apr_14__List_of_Managers.tex
\begin{longtable}{@{\extracolsep{\fill}} cclcc} 
  \caption{List of Manager Occupations and their Annual Wage and Education Requirements in our Sample.} 
  \label{tab:list of manager occupations} 
\\[-1.8ex]\hline 
\hline \\[-1.8ex] 
 & Code & Title & Wage & Education \\ 
\hline \\[-1.8ex]
\endhead
1 & 11-1011.00 & Chief Executives & \$ 170.5K & $7.540$ \\ 
2 & 11-1011.03 & Chief Sustainability Officers & \$ 170.5K & $6.920$ \\ 
3 & 11-9041.00 & Architectural and Engineering Managers & \$ 135.8K & $6.720$ \\ 
4 & 11-9041.01 & Biofuels/Biodiesel Technology and Product Devel... & \$ 135.8K & $6.480$ \\ 
5 & 11-2021.00 & Marketing Managers & \$ 124.2K & $6.680$ \\ 
6 & 11-3111.00 & Compensation and Benefits Managers & \$ 124.1K & $6.330$ \\ 
7 & 11-3021.00 & Computer and Information Systems Managers & \$ 123.8K & $5.550$ \\ 
8 & 11-2022.00 & Sales Managers & \$ 122.3K & $6.040$ \\ 
9 & 11-9121.00 & Natural Sciences Managers & \$ 122.1K & $8.130$ \\ 
10 & 11-9121.02 & Water Resource Specialists & \$ 122.1K & $6.860$ \\ 
11 & 11-9121.01 & Clinical Research Coordinators & \$ 122.1K & $6.060$ \\ 
12 & 11-3031.01 & Treasurers and Controllers & \$ 117.7K & $7.070$ \\ 
13 & 11-3031.02 & Financial Managers, Branch or Department & \$ 117.7K & $5.440$ \\ 
14 & 11-2011.00 & Advertising and Promotions Managers & \$ 116.3K & $5.210$ \\ 
15 & 11-3061.00 & Purchasing Managers & \$ 114.3K & $6.150$ \\ 
16 & 11-3131.00 & Training and Development Managers & \$ 112K & $6.630$ \\ 
17 & 11-3051.01 & Quality Control Systems Managers & \$ 109.6K & $6.030$ \\ 
18 & 11-3051.04 & Biomass Power Plant Managers & \$ 109.6K & $5.290$ \\ 
19 & 11-3051.03 & Biofuels Production Managers & \$ 109.6K & $4.980$ \\ 
20 & 11-3051.00 & Industrial Production Managers & \$ 109.6K & $4.920$ \\ 
21 & 11-3051.06 & Hydroelectric Production Managers & \$ 109.6K & $4.170$ \\ 
22 & 11-3051.02 & Geothermal Production Managers & \$ 109.6K & $4.060$ \\ 
23 & 11-3121.00 & Human Resources Managers & \$ 109.4K & $6.300$ \\ 
24 & 11-9033.00 & Education Administrators, Postsecondary & \$ 106.1K & $9.250$ \\ 
25 & 11-9111.00 & Medical and Health Services Managers & \$ 104.8K & $6.080$ \\ 
26 & 11-1021.00 & General and Operations Managers & \$ 103.8K & $4.920$ \\ 
27 & 11-9021.00 & Construction Managers & \$ 95.7K & $5.680$ \\ 
28 & 11-3071.03 & Logistics Managers & \$ 94.1K & $6.170$ \\ 
29 & 11-3071.02 & Storage and Distribution Managers & \$ 94.1K & $4.620$ \\ 
30 & 11-3071.01 & Transportation Managers & \$ 94.1K & $4.370$ \\ 
31 & 11-9032.00 & Education Administrators, Elementary and Second... & \$ 91.6K & $7.820$ \\ 
32 & 27-1011.00 & Art Directors & \$ 88.7K & $6.250$ \\ 
33 & 11-9039.01 & Distance Learning Coordinators & \$ 84.6K & $7.550$ \\ 
34 & 11-9039.02 & Fitness and Wellness Coordinators & \$ 84.6K & $6.580$ \\ 
35 & 15-2041.02 & Clinical Data Managers & \$ 83.7K & $6.100$ \\ 
36 & 11-9071.00 & Gaming Managers & \$ 81.6K & $3.680$ \\ 
37 & 11-9161.00 & Emergency Management Directors & \$ 79.8K & $6.120$ \\ 
38 & 11-9131.00 & Postmasters and Mail Superintendents & \$ 77K & $3.030$ \\ 
39 & 13-1011.00 & Agents and Business Managers of Artists, Perfor... & \$ 76.7K & $5.230$ \\ 
40 & 11-9013.01 & Nursery and Greenhouse Managers & \$ 76.6K & $4.960$ \\ 
41 & 11-9013.03 & Aquacultural Managers & \$ 76.6K & $4.420$ \\ 
42 & 11-9013.02 & Farm and Ranch Managers & \$ 76.6K & $3.880$ \\ 
43 & 19-1031.02 & Range Managers & \$ 67.3K & $5.950$ \\ 
44 & 11-9151.00 & Social and Community Service Managers & \$ 66.8K & $6.310$ \\ 
45 & 47-1011.03 & Solar Energy Installation Managers & \$ 66.1K & $3.680$ \\ 
46 & 25-9031.00 & Instructional Coordinators & \$ 64.6K & $7.700$ \\ 
47 & 27-2012.04 & Talent Directors & \$ 62.9K & $6.140$ \\ 
48 & 27-2012.05 & Technical Directors/Managers & \$ 62.9K & $5.770$ \\ 
49 & 27-2012.03 & Program Directors & \$ 62.9K & $5.080$ \\ 
50 & 27-2012.02 & Directors- Stage, Motion Pictures, Television, ... & \$ 62.9K & $4.570$ \\ 
51 & 11-9081.00 & Lodging Managers & \$ 60.9K & $4.890$ \\ 
52 & 11-9141.00 & Property, Real Estate, and Community Associatio... & \$ 60.2K & $5.040$ \\ 
53 & 11-9051.00 & Food Service Managers & \$ 56.7K & $2.540$ \\ 
54 & 39-4031.00 & Morticians, Undertakers, and Funeral Directors & \$ 56.5K & $4.810$ \\ 
55 & 27-2041.01 & Music Directors & \$ 56.2K & $8.210$ \\ 
56 & 11-9031.00 & Education Administrators, Preschool and Childca... & \$ 49.2K & $4.140$ \\ 
57 & 21-2021.00 & Directors, Religious Activities and Education & \$ 48.9K & $6$ \\ 
58 & 11-9199.03 & Investment Fund Managers & - & $7.410$ \\ 
59 & 11-9199.01 & Regulatory Affairs Managers & - & $6.500$ \\ 
60 & 15-1141.00 & Database Administrators & - & $6.440$ \\ 
61 & 11-9199.04 & Supply Chain Managers & - & $6.430$ \\ 
62 & 11-9199.07 & Security Managers & - & $6.150$ \\ 
63 & 11-9199.11 & Brownfield Redevelopment Specialists and Site M... & - & $6.120$ \\ 
64 & 11-2031.00 & Public Relations and Fundraising Managers & - & $6.100$ \\ 
65 & 11-9199.10 & Wind Energy Project Managers & - & $6.090$ \\ 
66 & 11-9199.09 & Wind Energy Operations Managers & - & $5.860$ \\ 
67 & 15-1199.09 & Information Technology Project Managers & - & $5.860$ \\ 
68 & 11-9061.00 & Funeral Service Managers & - & $5.710$ \\ 
69 & 11-9199.02 & Compliance Managers & - & $5.650$ \\ 
70 & 15-1142.00 & Network and Computer Systems Administrators & - & $5.580$ \\ 
71 & 15-1199.03 & Web Administrators & - & $5.350$ \\ 
72 & 11-9199.08 & Loss Prevention Managers & - & $4.950$ \\ 
73 & 39-1021.01 & Spa Managers & - & $4.220$ \\ 
74 & 11-3011.00 & Administrative Services Managers & - & $3.960$ \\ 
75 & 53-1021.01 & Recycling Coordinators & - & $3.890$ \\ 
\hline \\[-1.8ex] 
\end{longtable}

%% file: maincurrent.bib
@article{Hidalgo2009,
author = {Hidalgo, C{\'{e}}sar A and Hausmann, Ricardo},
doi = {10.1073/pnas.0900943106},
edition = {2009/06/22},
issn = {1091-6490},
journal = {Proceedings of the National Academy of Sciences of the United States of America},
language = {eng},
month = {jun},
number = {26},
pages = {10570--10575},
publisher = {National Academy of Sciences},
title = {{The building blocks of economic complexity}},
volume = {106},
year = {2009}
}

@article{Baldwin2014,
title = {Hidden structure: Using network methods to map system architecture},
journal = {Research Policy},
volume = {43},
number = {8},
pages = {1381-1397},
year = {2014},
issn = {0048-7333},
doi = {https://doi.org/10.1016/j.respol.2014.05.004},
url = {https://www.sciencedirect.com/science/article/pii/S0048733314001012},
author = {Carliss Baldwin and Alan MacCormack and John Rusnak}
}

@article{Saavedra2011,
author = {Saavedra, Serguei and Stouffer, Daniel B. and Uzzi, Brian and Bascompte, Jordi},
issn = {00280836},
journal = {Nature},
number = {7368},
pages = {233--235},
pmid = {21918515},
publisher = {Nature Publishing Group},
title = {{Strong contributors to network persistence are the most vulnerable to extinction}},
url = {http://dx.doi.org/10.1038/nature10433},
volume = {478},
year = {2011}
}

@article{Mariani2019,
archivePrefix = {arXiv},
arxivId = {1905.07593},
author = {Mariani, Manuel Sebastian and Ren, Zhuo Ming and Bascompte, Jordi and Tessone, Claudio Juan},
doi = {10.1016/j.physrep.2019.04.001},
eprint = {1905.07593},
issn = {03701573},
journal = {Physics Reports},
pages = {1--90},
title = {{Nestedness in complex networks: Observation, emergence, and implications}},
url = {https://www.sciencedirect.com/science/article/pii/S037015731930119X},
volume = {813},
year = {2019}
}

@article{Staniczenko2023,
author = {Staniczenko, Phillip P A and Panja, Debabrata},
doi = {10.1093/pnasnexus/pgad412},
issn = {2752-6542},
journal = {PNAS Nexus},
month = {dec},
number = {12},
pages = {pgad412},
title = {{Temporal origin of nestedness in interaction networks}},
url = {https://doi.org/10.1093/pnasnexus/pgad412},
volume = {2},
year = {2023}
}

@article{SergeiMaslov2002,
address = {Washington, DC},
archivePrefix = {arXiv},
arxivId = {cond-mat/0205380},
author = {Maslov, Sergei and Sneppen, Kim},
doi = {10.1126/science.1065103},
eprint = {0205380},
issn = {00368075},
journal = {Science},
number = {5569},
pages = {910--913},
pmid = {11988575},
primaryClass = {cond-mat},
publisher = {American Society for the Advancement of Science},
title = {{Specificity and stability in topology of protein networks}},
volume = {296},
year = {2002}
}

@article{Bastolla2009,
author = {Bastolla, Ugo and Fortuna, Miguel A. and Pascual-Garc{\'{i}}a, Alberto and Ferrera, Antonio and Luque, Bartolo and Bascompte, Jordi},
doi = {10.1038/nature07950},
issn = {00280836},
journal = {Nature},
number = {7241},
pages = {1018--1020},
pmid = {19396144},
title = {{The architecture of mutualistic networks minimizes competition and increases biodiversity}},
volume = {458},
year = {2009}
}

@article{Almeida-neto2008,
author = {Almeida-neto, M{\'{a}}rio and Guimar{\~{a}}es, Paulo and Jr, Paulo R Guimar{\~{a}}es and Loyola, Rafael D and Ulrich, Werner},
journal = {Oikos},
number = {8},
pages = {1227--1239},
title = {{A consistent metric for nestedness analysis in ecological systems: reconciling concept and measurement}},
volume = {117},
year = {2008}
}

@article{Stone1990,
author = {Stone, Lewi and Roberts, Alan},
doi = {10.1007/BF00317345},
issn = {1432-1939},
journal = {Oecologia},
number = {1},
pages = {74--79},
title = {{The checkerboard score and species distributions}},
url = {https://doi.org/10.1007/BF00317345},
volume = {85},
year = {1990}
}

@article{Bascompte2003,
author = {Bascompte, Jordi and Jordano, Pedro and Meli{\'{a}}n, Carlos J and Olesen, Jens M},
doi = {10.1073/pnas.1633576100},
journal = {Proceedings of the National Academy of Sciences},
month = {aug},
number = {16},
pages = {9383--9387},
publisher = {Proceedings of the National Academy of Sciences},
title = {{The nested assembly of plant–animal mutualistic networks}},
url = {https://doi.org/10.1073/pnas.1633576100},
volume = {100},
year = {2003}
}

@article{Saavedra2009,
author = {Saavedra, Serguei and Reed-Tsochas, Felix and Uzzi, Brian},
doi = {10.1038/nature07532},
issn = {00280836},
journal = {Nature},
number = {7228},
pages = {463--466},
pmid = {19052545},
publisher = {Nature Publishing Group},
title = {{A simple model of bipartite cooperation for ecological and organizational networks}},
volume = {457},
year = {2009}
}

@article{BenJ2009,
    author = {Jones, Benjamin F.},
    title = "{The Burden of Knowledge and the “Death of the Renaissance Man”: Is Innovation Getting Harder?}",
    journal = {The Review of Economic Studies},
    volume = {76},
    number = {1},
    pages = {283-317},
    year = {2009},
    month = {01},
    issn = {0034-6527},
    doi = {10.1111/j.1467-937X.2008.00531.x},
}

@article{Azoulay2020,
author = {Azoulay, Pierre and Jones, Benjamin F and Kim, J Daniel and Miranda, Javier},
doi = {10.1257/aeri.20180582},
journal = {American Economic Review: Insights},
number = {1},
pages = {65--82},
title = {{Age and High-Growth Entrepreneurship}},
url = {https://www.aeaweb.org/articles?id=10.1257/aeri.20180582},
volume = {2},
year = {2020}
}

@article{Harmand2015,
author = {Harmand, Sonia and Lewis, Jason E and Feibel, Craig S and Lepre, Christopher J and Prat, Sandrine and Lenoble, Arnaud and Bo{\"{e}}s, Xavier and Quinn, Rhonda L and Brenet, Michel and Arroyo, Adrian and Taylor, Nicholas and Cl{\'{e}}ment, Sophie and Daver, Guillaume and Brugal, Jean-Philip and Leakey, Louise and Mortlock, Richard A and Wright, James D and Lokorodi, Sammy and Kirwa, Christopher and Kent, Dennis V and Roche, H{\'{e}}l{\`{e}}ne},
doi = {10.1038/nature14464},
issn = {1476-4687},
journal = {Nature},
number = {7552},
pages = {310--315},
title = {{3.3-million-year-old stone tools from Lomekwi 3, West Turkana, Kenya}},
url = {https://doi.org/10.1038/nature14464},
volume = {521},
year = {2015}
}

@article{Kogan2021,
  title={Technology, vintage-specific human capital, and labor displacement: Evidence from linking patents with occupations},
  author={Kogan, Leonid and Papanikolaou, Dimitris and Schmidt, Lawrence DW and Seegmiller, Bryan},
  year={2021},
  issn = {1556-5068},
  journal = {SSRN},
  note = {SSRN 3585676},
  url = {http://dx.doi.org/10.2139/ssrn.3585676},
  institution={National Bureau of Economic Research}
}

@article{AndersonKatharineA2017Snam,
author = {Anderson, Katharine A},
issn = {1091-6490},
journal = {Proceedings of the National Academy of Sciences},
number = {48},
pages = {12720--12724},
publisher = {Proceedings of the National Academy of Sciences},
title = {{Skill networks and measures of complex human capital}},
volume = {114},
year = {2017}
}

@article{Alabdulkareem2018,
author = {Alabdulkareem, Ahmad and Frank, Morgan R. and Sun, Lijun and AlShebli, Bedoor and Hidalgo, C{\'{e}}sar and Rahwan, Iyad},
issn = {23752548},
journal = {Science Advances},
number = {7},
pages = {1--10},
title = {{Unpacking the polarization of workplace skills}},
volume = {4},
year = {2018}
}

@article{Althobaiti2022,
archivePrefix = {arXiv},
arxivId = {2204.07073},
author = {Althobaiti, Shahad and Alabdulkareem, Ahmad and Shen, Judy Hanwen and Rahwan, Iyad and Frank, Morgan and Moro, Esteban and Rutherford, Alex},
eprint = {2204.07073},
journal = {arXiv preprint arXiv:2204.07073},
title = {{Longitudinal Complex Dynamics of Labour Markets Reveal Increasing Polarisation}},
url = {https://arxiv.org/pdf/2204.07073v1.pdf},
year = {2022}
}

@article{Neffkeeaax3370,
author = {Neffke, Frank M H},
doi = {10.1126/sciadv.aax3370},
journal = {Science Advances},
number = {12},
publisher = {American Association for the Advancement of Science},
title = {{The value of complementary co-workers}},
url = {https://advances.sciencemag.org/content/5/12/eaax3370},
volume = {5},
year = {2019}
}

@article{Stephany2024,
author = {Stephany, Fabian and Teutloff, Ole},
doi = {https://doi.org/10.1016/j.respol.2023.104898},
issn = {0048-7333},
journal = {Research Policy},
number = {1},
pages = {104898},
title = {{What is the price of a skill? The value of complementarity}},
url = {https://www.sciencedirect.com/science/article/pii/S0048733323001828},
volume = {53},
year = {2024}
}

@article{Becker1962,
author = {Becker, Gary S.},
journal = {Journal of Political Economy},
number = {5},
pages = {9--49},
title = {{Investment in Human Capital : A Theoretical Analysis}},
volume = {70},
year = {1962}
}

@book{becker2009human, title={Human capital: A theoretical and empirical analysis, with special reference to education}, author={Becker, Gary S}, year={2009}, publisher={University of Chicago press} }

@book{williamson2007economic, title={The economic institutions of capitalism. Firms, markets, relational contracting}, author={Williamson, Oliver E}, year={2007}, publisher={Springer} }

@article{jacobson1993earnings, title={Earnings losses of displaced workers}, author={Jacobson, Louis S and LaLonde, Robert J and Sullivan, Daniel G}, journal={The American economic review}, pages={685--709}, year={1993}, publisher={JSTOR} }

@article{heckman2011economics,
  title={The economics of inequality: The value of early childhood education.},
  author={Heckman, James J},
  journal={American Educator},
  volume={35},
  number={1},
  pages={31},
  year={2011},
  publisher={ERIC}
}

@book{Mincer1974,
address = {New York},
author = {Mincer, Jacob},
isbn = {0870142658},
publisher = {National Bureau of Economic Research; distributed by Columbia University Press},
series = {Human behavior and social institutions, 2},
title = {{Schooling, experience, and earnings.}},
year = {1974}
}

@article{Gathmann2010,
author = {Gathmann, Christina and Sch{\"{o}}nberg, Uta},
doi = {10.1086/649786},
issn = {0734306X},
journal = {Journal of Labor Economics},
number = {1},
pages = {1--49},
title = {{How general is human capital? A task-based approach}},
volume = {28},
year = {2010}
}

@article{Poletaev2008,
author = {Poletaev, Maxim and Robinson, Chris},
doi = {10.1086/588180},
issn = {0734306X},
journal = {Journal of Labor Economics},
number = {3},
pages = {387--420},
title = {{Human capital specificity: Evidence from the Dictionary of Occupational Titles and Displaced Worker Surveys, 1984-2000}},
volume = {26},
year = {2008}
}

@book{Davidson1898book,
address = {New York},
author = {Davidson, John},
publisher = {G. P. Putnam's Sons},
title = {{The bargain theory of wages ...}},
year = {1898}
}

@incollection{Arrow1962,
author = {Arrow, Kenneth},
booktitle = {The Rate and Direction of Inventive Activity: Economic and Social Factors},
pages = {609--626},
publisher = {National Bureau of Economic Research, Inc.},
title = {{Economic Welfare and the Allocation of Resources for Invention}},
year = {1962}
}

@book{Goldin2008,
author = {Goldin, Claudia and Katz, Lawrence F.},
booktitle = {The Race between Education and Technology},
isbn = {9780674028678},
pages = {163--246},
title = {{The Race Between Technology \& Education}},
year = {2008}
}

@article{goos2007lousy,
  title={Lousy and lovely jobs: The rising polarization of work in Britain},
  author={Goos, Maarten and Manning, Alan},
  journal={The review of economics and statistics},
  volume={89},
  number={1},
  pages={118--133},
  year={2007},
  publisher={The MIT Press}
}

@article{Bertrand2009,
author = {Bertrand, Marianne and Goldin, Claudia and Katz, Lawrence F},
title = {{Dynamics of the gender gap for young professionals in the corporate and financial sectors}},
year = {2009}
}

@article{Canon2016,
author = {Canon, Maria and Golan, Limor},
journal = {The Regional Economist},
number = {July},
publisher = {Federal Reserve Bank of St. Louis},
title = {{Gender Pay Gap May Be Linked to Flexible and Irregular Hours}},
year = {2016}
}

@article{Goldin2015,
author = {Goldin, Claudia},
journal = {Center for American Progress},
title = {{Hours flexibility and the gender gap in pay}},
volume = {31},
year = {2015}
}

@article{Cha2014,
author = {Cha, Youngjoo and Weeden, Kim A},
issn = {0003-1224},
journal = {American Sociological Review},
number = {3},
pages = {457--484},
publisher = {Sage Publications Sage CA: Los Angeles, CA},
title = {{Overwork and the slow convergence in the gender gap in wages}},
volume = {79},
year = {2014}
}

@article{DelRio-Chanona2021,
author = {del Rio-Chanona, R Maria and Mealy, Penny and Beguerisse-D{\'{i}}az, Mariano and Lafond, Fran{\c{c}}ois and Farmer, J Doyne},
doi = {10.1098/rsif.2020.0898},
journal = {Journal of The Royal Society Interface},
month = {jun},
number = {174},
pages = {20200898},
publisher = {Royal Society},
title = {{Occupational mobility and automation: a data-driven network model}},
url = {https://doi.org/10.1098/rsif.2020.0898},
volume = {18},
year = {2021}
}

@article{Neffke2013,
author = {Neffke, Frank and Henning, Martin},
doi = {https://doi.org/10.1002/smj.2014},
issn = {0143-2095},
journal = {Strategic Management Journal},
month = {mar},
number = {3},
pages = {297--316},
publisher = {John Wiley \& Sons, Ltd},
title = {{Skill relatedness and firm diversification}},
url = {https://doi.org/10.1002/smj.2014},
volume = {34},
year = {2013}
}

@article{Frank2018,
author = {Frank, M.R. and Sun, L. and Cebrian, M. and Youn, H. and Rahwan, I.},
issn = {1742-5662},
journal = {Journal of the Royal Society Interface},
number = {139},
title = {{Small cities face greater impact from automation}},
url = {https://royalsocietypublishing.org/doi/10.1098/rsif.2017.0946},
volume = {15},
year = {2018}
}

@article{Autor2003,
author = {Autor, David H. and Levy, Frank and Murnane, Richard J.},
doi = {10.1162/003355303322552801},
issn = {00335533},
journal = {Quarterly Journal of Economics},
number = {4},
pages = {1279--1333},
title = {{The skill content of recent technological change: An empirical exploration}},
volume = {118},
year = {2003}
}

@article{Autor2014,
author = {Autor, David H},
doi = {10.1126/science.1251868},
journal = {Science},
month = {may},
number = {6186},
pages = {843--851},
publisher = {American Association for the Advancement of Science},
title = {{Skills, education, and the rise of earnings inequality among the “other 99 percent”}},
url = {https://doi.org/10.1126/science.1251868},
volume = {344},
year = {2014}
}

@article{Autor2013,
Author = {Autor, David H. and Dorn, David},
Title = {The Growth of Low-Skill Service Jobs and the Polarization of the US Labor Market},
Journal = {American Economic Review},
Volume = {103},
Number = {5},
Year = {2013},
Month = {August},
Pages = {1553-97},
URL = {https://www.aeaweb.org/articles?id=10.1257/aer.103.5.1553},
DOI = {10.1257/aer.103.5.1553}}

@techreport{Autor2022,
author = {Autor, David and Chin, Caroline and Salomons, Anna M and Seegmiller, Bryan},
publisher = {National Bureau of Economic Research},
title = {{New Frontiers: The Origins and Content of New Work, 1940–2018}},
year = {2022}
}

@article{SchwabeHenrik2020Awsa,
address = {United States},
author = {Schwabe, Henrik and Castellacci, Fulvio},
issn = {1932-6203},
journal = {PloS one},
number = {11},
pages = {e0242929--e0242929},
publisher = {Public Library of Science},
title = {{Automation, workers' skills and job satisfaction}},
volume = {15},
year = {2020}
}

@article{Frank2019,
author = {Frank, Morgan R and Autor, David and Bessen, James E and Brynjolfsson, Erik and Cebrian, Manuel and Deming, David J and Feldman, Maryann and Groh, Matthew and Lobo, Jos{\'{e}} and Moro, Esteban and Wang, Dashun and Youn, Hyejin and Rahwan, Iyad},
doi = {10.1073/pnas.1900949116},
journal = {Proceedings of the National Academy of Sciences},
month = {mar},
pages = {201900949},
title = {{Toward understanding the impact of artificial intelligence on labor}},
url = {http://www.pnas.org/content/early/2019/03/21/1900949116.abstract},
year = {2019}
}

@article{Moro2021,
author = {Moro, Esteban and Frank, Morgan R and Pentland, Alex and Rutherford, Alex and Cebrian, Manuel and Rahwan, Iyad},
doi = {10.1038/s41467-021-22086-3},
issn = {2041-1723},
journal = {Nature Communications},
number = {1},
pages = {1972},
title = {{Universal resilience patterns in labor markets}},
url = {https://doi.org/10.1038/s41467-021-22086-3},
volume = {12},
year = {2021}
}

@article{tacchella2012new,
  title={A new metrics for countries' fitness and products' complexity},
  author={Tacchella, Andrea and Cristelli, Matthieu and Caldarelli, Guido and Gabrielli, Andrea and Pietronero, Luciano},
  journal={Scientific reports},
  volume={2},
  number={1},
  pages={723},
  year={2012},
  publisher={Nature Publishing Group UK London}
}

@inproceedings{hidalgo2018principle,
  title={The principle of relatedness},
  author={Hidalgo, C{\'e}sar A and Balland, Pierre-Alexandre and Boschma, Ron and Delgado, Mercedes and Feldman, Maryann and Frenken, Koen and Glaeser, Edward and He, Canfei and Kogler, Dieter F and Morrison, Andrea and others},
  booktitle={Unifying Themes in Complex Systems IX: Proceedings of the Ninth International Conference on Complex Systems 9},
  pages={451--457},
  year={2018},
  organization={Springer}
}

@article{write1992,
  title={On the meaning and measurement of nestedness of species assemblages},
  author={Wright, D.H. and Reeves, J.H},
  number = {2},
  journal = {Oecologia},
  doi = {10.1007/BF00317469},
  pages = {416--428},
  volume = {92},
  year = {1992}
}

@article{BeckerG.S1992TDoL,
address = {Cambridge, Mass. [etc.]},
author = {Becker, G S and Murphy, K M},
issn = {1531-4650},
journal = {The Quarterly journal of economics},
number = {4},
pages = {1137--1160},
publisher = {Oxford University Press (OUP)},
title = {{The Division of Labor, Coordination Costs, and Knowledge}},
volume = {107},
year = {1992}
}

@article{Balland2020,
author = {Balland, Pierre-Alexandre and Jara-Figueroa, Cristian and Petralia, Sergio G and Steijn, Mathieu P A and Rigby, David L and Hidalgo, C{\'{e}}sar A},
doi = {10.1038/s41562-019-0803-3},
issn = {2397-3374},
journal = {Nature Human Behaviour},
title = {{Complex economic activities concentrate in large cities}},
url = {https://doi.org/10.1038/s41562-019-0803-3},
year = {2020}
}

@article{Lucas1988,
author = {Lucas, Robert E},
doi = {https://doi.org/10.1016/0304-3932(88)90168-7},
issn = {0304-3932},
journal = {Journal of Monetary Economics},
number = {1},
pages = {3--42},
title = {{On the mechanics of economic development}},
url = {http://www.sciencedirect.com/science/article/pii/0304393288901687},
volume = {22},
year = {1988}
}

@unpublished{Jo2020,
archivePrefix = {arXiv},
arxivId = {2002.07239},
author = {Jo, Woo Seong and Park, Jaehyuk and Luhur, Arthur and Kim, Beom Jun and Ahn, Yong-Yeol},
eprint = {2002.07239},
pages = {1--8},
title = {{Extracting hierarchical backbones from bipartite networks}},
url = {http://arxiv.org/abs/2002.07239},
year = {2020}
}

@article{Lin2022,
author = {Lin, Ken-Hou and Hung, Koit},
doi = {10.1086/719407},
issn = {0002-9602},
journal = {American Journal of Sociology},
month = {mar},
number = {5},
pages = {1551--1601},
publisher = {The University of Chicago Press},
title = {{The Network Structure of Occupations: Fragmentation, Differentiation, and Contagion}},
url = {https://doi.org/10.1086/719407},
volume = {127},
year = {2022}
}

@article{Hopson2021,
author = {Hopson, Amy},
journal = {Monthly Labor Review},
pages = {1--11},
title = {{Mapping Employment Projections and O*NET data: a methodological overview}},
url = {https://www.bls.gov/opub/mlr/2021/article/mapping-employment-projections-and-onet-data.htm#top},
year = {2021}
}

@article{Hidalgo2021,
author = {Hidalgo, C{\'{e}}sar A},
doi = {10.1038/s42254-020-00275-1},
issn = {2522-5820},
journal = {Nature Reviews Physics},
title = {{Economic complexity theory and applications}},
url = {https://doi.org/10.1038/s42254-020-00275-1},
year = {2021}
}

@article{Serrano6483,
author = {Serrano, M {\'{A}}ngeles and Bogu{\~{n}}{\'{a}}, Mari{\'{a}}n and Vespignani, Alessandro},
doi = {10.1073/pnas.0808904106},
issn = {0027-8424},
journal = {Proceedings of the National Academy of Sciences},
number = {16},
pages = {6483--6488},
publisher = {National Academy of Sciences},
title = {{Extracting the multiscale backbone of complex weighted networks}},
url = {https://www.pnas.org/content/106/16/6483},
volume = {106},
year = {2009}
}

@article{Borner2018,
author = {B{\"{o}}rner, Katy and Scrivner, Olga and Gallant, Mike and Ma, Shutian and Liu, Xiaozhong and Chewning, Keith and Wu, Lingfei and Evans, James A.},
doi = {10.1073/pnas.1804247115},
issn = {10916490},
journal = {Proceedings of the National Academy of Sciences},
number = {50},
pages = {12630--12637},
pmid = {30530667},
title = {{Skill discrepancies between research, education, and jobs reveal the critical need to supply soft skills for the data economy}},
url = {http://www.pnas.org/lookup/doi/10.1073/pnas.1804247115},
volume = {115},
year = {2018}
}

@article{DemingDavidJ2020EDCJ,
author = {Deming, David J and Noray, Kadeem},
issn = {0033-5533},
journal = {The Quarterly journal of economics},
number = {4},
pages = {1965--2005},
title = {{Earnings Dynamics, Changing Job Skills, and STEM Careers}},
volume = {135},
year = {2020}
}

@article{Deming2017,
author = {Deming, David J.},
doi = {10.1093/qje/qjx022},
issn = {15314650},
journal = {Quarterly Journal of Economics},
number = {4},
pages = {1593--1640},
title = {{The growing importance of social skills in the labor market}},
volume = {132},
year = {2017}
}

@article{Evans2024,
author = {Evans, David and Mason, Claire and Chen, Haohui and Reeson, Andrew},
doi = {10.1038/s41562-023-01788-2},
issn = {2397-3374},
journal = {Nature Human Behaviour},
number = {1},
pages = {32--42},
title = {{Accelerated demand for interpersonal skills in the Australian post-pandemic labour market}},
url = {https://doi.org/10.1038/s41562-023-01788-2},
volume = {8},
year = {2024}
}

@article{mcnerney2021bridging,
  title={Bridging the short-term and long-term dynamics of economic structural change},
  author={McNerney, James and Li, Yang and Gomez-Lievano, Andres and Neffke, Frank},
  journal={arXiv preprint arXiv:2110.09673},
  year={2021}
}

@unpublished{Tong2021,
author = {Tong, Di and Wu, Lingfei and Evans, James Allen},
institution = {arXiv},
title = {{Low-skilled Occupations Face the Highest Re-skilling Pressure}},
url = {https://arxiv.org/abs/2101.11505},
year = {2021}
}

@article{mcnerney2011, 
author = {James McNerney  and J. Doyne Farmer  and Sidney Redner  and Jessika E. Trancik },
title = {Role of design complexity in technology improvement},
journal = {Proceedings of the National Academy of Sciences},
volume = {108},
number = {22},
pages = {9008-9013},
year = {2011},
doi = {10.1073/pnas.1017298108},
URL = {https://www.pnas.org/doi/abs/10.1073/pnas.1017298108},
eprint = {https://www.pnas.org/doi/pdf/10.1073/pnas.1017298108}
}

@misc{yang2022scaling,
      title={Scaling and the Universality of Function Diversity Across Human Organizations}, 
      author={Vicky Chuqiao Yang and Christopher P. Kempes and Hyejin Youn and Sidney Redner and Geoffrey B. West},
      year={2022},
      eprint={2208.06487},
      archivePrefix={arXiv},
      primaryClass={physics.soc-ph}
}

@article{Hermo2022,
author = {Hermo, Santiago and P{\"{a}}{\"{a}}llysaho, Miika and Seim, David and Shapiro, Jesse M},
doi = {10.1093/qje/qjac022},
issn = {0033-5533},
journal = {The Quarterly Journal of Economics},
month = {nov},
number = {4},
pages = {2309--2361},
title = {{Labor Market Returns and the Evolution of Cognitive Skills: Theory and Evidence}},
url = {https://doi.org/10.1093/qje/qjac022},
volume = {137},
year = {2022}
}

@article{Kosack2018,
author = {Kosack, Stephen and Coscia, Michele and Smith, Evann and Albrecht, Kim and Barab{\'{a}}si, Albert-L{\'{a}}szl{\'{o}} and Hausmann, Ricardo},
doi = {10.1073/pnas.1803228115},
journal = {Proceedings of the National Academy of Sciences},
month = {11},
number = {46},
pages = {11748--11753},
publisher = {Proceedings of the National Academy of Sciences},
title = {{Functional structures of US state governments}},
url = {https://doi.org/10.1073/pnas.1803228115},
volume = {115},
year = {2018}
}

@article{Mones2012,
author = {Mones, Enys and Vicsek, Lilla and Vicsek, Tam{\'{a}}s},
doi = {10.1371/journal.pone.0033799},
issn = {1932-6203 (Electronic)},
journal = {PloS one},
language = {eng},
number = {3},
pages = {e33799},
pmid = {22470477},
title = {{Hierarchy measure for complex networks.}},
volume = {7},
year = {2012}
}

@article{Hausmann2011,
author = {Hausmann, Ricardo and Hidalgo, C{\'{e}}sar A},
doi = {10.1007/s10887-011-9071-4},
issn = {1573-7020},
journal = {Journal of Economic Growth},
number = {4},
pages = {309--342},
title = {{The network structure of economic output}},
url = {https://doi.org/10.1007/s10887-011-9071-4},
volume = {16},
year = {2011}
}

@article{Hidalgo2007,
author = {Hidalgo, C{\'{e}}sar A and Winger, B. and Barab{\'{a}}si, A. L. and Hausmann, R.},
isbn = {1095-9203 (Electronic)\n0036-8075 (Linking)},
issn = {00368075},
journal = {Science},
number = {5837},
pages = {482--487},
pmid = {17656717},
title = {{The product space conditions the development of nations}},
volume = {317},
year = {2007}
}

@article{Leung2014,
author = {Leung, Ming D},
doi = {10.1177/0003122413518638},
issn = {0003-1224},
journal = {American Sociological Review},
month = {jan},
number = {1},
pages = {136--158},
publisher = {SAGE Publications Inc},
title = {{Dilettante or Renaissance Person? How the Order of Job Experiences Affects Hiring in an External Labor Market}},
url = {https://doi.org/10.1177/0003122413518638},
volume = {79},
year = {2014}
}

@article{Liu2013,
author = {Liu, Yujia and Grusky, David B},
issn = {0002-9602},
journal = {The American journal of sociology},
number = {5},
pages = {1330--1374},
title = {{The Payoff to Skill in the Third Industrial Revolution}},
volume = {118},
year = {2013}
}

@article{OClery2021,
author = {O'Clery, Neave and Yıldırım, Muhammed Ali and Hausmann, Ricardo},
doi = {10.1038/s41467-021-21689-0},
issn = {2041-1723},
journal = {Nature Communications},
number = {1},
pages = {1479},
title = {{Productive Ecosystems and the arrow of development}},
url = {https://doi.org/10.1038/s41467-021-21689-0},
volume = {12},
year = {2021}
}

@incollection{henrich2015secret,
  title={The secret of our success},
  author={Henrich, Joseph},
  booktitle={The Secret of Our Success},
  year={2015},
  publisher={princeton University press}
}

@article{richerson1999complex,
  title={Complex societies: The evolutionary origins of a crude superorganism},
  author={Richerson, Peter J and Boyd, Robert},
  journal={Human nature},
  volume={10},
  pages={253--289},
  year={1999},
  publisher={Springer}
}

@book{Hidalgo2015,
address = {Boulder, UNITED STATES},
author = {Hidalgo, C{\'{e}}sar A},
isbn = {9780465039715},
publisher = {Basic Books},
title = {{Why Information Grows : The Evolution of Order, from Atoms to Economies}},
url = {http://ebookcentral.proquest.com/lib/uic/detail.action?docID=2039752},
year = {2015}
}

@book{MitchellMelanie2009,
address = {Oxford [England] ;},
author = {Mitchell, Melanie},
booktitle = {Oxford University Press},
isbn = {9780195124415},
publisher = {Oxford University Press},
title = {{Complexity : a guided tour}},
year = {2009}
}

@book{Peterson1999ONET,
address = {Place of publication not identified},
isbn = {1-55798-556-1},
publisher = {American Psychological Association},
title = {{An occupational information system for the 21st century : the development of O*NET}},
year = {1999}
}

@misc{yoon2023,
      title={What makes Individual I's a Collective We; Coordination mechanisms \& costs}, 
      author={Jisung Yoon and Chris Kempes and Vicky Chuqiao Yang and Geoffrey West and Hyejin Youn},
      year={2023},
      eprint={2306.02113},
      archivePrefix={arXiv},
      primaryClass={physics.soc-ph}
}

@article{Youn2016,
author = {Youn, Hyejin  and Bettencourt, Luís M. A.  and Lobo, José  and Strumsky, Deborah  and Samaniego, Horacio  and West, Geoffrey B. },
title = {Scaling and universality in urban economic diversification},
journal = {Journal of The Royal Society Interface},
volume = {13},
number = {114},
pages = {20150937},
year = {2016},
doi = {10.1098/rsif.2015.0937},
URL = {https://royalsocietypublishing.org/doi/abs/10.1098/rsif.2015.0937},
eprint = {https://royalsocietypublishing.org/doi/pdf/10.1098/rsif.2015.0937}
}

@article{Bettencourt2014,
author = {Bettencourt, Luís M. A. and Samaniego, Horacio and Youn, Hyejin},
title = {Professional diversity and the productivity of cities},
journal = {Scientific Reports},
volume = {4},
number = {1},
pages = {5393},
year = {2014},
doi = {10.1038/srep05393},
URL = {https://doi.org/10.1038/srep05393}
}

@article{Hong2020,
author = {Hong, Inho and Frank, Morgan R. and Rahwan, Iyad and Jung, Woo Sung and Youn, Hyejin},
doi = {10.1126/sciadv.aba4934},
issn = {23752548},
journal = {Science Advances},
number = {34},
pages = {1--7},
pmid = {32937361},
title = {{The universal pathway to innovative urban economies}},
volume = {6},
year = {2020}
}

@unpublished{Park2020,
author = {Park, Jaehyuk and Sun, Lijun and Youn, Hyejin},
year = 2020,
title = {{Industrial Topics in Urban Labor System}}
}

@article{Acemoglu2011,
archivePrefix = {arXiv},
arxivId = {arXiv:1011.1669v3},
author = {Acemoglu, Daron and Autor, David},
doi = {10.1016/S0169-7218(11)02410-5},
eprint = {arXiv:1011.1669v3},
isbn = {978-0-444-53452-1},
issn = {15734463},
journal = {Handbook of Labor Economics},
number = {PART B},
pages = {1043--1171},
pmid = {25246403},
title = {{Skills, tasks and technologies: Implications for employment and earnings}},
volume = {4},
year = {2011}
}

@article{Frey2017,
archivePrefix = {arXiv},
arxivId = {1602.03506},
author = {Frey, Carl Benedikt and Osborne, Michael A.},
doi = {10.1016/j.techfore.2016.08.019},
eprint = {1602.03506},
isbn = {9780198570509},
issn = {00401625},
journal = {Technological Forecasting and Social Change},
pages = {254--280},
pmid = {23961883},
title = {{The future of employment: How susceptible are jobs to computerisation?}},
volume = {114},
year = {2017}
}

@unpublished{Frank2022,
author = {Frank, Morgan R and Ahn, Yong-yeol and Moro, Esteban},
title = {{AI exposure predicts unemployment risk}},
year = {2022}
}

@book{Gamble2002,
address = {London},
author = {Gamble and Blackwell},
issn = {0012-3242},
journal = {Director (London, England : 1983)},
number = {6},
publisher = {Institute of Directors},
title = {{Knowledge Management: A state of the art guide}},
volume = {55},
year = {2002}
}

@article{gomez2016explaining,
  title={Explaining the prevalence, scaling and variance of urban phenomena},
  author={Gomez-Lievano, Andres and Patterson-Lomba, Oscar and Hausmann, Ricardo},
  journal={Nature Human Behaviour},
  volume={1},
  number={1},
  pages={0012},
  year={2016},
  publisher={Nature Publishing Group UK London}
}

@article{Glaeser1999,
author = {Glaeser, Edward L},
doi = {https://doi.org/10.1006/juec.1998.2121},
issn = {0094-1190},
journal = {Journal of Urban Economics},
number = {2},
pages = {254--277},
title = {{Learning in Cities}},
volume = {46},
year = {1999}
}

@article{Glaeser2001,
author = {Glaeser, Edward L. and Mare, David C.},
doi = {10.1086/319563},
issn = {0734-306X},
journal = {Journal of Labor Economics},
month = {apr},
number = {2},
pages = {316--342},
publisher = {The University of Chicago Press},
title = {{Cities and Skills}},
url = {https://doi.org/10.1086/319563},
volume = {19},
year = {2001}
}

@article{Wheeler2001,
author = {Wheeler, Christopher H.},
doi = {10.1086/322823},
issn = {0734-306X},
journal = {Journal of Labor Economics},
month = {oct},
number = {4},
pages = {879--899},
publisher = {The University of Chicago Press},
title = {{Search, Sorting, and Urban Agglomeration}},
url = {https://doi.org/10.1086/322823},
volume = {19},
year = {2001}
}

@article{Bacolod2009,
author = {Bacolod, Marigee and Blum, Bernardo S and Strange, William C},
doi = {https://doi.org/10.1016/j.jue.2008.09.003},
issn = {0094-1190},
journal = {Journal of Urban Economics},
number = {2},
pages = {136--153},
title = {{Skills in the city}},
url = {https://www.sciencedirect.com/science/article/pii/S0094119008001083},
volume = {65},
year = {2009}
}

@article{Acemoglu2020,
author = {Acemoglu, Daron and Restrepo, Pascual},
doi = {10.1093/cjres/rsz022},
issn = {17521386},
journal = {Cambridge Journal of Regions, Economy and Society},
number = {1},
pages = {25--35},
title = {{The wrong kind of AI? Artificial intelligence and the future of labour demand}},
volume = {13},
year = {2020}
}

@article{McNerney2022,
archivePrefix = {arXiv},
arxivId = {1810.07774},
author = {McNerney, James and Savoie, Charles and Caravelli, Francesco and Carvalho, Vasco M. and {Doyne Farmer}, J.},
doi = {10.1073/pnas.2106031118},
eprint = {1810.07774},
issn = {10916490},
journal = {Proceedings of the National Academy of Sciences of the United States of America},
number = {1},
pmid = {34949713},
title = {{How production networks amplify economic growth}},
url = {https://arxiv.org/abs/1810.07774},
volume = {119},
year = {2022}
}

@article{Pichler2023,
author = {Pichler, Anton and Diem, Christian and Brintrup, Alexandra and Lafond, Fran{\c{c}}ois and Magerman, Glenn and Buiten, Gert and Choi, Thomas Y and Carvalho, Vasco M and Farmer, J Doyne and Thurner, Stefan},
doi = {10.1126/science.adi7521},
journal = {Science},
month = {oct},
number = {6668},
pages = {270--272},
publisher = {American Association for the Advancement of Science},
title = {{Building an alliance to map global supply networks}},
url = {https://doi.org/10.1126/science.adi7521},
volume = {382},
year = {2023}
}

@misc{Flood2022,
author = {Flood, Sarah and King, Miriam and Rodgers, Renae and Ruggles, Steven and Warren, J. Robert and Westberry, Michael},
booktitle = {IPUMS},
doi = {https://doi.org/10.18128/D030.V10.0},
title = {{Integrated Public Use Microdata Series, Current Population Survey: Version 10.0 [dataset]}},
year = {2022}
}

@article{Deming2018,
author = {Deming, David and Kahn, Lisa B.},
doi = {10.1086/694106},
isbn = {201913:09:39},
issn = {0734306X},
journal = {Journal of Labor Economics},
month = {jan},
number = {S1},
pages = {S337--S369},
publisher = {The University of Chicago Press},
title = {{Skill Requirements across Firms and Labor Markets: Evidence from Job Postings for Professionals}},
url = {https://doi.org/10.1086/694106},
volume = {36},
year = {2018}
}

@article{Borghans2014,
author = {Borghans, Lex and Weel, Bas Ter and Weinberg, Bruce A.},
doi = {10.1177/001979391406700202},
issn = {2162271X},
journal = {ILR Review},
number = {2},
pages = {287--334},
title = {People skills and the labor-market outcomes of underrepresented groups},
volume = {67},
year = {2014}
}

@article{Weinberger2014,
author = {Weinberger, Catherine J},
journal = {The Review of Economics and Statistics},
number = {5},
pages = {849--861},
title = {{The Increasing Complementarity between Cognitive and Social Skills}},
volume = {96},
year = {2014}
}

@article{Lindqvist2011,
author = {Lindqvist, Erik and Vestman, Roine},
issn = {19457782, 19457790},
journal = {American Economic Journal: Applied Economics},
month = {may},
number = {1},
pages = {101--128},
publisher = {American Economic Association},
title = {{The Labor Market Returns to Cognitive and Noncognitive Ability: Evidence from the Swedish Enlistment}},
url = {http://www.jstor.org.turing.library.northwestern.edu/stable/25760248},
volume = {3},
year = {2011}
}

@article{Kuhn2005,
author = {Kuhn, Peter and Weinberger, Catherine},
doi = {10.1086/430282},
journal = {Journal of Labor Economics},
number = {3},
pages = {395--436},
title = {{Leadership skills and wages}},
volume = {23},
year = {2005}
}

@article{FergusonJohn-Paul2013,
address = {Los Angeles, CA},
author = {Ferguson, John Paul and Hasan, Sharique},
issn = {0001-8392},
journal = {Administrative Science Quarterly},
number = {2},
pages = {233--256},
publisher = {SAGE Publications},
title = {{Specialization and Career Dynamics: Evidence from the Indian Administrative Service}},
volume = {58},
year = {2013}
}

@article{Merluzzi2016,
author = {Merluzzi, Jennifer and Phillips, Damon J.},
doi = {10.1177/0001839215610365},
isbn = {0001839215610},
issn = {19303815},
journal = {Administrative Science Quarterly},
number = {1},
pages = {87--124},
title = {{The Specialist Discount: Negative Returns for MBAs with Focused Profiles in Investment Banking}},
volume = {61},
year = {2016}
}

@article{Byun2018,
author = {Byun, Heejung and Frake, Justin and Agarwal, Rajshree},
doi = {10.1002/smj.2790},
issn = {10970266},
journal = {Strategic Management Journal},
number = {7},
pages = {1803--1833},
title = {{Leveraging who you know by what you know: Specialization and returns to relational capital}},
volume = {39},
year = {2018}
}

@article{Fini2022,
author = {Fini, Riccardo and Jourdan, Julien and Perkmann, Markus and Toschi, Laura},
doi = {10.1287/orsc.2022.1610},
isbn = {0000000339636},
issn = {1047-7039},
journal = {Organization Science},
number = {April 2023},
title = {{A New Take on the Categorical Imperative: Gatekeeping, Boundary Maintenance, and Evaluation Penalties in Science}},
year = {2022}
}

@article{Byun2023,
author = {Byun, Heejung and Raffiee, Joseph},
doi = {10.1177/00018392221143762},
issn = {19303815},
journal = {Administrative Science Quarterly},
number = {1},
pages = {270--316},
title = {{Career Specialization, Involuntary Worker–Firm Separations, and Employment Outcomes: Why Generalists Outperform Specialists When Their Jobs Are Displaced*}},
volume = {68},
year = {2023}
}

@article{Wu2019Science,
author = {Wu, Lingfei and Wang, Dashun and Evans, James A},
doi = {10.1038/s41586-019-0941-9},
issn = {1476-4687},
journal = {Nature},
number = {7744},
pages = {378--382},
title = {{Large teams develop and small teams disrupt science and technology}},
url = {https://doi.org/10.1038/s41586-019-0941-9},
volume = {566},
year = {2019}
}

@article{Wuchty2007,
address = {Washington, DC},
author = {Wuchty, Stefan and Jones, Benjamin F and Uzzi, Brian},
doi = {10.1126/science.1136099},
issn = {0036-8075},
journal = {Science (American Association for the Advancement of Science)},
number = {5827},
pages = {1036--1039},
publisher = {American Association for the Advancement of Science},
title = {{Increasing Dominance of Teams in Production of Knowledge}},
volume = {316},
year = {2007}
}

@book{norris1998markov,
  title={Markov chains},
  author={Norris, James R},
  number={2},
  year={1998},
  publisher={Cambridge university press}
}

@Article{MarkovchainRPackage,
    title = {Discrete Time Markov Chains with R},
    author = {Giorgio Alfredo Spedicato},
    month = {07},
    year = {2017},
    journal = {The R Journal},
    url =
      {https://journal.r-project.org/archive/2017/RJ-2017-036/index.html},
    note = {R package version 0.6.9.7},
  }

@article{DUNBAR1992,
title = {Neocortex size as a constraint on group size in primates},
journal = {Journal of Human Evolution},
volume = {22},
number = {6},
pages = {469-493},
year = {1992},
issn = {0047-2484},
doi = {https://doi.org/10.1016/0047-2484(92)90081-J},
url = {https://www.sciencedirect.com/science/article/pii/004724849290081J},
author = {R.I.M. Dunbar}
}

@article{VanDam2021,
title = {{Correspondence analysis, spectral clustering and graph embedding: applications to ecology and economic complexity}},
author = {van Dam, Alje and Dekker, Mark and Morales-Castilla, Ignacio and Rodr{\'{i}}guez, Miguel and Wichmann, David and Baudena, Mara},
doi = {10.1038/s41598-021-87971-9},
isbn = {0123456789},
issn = {20452322},
journal = {Scientific reports},
number = {1},
pages = {8926},
pmid = {33903623},
publisher = {Nature Publishing Group UK},
url = {https://doi.org/10.1038/s41598-021-87971-9},
volume = {11},
year = {2021}
}

@article{Neal1995,
author = {Neal, Derek},
issn = {0734306X, 15375307},
journal = {Journal of Labor Economics},
month = {may},
number = {4},
pages = {653--677},
publisher = {[University of Chicago Press, Society of Labor Economists, NORC at the University of Chicago]},
title = {{Industry-Specific Human Capital: Evidence from Displaced Workers}},
url = {http://www.jstor.org.proxy.cc.uic.edu/stable/2535197},
volume = {13},
year = {1995}
}

@article{Parent2000,
author = {Parent, Daniel},
doi = {10.1086/209960},
issn = {0734306X, 15375307},
journal = {Journal of Labor Economics},
month = {may},
number = {2},
pages = {306--323},
publisher = {[The University of Chicago Press, Society of Labor Economists, NORC at the University of Chicago]},
title = {{Industry‐Specific Capital and the Wage Profile: Evidence from the National Longitudinal Survey of Youth and the Panel Study of Income Dynamics}},
url = {http://www.jstor.org.proxy.cc.uic.edu/stable/10.1086/209960},
volume = {18},
year = {2000}
}

@article{Xu2021,
author = {Xu, Weipan and Qin, Xiaozhen and Li, Xun and Chen, Haohui and Frank, Morgan and Rutherford, Alex and Reeson, Andrew and Rahwan, Iyad},
doi = {10.1057/s41599-021-00862-2},
issn = {2662-9992},
journal = {Humanities and Social Sciences Communications},
number = {1},
pages = {187},
title = {{Developing China's workforce skill taxonomy reveals extent of labor market polarization}},
url = {https://doi.org/10.1057/s41599-021-00862-2},
volume = {8},
year = {2021}
}

@article{NelsonDylan2022,
author = {Nelson, Dylan and Wilmers, Nathan and Zhang, Letian},
doi = {10.5465/AMBPP.2022.11043abstract},
journal = {Academy of Management Proceedings},
number = {1},
pages = {11043},
title = {{Job Upgrading and Earnings Growth for Non-college Workers}},
url = {https://doi.org/10.5465/AMBPP.2022.11043abstract},
volume = {2022},
year = {2022}
}

@article{Aeppli2022,
author = {Aeppli, Clem and Wilmers, Nathan},
doi = {10.1073/pnas.2204305119},
issn = {10916490},
journal = {Proceedings of the National Academy of Sciences of the United States of America},
number = {42},
pages = {1--7},
pmid = {36191177},
title = {{Rapid wage growth at the bottom has offset rising US inequality}},
volume = {119},
year = {2022}
}

@article{VanderWouden2023,
author = {van der Wouden, Frank and Youn, Hyejin},
doi = {https://doi.org/10.1016/j.respol.2022.104698},
issn = {0048-7333},
journal = {Research Policy},
number = {2},
pages = {104698},
title = {{The impact of geographical distance on learning through collaboration}},
url = {https://www.sciencedirect.com/science/article/pii/S0048733322002190},
volume = {52},
year = {2023}
}

@article{Tibshirani2001,
author = {Tibshirani, Robert and Walther, Guenther and Hastie, Trevor},
doi = {10.1111/1467-9868.00293},
isbn = {1369-7412},
issn = {1369-7412},
journal = {Journal of the Royal Statistical Society: Series B (Statistical Methodology)},
number = {2},
pages = {411--423},
pmid = {306526},
title = {{Estimating the number of clusters in a data set via the gap statistic}},
url = {http://doi.wiley.com/10.1111/1467-9868.00293},
volume = {63},
year = {2001}
}

@article{ROUSSEEUW198753,
author = {Rousseeuw, Peter J},
doi = {https://doi.org/10.1016/0377-0427(87)90125-7},
issn = {0377-0427},
journal = {Journal of Computational and Applied Mathematics},
pages = {53--65},
title = {{Silhouettes: A graphical aid to the interpretation and validation of cluster analysis}},
url = {https://www.sciencedirect.com/science/article/pii/0377042787901257},
volume = {20},
year = {1987}
}

@article{Bustos2022,
author = {Bustos, Sebasti{\'{a}}n and Yıldırım, Muhammed A},
doi = {https://doi.org/10.1016/j.respol.2020.104153},
issn = {0048-7333},
journal = {Research Policy},
number = {8},
pages = {104153},
title = {{Production Ability and economic growth}},
url = {https://www.sciencedirect.com/science/article/pii/S0048733320302286},
volume = {51},
year = {2022}
}

@article{Heiberger2021,
author = {Heiberger, Raphael H and {Munoz-Najar Galvez}, Sebastian and McFarland, Daniel A},
doi = {10.1177/00031224211056267},
issn = {0003-1224},
journal = {American Sociological Review},
month = {nov},
number = {6},
pages = {1164--1192},
publisher = {SAGE Publications Inc},
title = {{Facets of Specialization and Its Relation to Career Success: An Analysis of U.S. Sociology, 1980 to 2015}},
url = {https://doi.org/10.1177/00031224211056267},
volume = {86},
year = {2021}
}

@article{Teodoridis2018,
author = {Teodoridis, Florenta and Bikard, Micha{\"{e}}l and Vakili, Keyvan},
doi = {10.1177/0001839218793384},
issn = {0001-8392},
journal = {Administrative Science Quarterly},
month = {jul},
number = {4},
pages = {894--927},
publisher = {SAGE Publications Inc},
title = {{Creativity at the Knowledge Frontier: The Impact of Specialization in Fast- and Slow-paced Domains}},
url = {https://doi.org/10.1177/0001839218793384},
volume = {64},
year = {2018}
}

@article{Leahey2007,
author = {Leahey, Erin},
issn = {0003-1224},
journal = {American sociological review},
number = {4},
pages = {533--561},
publisher = {Sage Publications Sage CA: Los Angeles, CA},
title = {{Not by productivity alone: How visibility and specialization contribute to academic earnings}},
volume = {72},
year = {2007}
}

@article{Fortuna2010,
author = {Fortuna, Miguel A and Stouffer, Daniel B and Olesen, Jens M and Jordano, Pedro and Mouillot, David and Krasnov, Boris R and Poulin, Robert and Bascompte, Jordi},
doi = {https://doi.org/10.1111/j.1365-2656.2010.01688.x},
journal = {Journal of Animal Ecology},
number = {4},
pages = {811--817},
title = {{Nestedness versus modularity in ecological networks: two sides of the same coin?}},
url = {https://besjournals.onlinelibrary.wiley.com/doi/abs/10.1111/j.1365-2656.2010.01688.x},
volume = {79},
year = {2010}
}

@article{Atmar1993,
author = {Atmar, Wirt and Patterson, Bruce D},
doi = {10.1007/BF00317508},
issn = {1432-1939},
journal = {Oecologia},
number = {3},
pages = {373--382},
title = {{The measure of order and disorder in the distribution of species in fragmented habitat}},
url = {https://doi.org/10.1007/BF00317508},
volume = {96},
year = {1993}
}

@article{Park2019,
author = {Park, Jaehyuk and Wood, Ian B and Jing, Elise and Nematzadeh, Azadeh and Ghosh, Souvik and Conover, Michael D and Ahn, Yong-Yeol},
doi = {10.1038/s41467-019-11380-w},
issn = {2041-1723},
journal = {Nature Communications},
number = {1},
pages = {3449},
title = {{Global labor flow network reveals the hierarchical organization and dynamics of geo-industrial clusters}},
url = {https://doi.org/10.1038/s41467-019-11380-w},
volume = {10},
year = {2019}
}

@article{Gomez-Lievano2021,
archivePrefix = {arXiv},
arxivId = {1812.02842},
author = {Gomez-Lievano, Andres and Patterson-Lomba, Oscar},
doi = {10.1098/rsos.210670},
eprint = {1812.02842},
issn = {20545703},
journal = {Royal Society Open Science},
month = {sep},
number = {9},
pages = {210670},
publisher = {Royal Society},
title = {{Estimating the drivers of urban economic complexity and their connection to economic performance}},
url = {https://doi.org/10.1098/rsos.210670},
volume = {8},
year = {2021}
}

@article{Carneiro1986,
author = {Carneiro, Robert L},
doi = {10.1086/jar.42.3.3630039},
issn = {0091-7710},
journal = {Journal of Anthropological Research},
month = {oct},
number = {3},
pages = {355--364},
publisher = {The University of Chicago Press},
title = {{On the Relationship between Size of Population and Complexity of Social Organization}},
url = {https://doi.org/10.1086/jar.42.3.3630039},
volume = {42},
year = {1986}
}

@article{Carlile2002,
author = {Carlile, Paul R.},
doi = {10.1287/orsc.13.4.442.2953},
issn = {10477039},
journal = {Organization Science},
number = {4},
pages = {442--455+456},
title = {{A pragmatic view of knowledge and boundaries: Boundary objects in new product development}},
volume = {13},
year = {2002}
}

@misc{Murray2023,
address = {Cambridge},
author = {Murray, Fiona and Stern, Scott and Williams, Heidi L and Guzman, Jorge},
booktitle = {NBER Working Paper Series},
doi = {10.3386/w31541},
isbn = {0898-2937},
publisher = {National Bureau of Economic Research},
title = {{Accelerating Innovation Ecosystems: The Promise and Challenges of Regional Innovation Engines}},
year = {2023}
}

@article{Azoulay2021,
author = {Azoulay, Pierre and Greenblatt, Wesley H and Heggeness, Misty L},
doi = {https://doi.org/10.1016/j.respol.2021.104332},
issn = {0048-7333},
journal = {Research Policy},
number = {9},
pages = {104332},
title = {{Long-term effects from early exposure to research: Evidence from the NIH “Yellow Berets”}},
url = {https://www.sciencedirect.com/science/article/pii/S0048733321001311},
volume = {50},
year = {2021}
}

@article{KambourovGueorgui2013ACNO,
address = {New York, USA},
author = {Kambourov, Gueorgui and Manovskii, Iourii},
doi = {10.1017/S1365100510000350},
issn = {13651005},
journal = {Macroeconomic Dynamics},
number = {1},
pages = {172--194},
publisher = {Cambridge University Press},
title = {{A cautionary note on using (March) current population survey and panel study of income dynamics data to study worker mobility}},
volume = {17},
year = {2013}
}

@article{RotundoMaria2004Svgs,
address = {Oxford, UK},
author = {Rotundo, Maria and Sackett, Paul R},
issn = {0963-1798},
journal = {Journal of occupational and organizational psychology},
number = {2},
pages = {127--148},
publisher = {Blackwell Publishing Ltd},
title = {{Specific versus general skills and abilities: A job level examination of relationships with wage}},
volume = {77},
year = {2004}
}

@article{Jovanovic1997,
author = {Jovanovic, Boyan and Nyarko, Yaw},
doi = {https://doi.org/10.1016/S0167-2231(97)00012-2},
issn = {0167-2231},
journal = {Carnegie-Rochester Conference Series on Public Policy},
pages = {289--325},
title = {{Stepping-stone mobility}},
url = {https://www.sciencedirect.com/science/article/pii/S0167223197000122},
volume = {46},
year = {1997}
}

@article{Nedelkoska2015,
author = {Nedelkoska, Ljubica and Patt, Alexander and Ederer, Peer},
journal = {Available at SSRN},
title = {{Learning by problem solving}},
url = {https://papers.ssrn.com/sol3/papers.cfm?abstract_id=2673990},
year = {2015}
}

@article{Argote1990,
author = {Argote, Linda and Epple, Dennis},
issn = {00368075, 10959203},
journal = {Science},
month = {feb},
number = {4945},
pages = {920--924},
publisher = {American Association for the Advancement of Science},
title = {{Learning Curves in Manufacturing}},
url = {http://www.jstor.org/stable/2873885},
volume = {247},
year = {1990}
}

@article{WilkSteffanieL1995GtJC,
address = {Washington, DC},
author = {Wilk, Steffanie L and {Burris Desmarais}, Laura and Sackett, Paul R},
issn = {0021-9010},
journal = {Journal of applied psychology},
number = {1},
pages = {79--85},
publisher = {American Psychological Association},
title = {{Gravitation to Jobs Commensurate With Ability: Longitudinal and Cross-Sectional Tests}},
volume = {80},
year = {1995}
}

@article{Elliott2014,
author = {Elliott, Matthew and Golub, Benjamin and Jackson, Matthew O},
doi = {10.1257/aer.104.10.3115},
journal = {American Economic Review},
number = {10},
pages = {3115--3153},
title = {{Financial Networks and Contagion}},
url = {https://www.aeaweb.org/articles?id=10.1257/aer.104.10.3115},
volume = {104},
year = {2014}
}

@article{Acemoglu2012,
author = {Acemoglu, Daron and Carvalho, Vasco M and Ozdaglar, Asuman and Tahbaz-Salehi, Alireza},
doi = {https://doi.org/10.3982/ECTA9623},
issn = {0012-9682},
journal = {Econometrica},
month = {sep},
number = {5},
pages = {1977--2016},
publisher = {John Wiley & Sons, Ltd},
title = {{The Network Origins of Aggregate Fluctuations}},
url = {https://doi.org/10.3982/ECTA9623},
volume = {80},
year = {2012}
}

@article{Elliott2022,
author = {Elliott, Matthew and Golub, Benjamin and Leduc, Matthew V},
doi = {10.1257/aer.20210220},
journal = {American Economic Review},
number = {8},
pages = {2701--2747},
title = {{Supply Network Formation and Fragility}},
url = {https://www.aeaweb.org/articles?id=10.1257/aer.20210220},
volume = {112},
year = {2022}
}
